\documentclass[article]{IEEEtran}
\usepackage{graphicx}
\usepackage{grffile}
\usepackage{setspace}
\usepackage{longtable}
\usepackage{url}
\usepackage{cite}
\usepackage{multirow}
\usepackage{tabularx}
\usepackage{amsmath}
\usepackage{subcaption}
\usepackage{algorithm,algpseudocode}

\usepackage[usenames, dvipsnames]{color}

\title{Reconfiguration Algorithms for High Precision Communications in
  Time Sensitive Networks: Time-Aware Shaper Configuration with IEEE
  802.1Qcc (Extended Version) \thanks{Please direct correspondence to
    M.~Reisslein (reisslein@asu.edu).}}

\author{Ahmed Nasrallah, Venkatraman Balasubramanian, Akhilesh Thyagaturu, Martin Reisslein, and Hesham ElBakoury}

\begin{document}
\maketitle

\begin{abstract}
As new networking paradigms emerge for different networking
applications, e.g., cyber-physical systems, and different services are
handled under a converged data link technology, e.g., Ethernet,
certain applications with mission critical traffic cannot coexist on
the same physical networking infrastructure using traditional Ethernet
packet-switched networking protocols. The IEEE 802.1Q Time Sensitive
Networking (TSN) task group is developing protocol standards to
provide deterministic properties on Ethernet based packet-switched
networks. In particular, the IEEE 802.1Qcc, centralized management and
control, and the IEEE 802.1Qbv, Time-Aware Shaper, can be used to
manage and control scheduled traffic streams with periodic properties
along with best-effort traffic on the same network infrastructure. In
this paper, we investigate the effects of using the IEEE 802.1Qcc
management protocol to accurately and precisely configure TAS enabled
switches (with transmission windows governed by gate control lists
(GCLs) with gate control entries (GCEs)) ensuring ultra-low latency,
zero packet loss, and minimal jitter for scheduled TSN traffic. We
examine both a centralized network/distributed user
model (hybrid model) and a fully-distributed (decentralized) 802.1Qcc model
on a typical industrial control network with the goal of maximizing
scheduled traffic streams.
\end{abstract}

\begin{IEEEkeywords}
  Low-latency traffic, Protocol adaptation,
  Reconfiguration, Time Sensitive Networking (TSN).
\end{IEEEkeywords}

\section{Introduction}

\subsection{Motivation}

\subsubsection{Centralized User/Distributed (Hybrid) Model}		\label{motivation:hybrid:sec}
IEEE 802.1 Time Sensitive Networking (TSN) provides a standardized
framework of tools for providing deterministic ultra-low latency
(ULL), e.g., for industrial control applications, automotive
networking, and avionics communication
systems~\cite{finn2018introduction,nas2019ult}. In particular, the
IEEE 802.1Qbv Time Aware Shaper (TAS) has received extensive attention
as a key tool for achieving deterministic ULL network service. The TAS
operation requires careful planning of the synchronized time cycles
and the gate times that are allocated to the scheduled traffic (ST)
and the unscheduled best effort traffic (BE). The TAS parameter
settings specifying the timing characteristics (cycle time, gate slot
allocations) are also commonly referred to as the Qbv schedule or the
TAS schedule. For a given static networking scenario, the TAS
operation with a properly configured Qbv schedule can ensure the
deterministic ULL required by demanding industrial and automotive
applications~\cite{nas2019per}.

Modern network scenarios often involve dynamic changes with varied use
cases, such as changes in the network nodes and network topology, or
the traffic pattern. For instance, nodes or links may be dynamically
added or removed. Or, nodes may inject additional traffic flows or
traffic flows may terminate, or the latency requirements of flows may
change dynamically. Such dynamic changes have been included in the use
cases defined by IEC/IEEE 802.1 TSN TG~\cite{usecas2018}.

In a typical industrial environment, sensors that periodically or
sometimes sporadically send ambient measurements to a local gateway
require certain Quality of Service (QoS) guarantees. In such a
volatile and dynamic environment, new machinery that requires
prioritized execution (e.g., emergency cooling procedures or
maintenance tasks for network traffic tests) may be brought onto the
factory floor. To deal with such scenarios, the Time-Aware Shaper
(TAS) Gate Control Lists (GCLs) in coordination with the Network
Management Entities (NMEs), e.g., Centralized Network Configuration
(CNC), has to adapt to changing environment conditions by judiciously
applying reconfiguration such that stream deadlines, QoS, and total
stream utilization times (reported by a stream registration procedure)
are satisfied.

Similarly, in a disaster management networking scenario, a network may
have physically lost large amounts of network resources (links and
routes) but needs to manage the flows that are currently present. A
central orchestrator can prudently time share the reserved resources such
that a large number of streams can be completely serviced, i.e., the total
execution time for each stream is completed by the network.

Generally, in such dynamic networking scenarios, applying only
admission control will clearly guarantee (in accordance with a traffic
shaper) the QoS metrics (of the admitted flows).  However, for a given
static network configuration, the total number of admissible streams
may be well below the number of streams that seek network
service. Therefore, adding a dynamic reconfiguration strategy to
manage and configure the network appears to be a plausible and
attractive solution that intuitively should lower capital and
operational expenditures as it mitigates the over-provisioning of
network resources. The general idea for such an allocation scheme is
to control network access in a timely and orderly fashion such that a
maximum number of streams can be effectively serviced.

Our objective therefore is to maximize the number of admitted flows
(i.e., tasks or streams) in such a dynamically changing and volatile
environment whilst keeping the TSN QoS metric guarantees.  To the
best of our knowledge there are no prior detailed studies on a
%fluctuating volatile source or a
dynamic stream resource allocation and admission control policy in
conjunction with a network reconfiguration policy being executed while
flows are carried in a TAS time scheduled network. In this paper, we
focus on the IEEE 802.1Qbv~\cite{IEEE8021Qbv} enhancements and design
a reconfiguration framework taking inspiration from the IEEE
802.1Qcc~\cite{IEEE8021Qcc} standards for managing, configuring, and
reconfiguring a TSN network.

\subsubsection{Decentralized Model}
The IEEE 802.1Qcc standard specifies three models for configuring the
Time-Aware Shaper (TAS) gating schedules (GCL/GCE timing computation):
a fully-centralized model, a centralized network/distributed user
model (hybrid model, see Section~\ref{motivation:hybrid:sec}), and a
fully-distributed configuration model. The centralized model greatly
eases control and configuration messages sent across the network and
can precisely configure TAS schedules due to having complete knowledge
of the network and full capabilities of each bridge. However the
centralized model suffers from common disadvantages, such as a
single-point of failure, relatively large capital/operational
(CapEx/OpEx) expenditures (as the centralized control may be
superfluous in a small-scale network~\cite{chen2017rap}), and adding
unnecessary complexity to a small-scale network. Thus, a
fully-distributed configuration model (e.g., SRP over MRP or RAP over
LRP) may be attractive for some networks. The fully-distributed
configuration model avoids the added complexity and single point of
failure of a centralized management entity.  Moreover, Chen et
al.~\cite{chen2017rap} have argued that the centralized configuration
models can be an over-design for real-time applications with relaxed
latency requirement (order of magnitude of milliseconds). Chen et
al. have also argued that the distributed model is more
scalable. (However, studies of the fully distributed model with RAP
over LRP targeted typically applications with relatively relaxed
latency requirements.)

In the absence of a Centralized Network Configuration (CNC) node, the
TSN Task Group (TG) specifies the IEEE 802.1CS (Link-Local
Registration Protocol, LRP)~\cite{IEEE8021CS} standard for
registration and distribution of application configuration parameters
between point-to-point links targeting newly published TSN features. A
legacy protocol, such as the Stream Reservation Protocol
(SRP)~\cite{IEEE8021Qat} which is primarily used for AVB application,
is intended to serve as the main resource reservation and admission
control protocol. However, extending and porting the SRP to be
utilized for bridges that support TAS will not suffice since bandwidth
reservation cannot directly apply TAS's time slot reservation
natively. Therefore, the Resource Allocation Protocol, IEEE 802.1Qdd
(RAP)~\cite{chen2017rap}, has been proposed to apply a distributed
resource reservation that can exchange TSN features.

In our model, the switch computes the TAS time slot for all admitted
streams as follows. In the absence of admission control, we predefine
the TAS slot to be a minimum of 10\% and a maximum of 90\% of the
Cycle Time (CT), even when no streams are registered, so as to avoid
starvation of Best Effort (BE) traffic. With admission control, the
static predefinition, which can potentially waste resources if unused,
can be eliminated. Essentially, as streams get registered, we keep
track of the remaining load on each egress port until the load (which
depends on the slot size and CT) is negative (oversubscribed link).
Such a link over-subscription invokes a procedure call that increases
the slot time (by a step size of 1\%, or more fine-grained increments)
until the remaining load is positive. This procedure is iteratively
called until all registered streams and the new stream are
appropriately registered with sufficient Scheduled Traffic (ST) slot
time to transmit all frames during a single appropriately sized
CT. Note that the TAS time slot is defined as the portion of the CT
that is allocated to high-priority ST traffic.

Our proposed TAS configuration/reconfiguration, is designed for the
fully-distributed configuration model. In the distributed approach,
the GCE slot parameters are configured in a distributed manner by the
switches as per the distributed algorithm/procedure explained in
Sec.~\ref{tas:reconfig:prot}. In the centralized approach, the GCE
slot parameters are configured centrally by the CNC with the
Centralized User Configuration (CUC) node assisting in passing
end-station (source/sink, devices, etc.) capabilities and
parameters. Similarly, the ``hybrid'' model also utilizes the CNC for
configuration exchange and network side management (see
Section~\ref{motivation:hybrid:sec}).

Regarding the differences between the hybrid model and the fully
centralized model, the main network-side difference is the way the
User/Network configuration Information (UNI) is propagated. In the
fully centralized model, the sources communicate Control Data Traffic
(CDT) messages to/from the CUC node, instead of having each source
interact directly with the CNC. According to the 802.1Qcc standard,
the general advantage is that the computation complexity (especially
for industrial/automotive applications with computationally complex
I/O timing requiring detailed knowledge of the application's
software/hardware within each end station) can be tolerated by having
the CUC handle end station discovery, retrieving end station
capabilities and user requirements, and configuring TSN features in
end stations. The CNC is used to only manage and configure network
side components. In terms of benefits, the configuration delay is
potentially reduced, scheduling optimality increased, and hardware
performance overhead/complexity reduced with a shorter network
response time for networks involving command and control systems. The
development and investigation of a reconfiguration approach for the
fully centralized configuration model is left as future work.

Our study focuses on the centralized network/distributed user model
(hybrid model) and the fully-distributed (decentralized) configuration
model.  For brevity we refer to the centralized network/distributed
user model (hybrid model) also as the centralized model or the
centralized topology.  We refer to the fully-distributed
(decentralized) model also as the decentralized model or the
decentralized topology.

Our proposed TAS reconfiguration architecture maps and propagates
stream information and conducts dynamic TAS time slot reservations
(including GCL/GCE scheduling and provisioning) on local shared stream
database records to guarantee TSN QoS for all admitted streams and to
maximize stream admission using a reconfiguration strategy of TAS
gating schedules within each bridge in a fully decentralized IEEE
802.1Qcc model.

\subsection{Related Work} \label{tsn:rel:sec}
Raagaard et al.~\cite{raagaard2017fog} presents a heuristic algorithm
that reconfigures TAS switches according to runtime network
conditions. Feasible schedules are produced and forwarded using a
configuration agent (composed of a Centralized User Configuration
(CUC) and Centralized Network Configuration (CNC)). Raagaard et al's
model places emphasis on appearing and disappearing synthetic flows in
a fog computing platform that takes into account the flow's properties
and possible routes. Contrary to this approach our framework performs
flow maximization with optimal reconfiguration based on firm bandwidth
computation strategies at run-time. Further, we show the equilibrium
point for this algorithm.

Further related work that is complementary to
our study has been conducted by
Pop et al.~\cite{pop2018enabling}, Hackel et al.~\cite{hackel2019software},
Herlich et al.~\cite{herlich2016proof},
Nayak et al.~\cite{nayak2016time,nayak2017incremental,nayak2017routing},
and Kobzan et al.~\cite{kobzan2018secure}.

\subsection{Contributions}
We comprehensively evaluate the performance of TAS
for reconfigurations in the hybrid and fully distributed models
with respect to
network deployment parameters, such as, maximum window size for the
Gate Control List (GCL) repetition, gating ratio proportion, i.e.,
Gate Control Entry (GCE) proportion, to control delay perceived at the
receiving end, signaling impact on Scheduled Traffic (ST) and
Best-Effort traffic (BE) classes, and packet loss rate experienced at
the receiving end. In particular, we make the following contributions:
\begin{itemize}
\item[i)] We design a CNC interface for a TSN network to globally
  manage and configure TSN streams, including admission control and
  resource reservation.
\item[ii)] We integrate the CNC in the control plane with TAS in the
  data plane to centrally manage and shape traffic using the CNC as
  the central processing entity for flow schedules as more flows are
  added.
\item[iii)] We modify and test the model to operate in a distributed
  fashion, i.e., the control and data planes are combined.
\item[iv)] We evaluate each design approach and for a range of numbers of
  streams and sources with different TAS parameters. We show results
  based on admission ratios, network signaling overhead, and QoS
  metrics.
\end{itemize}

\subsection{Organization}
This article is organized as follows. Section~\ref{tsn:back:sec}
provides background information and an overview of related work on the
802.1 TSN standardization, focusing on the enhancements to scheduled
traffic and centralized management and
configuration. Section~\ref{tsn:design:sec} shows the complete
top-down design of the CNC (hybrid model) and main components that
achieve ultra-low latencies and guaranteed QoS for a multitude of ST
streams. Similarly, Section~\ref{tas:reconfig:prot} shows the approach
used in implementing the decentralized (fully distributed) TAS
reconfiguration model. The simulation setup as well as main
parameters and assumptions are given in Section~\ref{tsn:sim:setup} and
results are presented in Sections~\ref{eval:cent:sec} and~\ref{eval:dec:sec}. Finally conclusions and future work are outlined
in Section~\ref{concl:sec}.

\section{Background: IEEE 802.1 Time Sensitive Networking} \label{tsn:back:sec}

\subsection{IEEE 802.1Qbv: Time Aware Shaper (TAS)}
TAS's main operation is to schedule critical traffic streams in
reserved time-triggered windows. In order to prevent lower priority
traffic, e.g., best effort (BE) traffic, from interfering with the
scheduled traffic (ST) transmissions, ST windows are preceded by a
so-called guard band. TAS is applicable for Ultra Low Latency (ULL)
requirements but needs to have all time-triggered windows
synchronized, i.e., all bridges from sender to receiver must be
synchronized in time. TAS utilizes a gate driver mechanism that
opens/closes according to a known and agreed upon time schedule for
each port in a bridge. In particular, the Gate Control List (GCL)
represents Gate Control Entries (GCEs), i.e., a sequence of on and off
time periods that represent whether a queue is eligible to transmit or
not.

The frames of
a given egress queue are eligible for transmission according to the
GCL, which is synchronized in time through the 802.1AS time
reference. Frames are transmitted according to the GCL/GCE and
transmission selection decisions. Each individual software queue has
its own transmission selection algorithm, e.g., strict priority
queuing. Overall, the IEEE 802.1Qbv transmission selection transmits a
frame from a given queue with an open gate if: $(i)$ The queue
contains a frame ready for transmission, $(ii)$ higher priority
traffic class queues with an open gate do \textit{not} have a frame to
transmit, and $(iii)$ the frame transmission can be completed before
the gate closes for the given queue. Note that these transmission
selection conditions ensure that low-priority traffic is allowed to
\textit{start} transmission only if the transmission will \textit{be
  completed} by the start of the scheduled traffic window for
high-priority traffic. Thus, this transmission selection effectively
enforces a ``guard band'' to prevent low-priority traffic from
interfering with high-priority traffic~\cite{finn2018introduction}.

\subsection{IEEE 802.1Qcc: Centralized Management and Configuration}

\begin{figure}[t!]  \centering
	\includegraphics[width=3.5in]{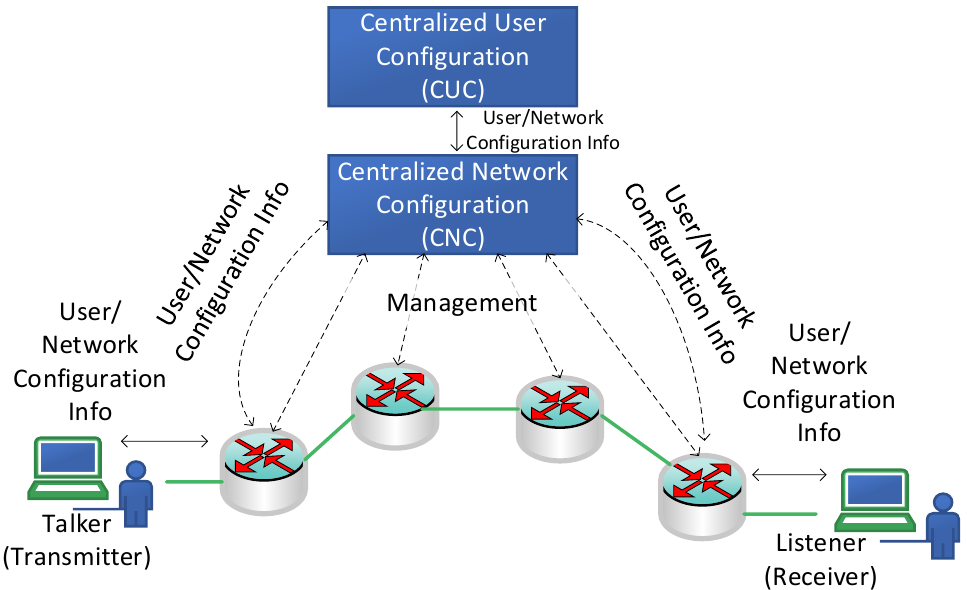} \vspace{-0.5cm}
	\caption{Illustration of Centralized Network Configuration (CNC): End stations interact with the network entities via the User-Network Interface (UNI). The CNC receives the requests, e.g., flow reservation requests, and provides corresponding management functions.  An optional CUC provides delay-optimized configuration, e.g., for closed-loop IACS applications.  The solid arrows represent the protocol, e.g., YANG or TLV, that is used as the UNI for exchanging configuration information between Talkers/Listeners (users) and Bridges (network).  The dashed arrows represent the protocol, e.g., YANG or TLV, that transfers configuration information between edge bridges and the CNC.} \label{fig_tsn_Qcc}
\end{figure}

IEEE 802.1Qcc~\cite{IEEE8021Qcc} provides a set of tools to globally
manage and control the network. In particular, IEEE~802.1Qcc enhances
the existing Stream Reservation Protocol (SRP) with a User Network
Interface (UNI) which is supplemented by a Centralized Network
Configuration (CNC) node, as shown in Fig.~\ref{fig_tsn_Qcc}. The UNI
provides a common method of requesting layer 2 services. Furthermore,
the CNC interacts with the UNI to provide a centralized means for
performing resource reservation, scheduling, and other types of
configuration via a remote management protocol, such as
NETCONF~\cite{enns2006netconf} or RESTCONF~\cite{bierman2017restconf};
hence, 802.1Qcc is compatible with the IETF YANG/NETCONF data modeling
language.

\section{Centralized Model Design and Framework Considerations} \label{tsn:design:sec}
This section presents our design methodology and main signaling
framework for the centralized network/distributed user
model (hybrid model). Our main goals behind designing the CNC is given by the
following constraints. Additionally, the CNC can be logically or
physically connected to the data-plane with in-band or out-of-band
management links.
\begin{enumerate}
\item Our focus is mainly on stream based network adaptation. By this
  technique, fluctuating streams (already registered streams and new
  incoming streams) and their requirements can be accommodated by the
  network dynamically based on a single remote procedure call to the
  CNC.
\item Identify and execute flow requirements by populating the
  registration table. The control plane resource orchestration is
  purely carried out by the monitoring of existing flows which have
  been satisfied.
\item Optimizing resource allocation (maximize admitted streams/flows)
  based on a bounded latency and network utilization, i.e., prioritize
  flows that request low network resources.
\end{enumerate}
Our main assumption to accurately apply admission control and,
consequently, reconfiguration, is that each source must define a flow
in terms of total resources needed (governed by the bandwidth
requirements) and the total time needed for the resource to be used
(which in our traffic model is termed as the resource utilization
time). Essentially, the CNC uses this information (which is tagged in
the Ethernet frame header) to determine whether a frame is admitted or
rejected.

\begin{figure}[t!] \centering
\includegraphics[width=3.3in]{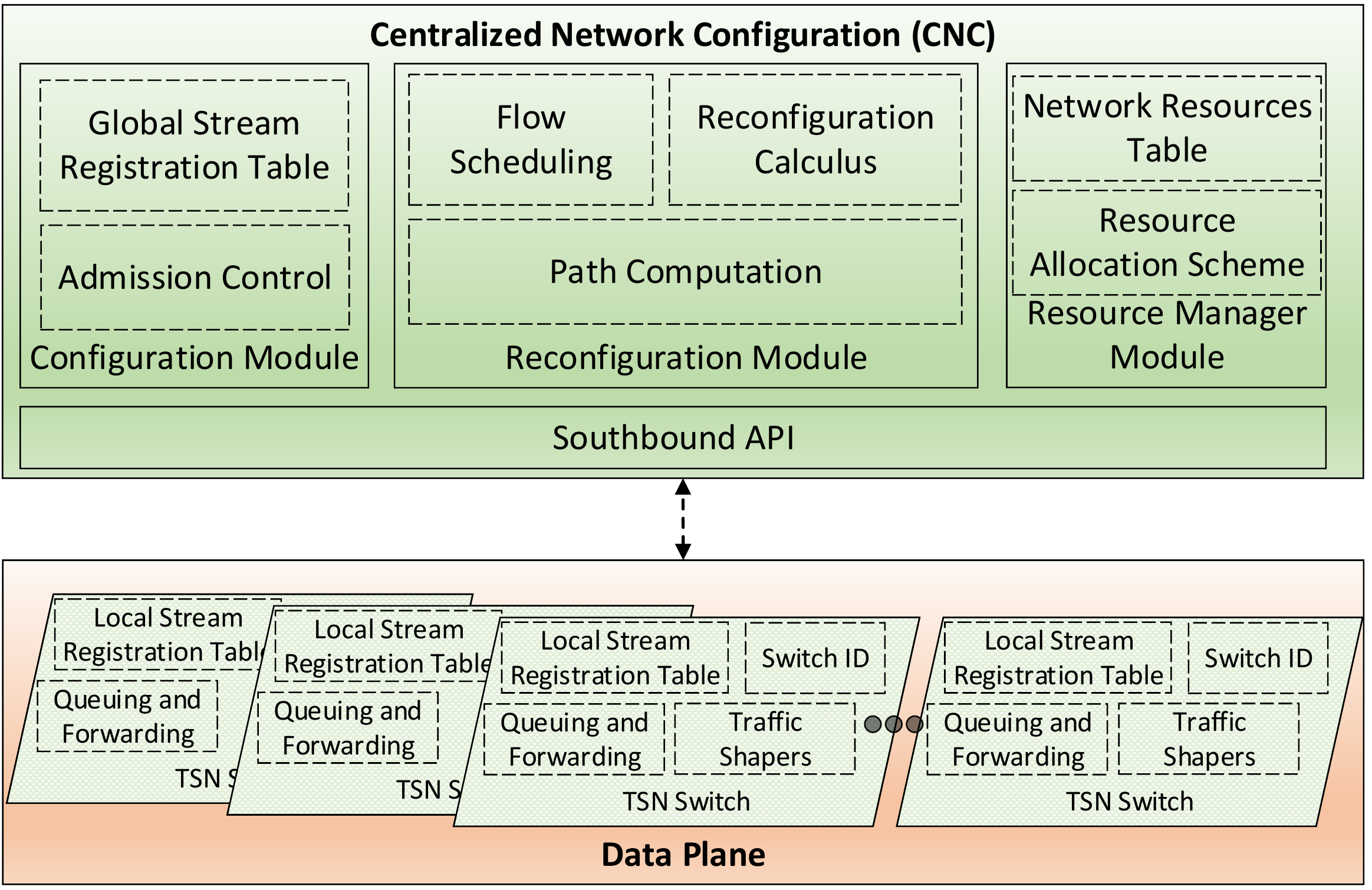} \vspace{0.2cm}
\caption{Network Management Entity Framework for TSN Switches:
  Centralized Network Configuration (CNC) is used to send and receive
  Control Data Traffic (CDT) to configure routing segments and network
  resources with the goal of maximizing network flows/streams.}
	\label{fig_cnc_reconfig}
\end{figure}

\subsection{Core Components}
Our design is split into two layers, Control Plane and Data Plane,
following the decoupling SDN paradigm, thereby inheriting the benefits
of the orthogonality of the two planes, as shown in
Fig.~\ref{fig_cnc_reconfig}.

\subsubsection{Configuration Module}
The configuration module is the main component that interacts with the
registered flows and network components. It includes the global stream
registration table which records all approved stream transmitting in
the network, and the admission control element that encapsulates and
decapsulates CDT headers and forwards the information to the necessary
module/element.

\paragraph{Global Stream Registration Table}
The source streams (devices/users) make a Remote Procedural Call (RPC)
via the stream registration interface for providing information that
can be mapped as a unique tuple structure identification $<Flow ID,
Bridge Gateway>$. Upon receiving the registration packet, i.e.,
Control Data Traffic (CDT), the CNC determines whether the new stream
can be accepted in its stream table. To guarantee the QoS for all
registered streams, admission control principles are applied to all
streams according to the stream's path, required network resources,
and available resources.

\paragraph{Admission Control}
The admission control element is the first element that the new
streams interacts with. The admission control element under the
configuration module globally manages all streams transmitting in the
TSN domain governed by the CNC. The admission control element extracts
the necessary information from the CDT packet and forwards the
information according to the CDT type. The CNC apples several steps to
decide whether to accept or reject the stream transmission request.
\begin{enumerate}
\item The CNC checks the destination address(es) of the stream and
  consults its resource manager module for network resources available
  on the new stream's path, which is computed based on the path
  computation element within the CNC.
\item According to the bandwidth required for the new stream
  (calculated at the bridge gateway for the new stream), all links on
  the path are checked to see if enough bandwidth is available for the
  new stream.
\item In the event that not enough resources are available, the CNC
  applies the TAS reconfiguration module to identify the bottleneck
  link(s) and to check whether the gating ratio can be increased for that
  specific traffic class whose current resource utilization and
  deadline will not cause a late deadline by being added to the TAS
  slot reservation.
\end{enumerate}

\subsubsection{Reconfiguration Module}
The reconfiguration module includes the flow scheduling element (for
our network model, the Time-Aware Shaper (TAS) is used in the data
plane), the reconfiguration calculus element which optimizes flow
registration according to each stream's total resource utilization and
flow deadlines, and finally the path computation element which defines
the path for all stream according to the QoS constraint.

\paragraph{Flow Scheduling}
The flow scheduling element currently takes the Time-Aware Shaper into
consideration. Due to the TAS's requirements on time synchronization
between network components (switches, hosts, etc.), the CNC follows
the same principle of scheduling flows according to a known timescale
(initially set to be 50~$\mu$s in our network model). The CNC then
passes on this time synchronization information to the TSN enabled
switches within its domain. Any approved streams will transmit frames
according to the time scale specified by the flow scheduler in the
CNC.

\paragraph{Reconfiguration Calculus}
In addition to centrally managing resources and providing admission
control policies to the network, the CNC can invoke the TAS
reconfiguration strategy with the goal of borrowing BE time
slots for pending ST traffic streams. This element consults the
resource manager module on the bottleneck link and checks whether the
added stream will oversubscribe the link. The TAS reconfiguration
incrementally (1\% of total CT) increases the traffic class slot time
and reserves it for the new stream.

\paragraph{Path Computation}
For large scale and complex LAN/MAN topologies, it is often required
to supplement streams with equal cost paths in the event of a path
disruption (e.g., link failure, stream saturation, and explicit
congestion). The CNC's path computation element is tasked with finding
such paths as a fail-over approach to avoid any violations to any
stream's QoS. Presently, our model has a rudimentary application of
path computation, i.e., it is defined statically for all core network
components (shortest path), since the main emphasis was on
reconfiguration based on stream characteristics as defined by the
source.

\subsubsection{Resource Manager Module}
The resource manager module centrally manages all network resources
within the CNC's domain. It includes the network resource table that
records all streams' usage of resources, and the resource allocation
scheme element to which we delegate the task of calculating the required
network resources for a given stream according to an allocation
scheme.

\paragraph{Network Resource Table}
To remove certain overheads on the configuration module, the network
resource table operates in tandem with the global stream registration
table to accurately determine the required network resources (mainly
bandwidth for our traffic model). It classifies streams based on
periodic stream properties. Any stream that has been approved by the
CNC has a record attached to it in the network resource table.

\paragraph{Resource Allocation Scheme}
Several allocation schemes can be implemented for all traffic classes
defined in the network. For periodic streams, the time slot given by
the flow scheduler (according to the TAS Cycle Time and number of
traffic classes) and the data rate defined by the source is used to
calculate the required bandwidth for each link on the path to the
destination (i.e., sink).

%For sporadic streams, the self-similar traffic model indicates an average data rate as defined by the
%source which is used to allocate the bandwidth required at each link on the path.

\subsubsection{Data Plane}
The data plane contains all core switches. Any TSN switch interfaced
by the CNC is given a switch ID and has a local stream registration
table. The remaining switch elements compose the forwarding and
queuing operation with several traffic shapers (802.1Qbv TAS in our
network model).

\paragraph{Local Stream Registration Table}
This data plane registry contains the subset of source streams that
are established for the corresponding bridge gateway and attached
sources to each port. The CNC delegates some control to the bridge
gateway to instruct and alert sources of any new network conditions
and explicit changes.

\paragraph{Traffic Shaper --- Time-Aware Shaper}
The traffic shaper is the main shaping and scheduling mechanism that
controls the gating schedules for all the traffic classes within the
TSN domain. All bridges are synchronized to the same gating schedule
GCL Cycle Time (CT) given by the CNC's flow schedule element (CT
indicates the time period for the GCL to repeat).

\section{Decentralized Model Design and Framework Considerations} \label{tas:reconfig:prot}

\begin{figure}[t!] \centering
\includegraphics[width=3.5in]{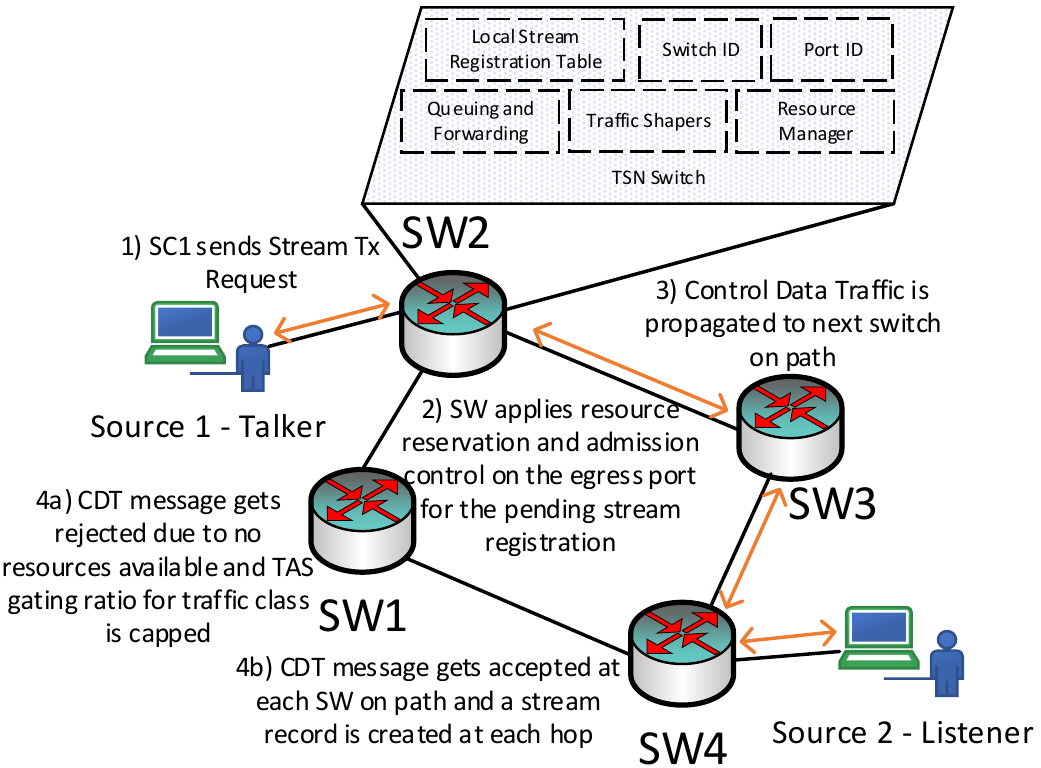} \vspace{-0.2cm}
\caption{A TSN fully distributed configuration model example
  illustrating the general strategy and logic of each TSN switch with
  TAS support. In the absence of a CNC to centrally manage network
  parameters, each switch performs admission control and resource
  reservation (according to the TAS time slot load) and propagates the
  information to the next hop on the stream path. A single rejection on
  one hop terminates the forwarding of the CDT, and sends another CDT
  on the reverse path indicating the stream rejection outcome. If all
  switches on the path accept the stream, then the source is notified
  of the stream acceptance outcome and can begin forwarding in the
  next TAS cycle. In our model, CDT traffic has higher priority than
  non-CDT traffic (including ST). The formal definition of the CDT
  traffic is left for future work. }
	\label{fig_distributed_reconfig}
\end{figure}

This section presents our design methodology and framework for the TAS
reconfiguration in the decentralized (fully distributed) model. Our
current proposed architecture generally follows the steps enumerated
below and illustrated in Fig.~\ref{fig_distributed_reconfig}. Our
description focuses on the additions to the design of RAP over LRP,
e.g., TAS slot computation/reservations.

\begin{enumerate}
	\item At each egress port (Port Identifier, PID), the TSN switch maintains a local stream registration table that includes information, such as flow ID, gateway (i.e., the first bridge that a talker is connected to), destination address(es), the traffic injection rate per GCL cycle time, and the calculated port bandwidth requirement. The traffic injection rate is not computed, rather the traffic injection rate is reported by the source (talker) to the network devices. It mainly indicates the bandwidth requirements of a stream. Bandwidth for a bridge egress port needed for a stream is computed using the ST injection rate (or ST rate), the average packet size, and the bridge TAS timing configuration (e.g., the CT and current traffic	class slot time). This information is carried and communicated between bridges using the CDT packet type identifier (or message type).
	\item A source (talker) can send a stream transmission request, i.e., a CDT message of type Stream Transmission Request to	register its stream and use the TSN service for scheduled traffic.
	\item Each switch maintains a resource manager module for each port. If the newly incoming stream is accepted (due to available resources and TAS slot space). The TAS slot size for a specific traffic class is governed by the CT and traffic class gating ratio (in time). The TAS ST slot can be configured/reconfigured according to stream requests and terminations. The stream registration message is then propagated towards the next switch, and a map is maintained for the stream (and any other streams) pending approval.
	\item If accepted by the last switch, then the stream registration record is added to the local stream registration table, and bandwidth resources are allocated for the stream and TAS slot space is modified (if necessary) on the reverse path. The main reason for allocating the resources in the reverse path is as follows. If we allocate the resources in the forward direction but a switch in the next hop rejects the stream (due to lack of resources),	then we have to release the resources reserved earlier for the stream. Therefore, we avoid the allocation until all hops provide	assurance that the stream will be accommodated.
	\item Each switch receiving the pending registration message adds the stream record to its local table, allocates the necessary resources and TAS slot reservation, and propagates the registration message towards the source gateway.
	\item The source gateway receives the pending stream registration message and similarly allocates the resources and finally sends an approval granted message towards the source which prompts the source to start sending data in the next available TAS cycle.
\end{enumerate}

While the previous enumerated points provide an overview of the procedural approach used, a key question remains, namely what happens if frames belonging to a given stream arrive after the gate for it's traffic class queue has been closed? Generally, the way the stream traffic end station (source/talker) operates is by synchronizing to the switch's TAS ST slot time for the class the stream belongs to (in	our case, it is ST) after the CDT stream registration is complete. Therefore, it will always send data traffic at the beginning of the ST slot. After the data traffic for the ST stream arrives at the first switch (where the gate for the ST queue will definitely be open due to source-network synchronization), the new ST traffic is queued at the back of the queue awaiting transmission. Since registration and resource allocation for all ST streams is enforced, the computed ST slot size will be large enough to transmit all ST packets (from all streams being serviced through that specific switch's egress port) queued that arrived from different sources during the cycle time. The delay bounds (a data traffic stream arriving before a specific time and not necessarily at the right time) is guaranteed since the cycle time for each switch along the path is configured to be large enough to transmit all ST frames registered on the multi hop path (up to 5 hops). Some potential ST traffic sent on cycle time $(t)$ can be transmitted at cycle time $(t+1)$, but would be blocked by some subsequent slots in the first cycle time belonging to different traffic classes (in our case, it is BE) leading to higher delays but still within tolerated delays since the cycle time is usually much smaller than the max delay bounds or threshold defined (e.g., 50 microseconds cycle time and 100 microseconds delay bounds).

In our simplified example in Fig.~\ref{fig_distributed_reconfig}, each switch, upon initialization and with admission control, provisions all the CT to BE traffic. When the first stream transmission request occurs, the switches exchange the CDT message and start building the local stream registration table. A switch builds the GCE slot time for ST by mapping each stream request to its internal stream registration table upon granting the stream	approval. The switch builds the ST slot time iteratively by	following each stream admission/termination. The switch passes each set of stream configuration parameters (UNI) through the 802.1Qcc registration and reservation protocol.

ST traffic stream load is viewed in two main ways in our implementation. $i)$ ST injection rate, and $ii)$ resource utilization time. The ST injection rate corresponds to the number of packets sent from an ST stream in one TAS cycle time, while the resource utilization time is the time that the stream reports to the network on how long the resources reserved will be used. The bandwidth requirement for a new stream request is calculated by taking the reported ST injection rate of the stream and multiplying that with the packet size for the stream. That quantity is then divided by the TAS slot time (at a specific switch). Essentially, an increase in the slot time corresponds to a decrease in the bandwidth requirement (since more time is given) and vice versa. Note that this only works for streams with periodic data transmissions and not for sporadic data streams since we cannot calculate how much time is needed for a sporadic data transmission. This procedure is only for ST traffic; BE traffic is transmitted with the leftover slot time after the ST traffic has been serviced.

\subsection{Core Components}

\begin{figure}[t!] \centering
	\includegraphics[width=3.5in]{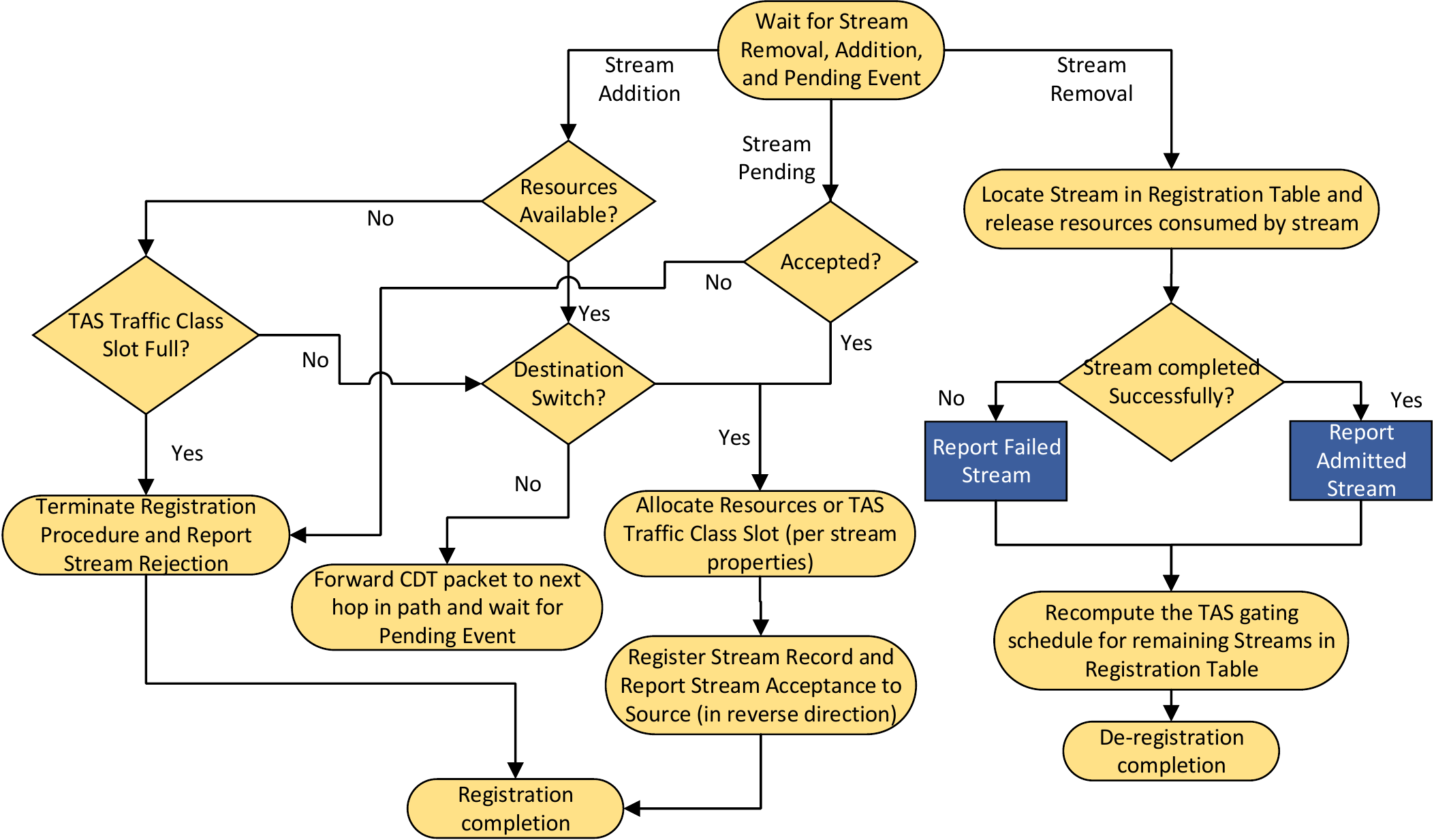} \vspace{-0.2cm}
	\caption{The main logical steps performed by each switch along the stream's path are shown to apply stream registration and reservation. Each switch generally waits for an event (addition, removal, or pending) for each stream. For instance, a stream removal is usually based on the resource utilization time (stream life time) that was specified at stream establishment. The bridges that allocated resources for the stream can remove the stream after the stream life time has expired. For the cases of stream addition or pending, the event is the CDT message received (whether in the forward or reverse direction). Towards completing (finalizing, confirming) a stream reservation (registration), the pending event is the event for a CDT message in the reverse direction where each switch (not the last switch) waits for the approval (confirmation of reservation) of the next-hop switch.}
	\label{fig_stream_registration}
\end{figure}

This section outlines the main components required to successfully implement stream admission control and resource reservation within switches that support the TAS traffic shaper in a distributed fashion. Fig.~\ref{fig_stream_registration} illustrates the typical registration/reservation procedure for all streams within the TSN domain.

\subsubsection{Admission Control}
The admission control element extracts the necessary information from the CDT packet and forwards the information according to the CDT type. The switch forwarding mechanism applies several steps to accurately decide whether to accept or reject the stream transmission request Note that the stream transmission request corresponds to a CDT message. In particular, the switch consults the
resource manager module to check if enough resources (bandwidth) is available for the new stream that is calculated by the reported traffic injection rate, the maximum cycle time, and the traffic class's TAS slot time. A given stream's bandwidth requirement is calculated by multiplying the ST injection rate with the average packet size and dividing by the current ST slot size. Note that the traffic class TAS slot time is the time during which the TAS gate is open to transmit frames belonging to the considered class.  Also note that all GCEs are executed during each CT. If the CT is smaller than the aggregate of the GCEs, then we need to either increase the CT or reject streams that cause the exceedance of the CT.

\subsubsection{Flow Scheduling}
This element currently takes the Time-Aware Shaper into consideration. Due to the TAS's requirements on time synchronization between network components (switches, hosts, etc.), all switches/hosts follow the TAS GCL timescale cycle time (e.g., 50~$\mu$s). Depending on the number of traffic classes supported, the TAS cycle time can be divided into appropriate slots for each traffic class load. The TAS CT is divided among all the traffic classes (in our evaluation model, we consider two traffic classes, BE and ST). Currently, in our evaluations, the CT is initially predefined to 50 microseconds. Note that the CT could be changed/configured dynamically. The dynamic adaptation of the CT with respect to new stream additions, application specifications, or other events is a topic for future work.

\subsubsection{Stream Registration Table}
Stream creation follows a Poisson process with a mean duration. Different scenarios with varying mean duration enables analysis of how reconfiguration works in multiple settings. The stream life-time is defined by the duration. For example, how the switches in the path are aware of the Poisson parameter from the edge switch. All GCE entries and Queues are based purely on this slots allocated based on the Poisson parameter. Although, our approach does not intend to minimize delay, we follow a delay bound strategy that leads to finding the limiting value. Implicitly, the limiting value can be minimized that can provide a minimized delay. Additionally, the stream registration table contains the characteristics of the source streams that are established for the corresponding bridge egress port. Each record gets populated (if accepted) on the reverse path taken by the stream's registration message (after reaching the destination switch).

\subsubsection{Traffic Shaper - Time-Aware Shaper}
The main shaping and scheduling mechanism that controls the gating schedules for all the traffic classes within the TSN domain. All bridges are synchronized to the same gating schedule GCL CT that is initially predefined by network administrator. Ideally, we want the CT to be large enough for all streams from all
traffic classes to be accommodated and small enough that the all streams QoS fits the delay requirements.
In our current evaluations, the CT is predefined at the widely used 50 microseconds.

\subsubsection{Reconfiguration Calculus}
In essence, the reconfiguration (dynamic configuration) of the TAS schedules (switch GCL/GCE) for each egress port is dynamically invoked according to two principle events, $i)$ adding a new stream, and $ii)$
removing an existing stream. The switch's gating ratio (for a particular stream belonging to a defined traffic class) reports certain parameters (e.g., packet injection rate, maximum packet size, latency requirement, deadline, application response time, etc.) which are then used to check if enough slot time is available (which corresponds to attempting bandwidth reservation). In the event that no slots are  available, the GCE slot size is recomputed (according to the registered stream properties within the registration table) generally by allocating more resources from Best Effort Traffic.
The stream life time is reported by the source to the network as user/network information (UNI). Each UNI is propagated by each switch along the path which allows the switch to register the stream and store the stream's resource utilization time, (stream life-time), among other critical information. Any information	pertaining to the UNI of a stream is recorded in the stream	registration table. In terms of GCEs for TAS with support of ST and BE traffic classes, only two GCEs within a GCL (1/0 (ST/BE) for the	first entry and 0/1 (BE/ST) for the second) are necessary with a total of three outbound queues for each egress channel port in a TSN switch. Two queues for each traffic class, and another queue for CDT traffic (signaling traffic). Upon initialization, each switch allocates $0$ resources to ST, and BE therefore gets all the leftover CT slot. As streams get registered, the ST slot time is recomputed
(according to the stream packet size, ST injection rate, and current slot time, if the slot exists). If the stream is the first stream to the switch, i.e., ST slot is $0$, then the ST slot size is defined
to be 1\% of the CT at a minimum by borrowing necessary time slot from BE.

\subsubsection{Path Computation}
While this module is fundamentally necessary in any switch (in a decentralized/distributed network), we manually define static routing tables for destination addresses and associated ports on each switch.
Essentially, we assume a manual procedure to compute paths, i.e., we assume that there is a path computation module that is used in both centralized and distributed configuration models. We make this assumption to simplify operations and place emphasis on the TAS reconfiguration technique.

\subsubsection{Network Resource Table}
To remove certain overheads on the configuration procedure, the network resource table operates in tandem with the stream registration table to accurately determine the required network resources (mainly bandwidth for our traffic model) per switch egress port. It classifies streams based on periodic and sporadic streams properties, though currently the focus is on periodic ST streams. Any approved stream by the switch has a record attached to it in the network resource table, located within each switch, which can be called to compute and store current and remaining link/port loads for each switch. Each egress port has a network resource table. More details on the network resource table will be provided in the next iteration of this document.

\section{Performance Evaluation}  \label{tsn:sim:setup}

\begin{figure}[t!]  \centering
\includegraphics[width=3.3in]{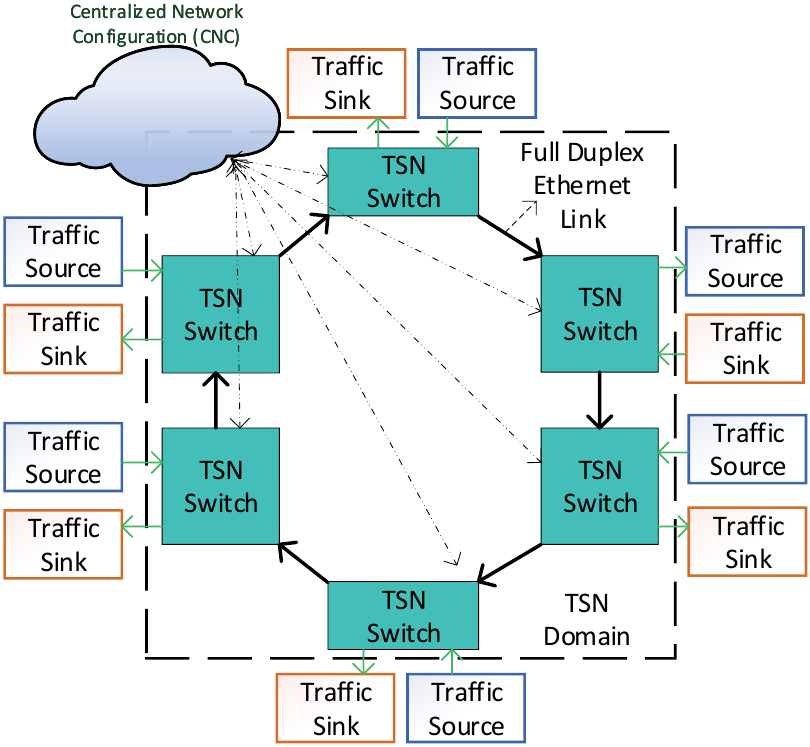} \vspace{-0.2cm}
\caption{Industrial control loop
  topology~\cite{guck2016function}. Each source generates stream data
  with varying hop counts and packets rates unidirectionally or
  bidirectionally across the six switches ultimately destined to a
  sink} \label{fig_tsn_indLoop}
\end{figure}

\subsection{System Overview and Simulation Setup}  \label{tsn:eval:sec}
This section explains the simulation setup and model. Furthermore, the
topology and simulation scenarios will be presented. Throughout, we
employ the OMNet++~\cite{varga2008overview} simulation environment.

\begin{table}
	\caption{Simulation Parameters}
	\begin{tabularx}{\columnwidth}{|X|X|X|}
		\hline
		\multicolumn{1}{|c|}{\textbf{Key}} & \multicolumn{1}{|c|}{\textbf{Symbol}} & \multicolumn{1}{c|}{\textbf{Value}} \\
		\hline
		Simulation Duration & $Sim_{limit}$ &  $100$ seconds \\
		\hline
		Initialized Cycle Time & $GCL_{CT}$ & $50~\mu$s \\
		\hline
		Initialized Gating Ratio & $ST^{R}_{init}$ & $20$\% (i.e., 10~$\mu$s) \\
		\hline
		Average Streams per Second & $\pi$ & $1 - 20$ \\
		\hline
		Average stream duration & $\tau$ & $2 - 5$ seconds \\
		\hline
		Number of Frames per Cycle & $\gamma$ & $1$ \\
		\hline
		BE Traffic Intensity & $\rho_{L}$ & $0.1, 1.0, 2.0$ (low, mid, and high) \\
		\hline
		ST sources & $S$ & $6$ \\
		\hline
		Hurst Parameter & $H$ & $0.5$ \\
		\hline
		Queue Size & $Q_{size}$ & $512$~Kb
		%\note{Should this be $100$~Kb since we want the nax delay in the queue to be $100~\mu$s (end to end) following
		%the max queuing delay equation under the simple queuing model \begin{equation} Delay \leq B \div R
		%\end{equation} where $B$ is the size of the queue and $R$ is the channel capacity}
		\\
		\hline
	\end{tabularx}
	\label{table: simulation parameters}
\end{table}

\begin{table}[t!]
	\centering
	\footnotesize
	\caption{Traffic proportions relative to number of hops for the industrial control loop topology}
	\label{hops}
	\begin{tabular}{cccccc}
		\textbf{Hops \#} & 1 & 2 & 3 & 4 & 5 \\
		\textbf{Range Distribution} & 20 \% & 20 \% & 20 \% & 20 \% & 20 \% \\
	\end{tabular}
\end{table}

\subsubsection{Network Model}	\label{tas:sec:net}
The network topology is modeled around an industrial control loop
topology that consists of six core switches in a ring topology
connected to the CNC as shown in Fig.~\ref{fig_tsn_indLoop}. Each
switch-to-switch link operates as a full-duplex Ethernet link with a
capacity (transmission bitrate) $R = 1$~Gbps. Each switch can act as a
gateway for a number of traffic sources and one sink. The distance
between two successive switches along the ring is fixed to 100~m and
the switch-to-switch propagation delay is set accordingly to
0.5~$\mu$s. All switches are configured to use 802.1Qbv TAS as the
traffic shaper for each switch-to-switch egress port whose flow
schedule (ST gating ratio and cycle time) is configured by the CNC in
the centralized (hybrid) model and independently in the decentralized
(fully distributed) model.

\subsubsection{Traffic Model}		\label{tas:sec:traffic}
We consider periodic (pre-planned) traffic and sporadic self-similar
Poisson traffic for ST traffic and for BE traffic, respectively. To
emulate dynamic conditions in the network, we employ several
distributed ST sources that generate $\pi$ ST streams according to the
network and traffic parameters shown in Table~\ref{table: simulation
  parameters}. The stream generation follows a Poisson distribution
where $\pi$ represents the average number of generated streams per
second. Each stream within a source injects packet traffic with packet
size $64$~bytes and has an ST traffic injection rate that is uniformly
distributed with a value ($\gamma$) statically defined at stream
creation time and a destination address assigned by the number of
switch-to-switch hops as shown in Table.~\ref{hops}. Furthermore, at
the stream creation time, each stream is given a start time (usually
the current runtime), and a finish time based on $\tau$. We consider
admission as the completion of the flow from start to the finish time
reported by the source. Each source is attached to a core TSN switch
gateway (first hop switch). While the TSN switches operate with time
synchronization, the ST sources (outside the TSN domain) do not need
to be synchronized. Therefore, the gateways can in the worst case
delay ST traffic by a maximum of 1 cycle times. However, note that the
ST traffic follows an isochronous traffic class as specified by
IEC/IEEE 60802 where the sources are synchronized with the network
after stream registration is initiated.

\subsection{Centralized Model Evaluation}	\label{eval:cent:sec}
In evaluating the proposed solution described in
Section~\ref{tsn:design:sec}, we consider both periodic and sporadic
sources for ST and BE traffic, as described in
Section~\ref{tas:sec:traffic}. We evaluated the CNC with TAS shaper on
the industrial control loop for the unidirectional and bi-directional
topologies and results are collected for tests following the simulation
parameters shown in Table~\ref{table: simulation parameters}.

\subsubsection{Unidirectional Ring Topology}

\begin{figure} [t!] \centering
	\includegraphics[width=3.3in]{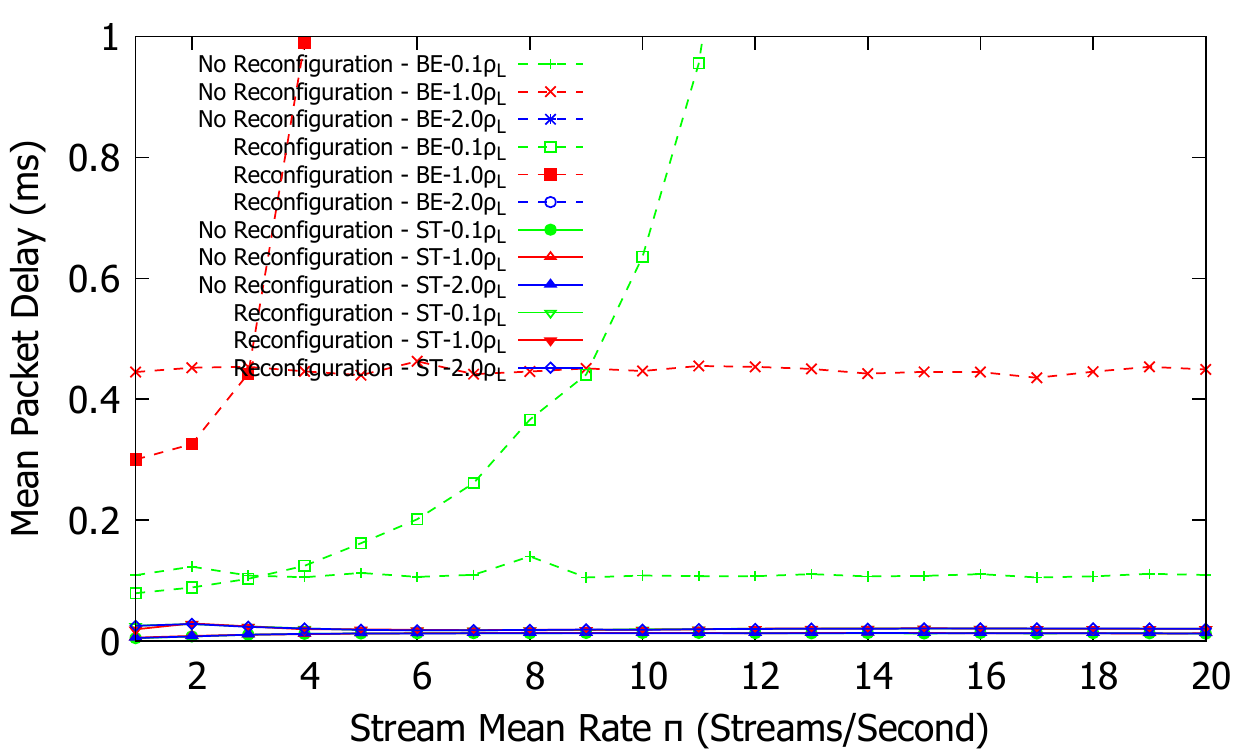}
	\caption{Centralized Unidirectional Topology: Mean end-to-end delay for ST and BE traffic for $\tau = 2$ under different BE loads~$\rho_{L}$, ST stream rates~$\pi$, and initialized gating ratio of $20\%$.}
	\label{fig_delay_2}
\end{figure}

\begin{figure} [t!] \centering
	\includegraphics[width=3.3in]{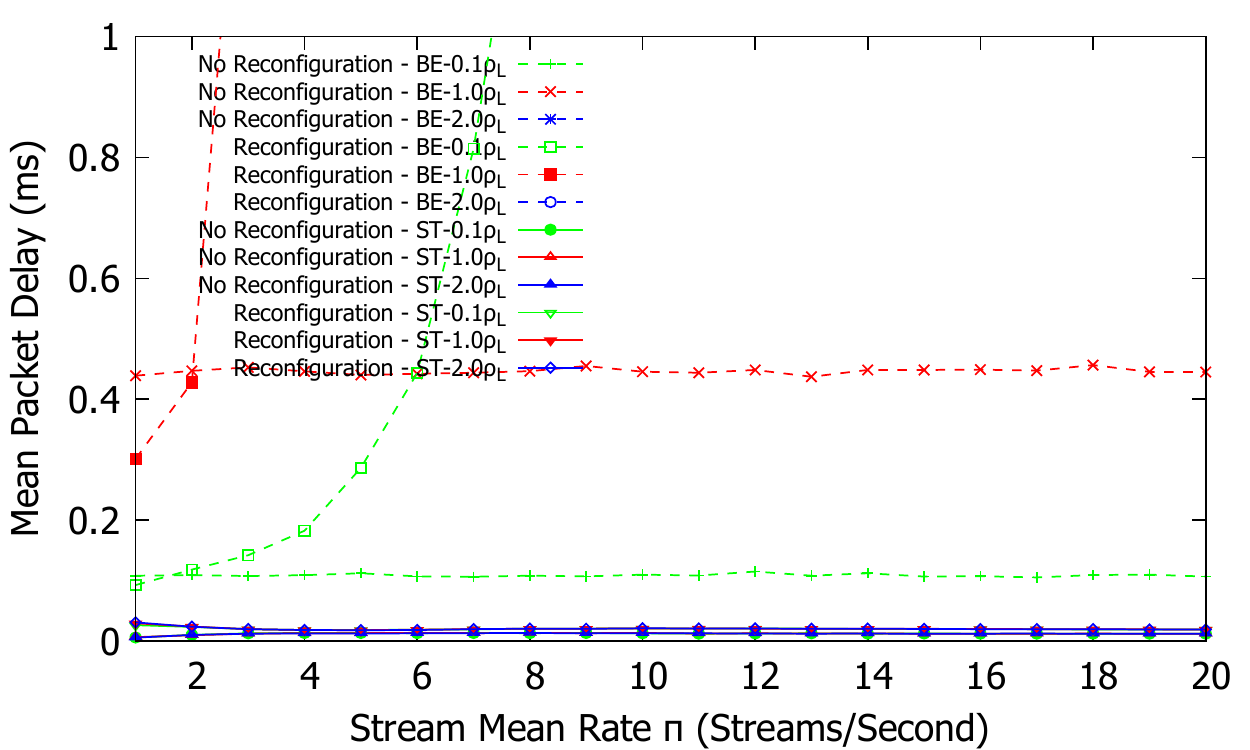}
	\caption{Centralized Unidirectional Topology: Mean end-to-end delay for ST and BE traffic for $\tau = 3$ under different BE loads~$\rho_{L}$, ST stream rates~$\pi$, and initialized gating ratio of $20\%$.}
	\label{fig_delay_3}
\end{figure}

\begin{figure} [t!] \centering
	\includegraphics[width=3.3in]{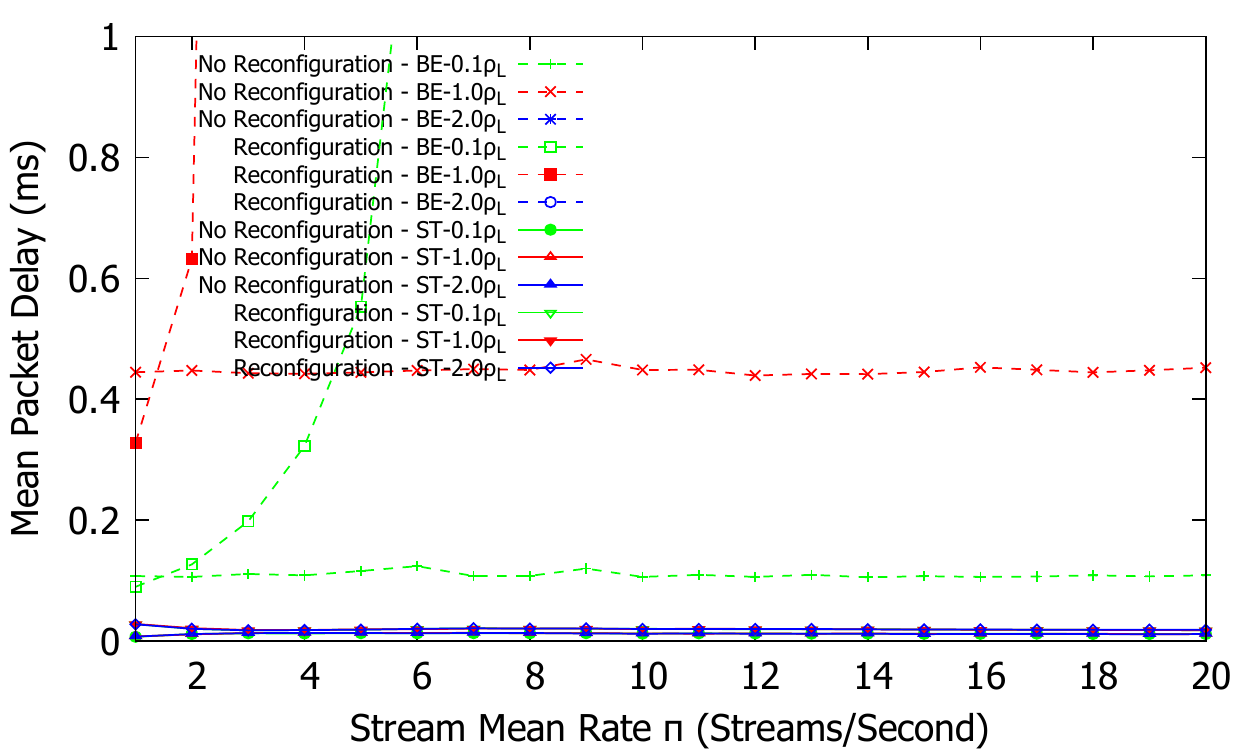}
	\caption{Centralized Unidirectional Topology: Mean end-to-end delay for ST and BE traffic for $\tau = 4$ under different BE loads~$\rho_{L}$, ST stream rates~$\pi$, and initialized gating ratio of $20\%$.}
	\label{fig_delay_4}
\end{figure}

\begin{figure} [t!] \centering
	\includegraphics[width=3.3in]{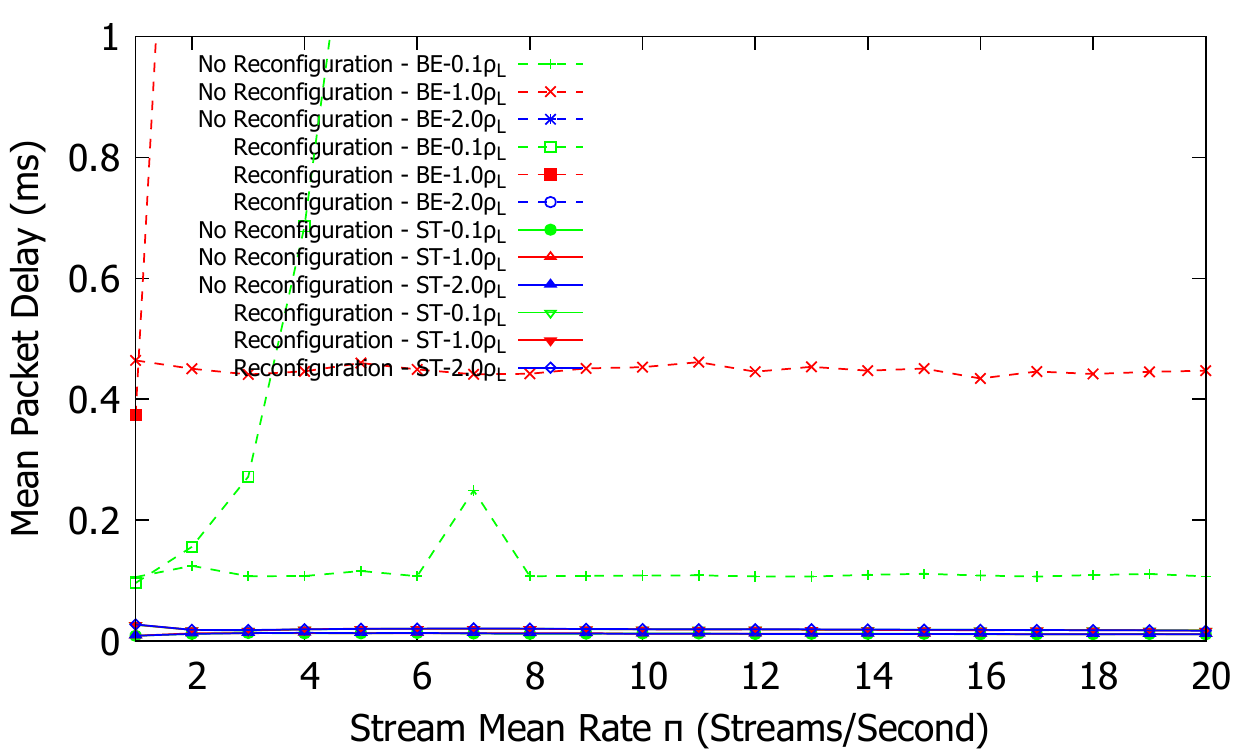}
	\caption{Centralized Unidirectional Topology: Mean end-to-end delay for ST and BE traffic for $\tau = 5$ under different BE loads~$\rho_{L}$, ST stream rates~$\pi$, and initialized gating ratio of $20\%$.}
	\label{fig_delay_5}
\end{figure}

Figs.~\ref{fig_delay_2}--\ref{fig_delay_5} show the average mean
delay for ST traffic and for BE traffic for the centralized
unidirectional ring topologies. The average delays in general are low
and stable for both BE and ST traffic. Since the CNC manages the ST
traffic streams and therefore guarantees the bandwidth rates needed to
transmit across a single switch hop in one CT, the ST delays are less
than $100~\mu$s for all $\tau$ values. The ST delays with
reconfiguration active at the CNC experience higher delays than ``No
Reconfiguration'' since we essentially push more frames into the
network that increase the queuing delay.
%%%%Note that we do not take into account the queuing delay.
%%%%Modeling the queuing delay and using it to guarantee max delays
%%%%is left for future work.
BE traffic
experiences much higher delays than ST. With the ``No Reconfiguration''
approach, the BE traffic delay is near constant since the gating ratio is
left unchanged throughout the simulation. For the test with
reconfiguration, the BE mean delay increases dramatically since we
tend to accept and exhaust the CT with ST streams over the course of
the simulation run.

\begin{figure} [t!] \centering
	\begin{subfigure}{\columnwidth} \centering
		\includegraphics[width=3.3in]{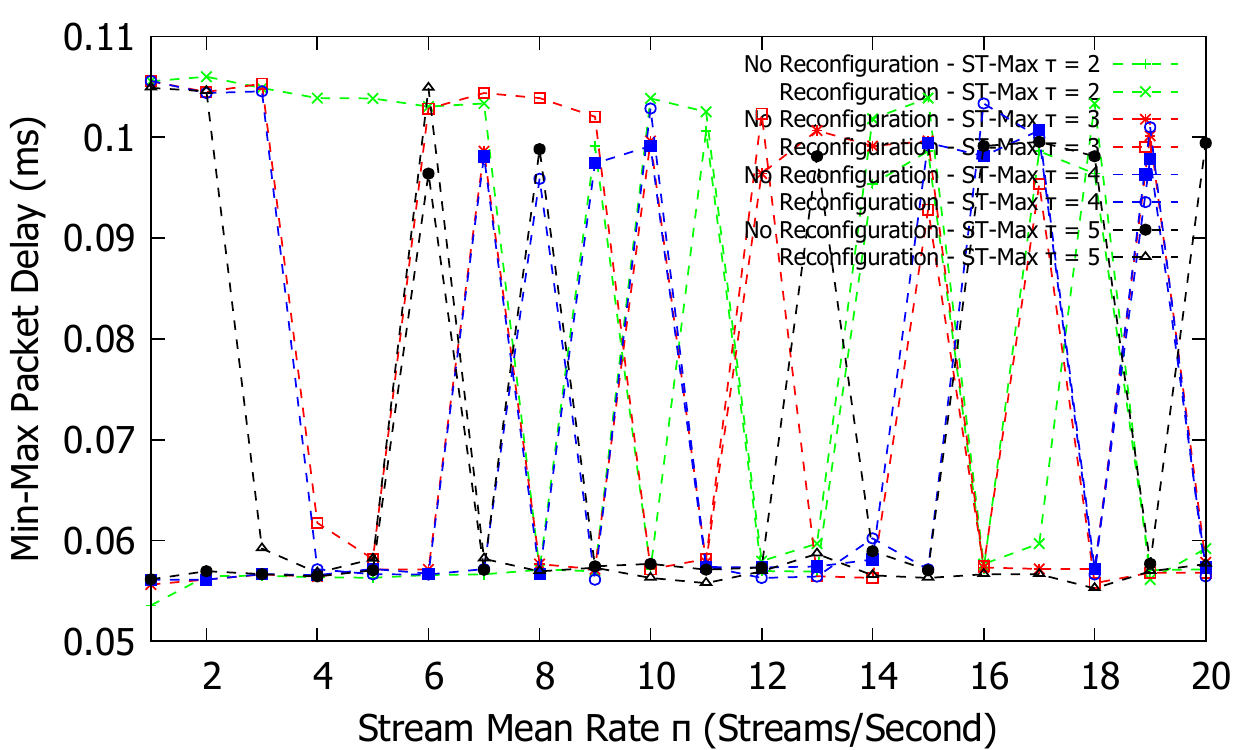}
		\caption{Low $\rho_{L}$}
	\end{subfigure}
	\begin{subfigure}{\columnwidth} \centering
		\includegraphics[width=3.3in]{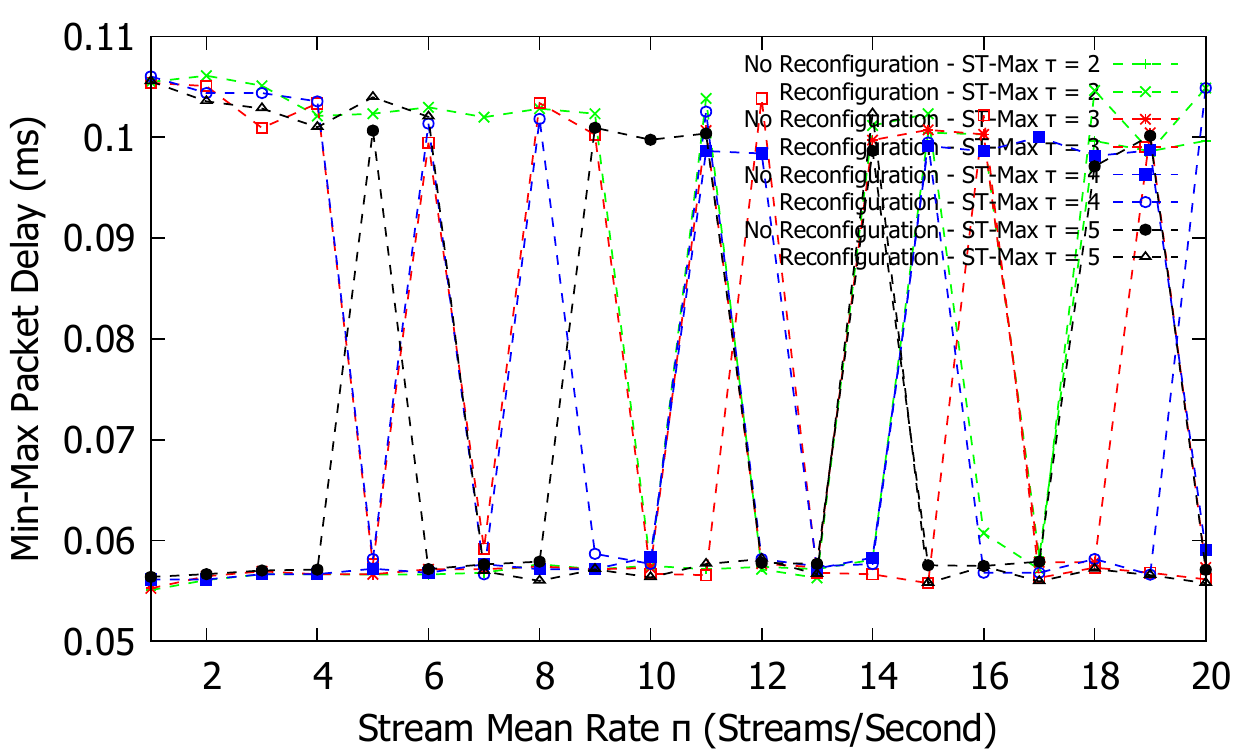}
		\caption{Mid $\rho_{L}$}
	\end{subfigure}
	\begin{subfigure}{\columnwidth} \centering
		\includegraphics[width=3.3in]{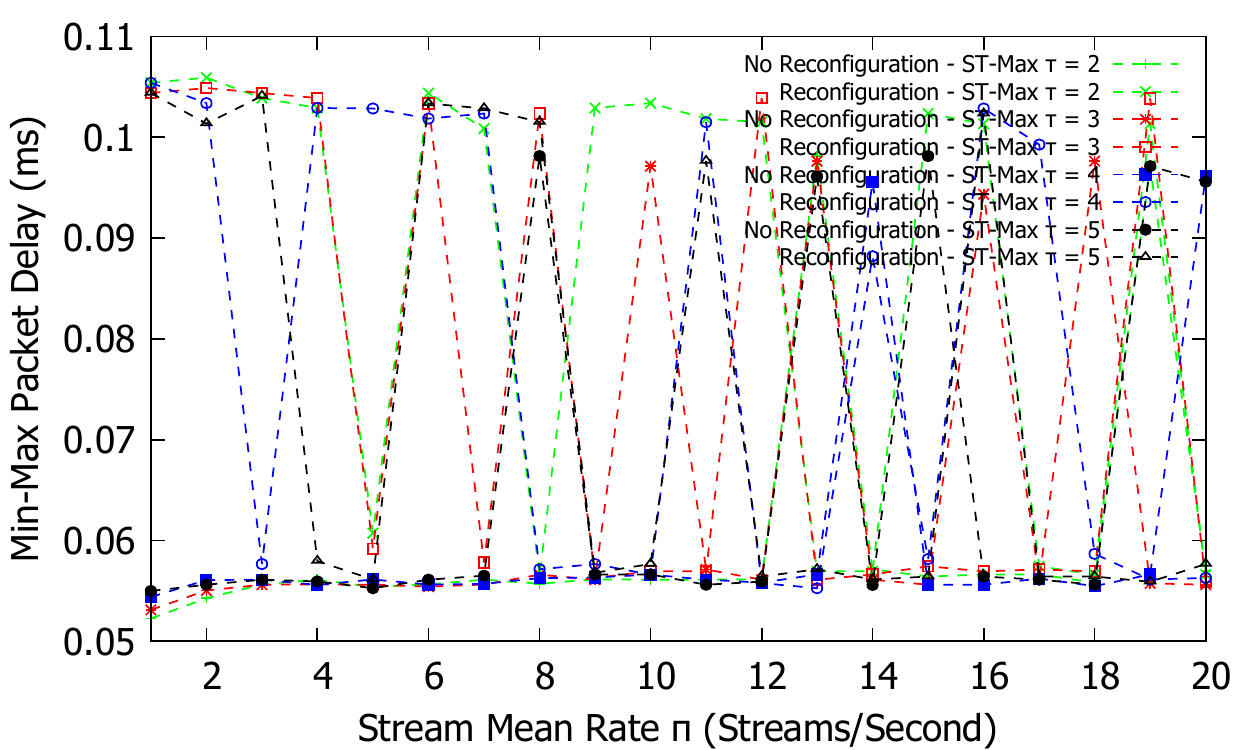}
		\caption{High $\rho_{L}$}
	\end{subfigure}
	\caption{Centralized Unidirectional Topology: Max delay as a results of TAS with centralized configuration (CNC) management entity.}
	\label{fig_maxDelay}
\end{figure}

As mentioned in the introduction section, TSN needs to guarantee and
bound the maximum delay in order to deterministically forward traffic
across a TSN domain. Fig.~\ref{fig_maxDelay} shows the maximum delay
evaluation for ST traffic. For the unidirectional ring topology with a
maximum of five hop streams, the reconfiguration approach maximum
delay is bounded at $0.105$~ms, while for the ``No reconfiguration''
approach, the max delay is bounded at nearly $60\mu$~s. For TAS and
the CNC's registration and reservation procedure, the guarantee is
applied to bandwidth as a share of the egress port using time division
multiplexing.
%%%%Therefore, the queuing delay is not currently addressed. However,
With the parameters chosen empirically, the
maximum delays is capped to approximately $100\mu$~s which is ideal
for the topology chosen and critical ST traffic.

\begin{figure} [t!] \centering
	\begin{subfigure}{\columnwidth} \centering
		\includegraphics[width=3.3in]{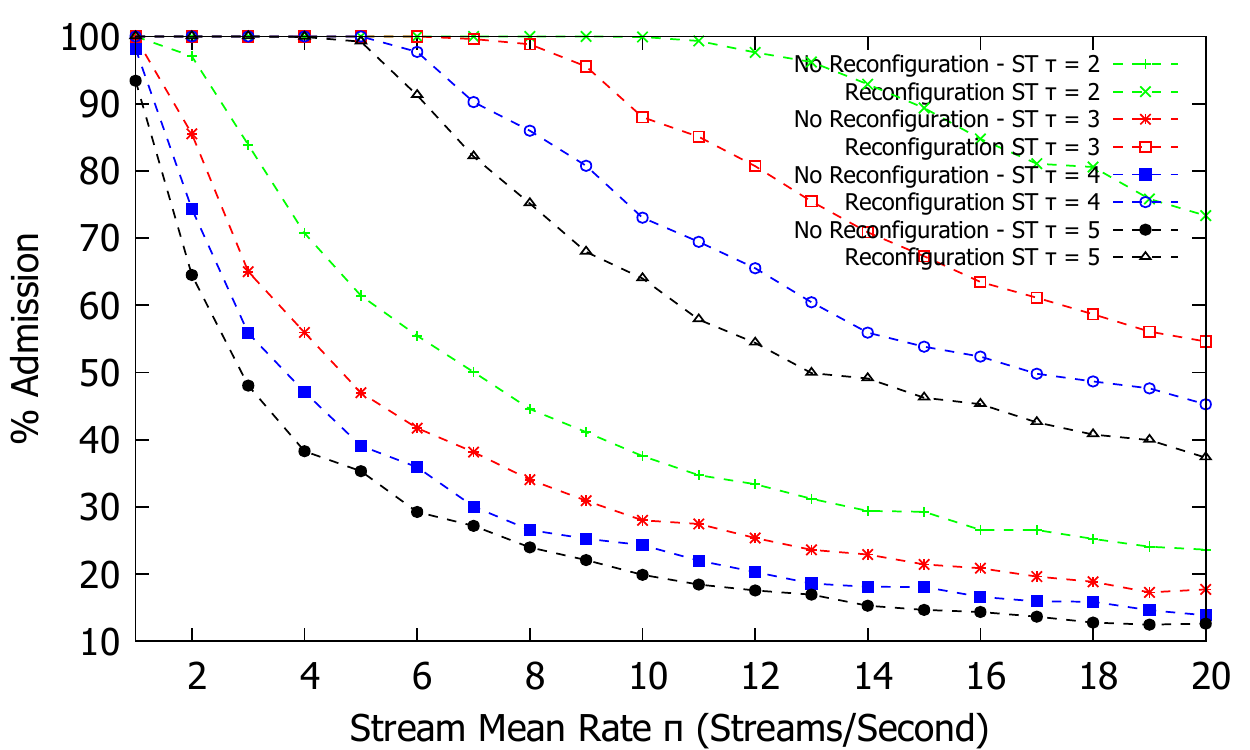}
		\caption{Low $\rho_{L}$}
	\end{subfigure}
	\begin{subfigure}{\columnwidth} \centering
		\includegraphics[width=3.3in]{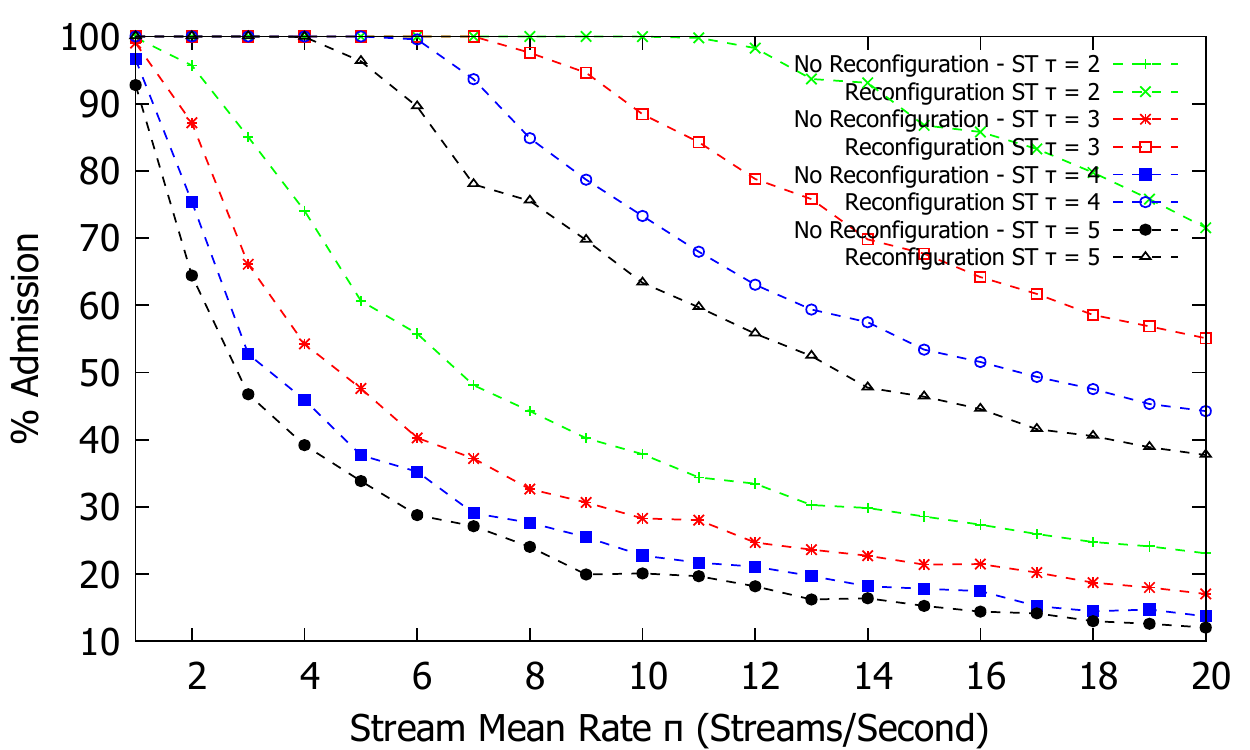}
		\caption{Mid $\rho_{L}$}
	\end{subfigure}
	\begin{subfigure}{\columnwidth} \centering
		\includegraphics[width=3.3in]{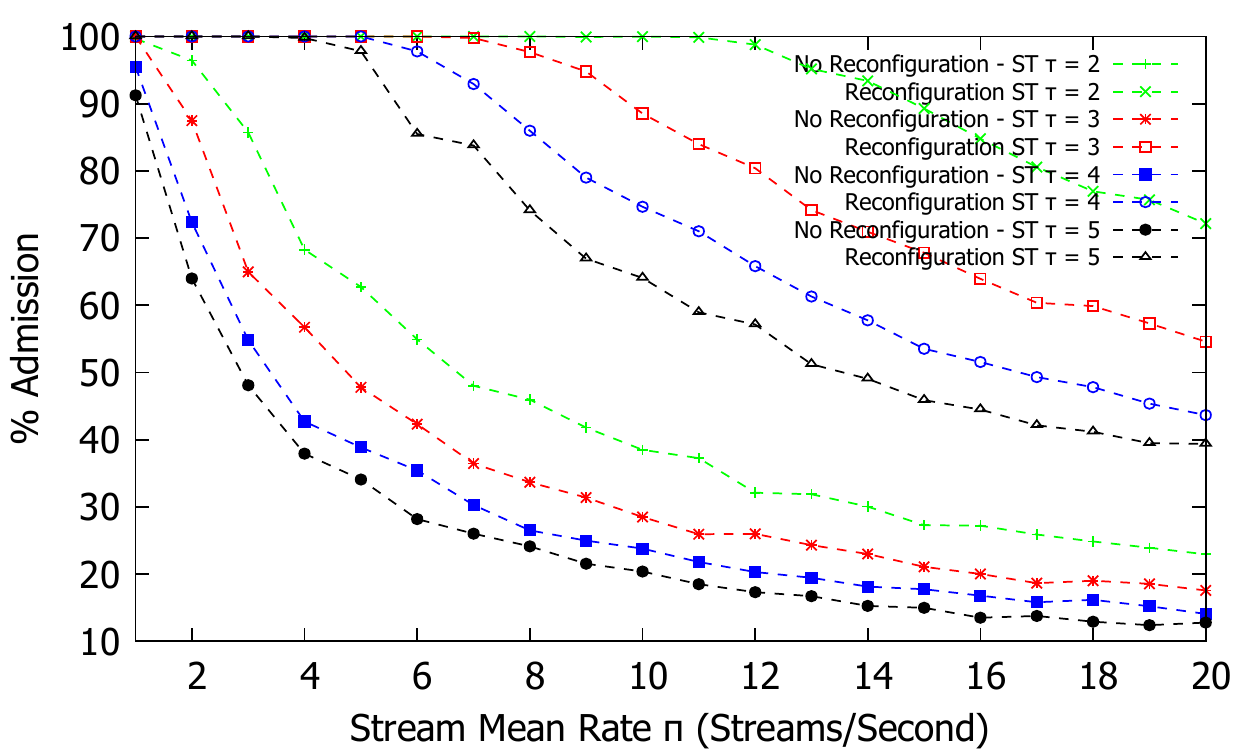}
		\caption{High $\rho_{L}$}
	\end{subfigure}
	\caption{Centralized Unidirectional Topology: Stream Admission as a results of TAS with centralized configuration (CNC) management entity.}
	\label{fig_admin}
\end{figure}

While QoS metrics are important, another factor that determines the
performance gains is the admission ratio for the
system. Fig.~\ref{fig_admin} shows the stream admission ratio that
results for both reconfiguration and no reconfiguration. In general,
each generated stream needs a data rate of about $10.24$~Mbps for a
50~$\mu$s CT for each egress port on the stream's path with one ST
packet rate per CT and fixed packet size of $64$~B. With an egress
port channel capacity of $R=10^{9}$, approximately $100$ streams can
be guaranteed. Compared to the ``no reconfiguration'' approach, the
reconfiguration significantly improves the admission rates at the cost
of higher BE traffic delays since the ST slot borrows BE time slots to
accommodate the ST streams.

\begin{figure} [t!] \centering
	\begin{subfigure}{\columnwidth} \centering
		\includegraphics[width=3.3in]{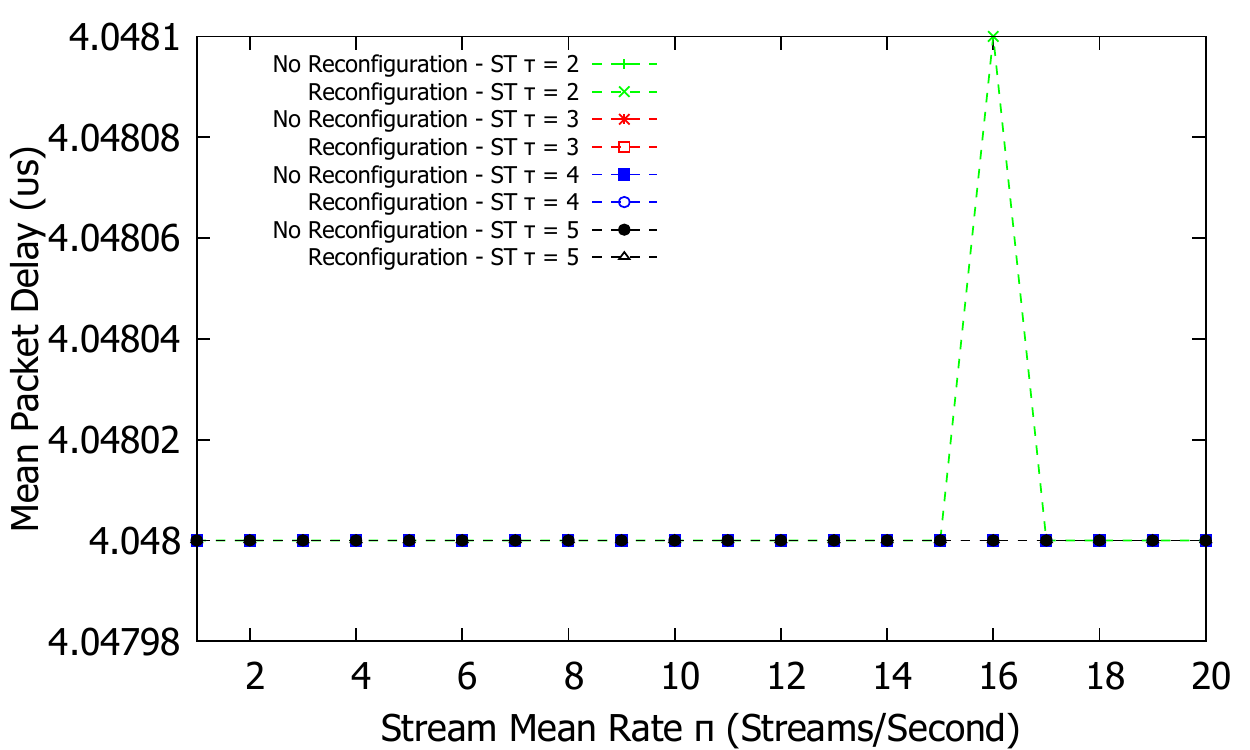}
		\caption{Low $\rho_{L}$}
	\end{subfigure}
	\begin{subfigure}{\columnwidth} \centering
		\includegraphics[width=3.3in]{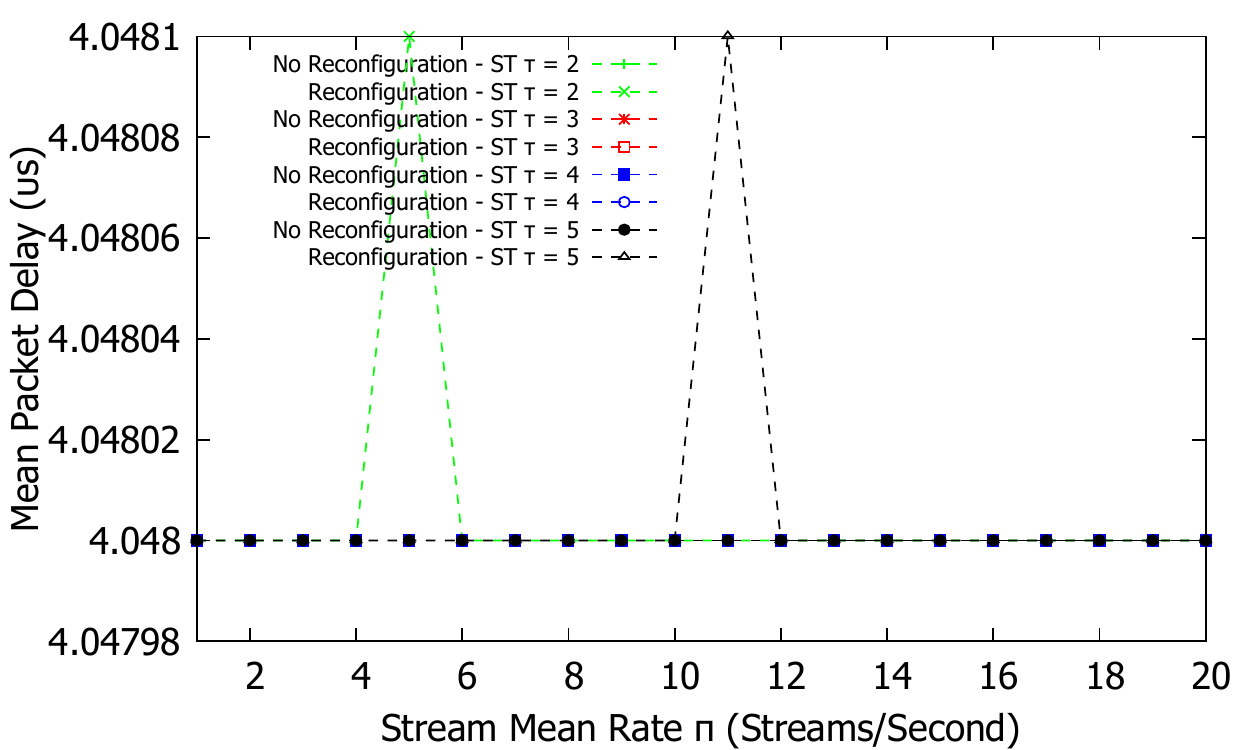}
		\caption{Mid $\rho_{L}$}
	\end{subfigure}
	\begin{subfigure}{\columnwidth} \centering
		\includegraphics[width=3.3in]{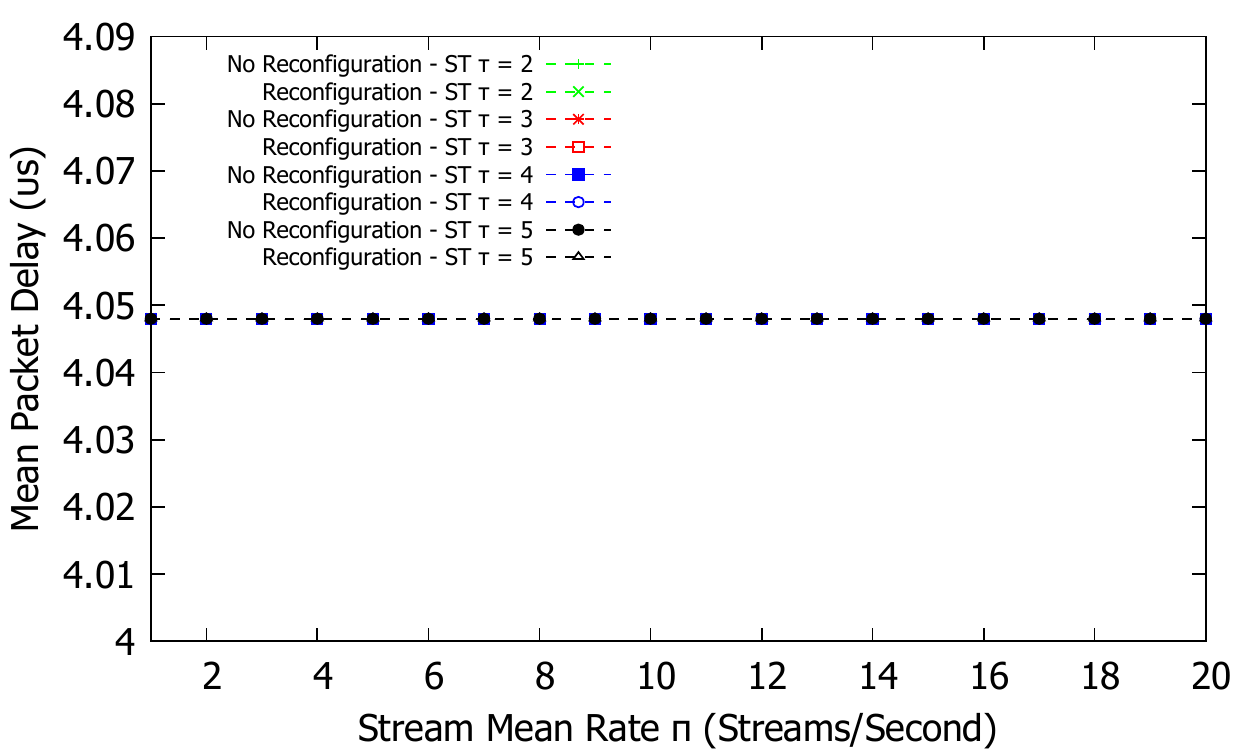}
		\caption{High $\rho_{L}$}
	\end{subfigure}
	\caption{Centralized Unidirectional Topology: Stream Signaling delay as a results of TAS with centralized configuration (CNC) management entity.}
	\label{fig_signalDelay}
\end{figure}

CDT traffic that requests transmission guarantees from the CNC
experiences some delay before being either admitted or
rejected. Fig.~\ref{fig_signalDelay} shows the average signaling
latency for ST stream registration. Since the control plane is out-of
band from the data plane within the TSN domain, the delay is constant
throughout the simulation run.

\begin{figure} [t!] \centering
	\begin{subfigure}{\columnwidth} \centering
		\includegraphics[width=3.3in]{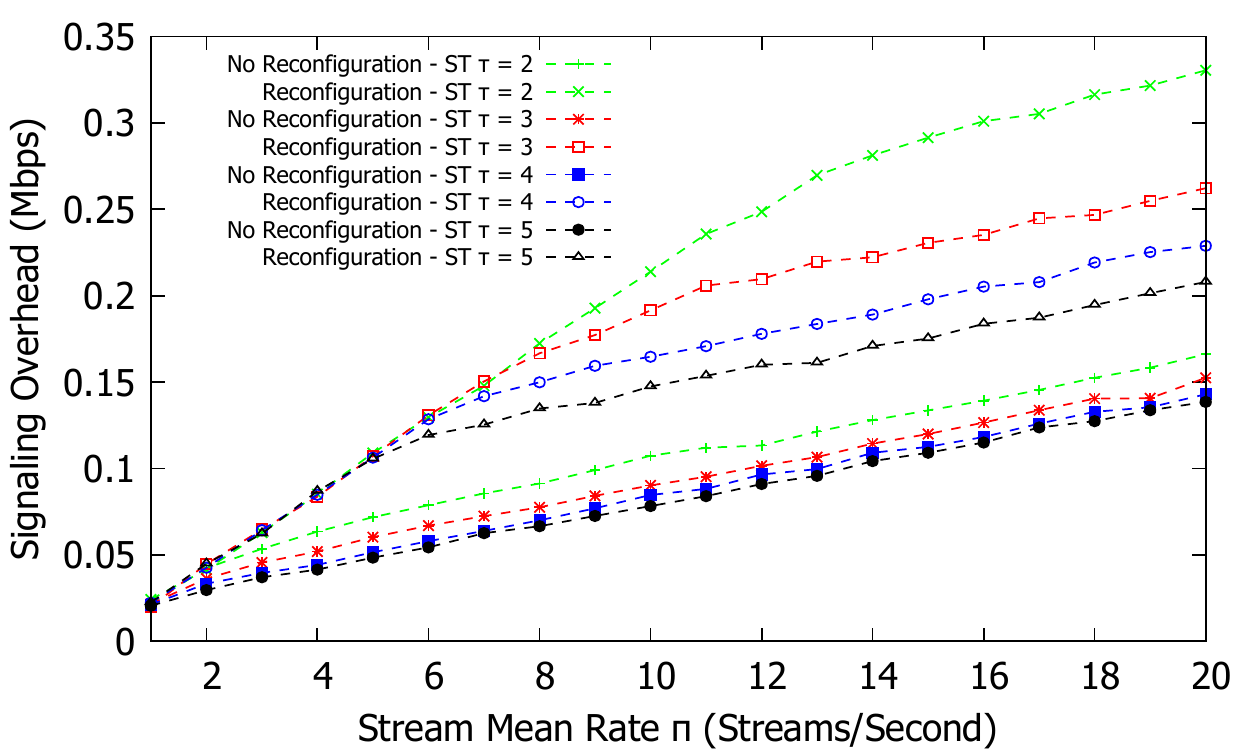}
		\caption{Low $\rho_{L}$}
	\end{subfigure}
	\begin{subfigure}{\columnwidth} \centering
		\includegraphics[width=3.3in]{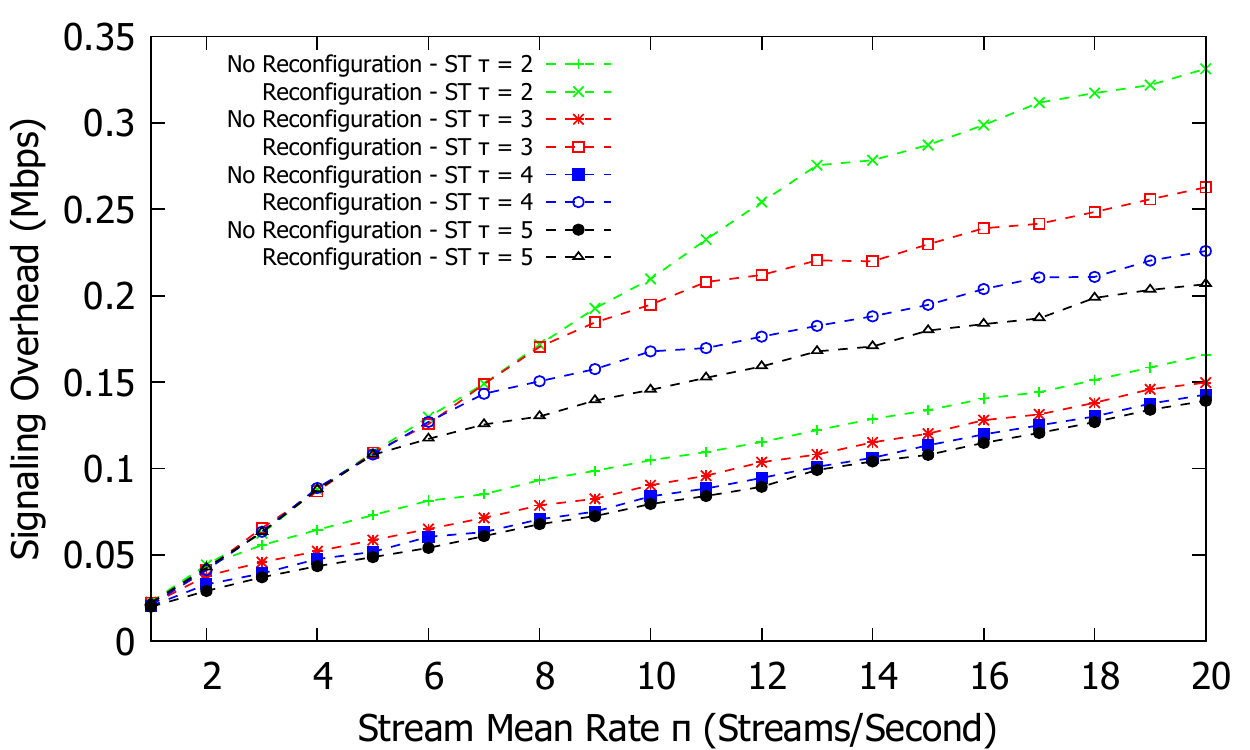}
		\caption{Mid $\rho_{L}$}
	\end{subfigure}
	\begin{subfigure}{\columnwidth} \centering
		\includegraphics[width=3.3in]{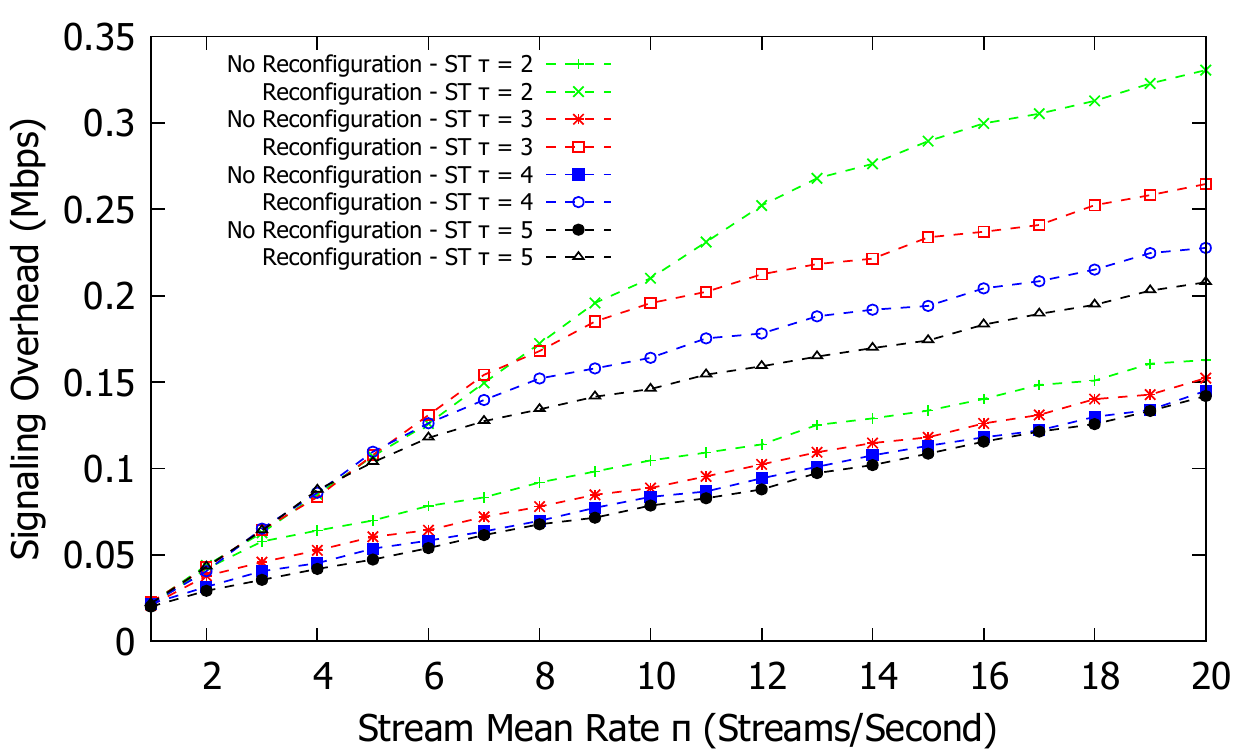}
		\caption{High $\rho_{L}$}
	\end{subfigure}
	\caption{Centralized Unidirectional Topology: Stream average signaling Overhead as a results of TAS with centralized configuration (CNC) management entity.}
	\label{fig_signalOverhead}
\end{figure}

Stream registration and reservation introduces some control plane
overhead. Fig.~\ref{fig_signalOverhead} shows the signaling
performance overhead. More specifically, the overhead is measured at
the CNC or both incoming and outgoing control (CDT)
traffic. Generally, the reconfiguration approach introduces more
signaling overhead; however, Ethernet generally has large bandwidths,
thus the CDT traffic rates are minimal compared to the link
capacities. Furthermore, when $\tau = 2$, we observe higher signaling
overhead due to accepting larger numbers of streams (rejections are
inexpensive compared to acceptance) both with and without
reconfiguration.

\begin{figure} [t!] \centering
	\begin{subfigure}{\columnwidth} \centering
		\includegraphics[width=3.3in]{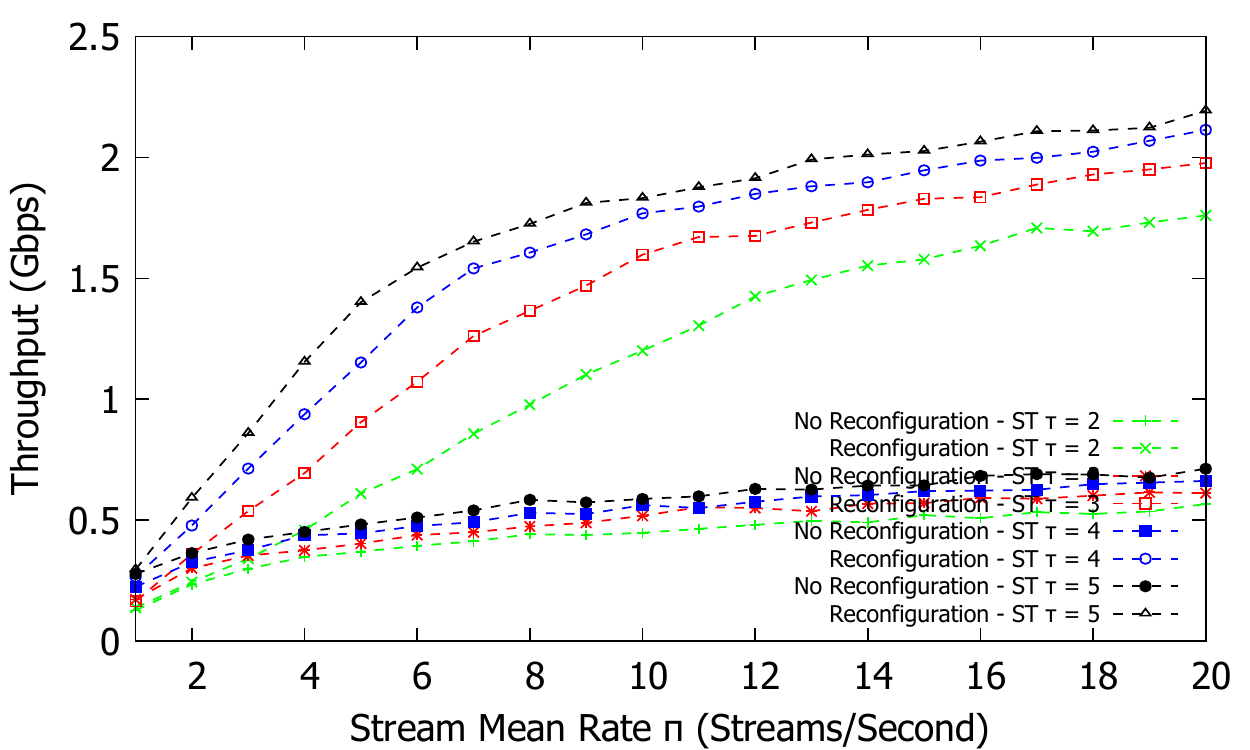}
		\caption{Low $\rho_{L}$}
	\end{subfigure}
	\begin{subfigure}{\columnwidth} \centering
		\includegraphics[width=3.3in]{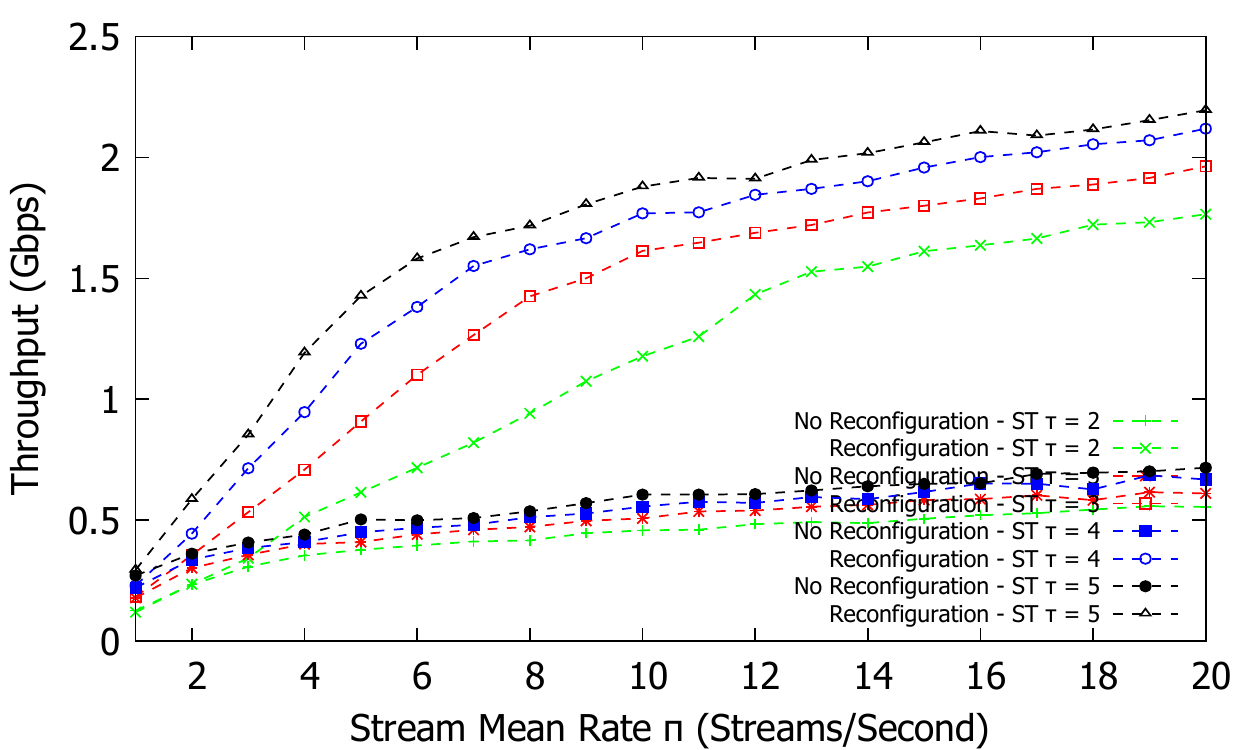}
		\caption{Mid $\rho_{L}$}
	\end{subfigure}
	\begin{subfigure}{\columnwidth} \centering
		\includegraphics[width=3.3in]{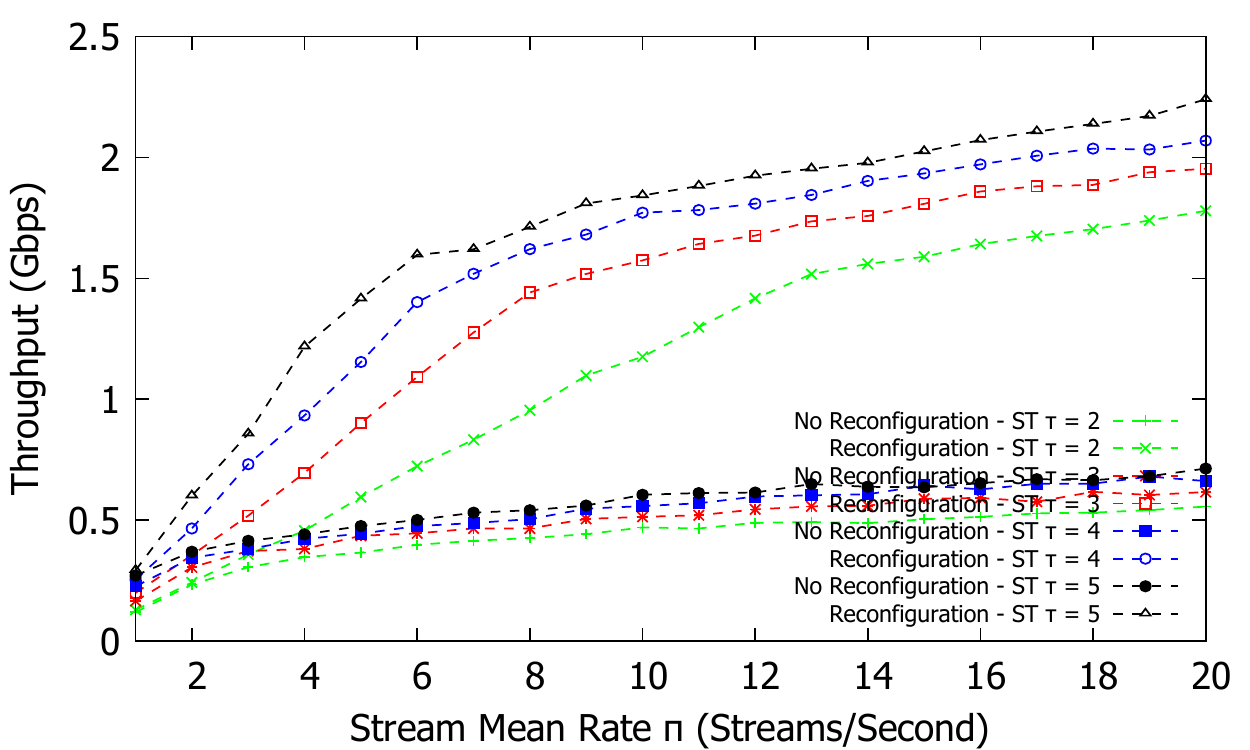}
		\caption{High $\rho_{L}$}
	\end{subfigure}
	\caption{Centralized Unidirectional Topology: ST Total average throughput measured at the sink as a results of TAS with centralized configuration (CNC) management entity.}
	\label{fig_avgTput_ST}
\end{figure}

\begin{figure} [t!] \centering
	\begin{subfigure}{\columnwidth} \centering
		\includegraphics[width=3.3in]{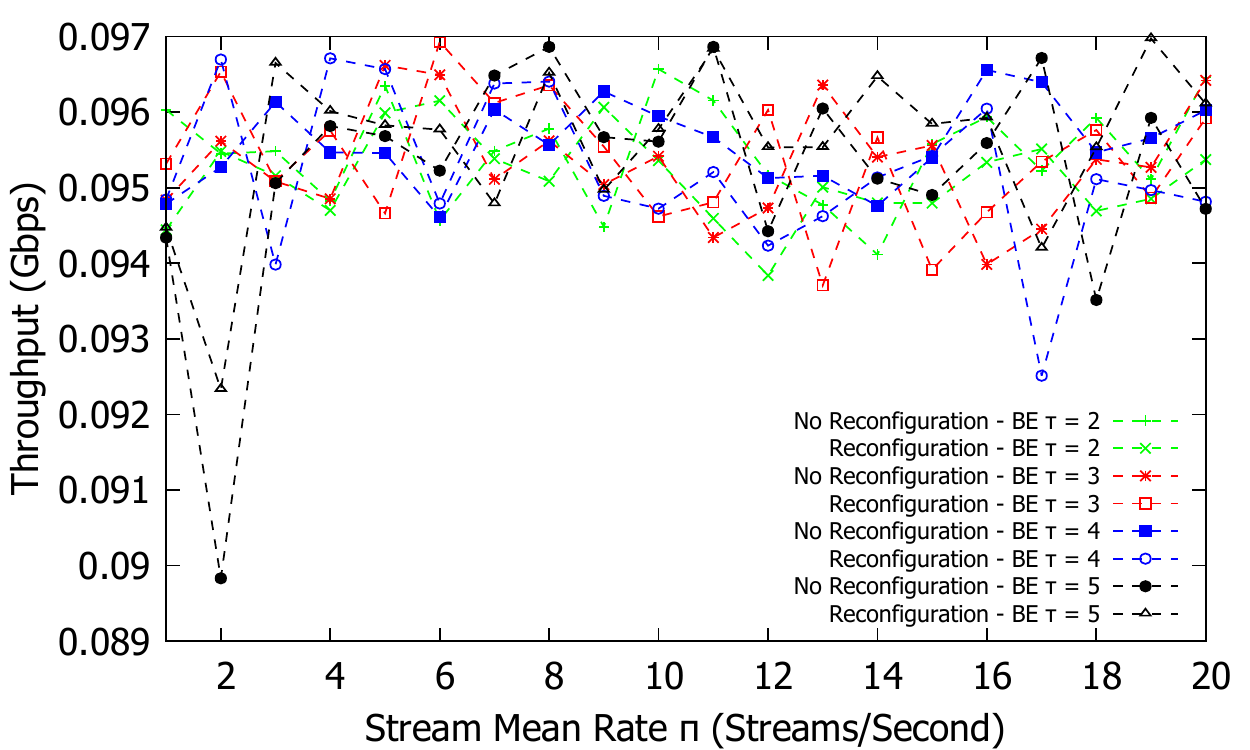}
		\caption{Low $\rho_{L}$}
	\end{subfigure}
	\begin{subfigure}{\columnwidth} \centering
		\includegraphics[width=3.3in]{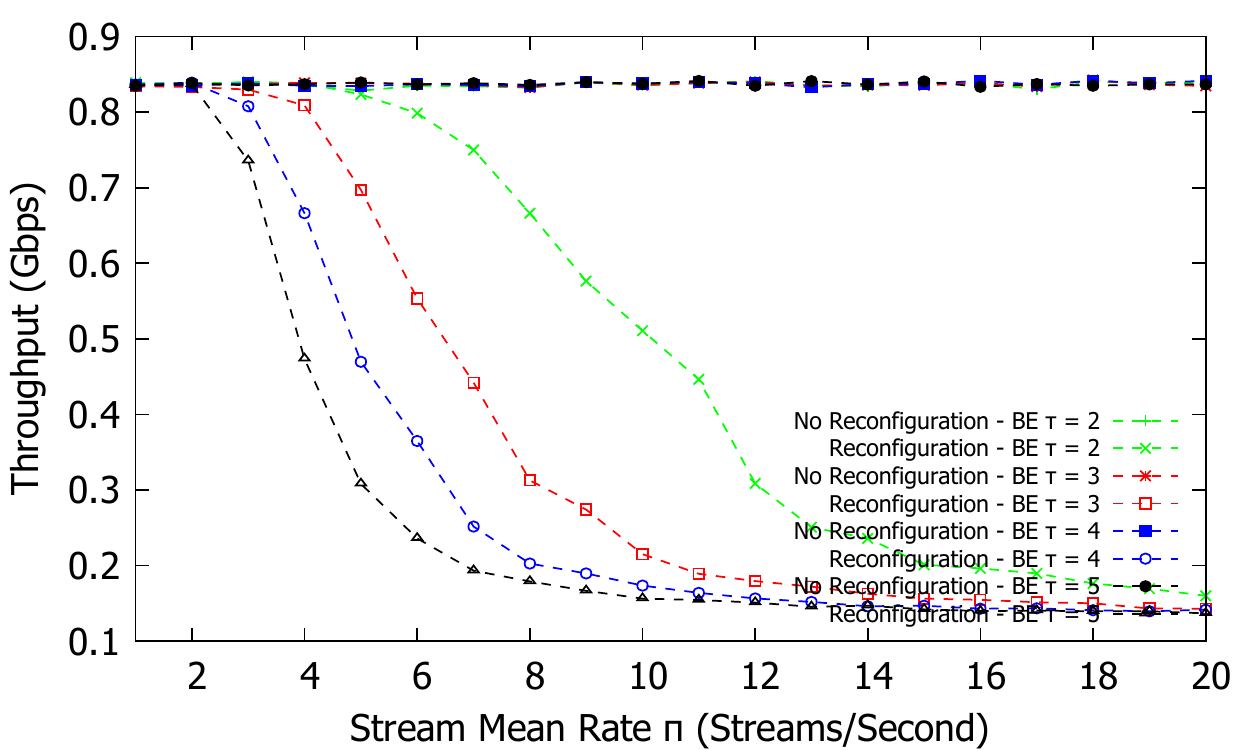}
		\caption{Mid $\rho_{L}$}
	\end{subfigure}
	\begin{subfigure}{\columnwidth} \centering
		\includegraphics[width=3.3in]{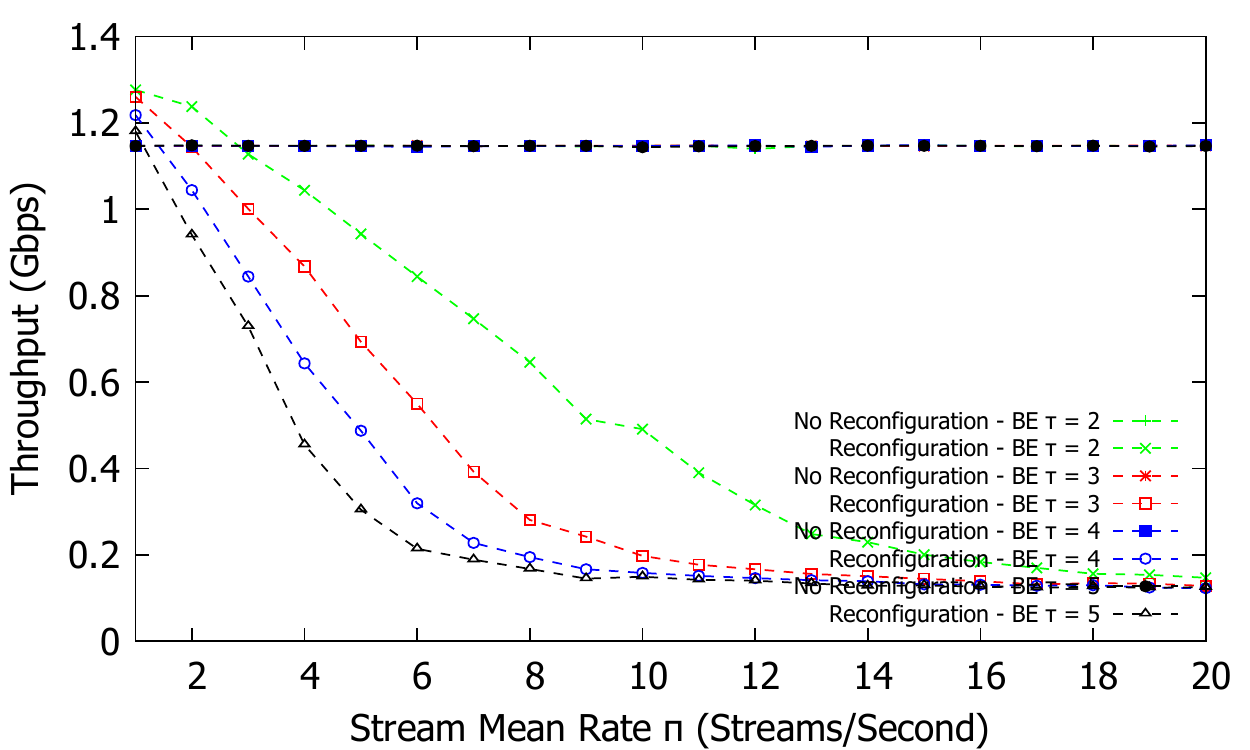}
		\caption{High $\rho_{L}$}
	\end{subfigure}
	\caption{Centralized Unidirectional Topology: BE Total average throughput measured at the sink as a results of TAS with centralized configuration (CNC) management entity.}
	\label{fig_avgTput_BE}
\end{figure}

Observing the network's throughput gain, Fig.~\ref{fig_avgTput_ST} and
Fig.~\ref{fig_avgTput_BE} show the average throughput measured at the
sink for both ST and BE traffic. While the total throughput that can
be achieved is around $6$~Gbps, the maximum throughput allowed is
around $2$~Gbps since some switches get bottlenecked faster than other
switches which restricts the addition of more flows. For the ``no
reconfiguration'' approach, BE traffic is varied between $0.1$ traffic
intensity to $2.0$. Fig.~\ref{fig_avgTput_BE}[a]--[b] shows an average
throughput of $0.1$~Gbps and $1.0$~Gbps, respectively, which is expected
and no frames are dropped. However, as the load reaches $2.0$~Gbps and
more ST streams are accepted, i.e., the BE traffic time slot is shortened,
BE traffic starts to suffer and caps at around $1.2$~Gbps, as shown in
Fig.~\ref{fig_avgTput_BE}[c]. With reconfiguration, BE tends to suffer
more since we shift the time slot of BE to ST (maximum of $90\%$) and
the throughput drops to $0.1$~Gbps as shown in
Fig.~\ref{fig_avgTput_BE}[b][c].

\begin{figure} [t!] \centering
	\begin{subfigure}{\columnwidth} \centering
		\includegraphics[width=3.3in]{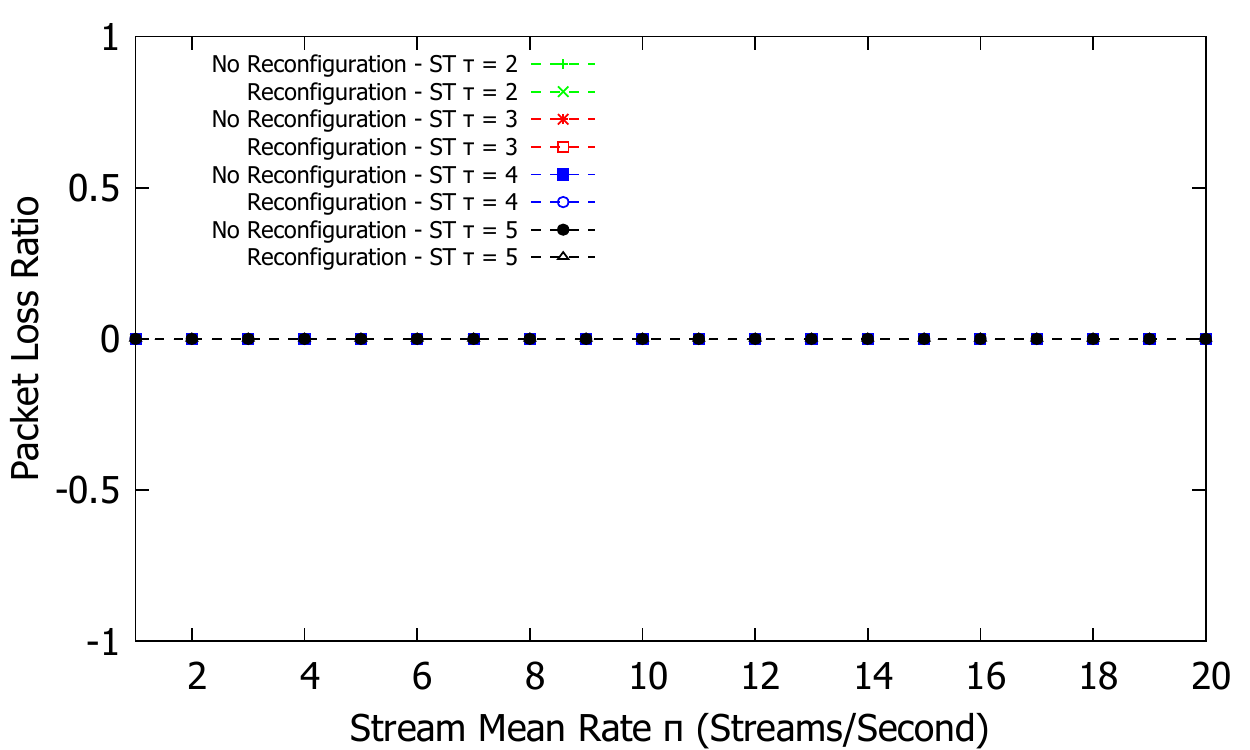}
		\caption{Low $\rho_{L}$}
	\end{subfigure}
	\begin{subfigure}{\columnwidth} \centering
		\includegraphics[width=3.3in]{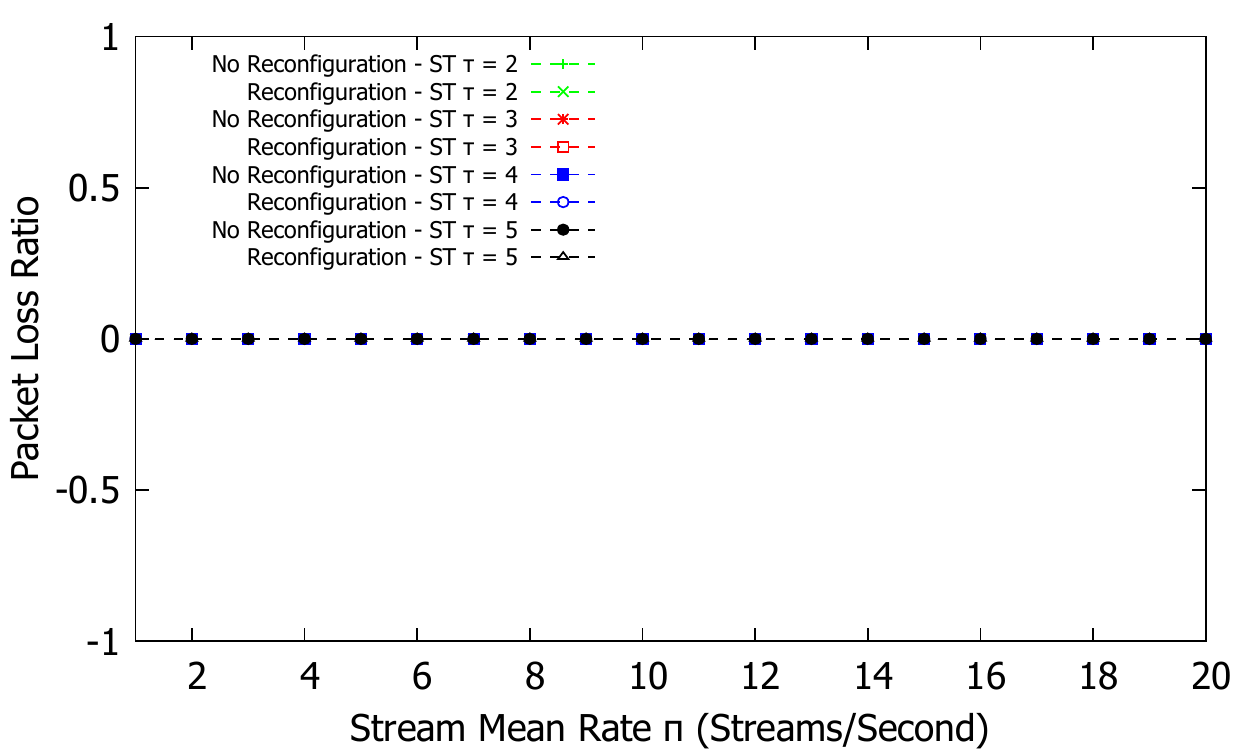}
		\caption{Mid $\rho_{L}$}
	\end{subfigure}
	\begin{subfigure}{\columnwidth} \centering
		\includegraphics[width=3.3in]{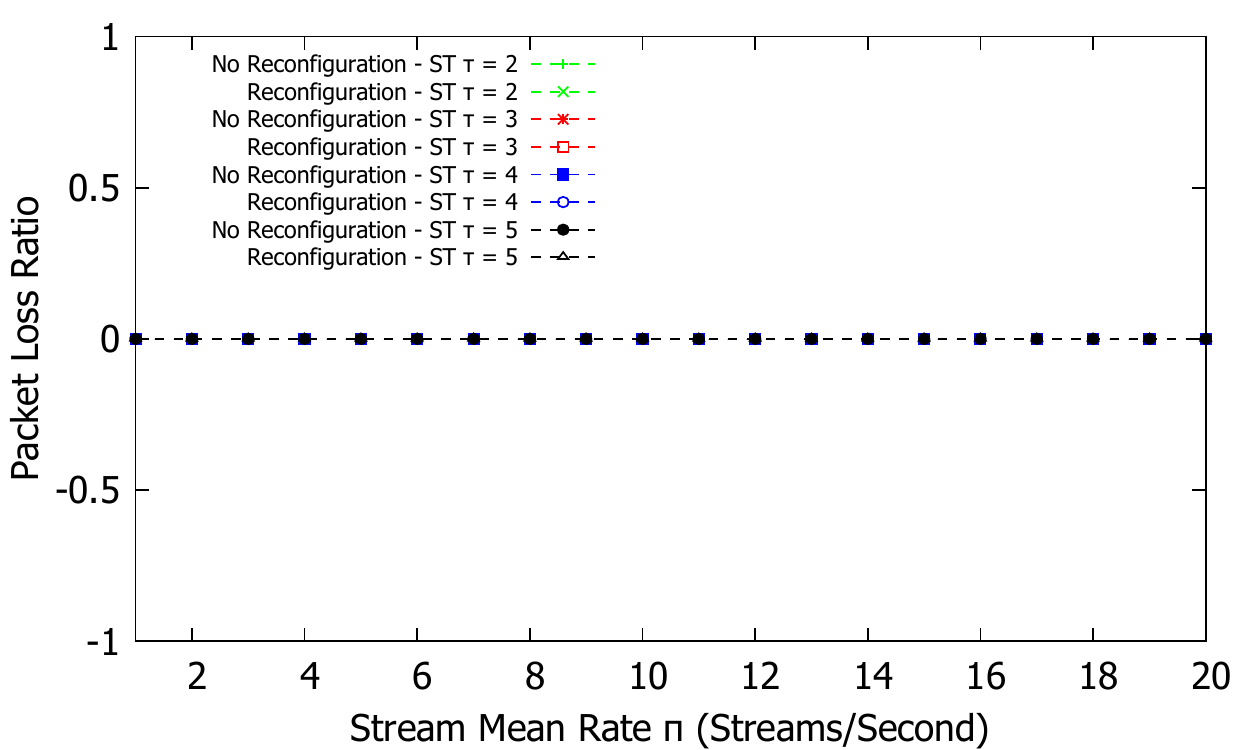}
		\caption{High $\rho_{L}$}
	\end{subfigure}
	\caption{Centralized Unidirectional Topology: ST Frame loss ratio as a results of TAS with centralized configuration (CNC) management entity.}
	\label{fig_lossProb_ST}
\end{figure}

\begin{figure} [t!] \centering
	\begin{subfigure}{\columnwidth} \centering
		\includegraphics[width=3.3in]{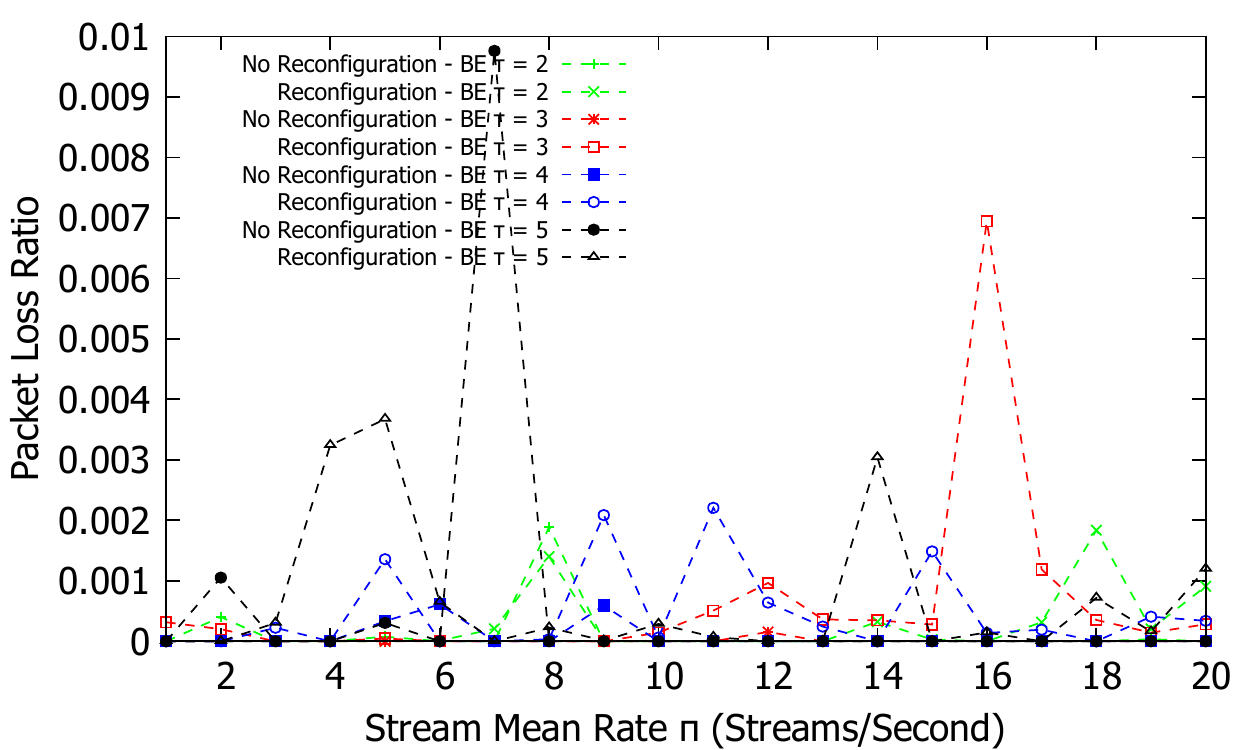}
		\caption{Low $\rho_{L}$}
	\end{subfigure}
	\begin{subfigure}{\columnwidth} \centering
		\includegraphics[width=3.3in]{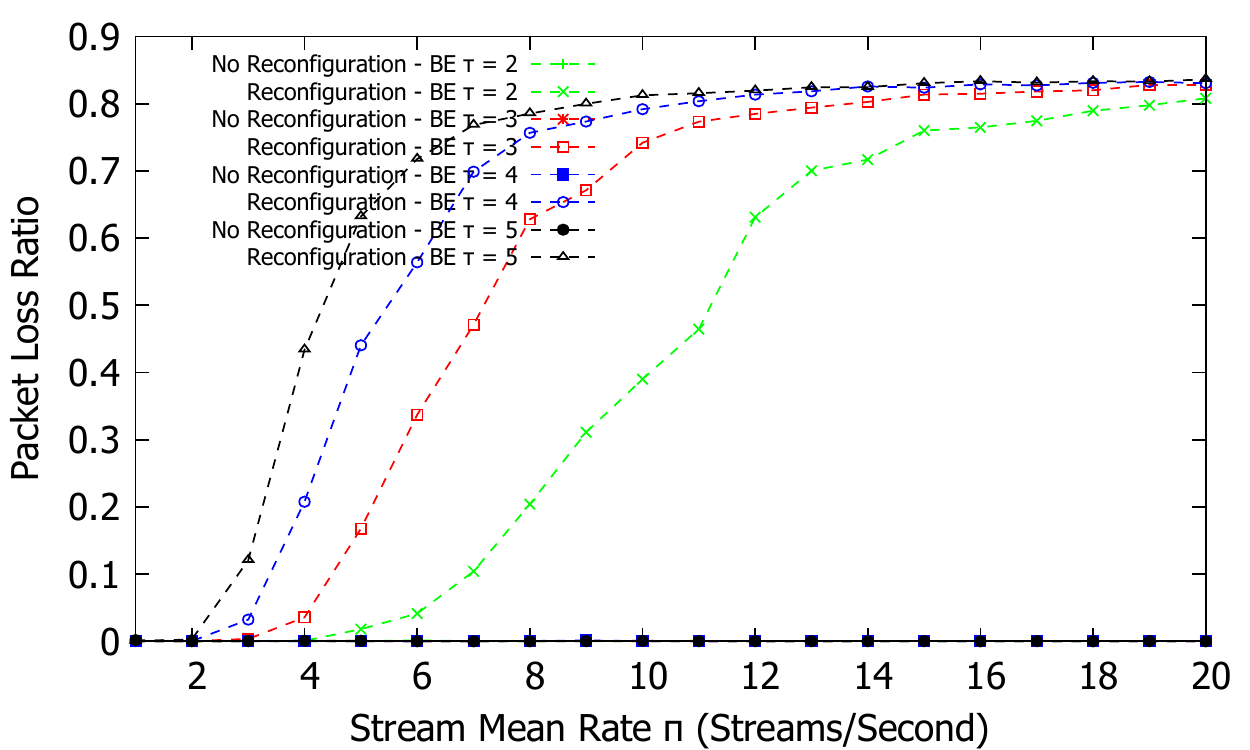}
		\caption{Mid $\rho_{L}$}
	\end{subfigure}
	\begin{subfigure}{\columnwidth} \centering
		\includegraphics[width=3.3in]{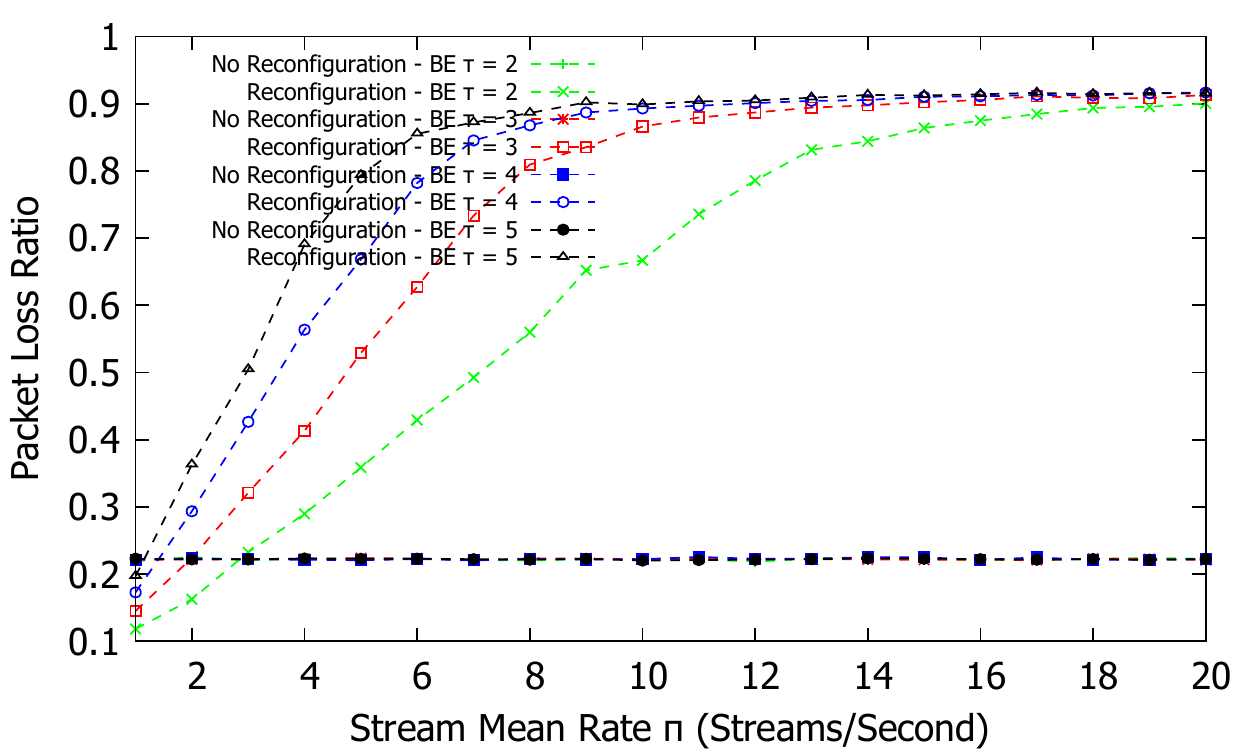}
		\caption{High $\rho_{L}$}
	\end{subfigure}
	\caption{Centralized Unidirectional Topology: BE Frame loss ratio as a results of TAS with centralized configuration (CNC) management entity.}
	\label{fig_lossProb_BE}
\end{figure}

To show the performance of the CNC management of streams,
Fig.~\ref{fig_lossProb_ST} and Fig.~\ref{fig_lossProb_BE} show the
packet loss ratio for ST and BE traffic in the network. Since the CNC
manages only ST streams, the TSN guarantees (which include zero packet
loss since retransmissions are in general too expansive for ST
traffic) are only valid for ST streams. For BE under the
reconfiguration approach, as the load for ST traffic increases, the
packet loss increases as well. For the ``no reconfiguration''
approach, the loss typically is constant even for high loads of BE
traffic.

Overall, the reconfiguration approach certainly provides a means to
manage and to ensure that the number of ST streams is maximized
according to the link capacity. However, we observe from the results
and the topology used that any bottleneck switch can generally reduce
the link utilization which can significantly drop the throughput, even
if the delay and loss are guaranteed. Selecting different paths (if
one exists) and modeling the queue to ensure maximum ST stream delays
can help potential bottleneck links and increase throughput throughout
the network.

\subsubsection{Bi-Directional Ring Topology}

\begin{figure} [t!] \centering
	\includegraphics[width=3.3in]{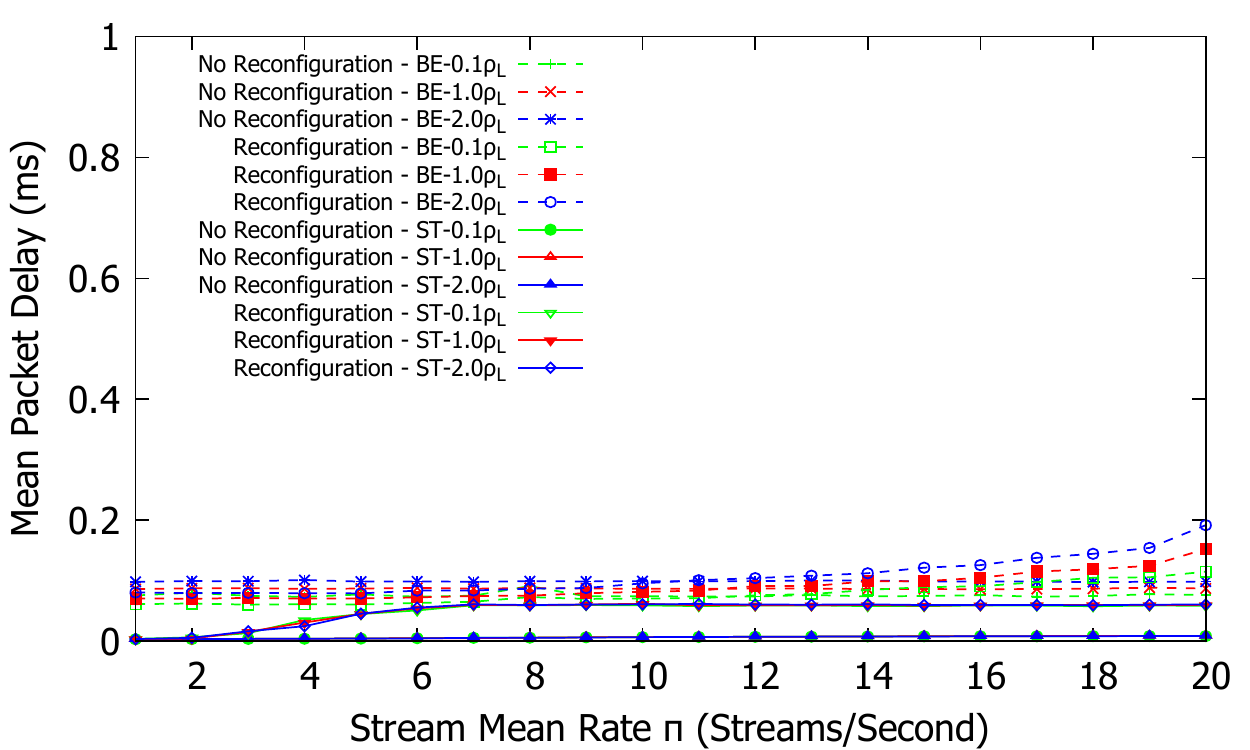}
	\caption{Centralized Bi-directional Topology: Mean end-to-end delay for ST and BE traffic for $\tau = 2$ under different BE loads~$\rho_{L}$, ST stream rates~$\pi$, and initialized gating ratio of $20\%$.}
	\label{fig_delay_2_bi}
\end{figure}

\begin{figure} [t!] \centering
	\includegraphics[width=3.3in]{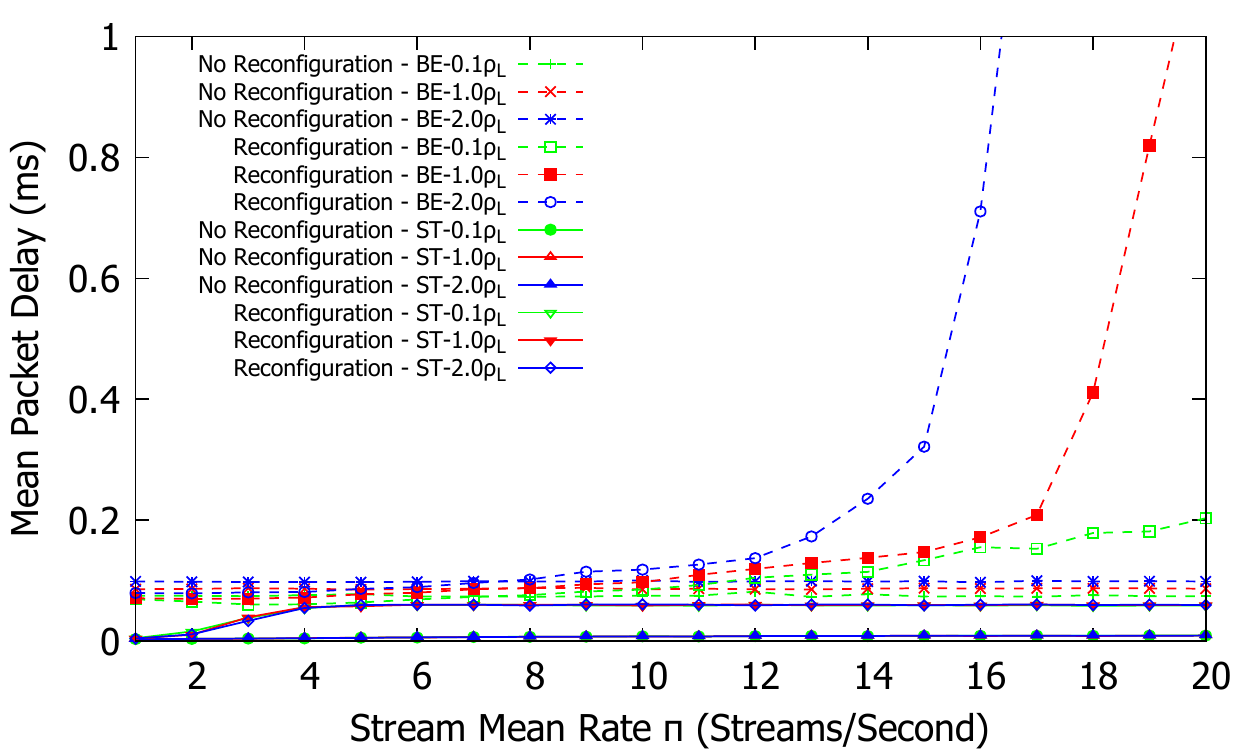}
	\caption{Centralized Bi-directional Topology: Mean end-to-end delay for ST and BE traffic for $\tau = 3$ under different BE loads~$\rho_{L}$, ST stream rates~$\pi$, and initialized gating ratio of $20\%$.}
	\label{fig_delay_3_bi}
\end{figure}

\begin{figure} [t!] \centering
	\includegraphics[width=3.3in]{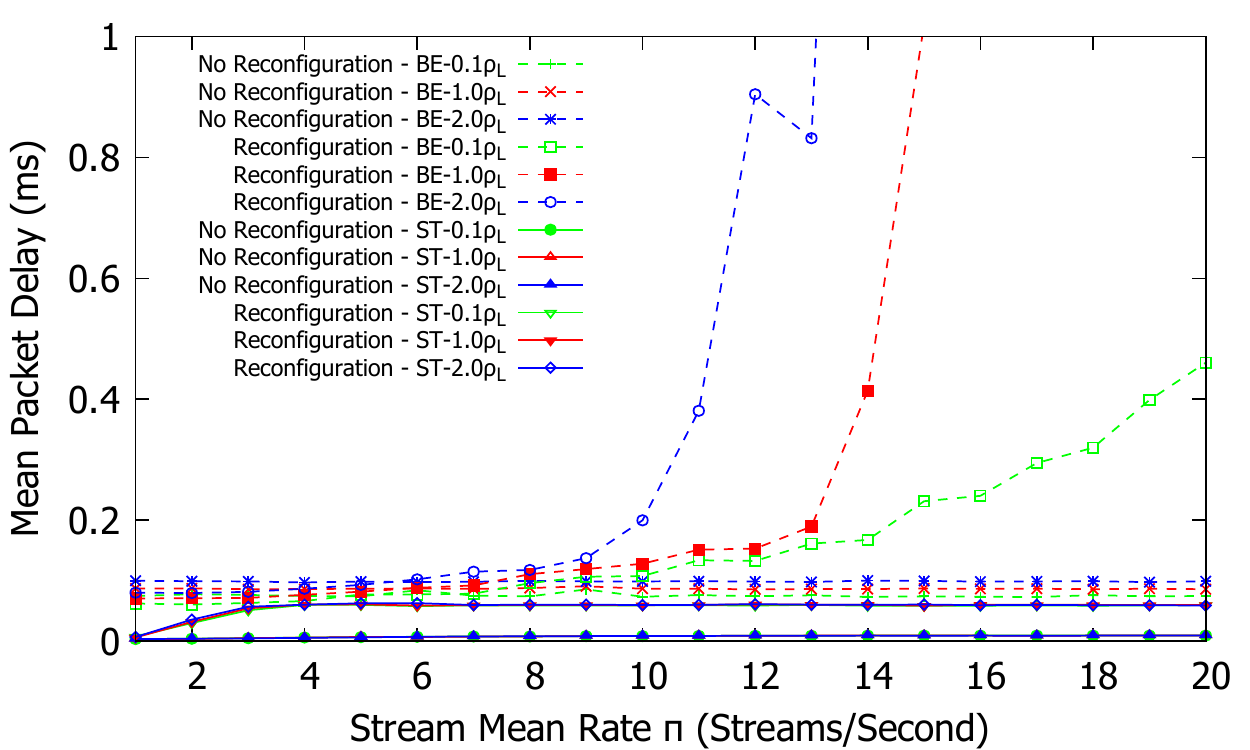}
	\caption{Centralized Bi-directional Topology: Mean end-to-end delay for ST and BE traffic for $\tau = 4$ under different BE loads~$\rho_{L}$, ST stream rates~$\pi$, and initialized gating ratio of $20\%$.}
	\label{fig_delay_4_bi}
\end{figure}

\begin{figure} [t!] \centering
	\includegraphics[width=3.3in]{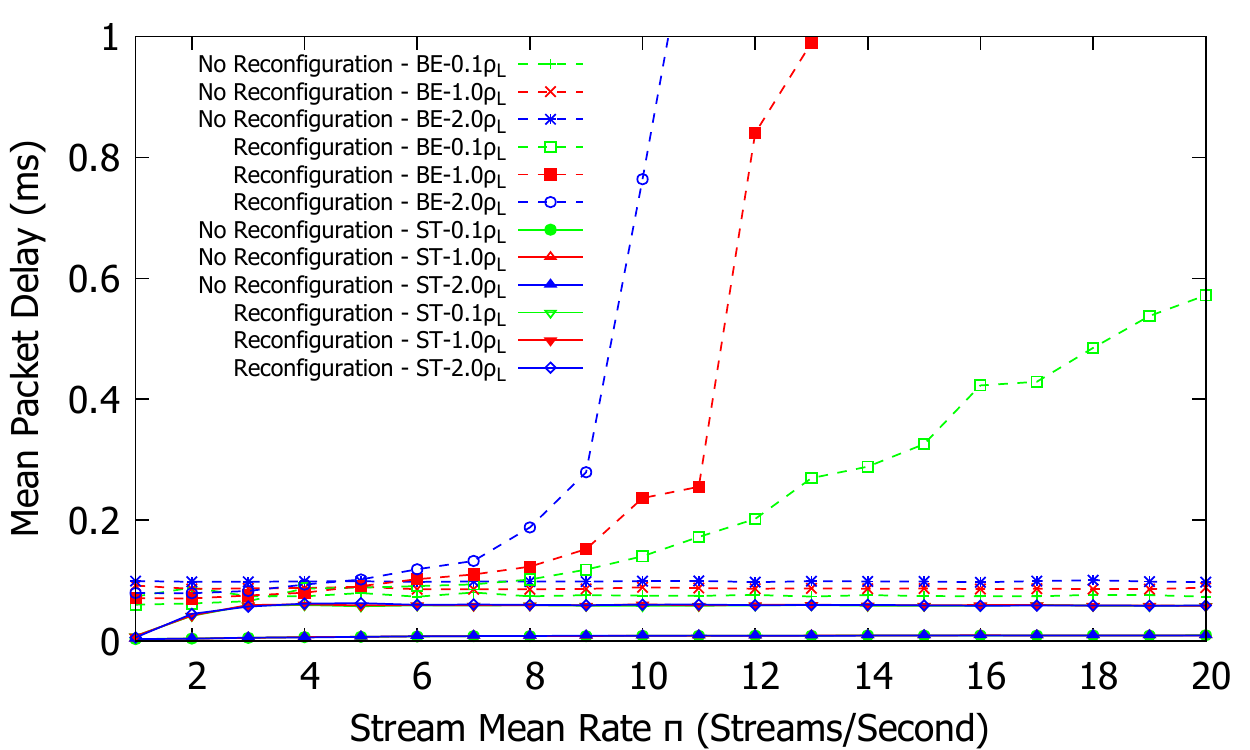}
	\caption{Centralized Bi-directional Topology: Mean end-to-end delay for ST and BE traffic for $\tau = 5$ under different BE loads~$\rho_{L}$, ST stream rates~$\pi$, and initialized gating ratio of $20\%$.}
	\label{fig_delay_5_bi}
\end{figure}

The unidirectional ring topology certainly simplifies the complexities of calculating the ST slot window. However, to show the a more pronounced gain in stream utilization and admission, a bi-directional ring topology is used with static shortest path routes. The two port switch now has two paths to the destination according to the hop count specified in Table.~\ref{hops}, where about $60\%$ of streams take one port and the rest ($40\%$) take the other port. Note that the edge links (switch to sink and source to switch) are given higher link capacities to avoid congestion at the edges where the CNC currently does not control (at least $2$~Gbps for the bi-directional ring). Fig.~\ref{fig_delay_2_bi} - \ref{fig_delay_5_bi} shows the average mean delay evaluation for both ST and BE traffic under different stream lifetime values, $\tau$. Compared to the unidirectional topology, the bi-directional provides significantly better delay results since an extra port with full-duplex link support now provides extra capacity to service streams giving more slot reservations to BE even at high ST stream load.

\begin{figure} [t!] \centering
	\begin{subfigure}{\columnwidth} \centering
		\includegraphics[width=3.3in]{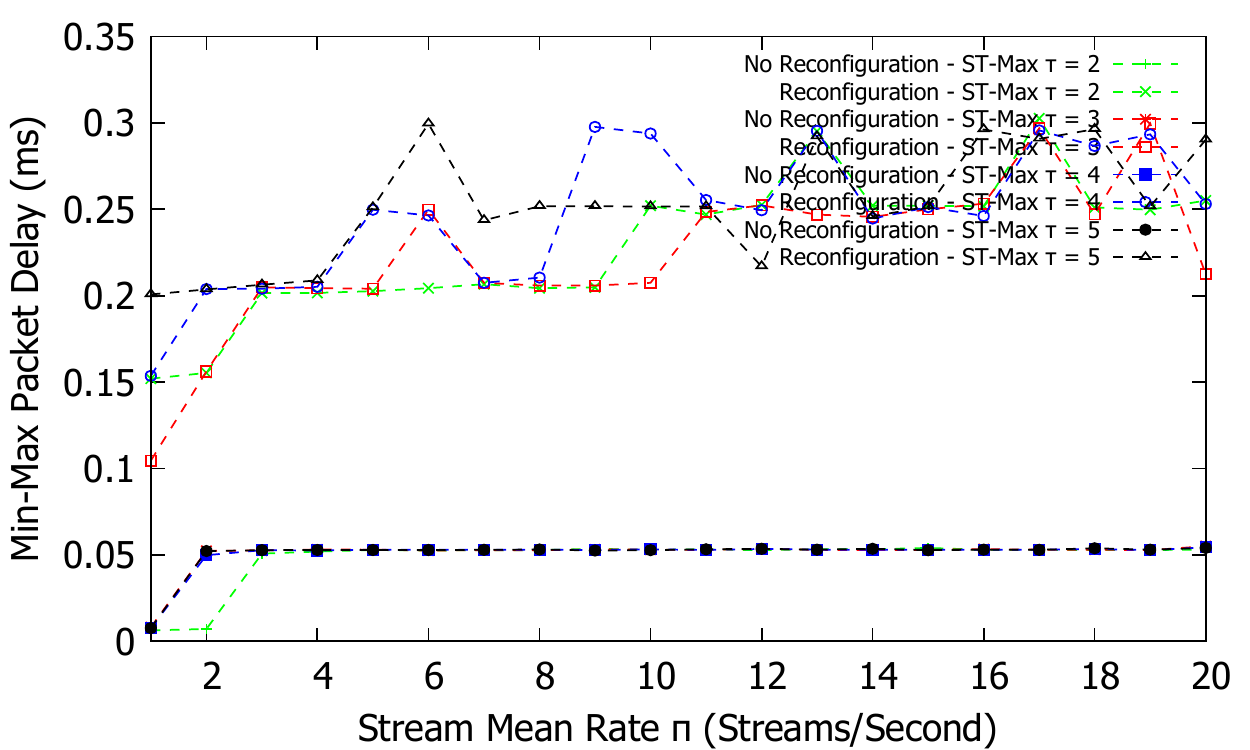}
		\caption{Low $\rho_{L}$}
	\end{subfigure}
	\begin{subfigure}{\columnwidth} \centering
		\includegraphics[width=3.3in]{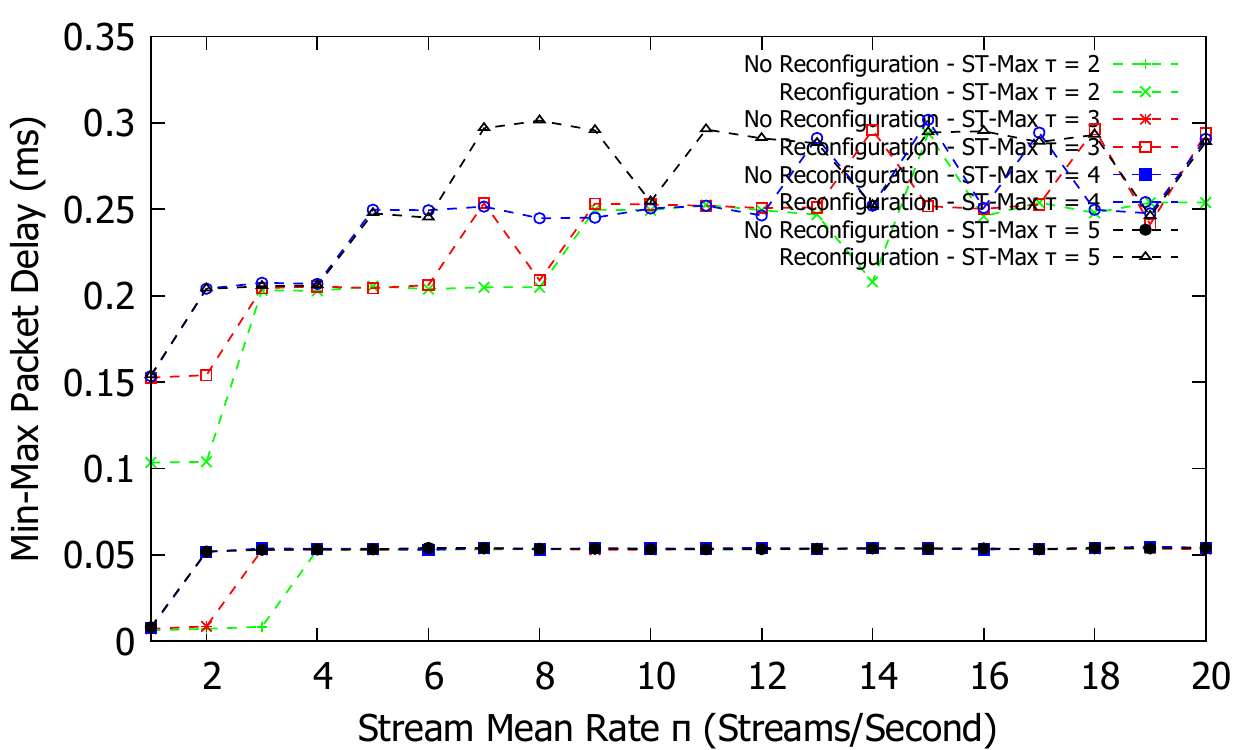}
		\caption{Mid $\rho_{L}$}
	\end{subfigure}
	\begin{subfigure}{\columnwidth} \centering
		\includegraphics[width=3.3in]{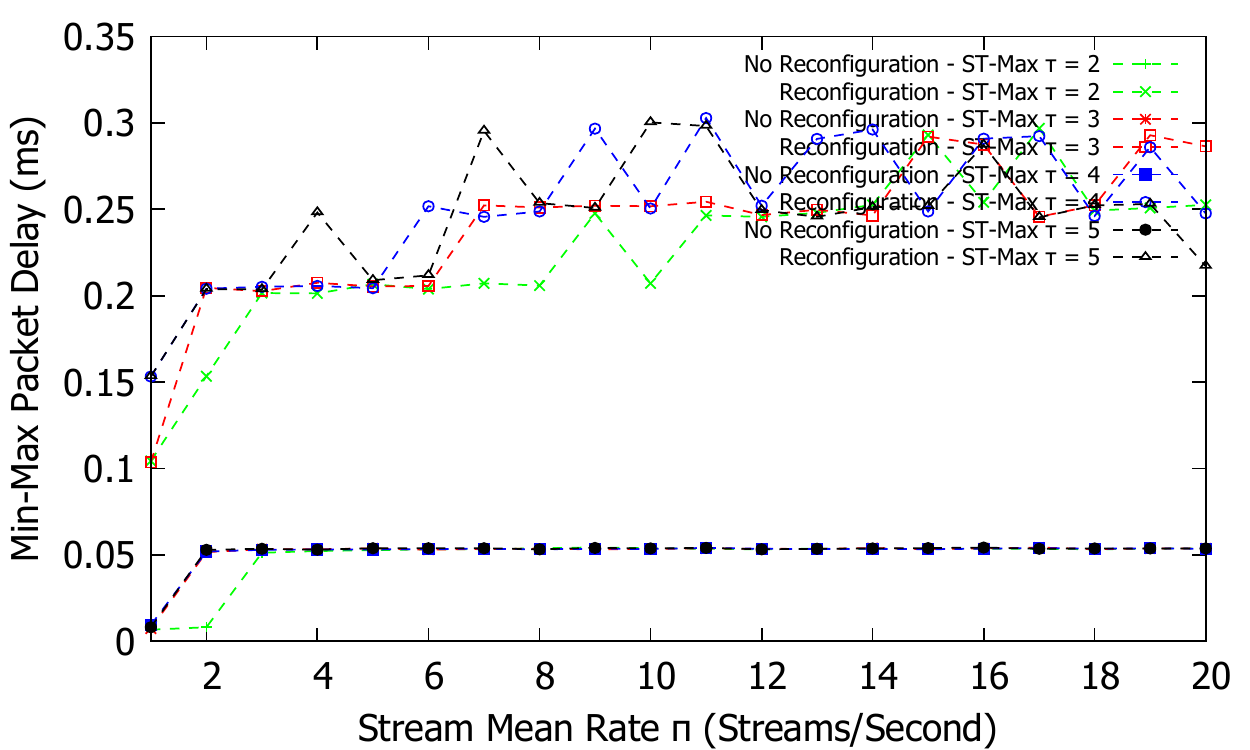}
		\caption{High $\rho_{L}$}
	\end{subfigure}
	\caption{Centralized Bi-directional Topology: Max delay as a results of TAS with centralized configuration (CNC) management entity.}
	\label{fig_maxDelay_bi}
\end{figure}

In terms of maximum delays, the bi-directional topology configuration produces higher than expected max delays due to the increasing ST stream acceptance without taking into account queue delays. For the bi-directional topology tests, the queue sizes were left the same (see Table.~\ref{table: simulation parameters}) and the CNC typically guarantees an upper bound delay per cycle per hop, i.e., each switch hop constitutes a max delay of one CT (or $50\mu$s in our tests). Fig.~\ref{fig_maxDelay_bi} shows the maximum delay evaluation for ST traffic for the bi-directional ring topology with CNC present. With the ``no reconfiguration'' approach, and since the ST slot size is kept at the initialized value ($20\%$ of CT or $10~\mu$s), the maximum delay is constant at $50~\mu$s. However, if reconfiguration is active at the CNC, then the maximum delay is bounded at $300~\mu$s due to the high admission rate and larger queuing delay per cycle. Note that the maximum hop traversal for any ST stream is kept at $3$~hops due to the bi-directional topology.

\begin{figure} [t!] \centering
	\begin{subfigure}{\columnwidth} \centering
		\includegraphics[width=3.3in]{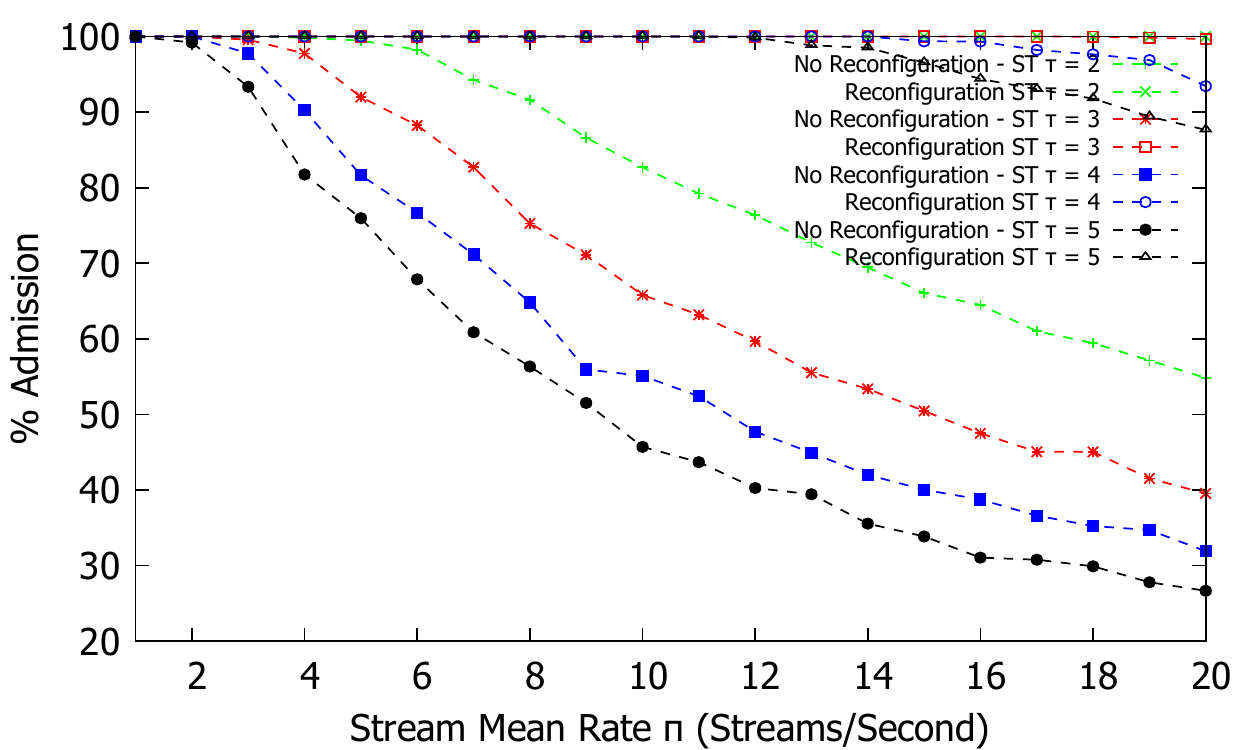}
		\caption{Low $\rho_{L}$}
	\end{subfigure}
	\begin{subfigure}{\columnwidth} \centering
		\includegraphics[width=3.3in]{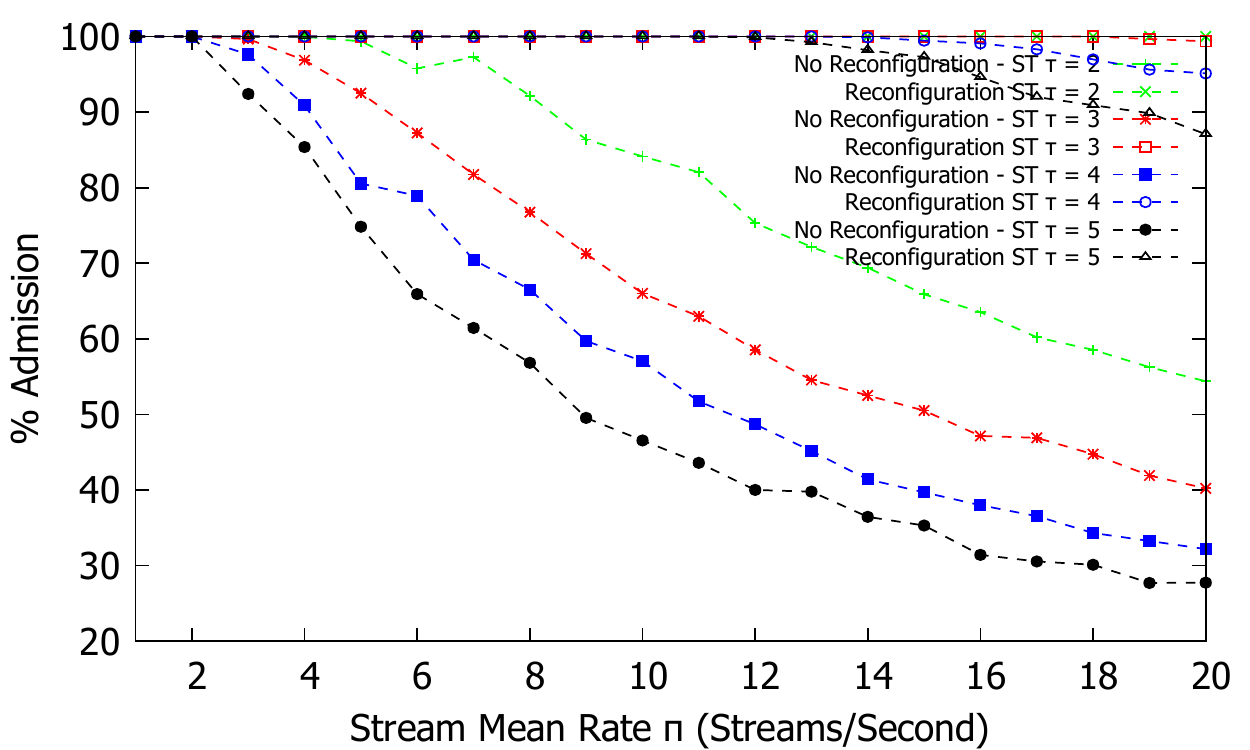}
		\caption{Mid $\rho_{L}$}
	\end{subfigure}
	\begin{subfigure}{\columnwidth} \centering
		\includegraphics[width=3.3in]{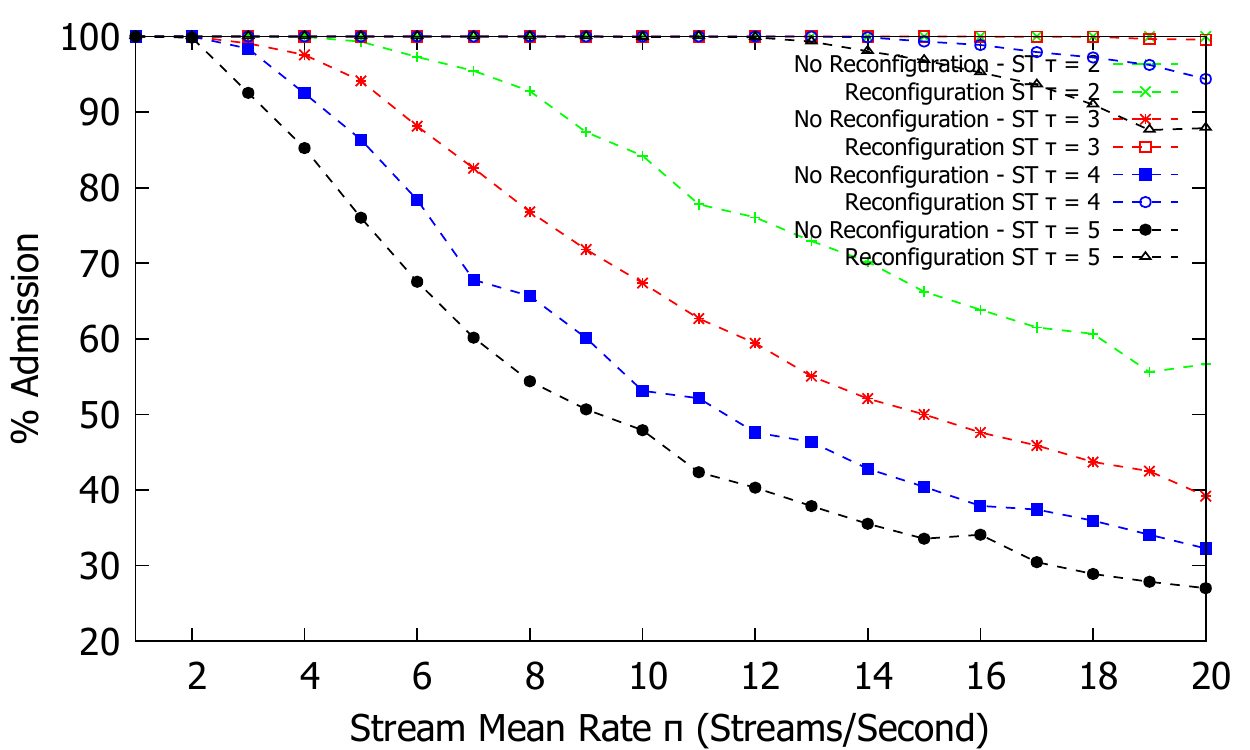}
		\caption{High $\rho_{L}$}
	\end{subfigure}
	\caption{Centralized Bi-directional Topology: Stream Admission as a results of TAS with centralized configuration (CNC) management entity.}
	\label{fig_admin_bi}
\end{figure}

While the max delay suffers due to not taking into account queuing delay in general, the admission rate is much higher (by about $40\%$) at high ST loads. Fig.~\ref{fig_admin_bi} shows the stream admission ratio results. With $\pi = 20$ and $\tau = 5$, the admission rate is close to $90\%$ for the bi-directional topology with reconfiguration active at the CNC. In contrast, the ``no reconfiguration'' approach improves slightly (by about $20\%$) compared to the unidirectional ring since the initialized gating ratio is too restrictive and can largely underutilize the link. Note that while we show the admission rates for different BE loads, $\rho_{L}$, the admission ratio does not change since TAS effectively segments the traffic at the egress switch/port, i.e., BE traffic does not block at ST traffic.

\begin{figure} [t!] \centering
	\begin{subfigure}{\columnwidth} \centering
		\includegraphics[width=3.3in]{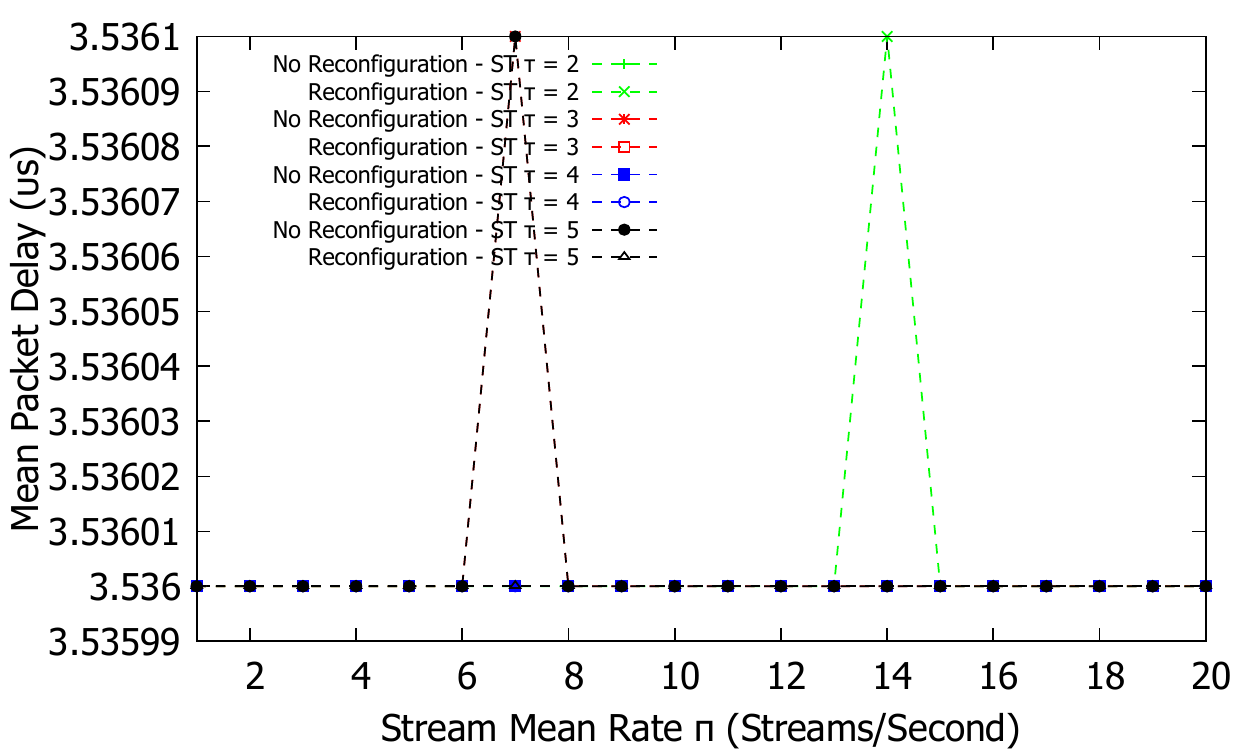}
		\caption{Low $\rho_{L}$}
	\end{subfigure}
	\begin{subfigure}{\columnwidth} \centering
		\includegraphics[width=3.3in]{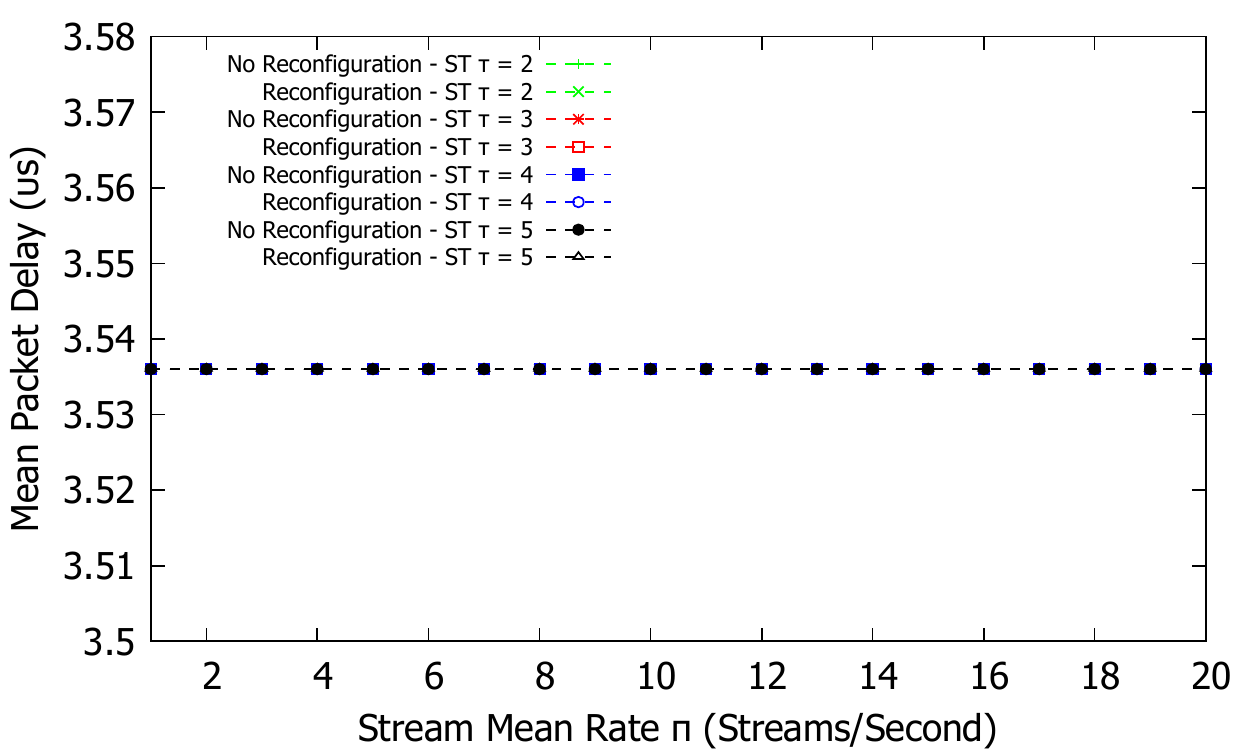}
		\caption{Mid $\rho_{L}$}
	\end{subfigure}
	\begin{subfigure}{\columnwidth} \centering
		\includegraphics[width=3.3in]{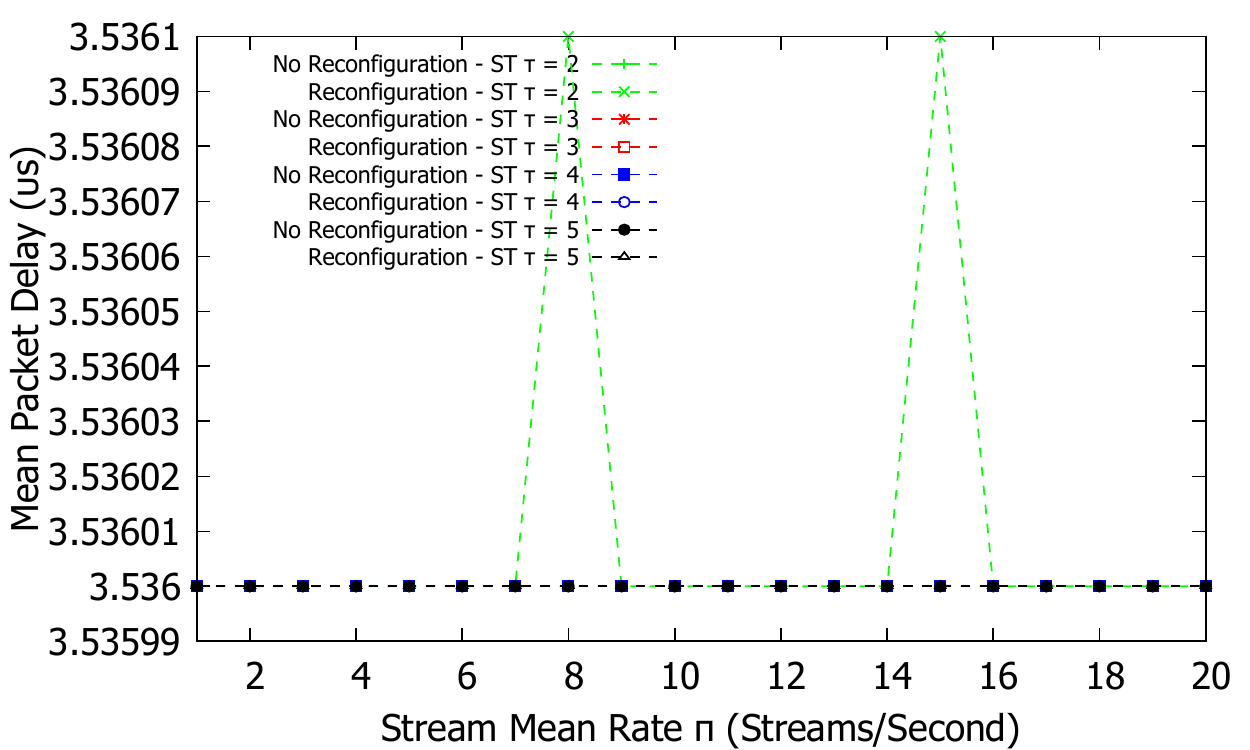}
		\caption{High $\rho_{L}$}
	\end{subfigure}
	\caption{Centralized Bi-directional Topology: Stream Signaling delay as a results of TAS with centralized configuration (CNC) management entity.}
	\label{fig_signalDelay_bi}
\end{figure}

Similar to the unidirectional ring, the bidirectional ring topology provides constant signaling delay due to the CNC out-of band signaling channels. Fig.~\ref{fig_signalDelay_bi} shows the signaling delay for ST stream registration. Note that the average signaling delay is lower than in the unidirectional ring since the edge links, specifically the source to switch link, is larger than in the unidirectional ring hence the transmission delay is shorter.

\begin{figure} [t!] \centering
	\begin{subfigure}{\columnwidth} \centering
		\includegraphics[width=3.3in]{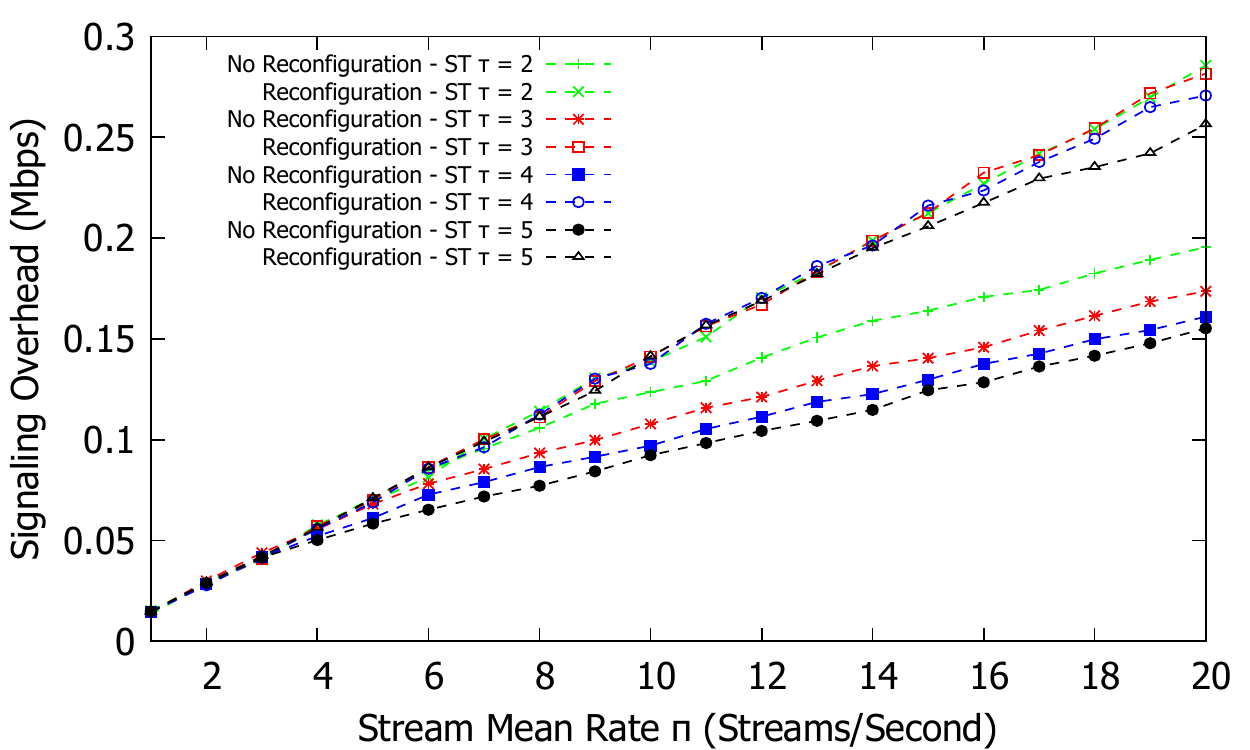}
		\caption{Low $\rho_{L}$}
	\end{subfigure}
	\begin{subfigure}{\columnwidth} \centering
		\includegraphics[width=3.3in]{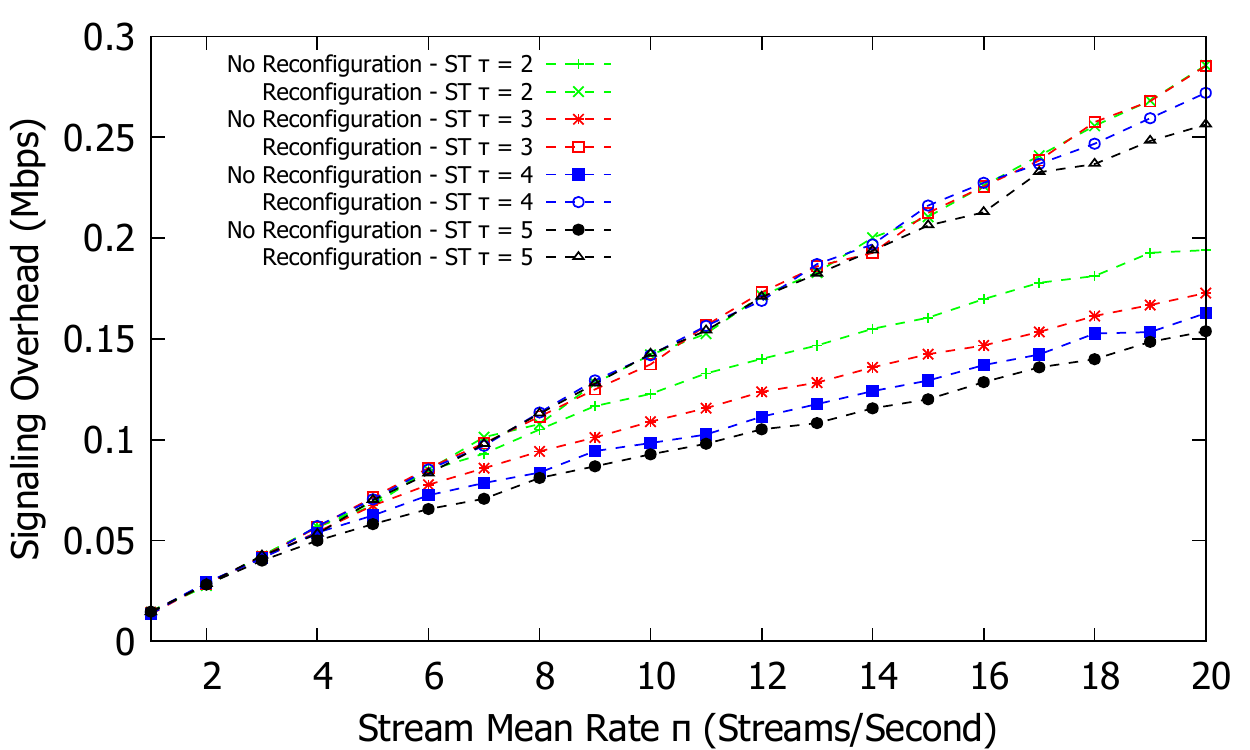}
		\caption{Mid $\rho_{L}$}
	\end{subfigure}
	\begin{subfigure}{\columnwidth} \centering
		\includegraphics[width=3.3in]{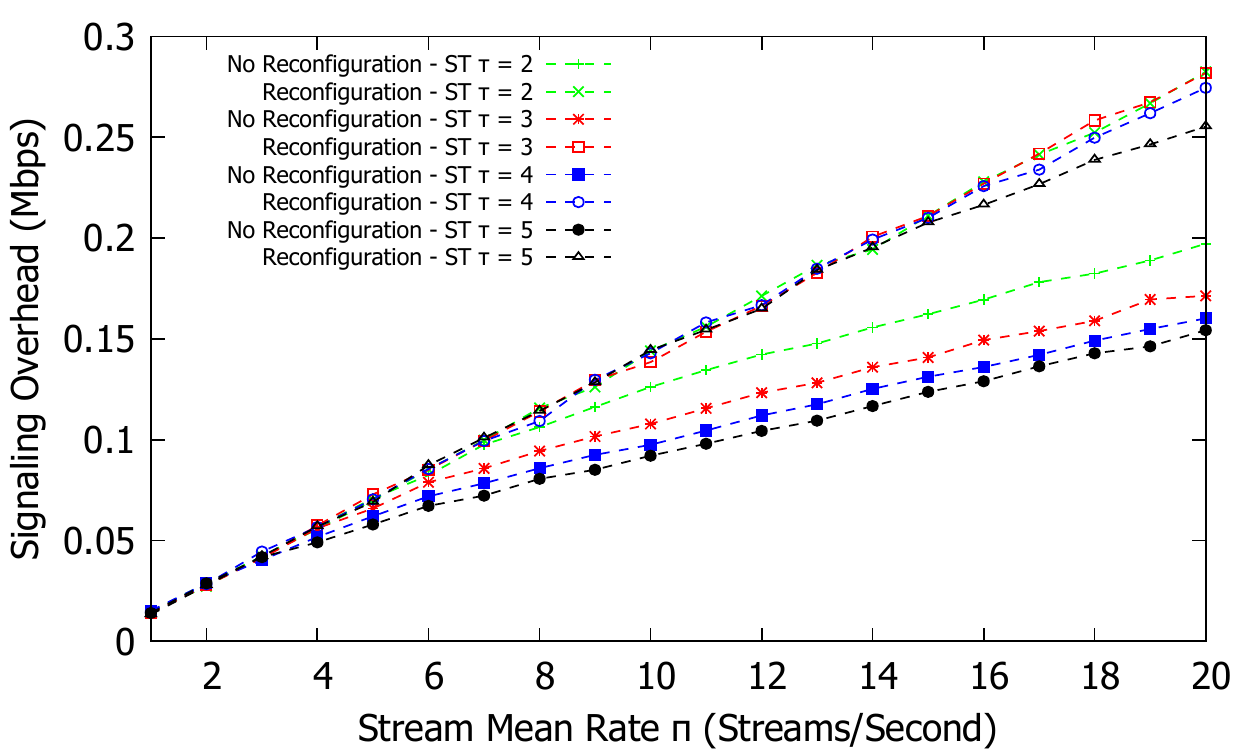}
		\caption{High $\rho_{L}$}
	\end{subfigure}
	\caption{Centralized Bi-directional Topology: Stream average signaling Overhead as a results of TAS with centralized configuration (CNC) management entity.}
	\label{fig_signalOverhead_bi}
\end{figure}

Since the topology is effectively the same (albeit having another port to the switch), the signaling overhead in general is very similar to the unidirectional topology. Fig.~\ref{fig_signalOverhead_bi} shows the signaling performance overhead. Note that while the hop traversal is reduced (since the stream can take one of two paths to the destination governed by hop traversal), the number of sent and received CDT frames are the same in general. Clearly, similar to the unidirectional topology, the reconfiguration approach generates more CDT traffic. Note that rejections in general are less costly in terms of sent and received frames in the network. Therefore, the higher the admission rate, the more overhead is observed in the control plane, though based on Fig.~\ref{fig_signalOverhead_bi}, the overall overhead is not even close to $1$~Mbps and therefore is much lower compared to the channel capacity.

\begin{figure} [t!] \centering
	\begin{subfigure}{\columnwidth} \centering
		\includegraphics[width=3.3in]{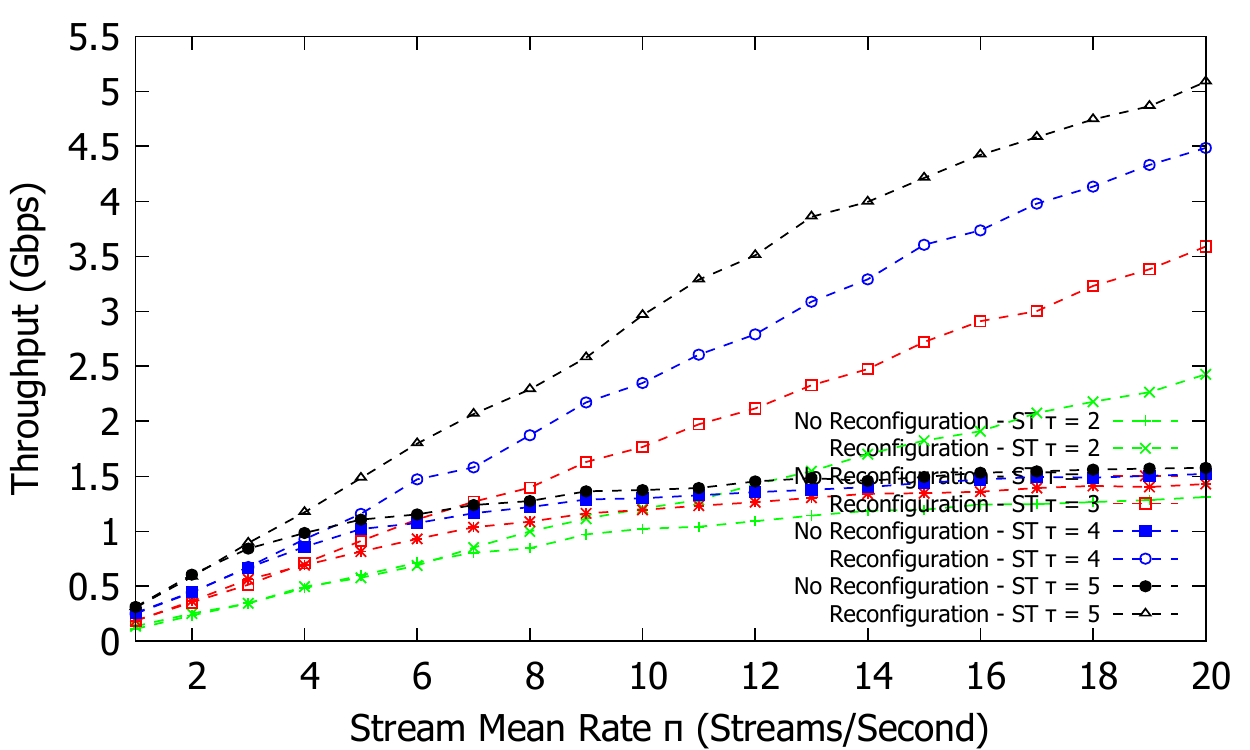}
		\caption{Low $\rho_{L}$}
	\end{subfigure}
	\begin{subfigure}{\columnwidth} \centering
		\includegraphics[width=3.3in]{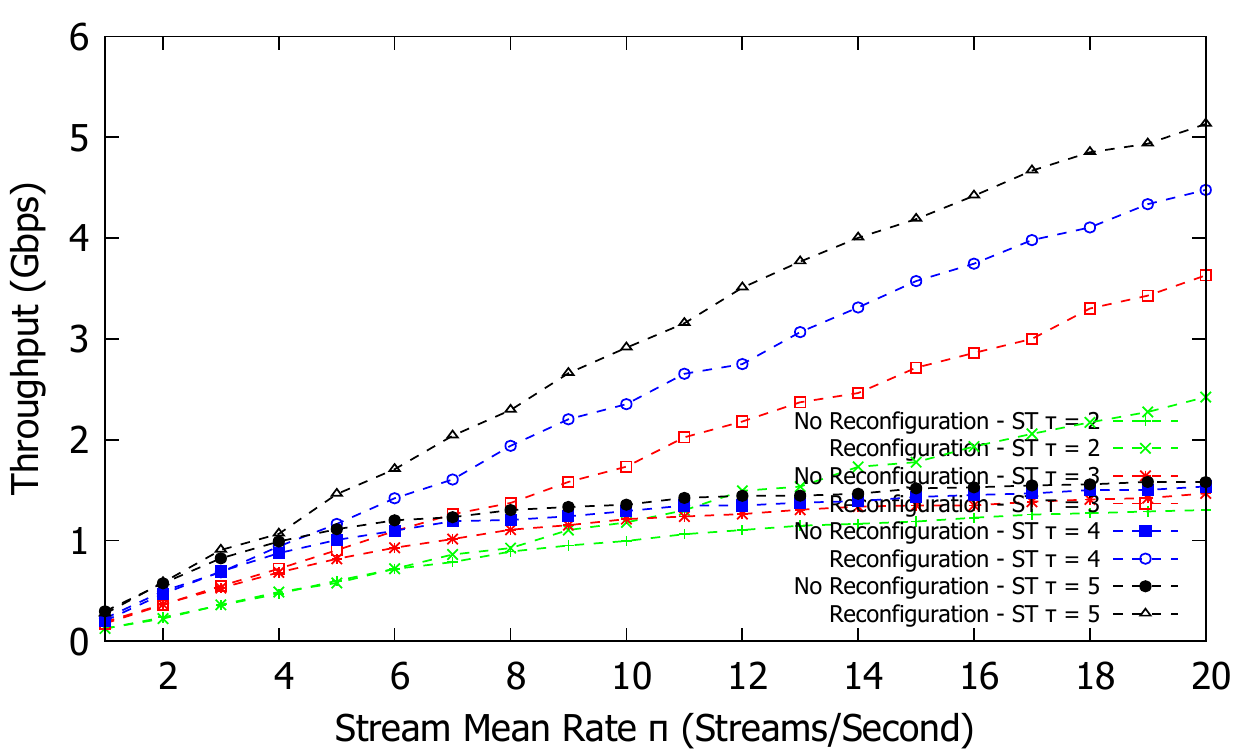}
		\caption{Mid $\rho_{L}$}
	\end{subfigure}
	\begin{subfigure}{\columnwidth} \centering
		\includegraphics[width=3.3in]{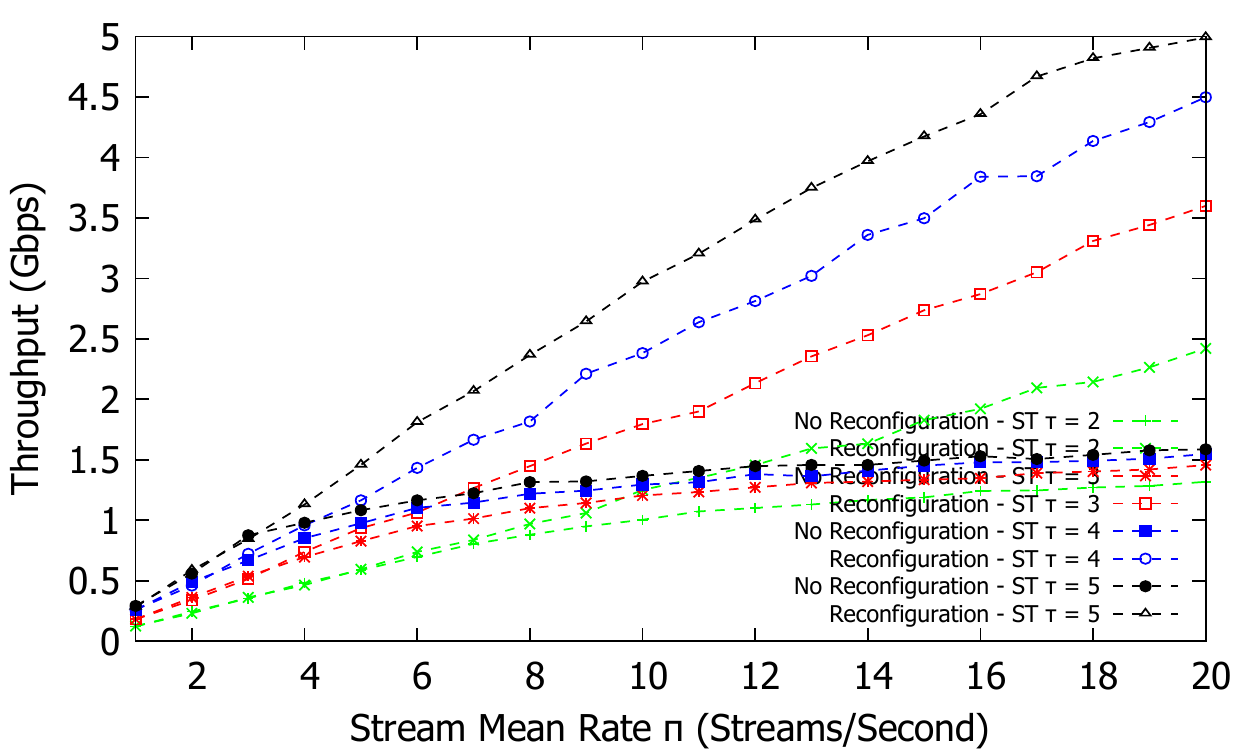}
		\caption{High $\rho_{L}$}
	\end{subfigure}
	\caption{Centralized Bi-directional Topology: ST Total average throughput measured at the sink as a results of TAS with centralized configuration (CNC) management entity.}
	\label{fig_avgTput_ST_bi}
\end{figure}

\begin{figure} [t!] \centering
	\begin{subfigure}{\columnwidth} \centering
		\includegraphics[width=3.3in]{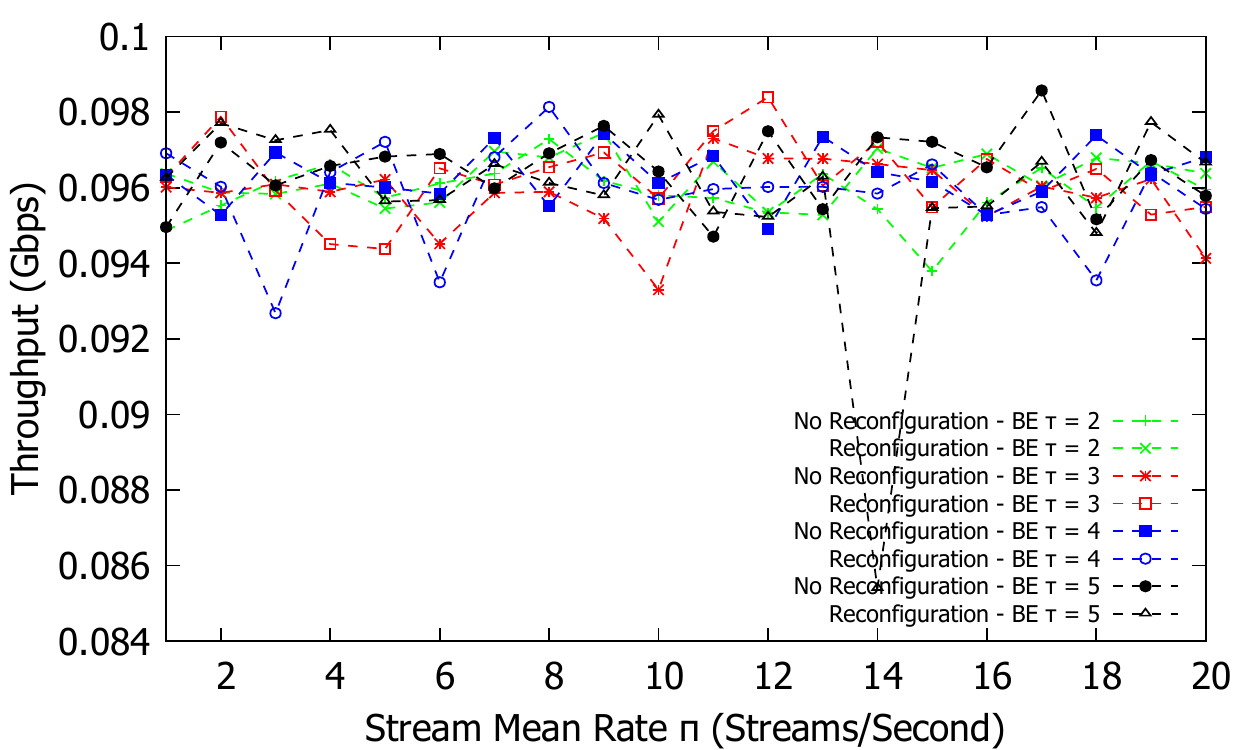}
		\caption{Low $\rho_{L}$}
	\end{subfigure}
	\begin{subfigure}{\columnwidth} \centering
		\includegraphics[width=3.3in]{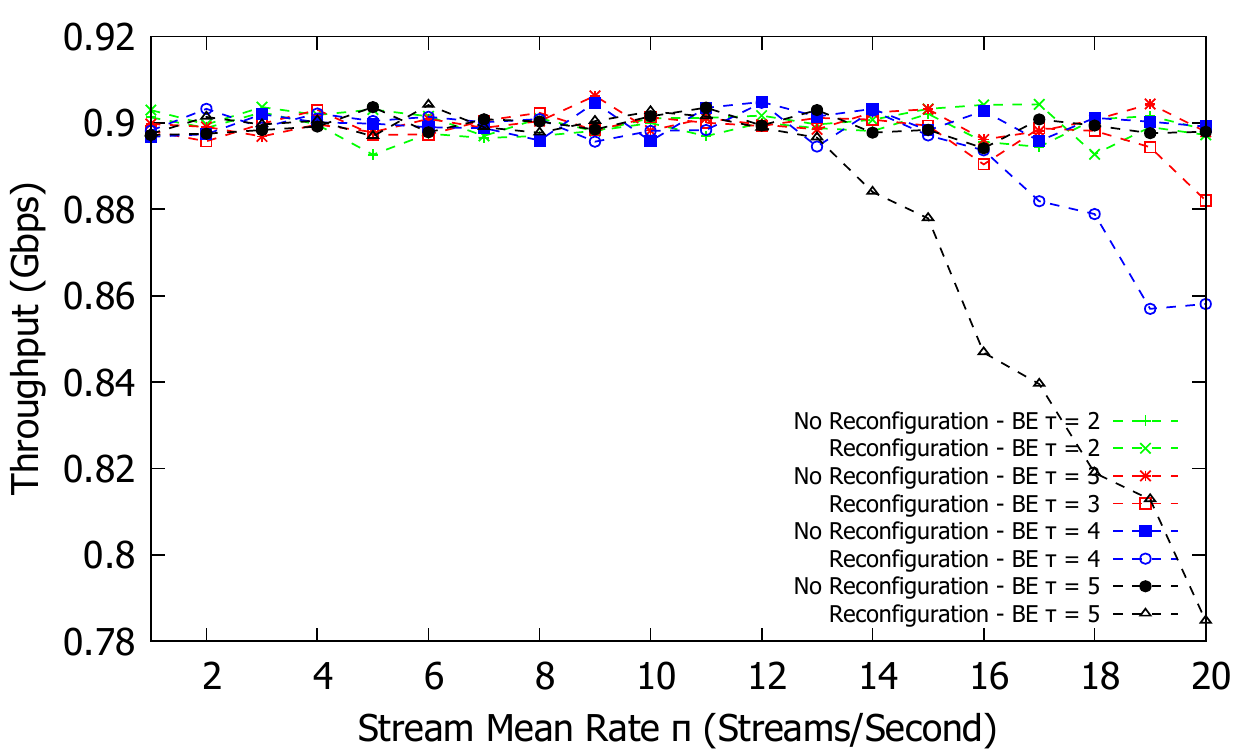}
		\caption{Mid $\rho_{L}$}
	\end{subfigure}
	\begin{subfigure}{\columnwidth} \centering
		\includegraphics[width=3.3in]{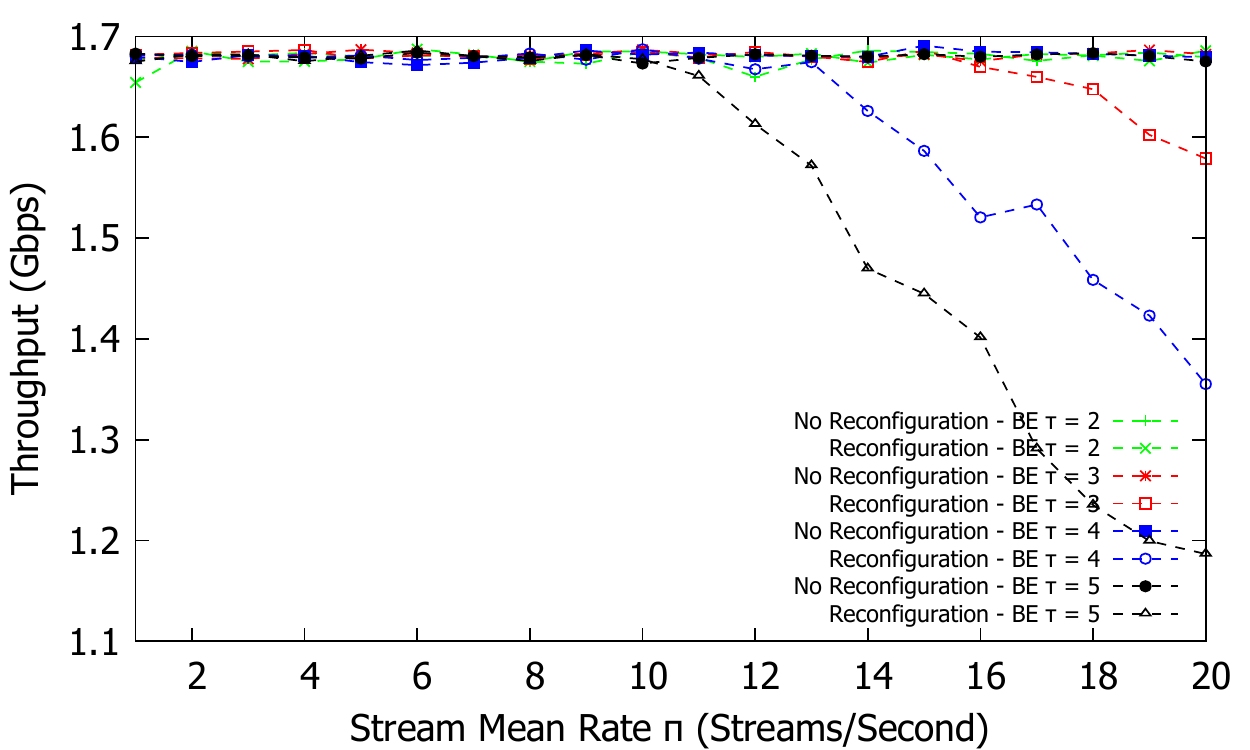}
		\caption{High $\rho_{L}$}
	\end{subfigure}
	\caption{Centralized Bi-directional Topology: BE Total average throughput measured at the sink as a results of TAS with centralized configuration (CNC) management entity.}
	\label{fig_avgTput_BE_bi}
\end{figure}

In terms of overall throughput, Fig.~\ref{fig_avgTput_ST_bi} and Fig.~\ref{fig_avgTput_BE_bi} shows the average throughput measured at the sink for both ST and BE traffic for the bi-directional ring topology. Compared to the unidirectional ring, the throughput for the bi-directional ring is much higher (sink maximum capacity is around $12$~Gbps from all switch to sink channels).

\begin{figure} [t!] \centering
	\begin{subfigure}{\columnwidth} \centering
		\includegraphics[width=3.3in]{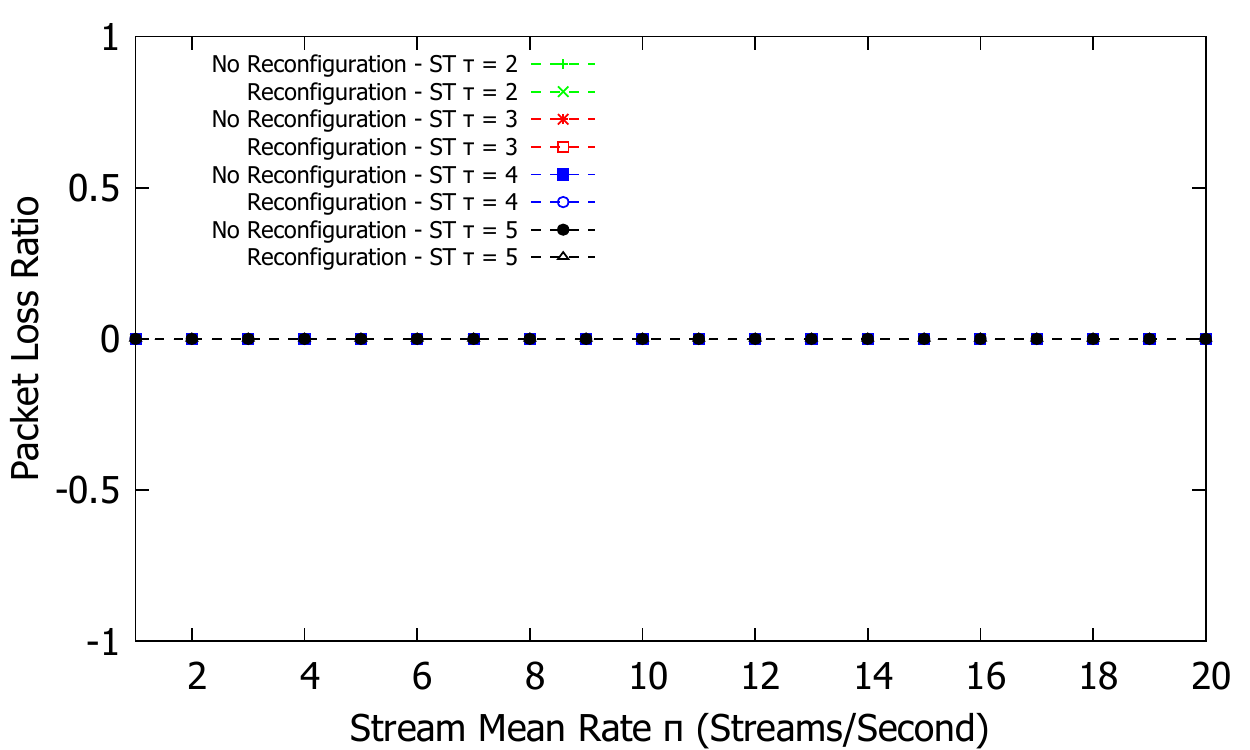}
		\caption{Low $\rho_{L}$}
	\end{subfigure}
	\begin{subfigure}{\columnwidth} \centering
		\includegraphics[width=3.3in]{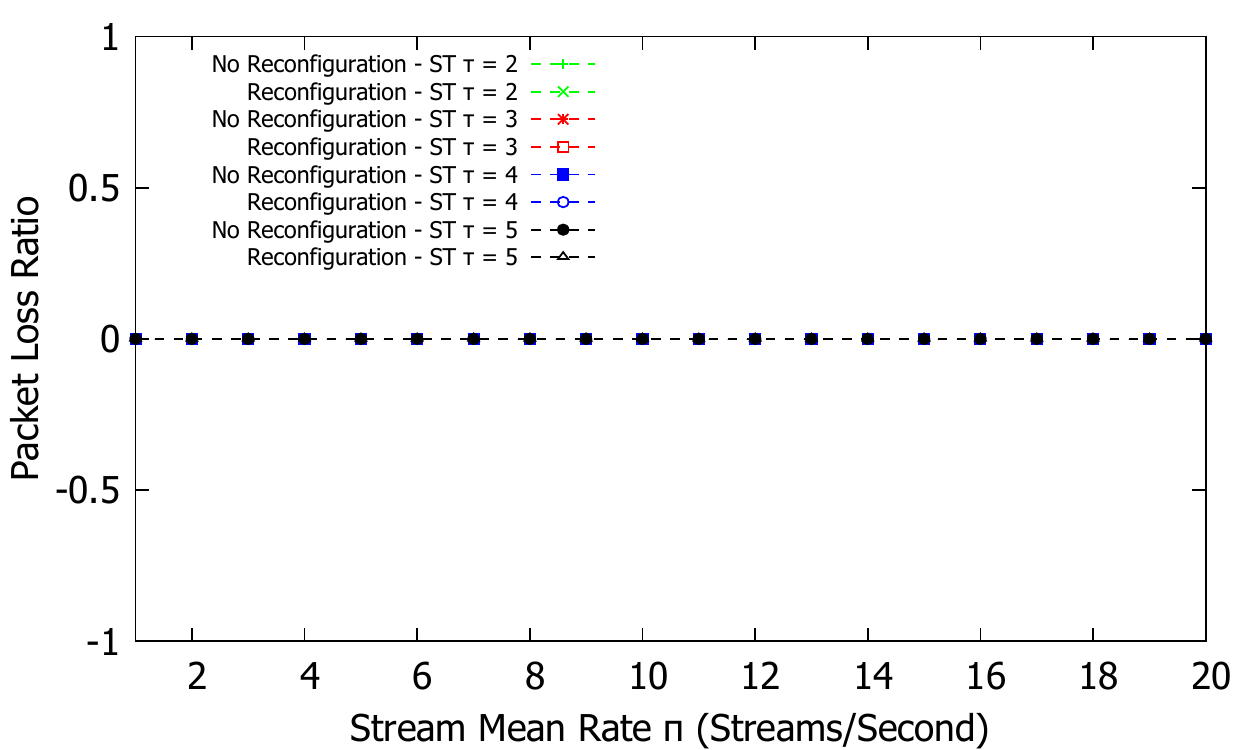}
		\caption{Mid $\rho_{L}$}
	\end{subfigure}
	\begin{subfigure}{\columnwidth} \centering
		\includegraphics[width=3.3in]{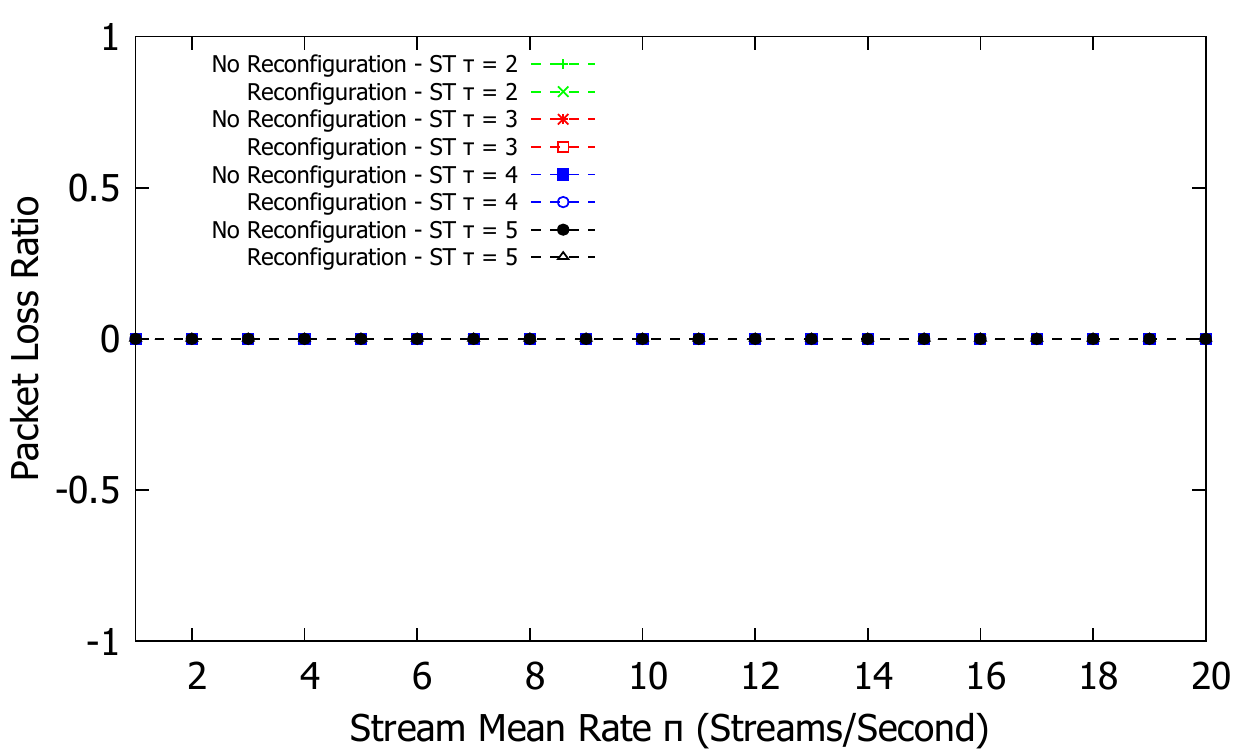}
		\caption{High $\rho_{L}$}
	\end{subfigure}
	\caption{Centralized Bi-directional Topology: ST Frame loss ratio as a results of TAS with centralized configuration (CNC) management entity.}
	\label{fig_lossProb_ST_bi}
\end{figure}

\begin{figure} [t!] \centering
	\begin{subfigure}{\columnwidth} \centering
		\includegraphics[width=3.3in]{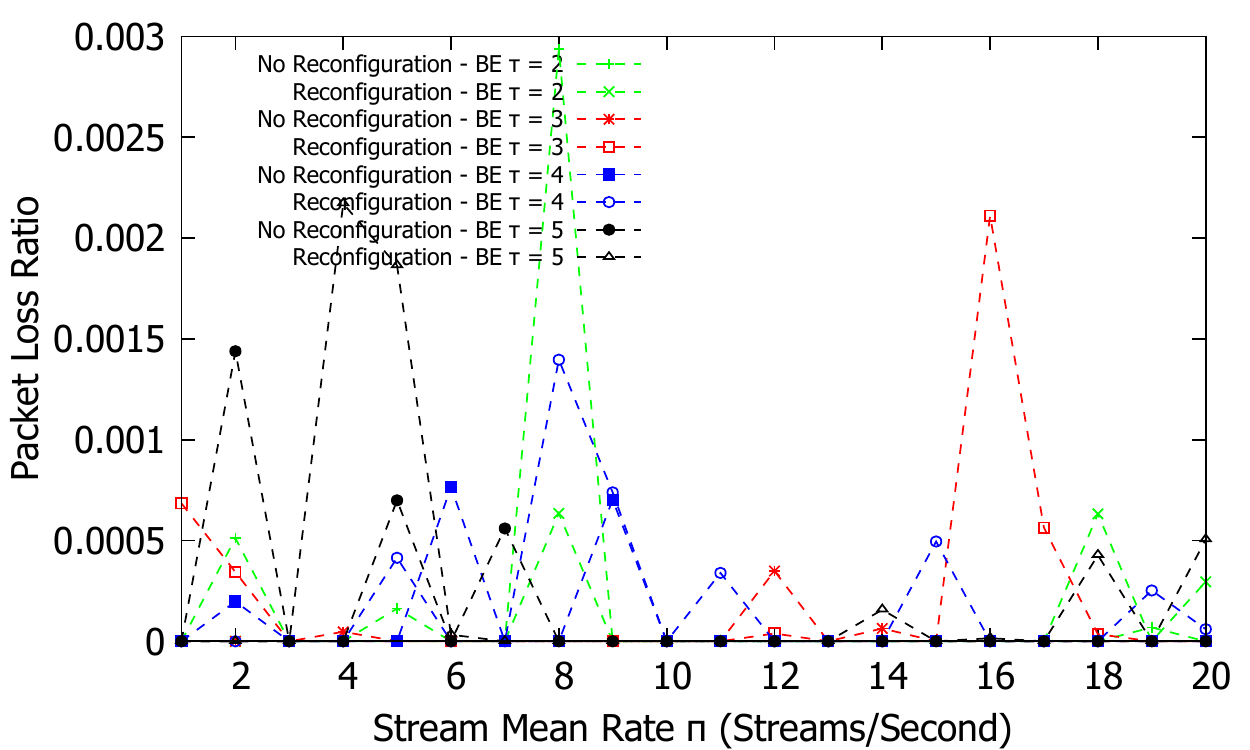}
		\caption{Low $\rho_{L}$}
	\end{subfigure}
	\begin{subfigure}{\columnwidth} \centering
		\includegraphics[width=3.3in]{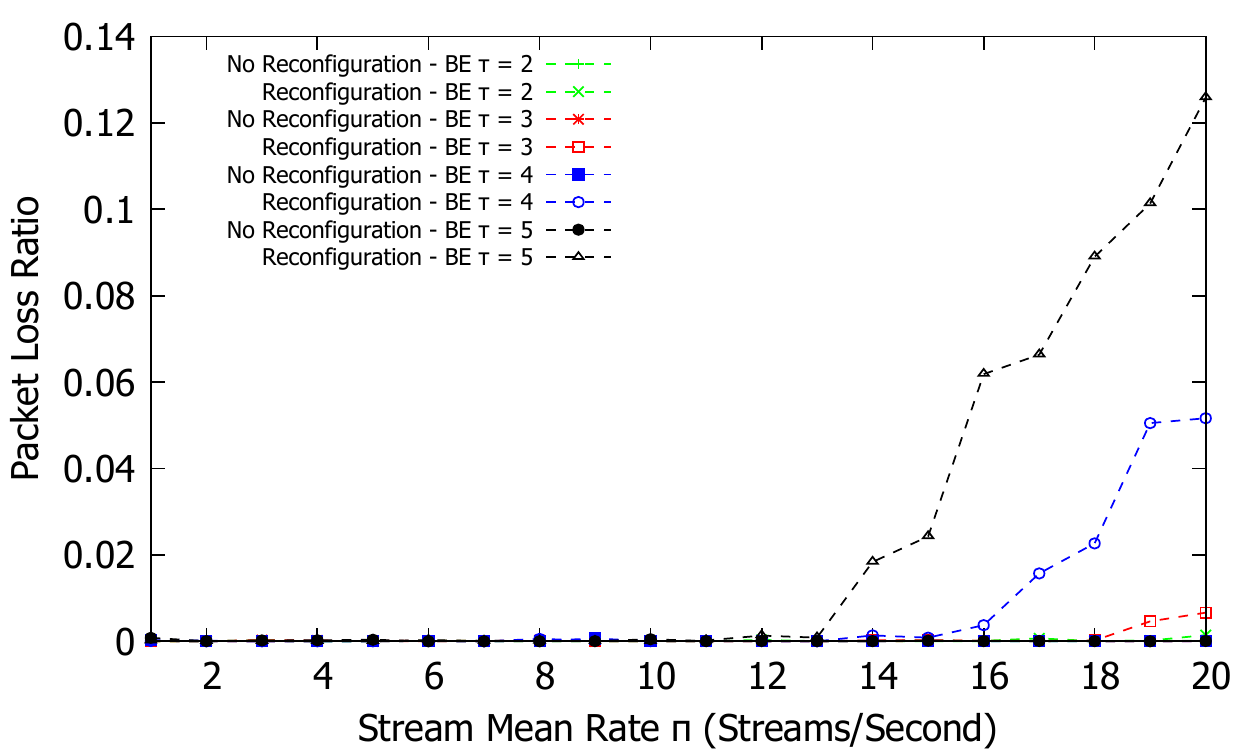}
		\caption{Mid $\rho_{L}$}
	\end{subfigure}
	\begin{subfigure}{\columnwidth} \centering
		\includegraphics[width=3.3in]{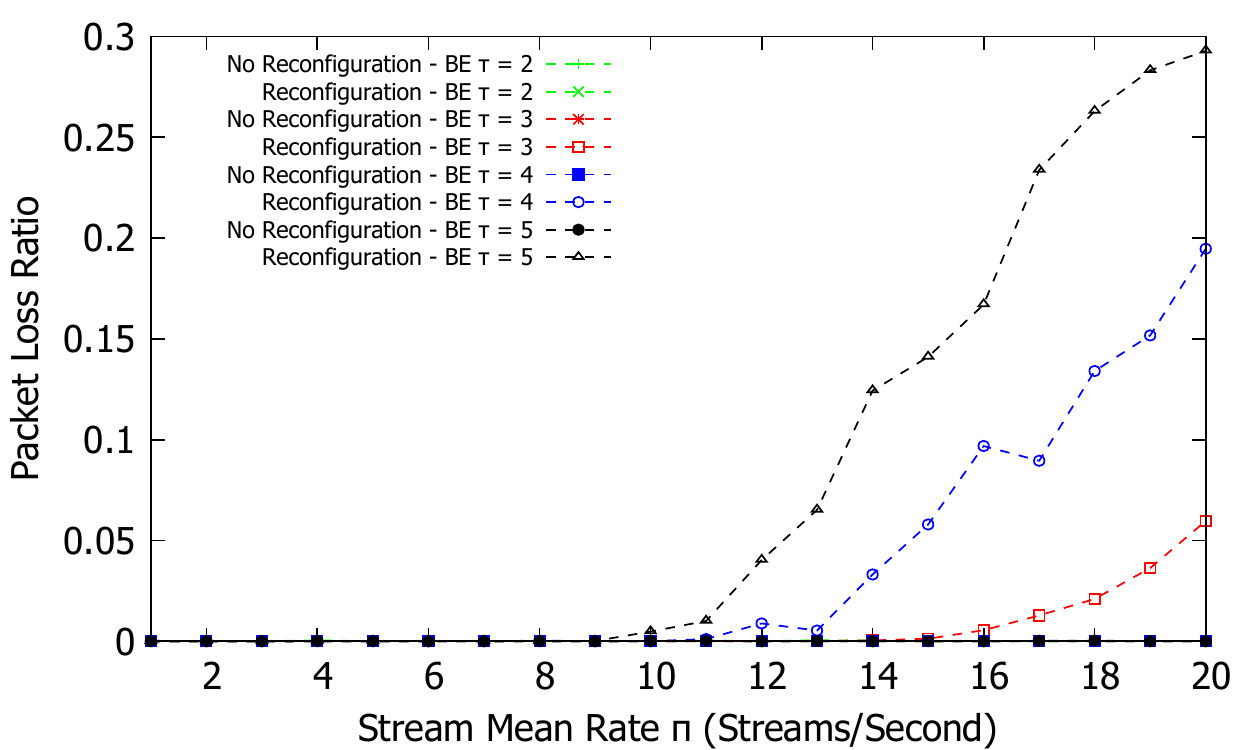}
		\caption{High $\rho_{L}$}
	\end{subfigure}
	\caption{Centralized Bi-directional Topology: BE Frame loss ratio as a results of TAS with centralized configuration (CNC) management entity.}
	\label{fig_lossProb_BE_bi}
\end{figure}

Similar to the unidirectional ring topology, bi-directional topology achieves zero loss to ST streams while significantly improving the loss rate for BE traffic. Fig.~\ref{fig_lossProb_ST_bi} and Fig.~\ref{fig_lossProb_BE_bi} shows the packet loss ratio for ST and BE traffic in the network for the bi-directional ring topology. Maximum BE loss as high BE traffic intensity, $\rho_{L} = 2.0$, is around $30\%$ which is a significant reduction from the unidirectional topology (of around $90\%$).

In contrast to the unidirectional topology, the bi-directional topology with reconfiguration active at the CNC achieves improved QoS metrics and admission rates. However, without modeling the queue characteristics and guaranteeing queuing delay, it is difficult to guarantee maximum delays. Modeling the ST queue at each egress port in conjunction with TAS configuration properties grants the possibility to produce deterministic forwarding plane at the data plane for ST streams. This part is left for the next iteration of this report.

\subsection{Decentralized Model Evaluation}	\label{eval:dec:sec}
In evaluating the proposed solution describes in more detail in section~\ref{tas:reconfig:prot}, we consider both periodic and sporadic sources for ST and BE traffic as discussed in section~\ref{tas:sec:traffic} respectively. We evaluated network with TAS shaper on the industrial control loop unidirectional and bi-directional topology and results are collected for tests following the simulation parameters shown in Table.\ref{table: simulation parameters}.

\subsubsection{Unidirectional Ring Topology}

\begin{figure} [t!] \centering
	\includegraphics[width=3.3in]{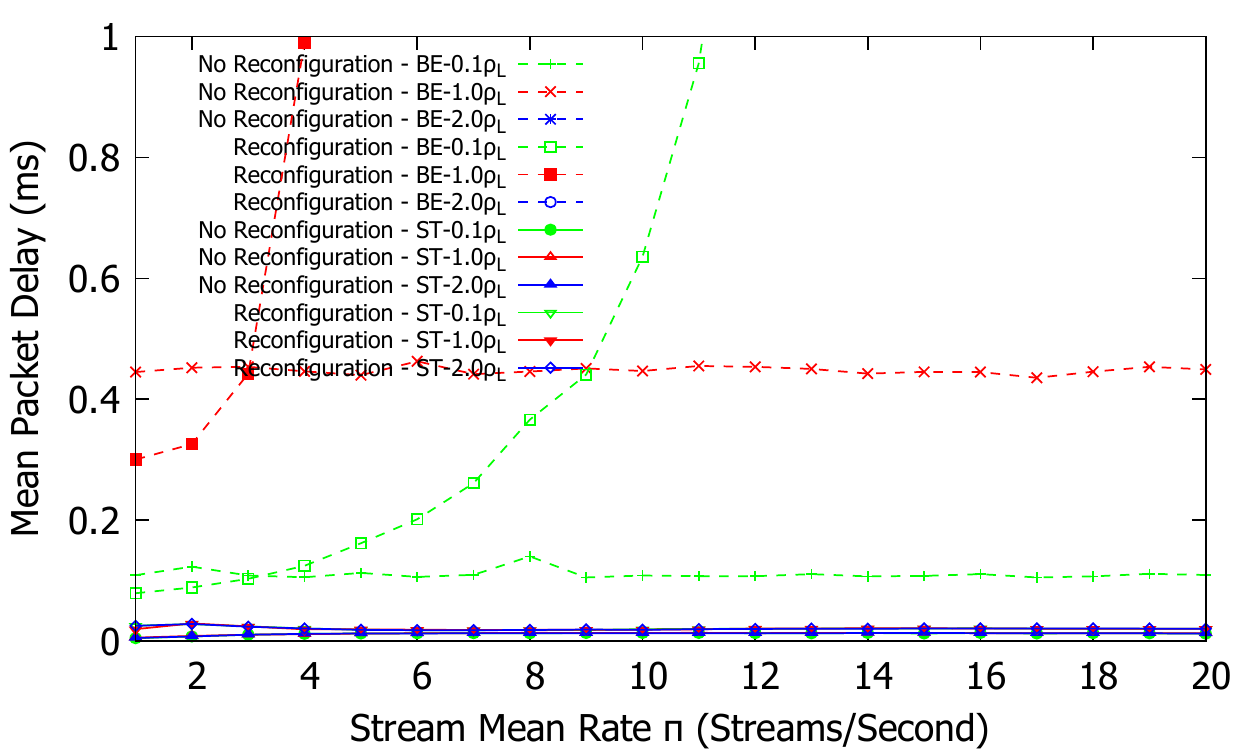}
	\caption{Decentralized Unidirectional Topology: Mean end-to-end delay for ST and BE traffic for $\tau = 2$ under different loads~$\rho_{L}$, mean traffic rates~$\pi$, and initialized gating ratio of $20\%$.}
	\label{fig_delay_2_dec}
\end{figure}

\begin{figure} [t!] \centering
	\includegraphics[width=3.3in]{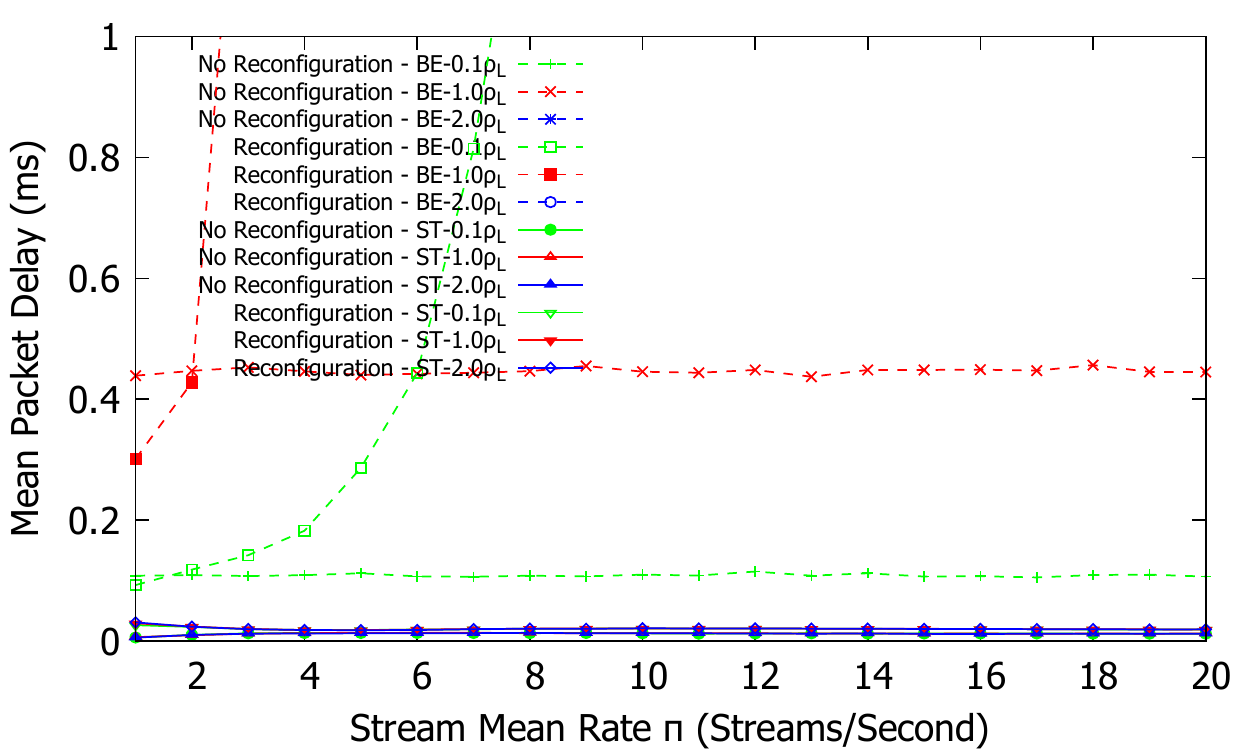}
	\caption{Decentralized Unidirectional Topology: Mean end-to-end delay for ST and BE traffic for $\tau = 3$ under different loads~$\rho_{L}$, mean traffic rates~$\pi$, and initialized gating ratio of $20\%$.}
	\label{fig_delay_3_dec}
\end{figure}

\begin{figure} [t!] \centering
	\includegraphics[width=3.3in]{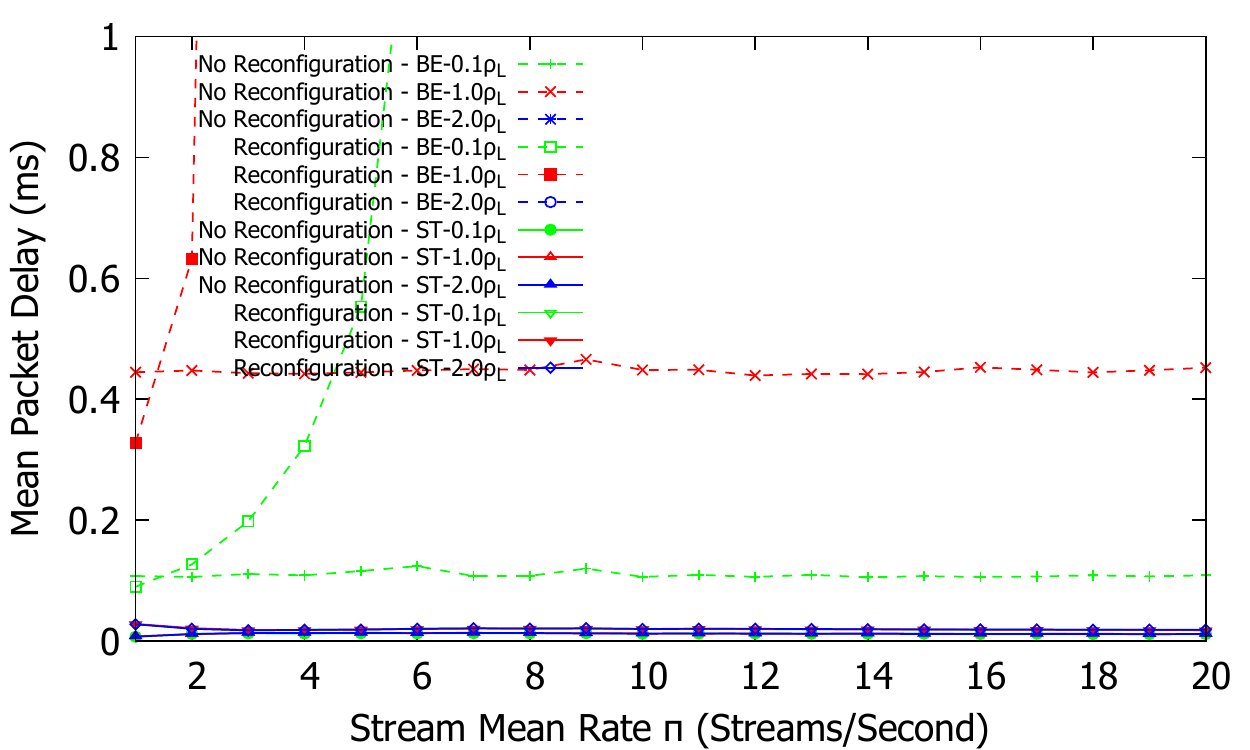}
	\caption{Decentralized Unidirectional Topology: Mean end-to-end delay for ST and BE traffic for $\tau = 4$ under different loads~$\rho_{L}$, mean traffic rates~$\pi$, and initialized gating ratio of $20\%$.}
	\label{fig_delay_4_dec}
\end{figure}

\begin{figure} [t!] \centering
	\includegraphics[width=3.3in]{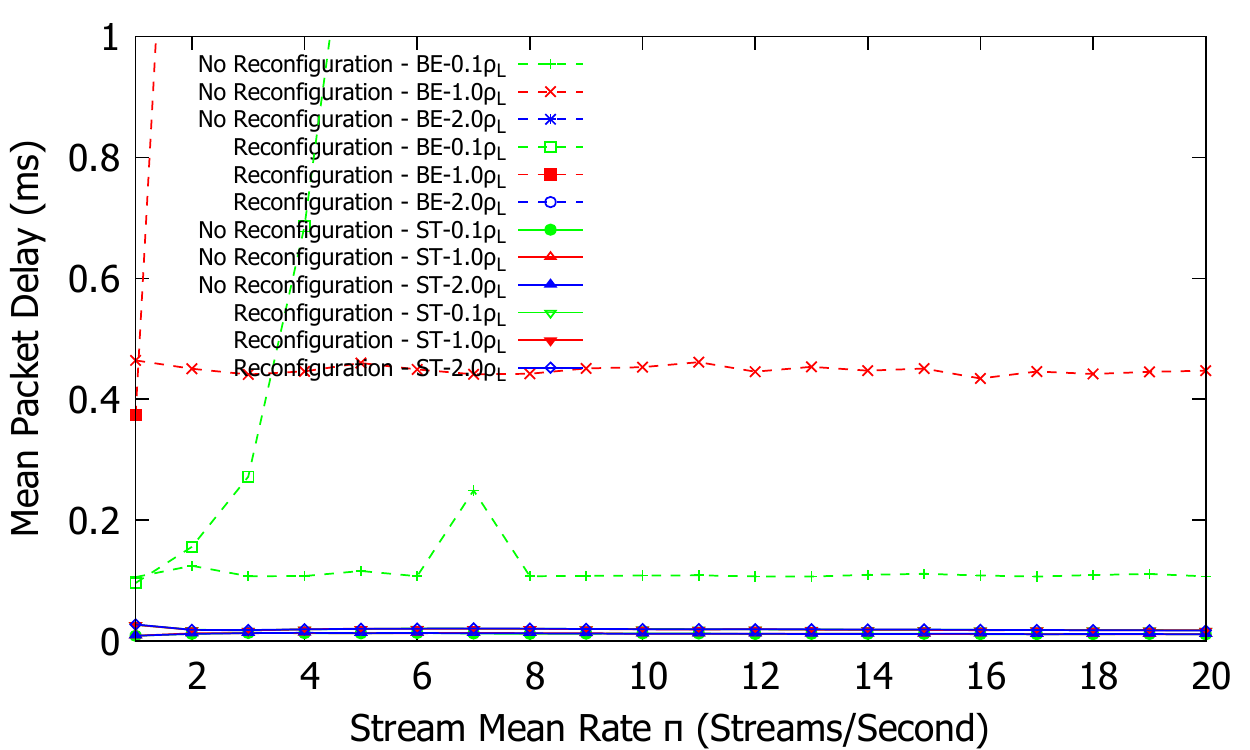}
	\caption{Decentralized Unidirectional Topology: Mean end-to-end delay for ST and BE traffic for $\tau = 5$ under different loads~$\rho_{L}$, mean traffic rates~$\pi$, and initialized gating ratio of $20\%$.}
	\label{fig_delay_5_dec}
\end{figure}

Decentralized model essentially transfers some of the CNC functions (e.g., TAS reconfiguration and resource reservation modules) from the centralized model down to the data plane switch's egress TAS enabled ports. The main difference between the centralized and decentralized models is the signaling performance which is now in-band and can affect data traffic. Fig.~\ref{fig_delay_2_dec} - \ref{fig_delay_5_dec} shows the average mean delay evaluation for both ST and BE traffic. While the CDT traffic is in-band, the average delay is about the same as the centralized topology average delay in Fig.~\ref{fig_delay_2} - \ref{fig_delay_5}. Typically, the ST stream's average delay is minimal to near constant for both the reconfiguration and ``no reconfiguration'' approaches. For BE, the ``no reconfiguration'' approach produces constant average delay for each BE $\rho_{L}$ traffic intensity. When used with reconfiguration, the mean delay start to democratically increase since BE traffic time slots are begin reserved for ST streams.

\begin{figure} [t!] \centering
	\begin{subfigure}{\columnwidth} \centering
		\includegraphics[width=3.3in]{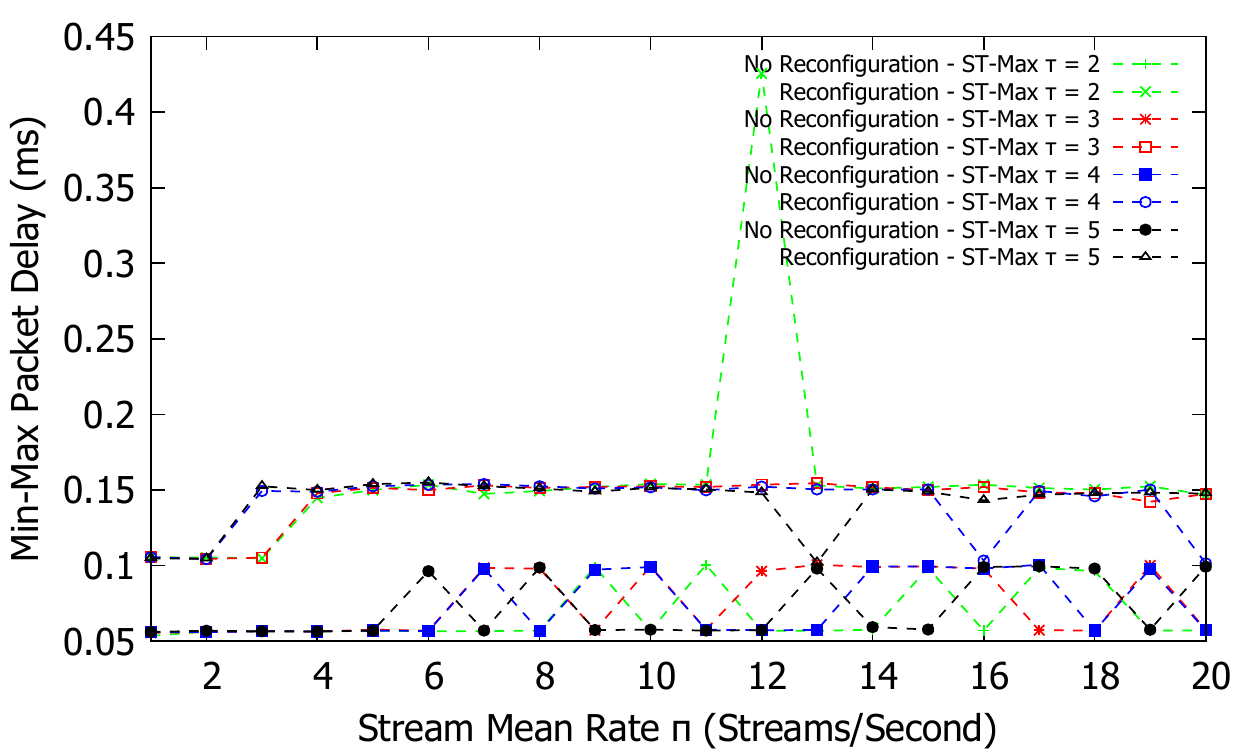}
		\caption{Low $\rho_{L}$}
	\end{subfigure}
	\begin{subfigure}{\columnwidth} \centering
		\includegraphics[width=3.3in]{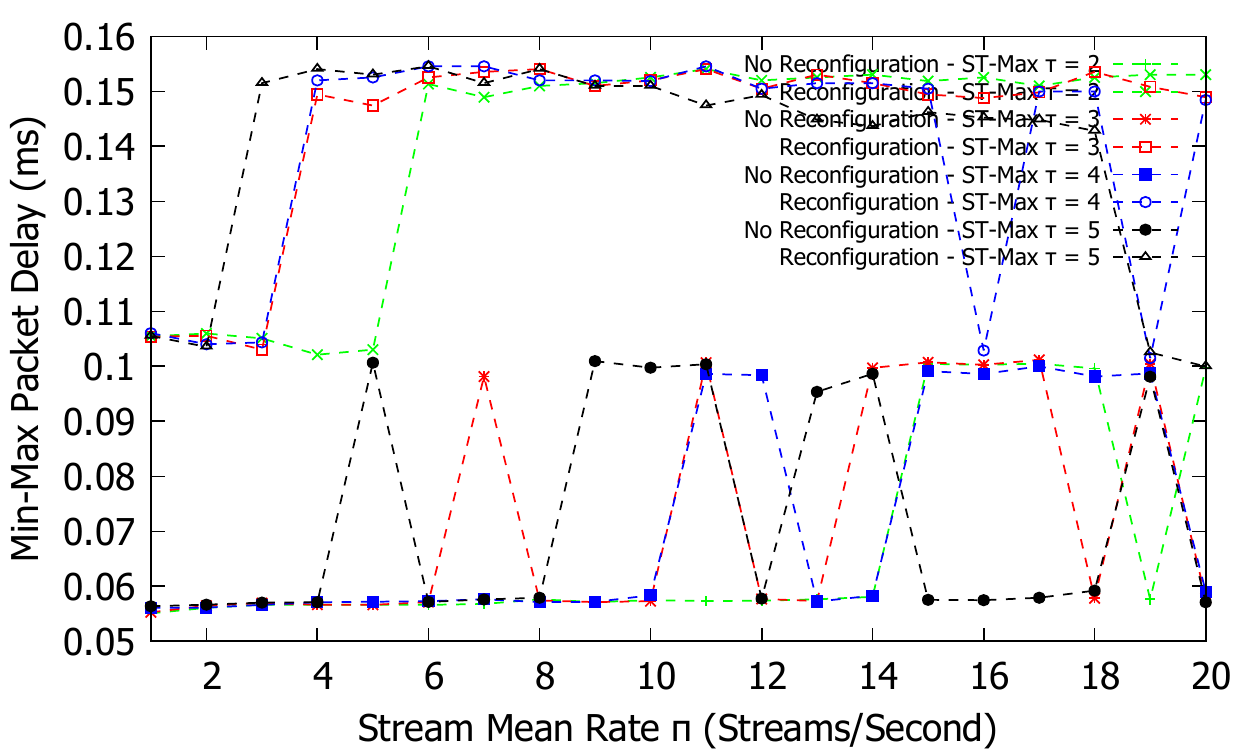}
		\caption{Mid $\rho_{L}$}
	\end{subfigure}
	\begin{subfigure}{\columnwidth} \centering
		\includegraphics[width=3.3in]{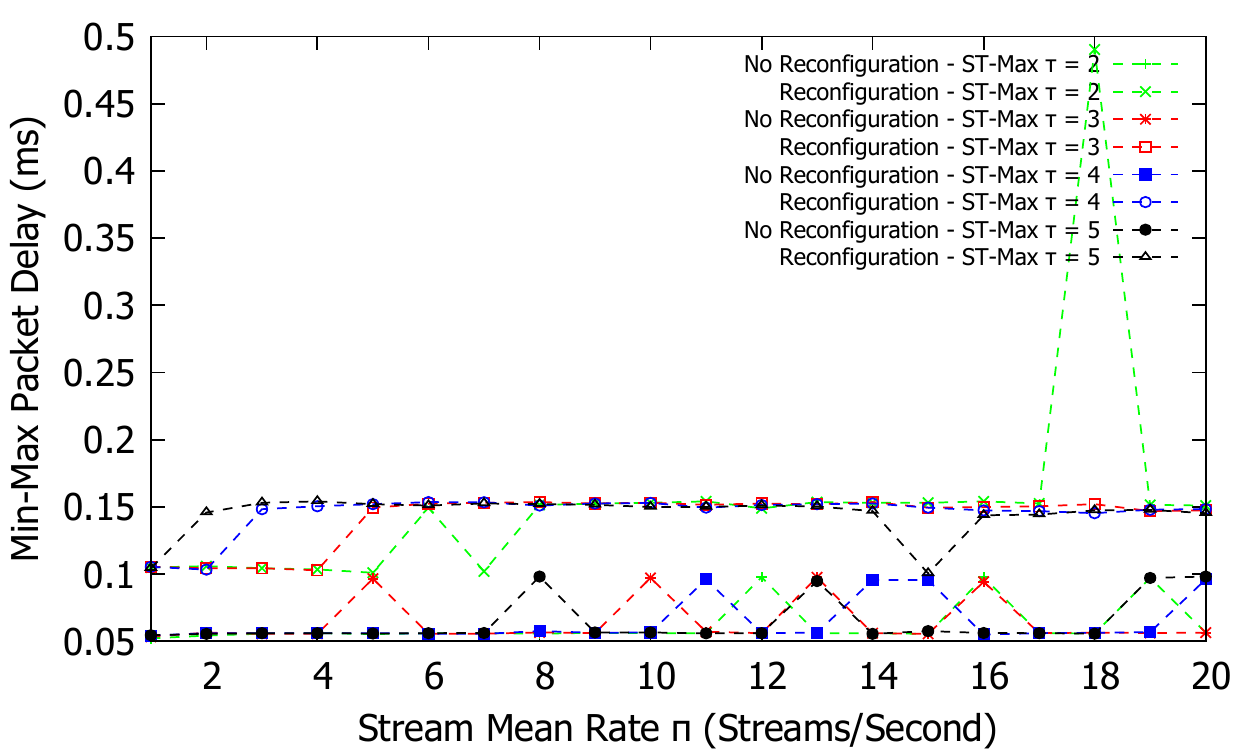}
		\caption{High $\rho_{L}$}
	\end{subfigure}
	\caption{Decentralized Unidirectional Topology: Max delay as a results of TAS.}
	\label{fig_maxDelay_dec}
\end{figure}

In terms of maximum delays under the unidirectional topology using the decentralized model, Fig.~\ref{fig_maxDelay_dec} shows the maximum delay evaluation for ST traffic. In contrast to the average delay, the maximum delay does get affected by in-band CDT traffic. In the decentralized model, the CDT traffic is given the highest priority above both ST and BE traffic. Therefore, the maximum delays can reach about $50~\mu$s more than the ST threshold of $100~\mu$s.

\begin{figure} [t!] \centering
	\begin{subfigure}{\columnwidth} \centering
		\includegraphics[width=3.3in]{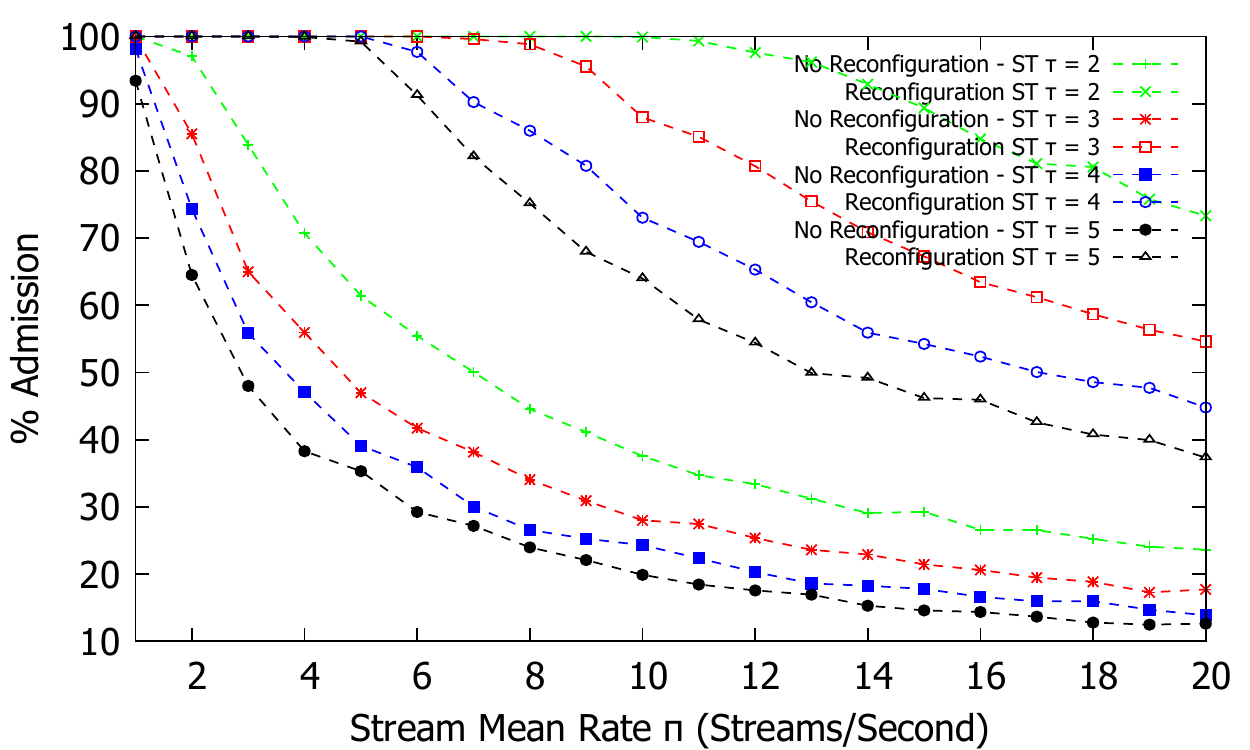}
		\caption{Low $\rho_{L}$}
	\end{subfigure}
	\begin{subfigure}{\columnwidth} \centering
		\includegraphics[width=3.3in]{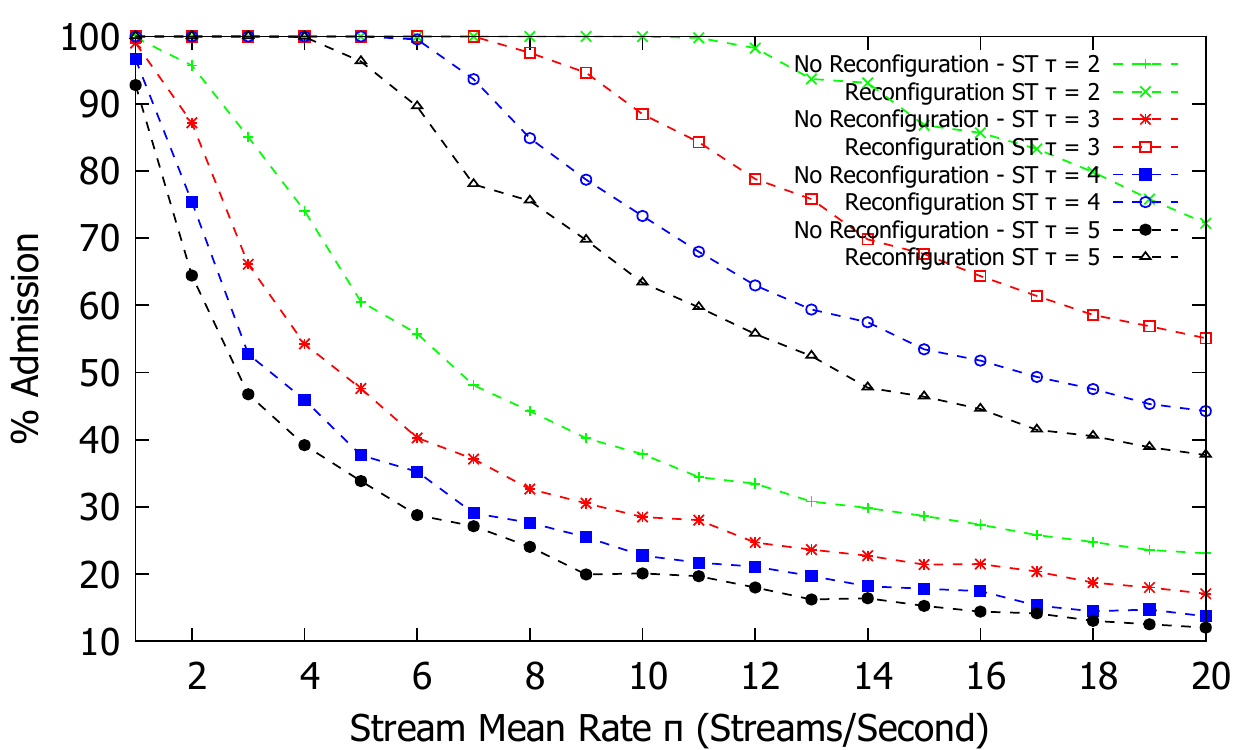}
		\caption{Mid $\rho_{L}$}
	\end{subfigure}
	\begin{subfigure}{\columnwidth} \centering
		\includegraphics[width=3.3in]{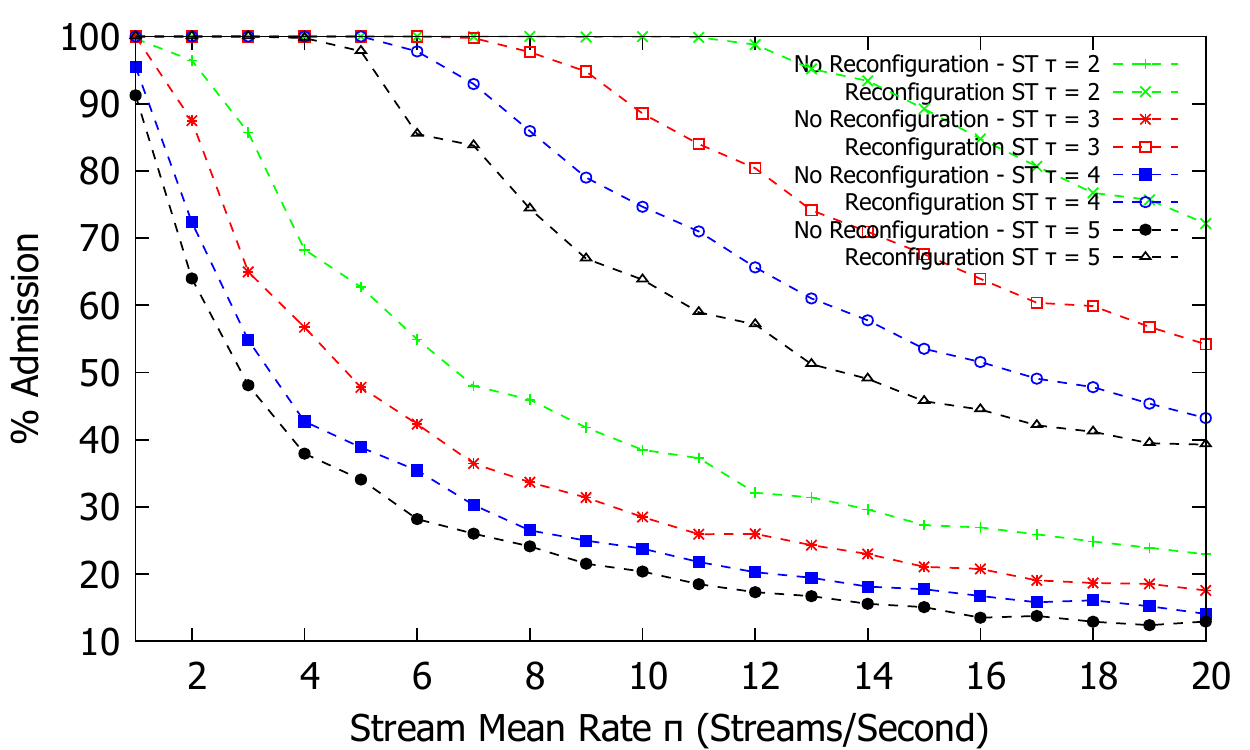}
		\caption{High $\rho_{L}$}
	\end{subfigure}
	\caption{Decentralized Unidirectional Topology: Stream admission as a results of TAS.}
	\label{fig_admin_dec}
\end{figure}

The admission rate is very similar to the centralized model since the network parameters used are identical. Fig.~\ref{fig_admin_dec} shows the stream admission ratio results.

\begin{figure} [t!] \centering
	\begin{subfigure}{\columnwidth} \centering
		\includegraphics[width=3.3in]{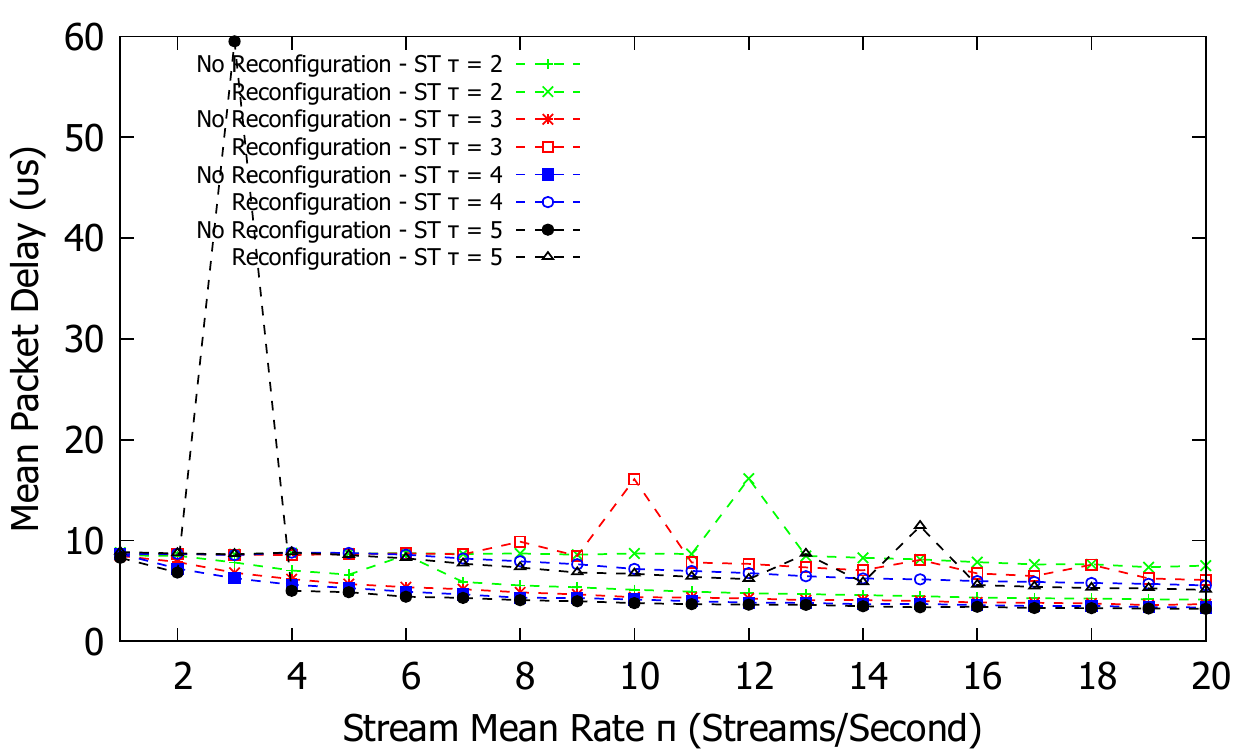}
		\caption{Low $\rho_{L}$}
	\end{subfigure}
	\begin{subfigure}{\columnwidth} \centering
		\includegraphics[width=3.3in]{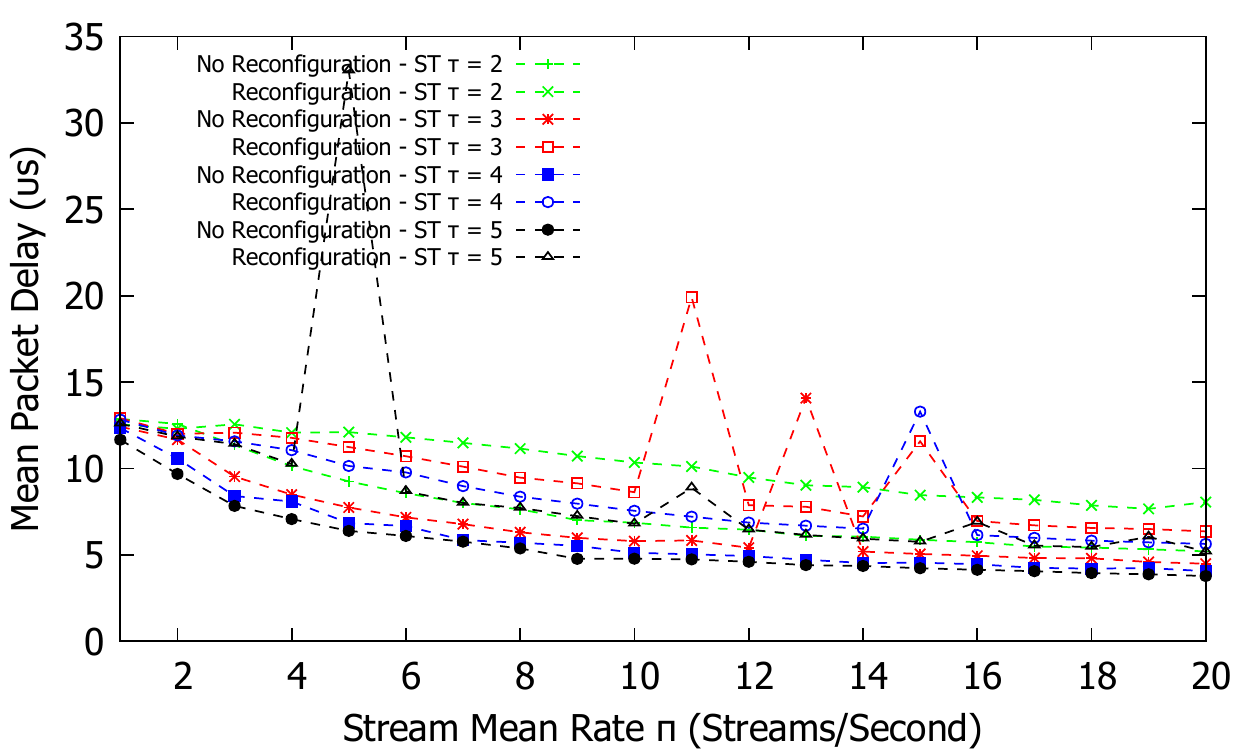}
		\caption{Mid $\rho_{L}$}
	\end{subfigure}
	\begin{subfigure}{\columnwidth} \centering
		\includegraphics[width=3.3in]{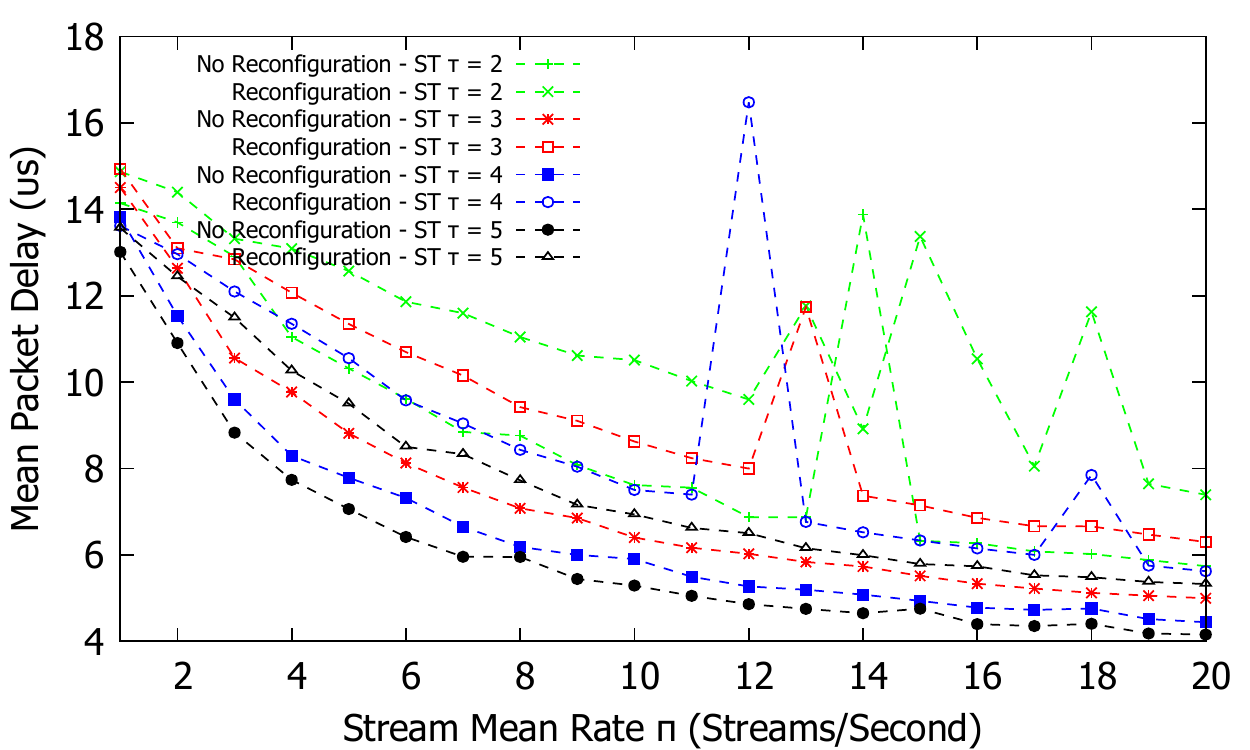}
		\caption{High $\rho_{L}$}
	\end{subfigure}
	\caption{Decentralized Unidirectional Topology: Stream average signaling delay as a results of TAS.}
	\label{fig_signalDelay_dec}
\end{figure}

In contrast to the centralized model, the decentralized model's in-band CDT traffic implies varied stream signaling delays as shown in Fig.~\ref{fig_signalDelay_dec} which shows the signaling delay for ST stream registration. As the number of streams generated ($\pi$) increase, the overall average signaling delay decreases which is due to the increased rejections as more streams attempt to request network resources. In the decentralized model, a rejection by an intermediate bottlenecked switch implies a termination of the CDT traffic and a notification to any previous pending stream records to cancel the potential reservation and eventually notify the source of the rejection. If this rejection happens closer to the source in the CDT registration procedure, then the average delay will be much shorter when compared to an stream acceptance. In general, the average stream signaling delay is in the order of microseconds which is reasonable for most industrial control systems applications.

\begin{figure} [t!] \centering
	\begin{subfigure}{\columnwidth} \centering
		\includegraphics[width=3.3in]{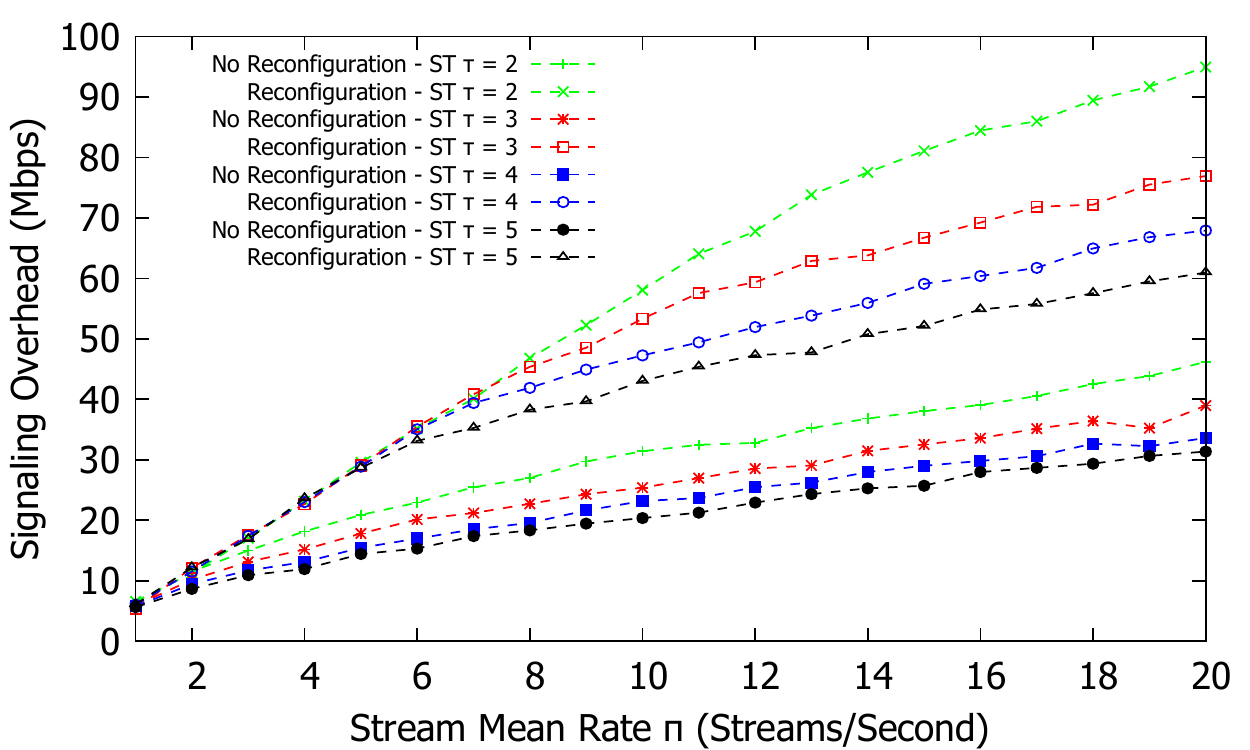}
		\caption{Low $\rho_{L}$}
	\end{subfigure}
	\begin{subfigure}{\columnwidth} \centering
		\includegraphics[width=3.3in]{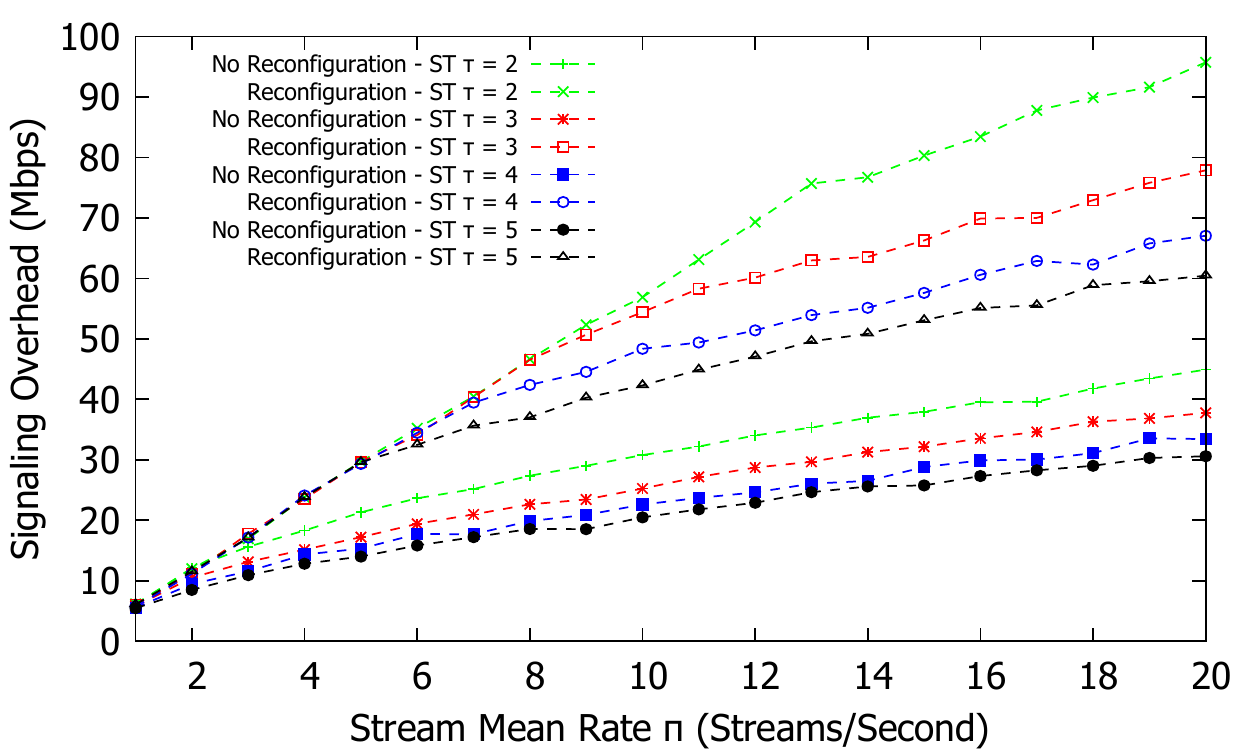}
		\caption{Mid $\rho_{L}$}
	\end{subfigure}
	\begin{subfigure}{\columnwidth} \centering
		\includegraphics[width=3.3in]{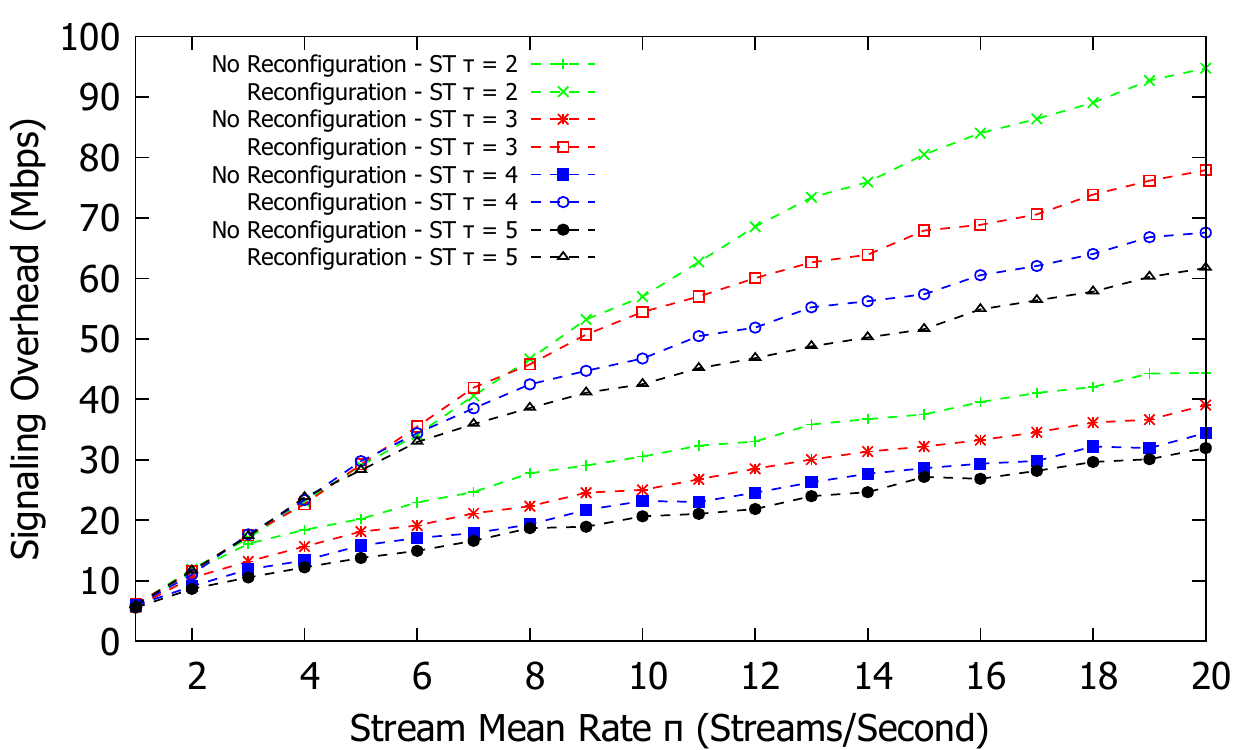}
		\caption{High $\rho_{L}$}
	\end{subfigure}
	\caption{Decentralized Unidirectional Topology: Stream Signaling Overhead as a results of TAS.}
	\label{fig_signalOverhead_dec}
\end{figure}

Generally, the decentralized model produced greater signaling overhead than the centralized model since CDT traffic is measured at each data traffic port for incoming and outgoing as shown in Fig.~\ref{fig_signalOverhead_dec}. Analogous to the signaling delay, the more ST streams accepted, the more overhead is observed. Therefore, as $\tau$ increases and consequently, the more rejections occur, the lower the overhead. As shown in Fig.~\ref{fig_signalOverhead_dec}, the results for the signaling overhead with reconfiguration shows a more pronounced difference with varied $\tau$ values.

\begin{figure} [t!] \centering
	\begin{subfigure}{\columnwidth} \centering
		\includegraphics[width=3.3in]{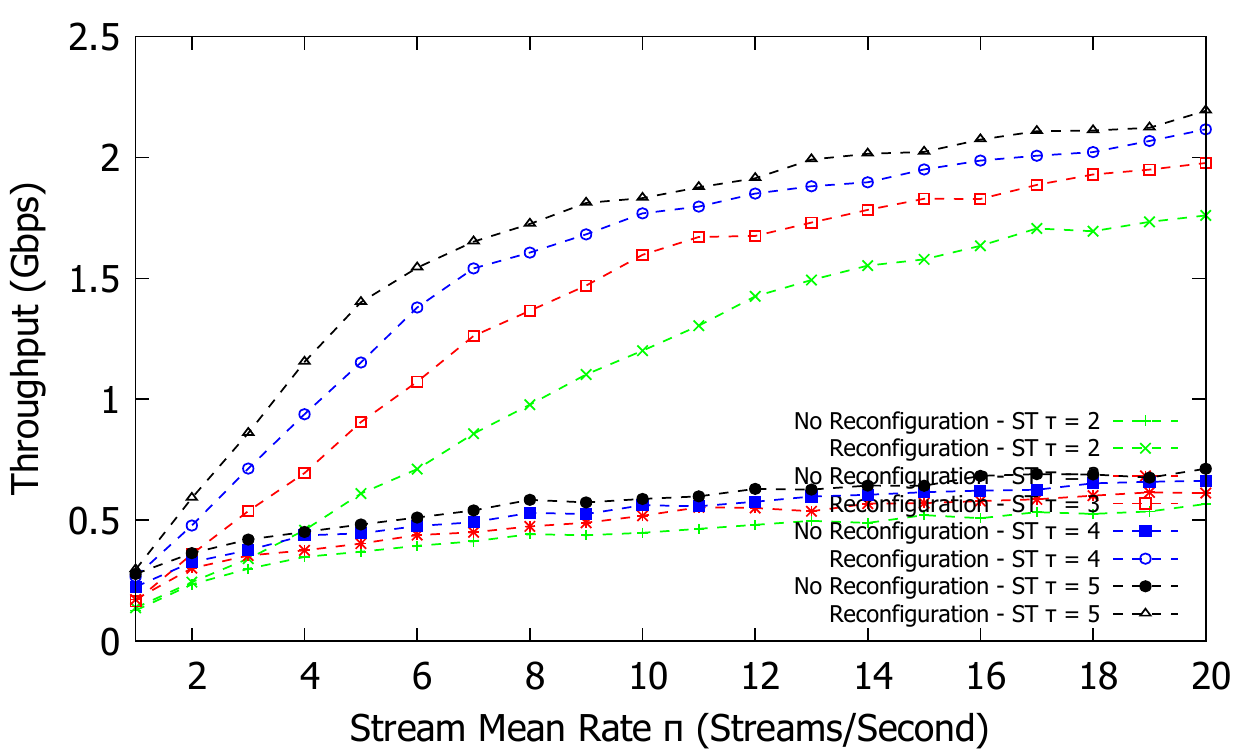}
		\caption{Low $\rho_{L}$}
	\end{subfigure}
	\begin{subfigure}{\columnwidth} \centering
		\includegraphics[width=3.3in]{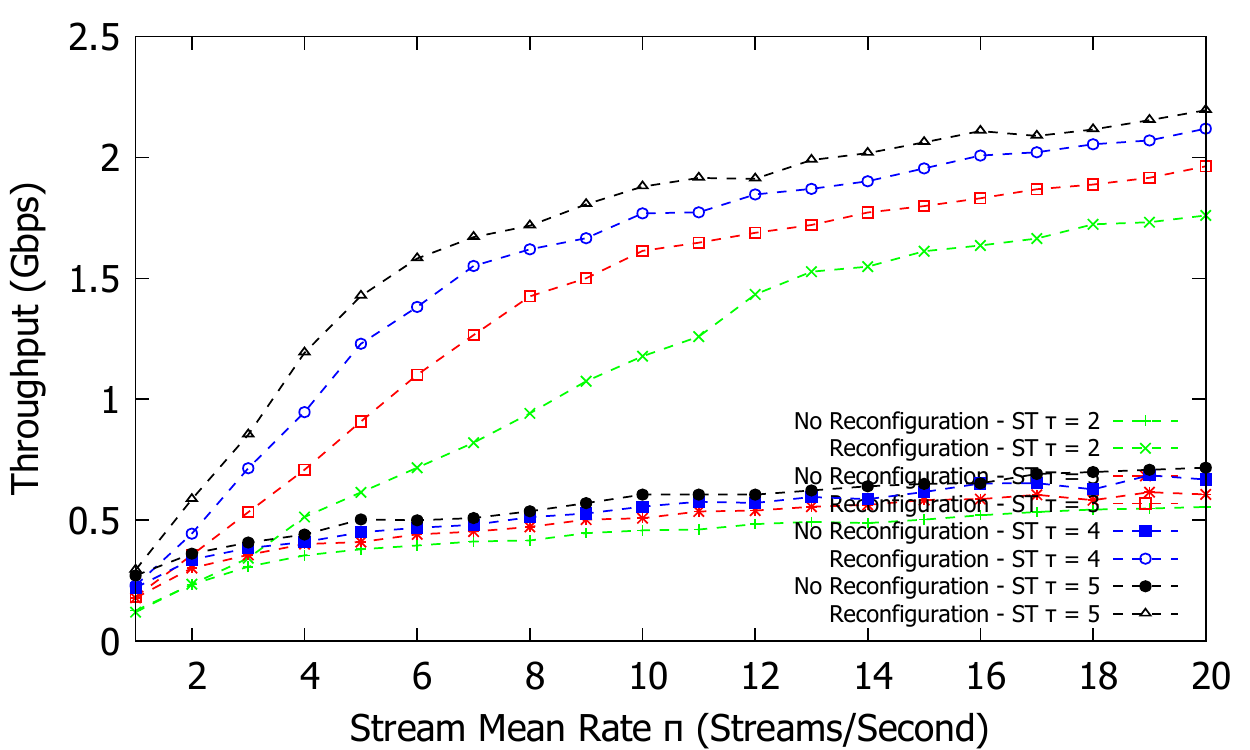}
		\caption{Mid $\rho_{L}$}
	\end{subfigure}
	\begin{subfigure}{\columnwidth} \centering
		\includegraphics[width=3.3in]{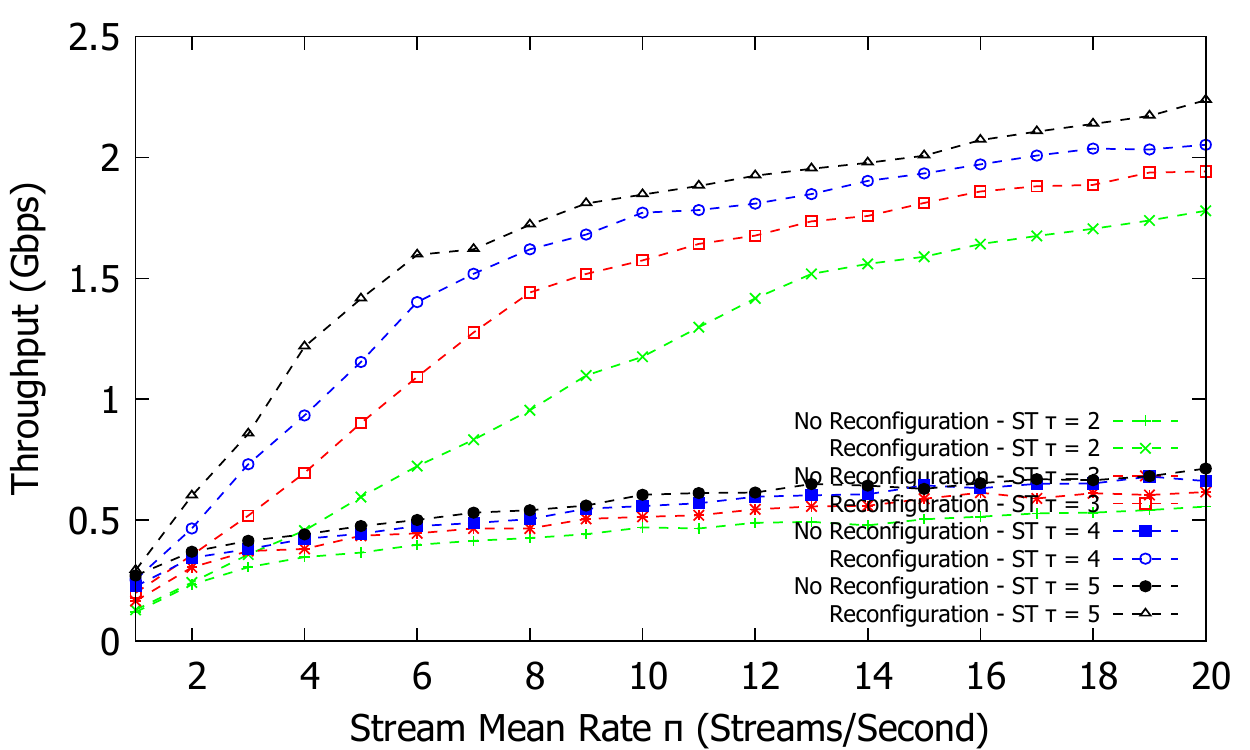}
		\caption{High $\rho_{L}$}
	\end{subfigure}
	\caption{Decentralized Unidirectional Topology: ST Total average throughput measured at the sink as a results of TAS.}
	\label{fig_avgTput_ST_dec}
\end{figure}

\begin{figure} [t!] \centering
	\begin{subfigure}{\columnwidth} \centering
		\includegraphics[width=3.3in]{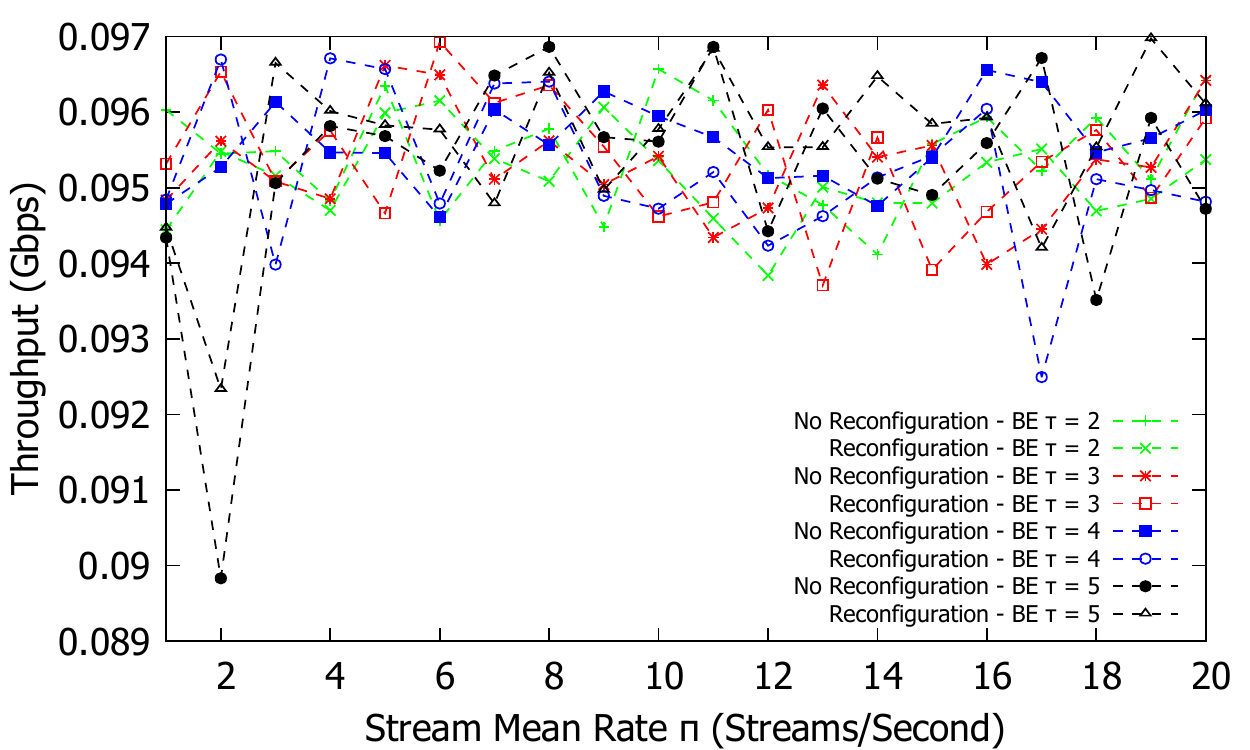}
		\caption{Low $\rho_{L}$}
	\end{subfigure}
	\begin{subfigure}{\columnwidth} \centering
		\includegraphics[width=3.3in]{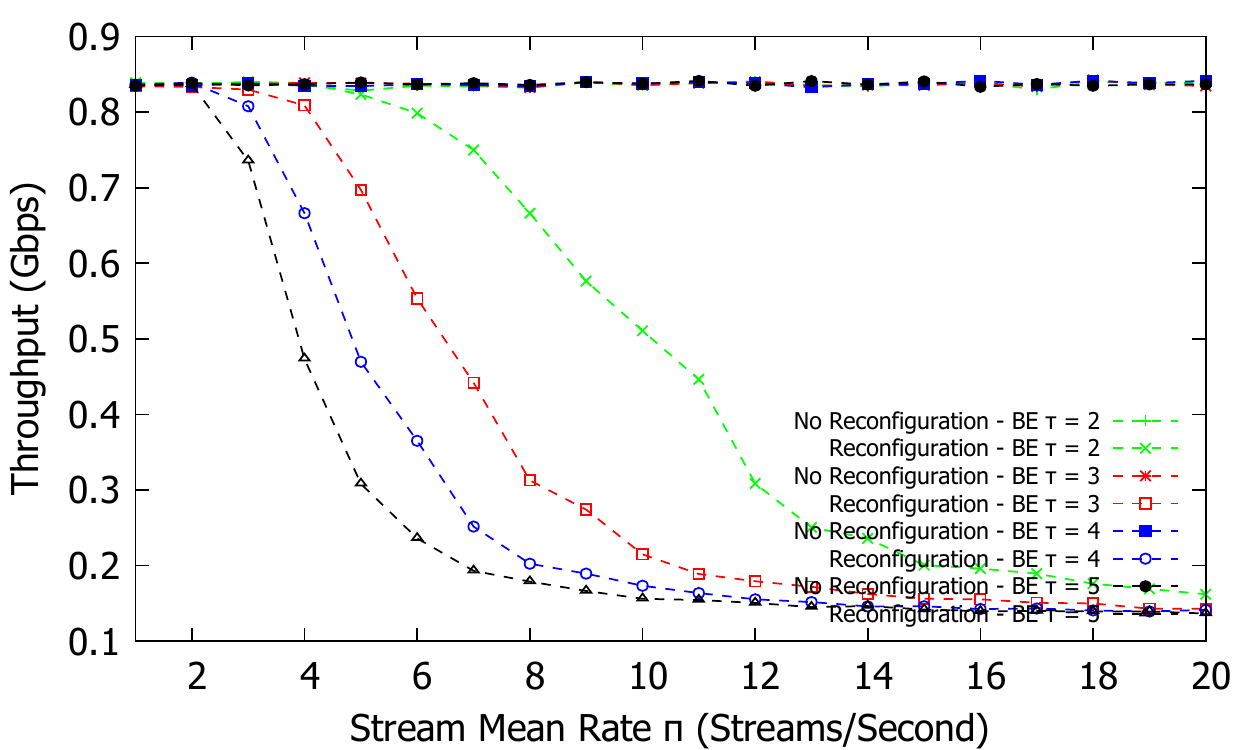}
		\caption{Mid $\rho_{L}$}
	\end{subfigure}
	\begin{subfigure}{\columnwidth} \centering
		\includegraphics[width=3.3in]{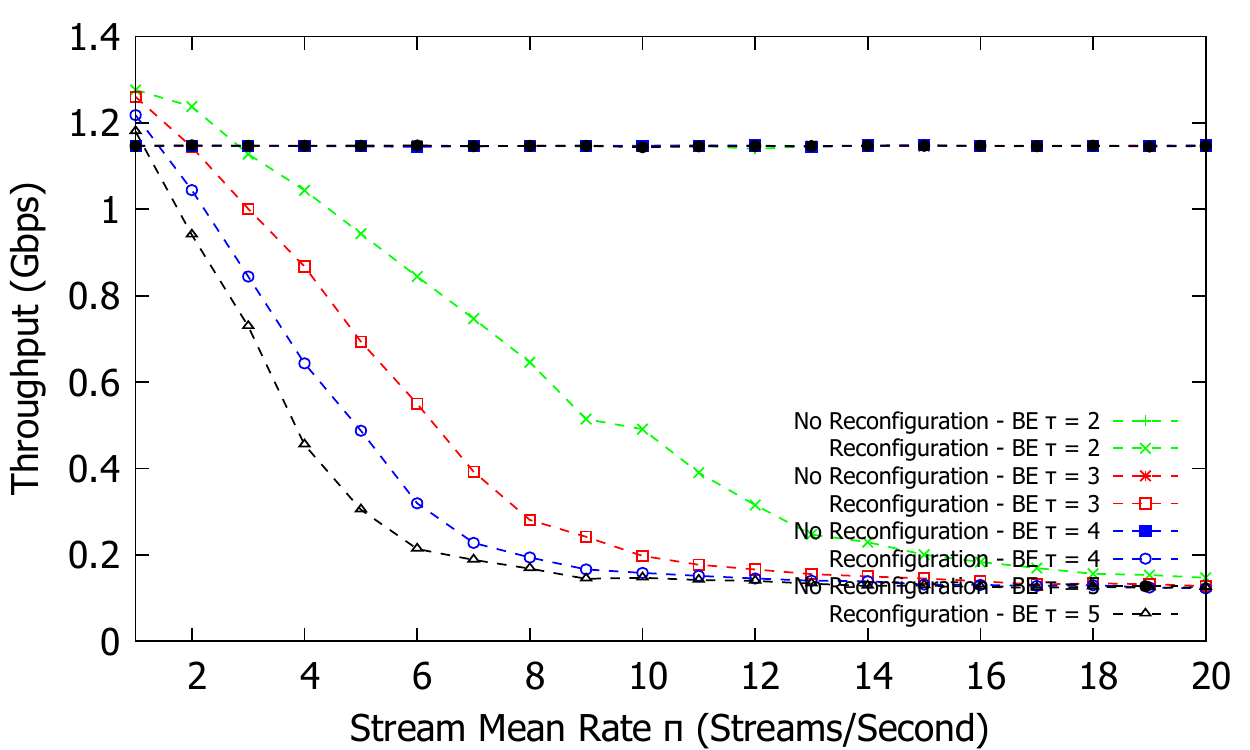}
		\caption{High $\rho_{L}$}
	\end{subfigure}
	\caption{Decentralized Unidirectional Topology: BE Total average throughput measured at the sink as a results of TAS.}
	\label{fig_avgTput_BE_dec}
\end{figure}

Throughput results are generally the same when compared to the unidirectional centralized model. Fig.~\ref{fig_avgTput_ST_dec} and Fig.~\ref{fig_avgTput_BE_dec} shows the average throughput measured at the sink for both ST and BE traffic.

\begin{figure} [t!] \centering
	\begin{subfigure}{\columnwidth} \centering
		\includegraphics[width=3.3in]{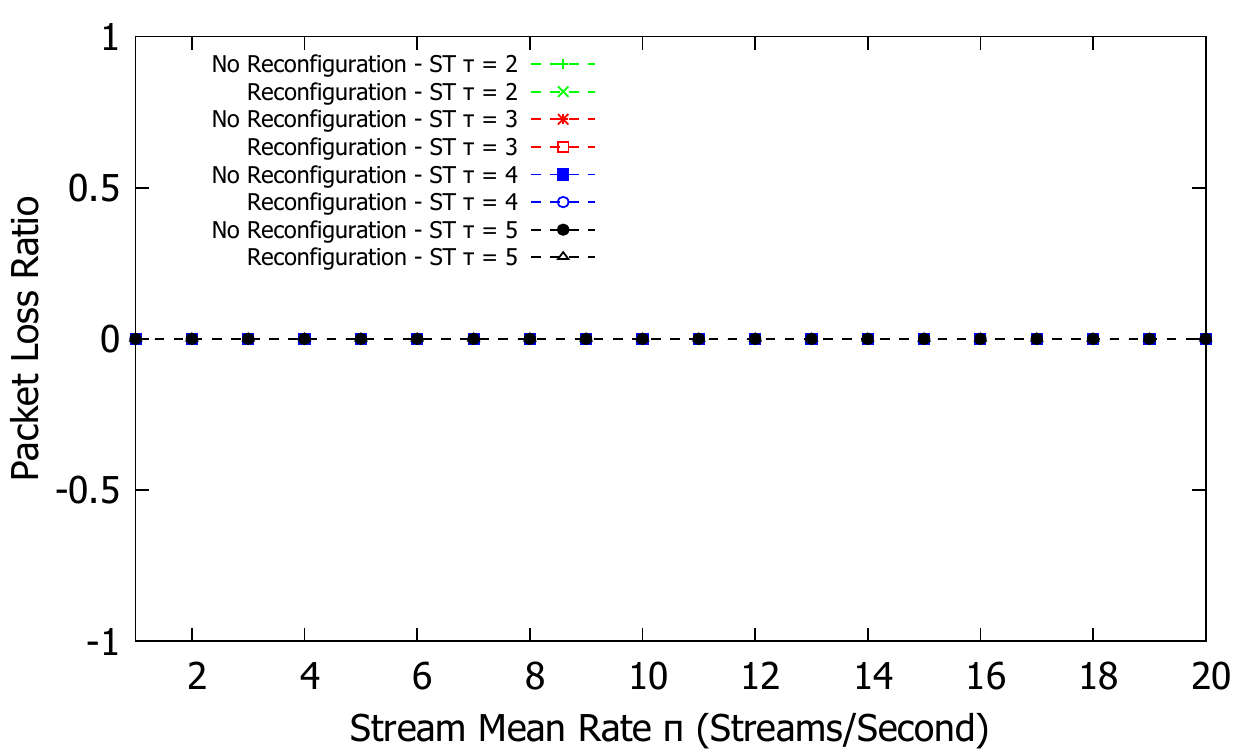}
		\caption{Low $\rho_{L}$}
	\end{subfigure}
	\begin{subfigure}{\columnwidth} \centering
		\includegraphics[width=3.3in]{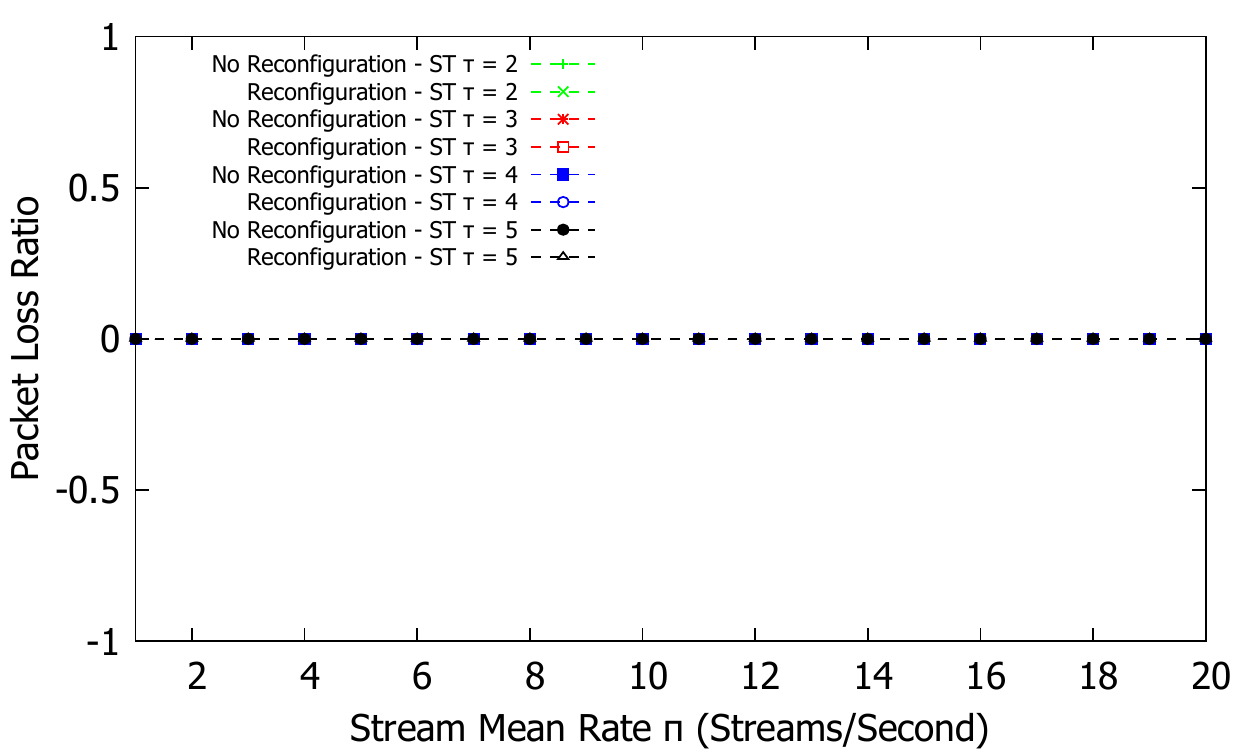}
		\caption{Mid $\rho_{L}$}
	\end{subfigure}
	\begin{subfigure}{\columnwidth} \centering
		\includegraphics[width=3.3in]{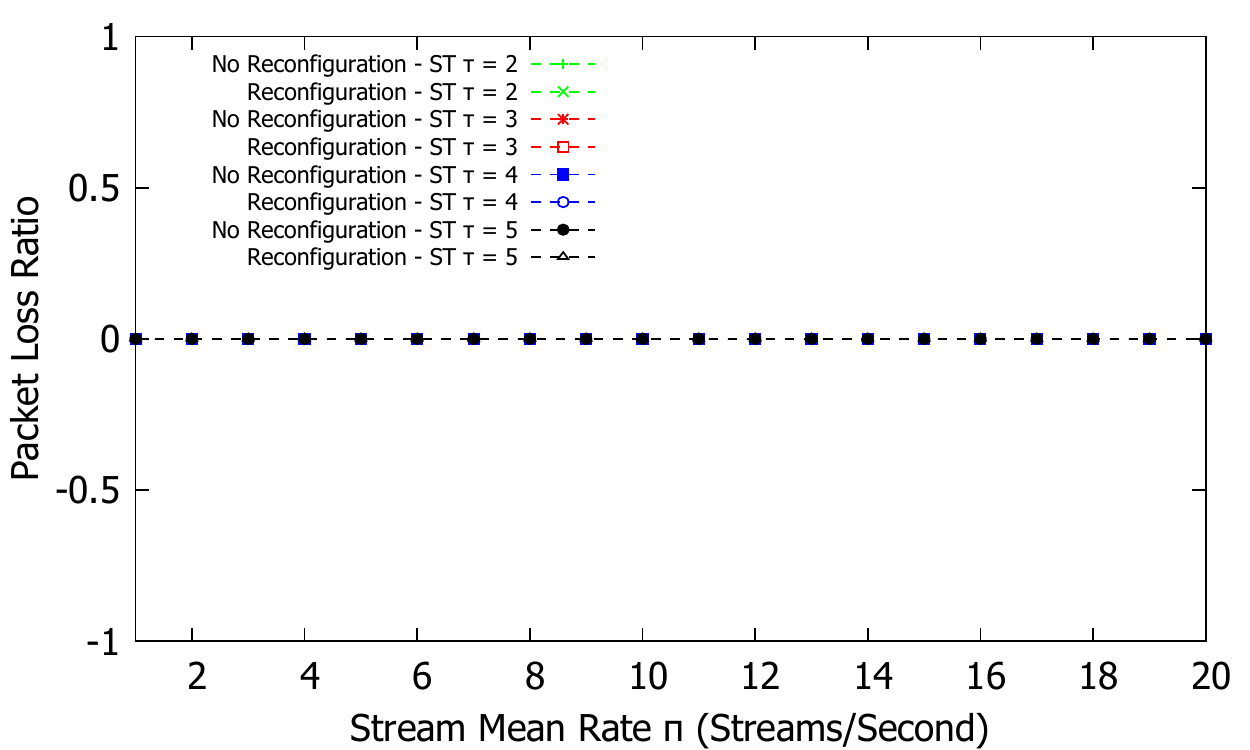}
		\caption{High $\rho_{L}$}
	\end{subfigure}
	\caption{Decentralized Unidirectional Topology: ST Frame loss ratio as a results of TAS.}
	\label{fig_lossProb_ST_dec}
\end{figure}

\begin{figure} [t!] \centering
	\begin{subfigure}{\columnwidth} \centering
		\includegraphics[width=3.3in]{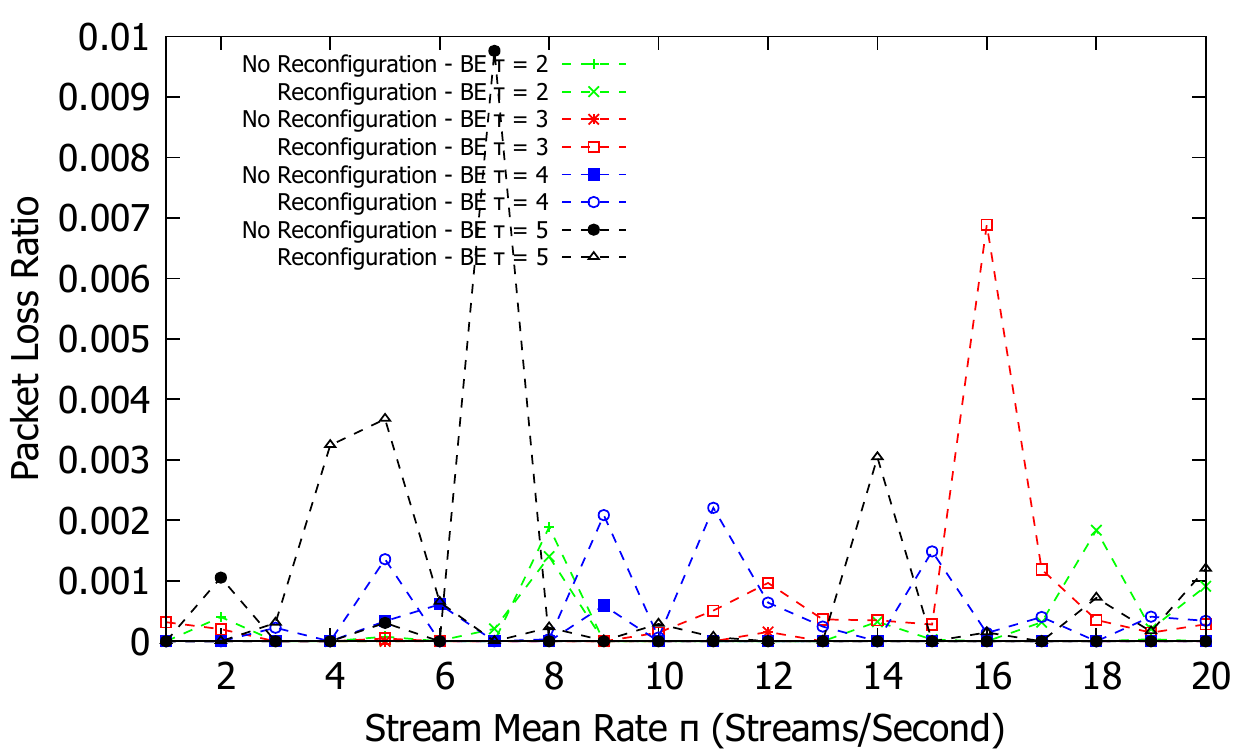}
		\caption{Low $\rho_{L}$}
	\end{subfigure}
	\begin{subfigure}{\columnwidth} \centering
		\includegraphics[width=3.3in]{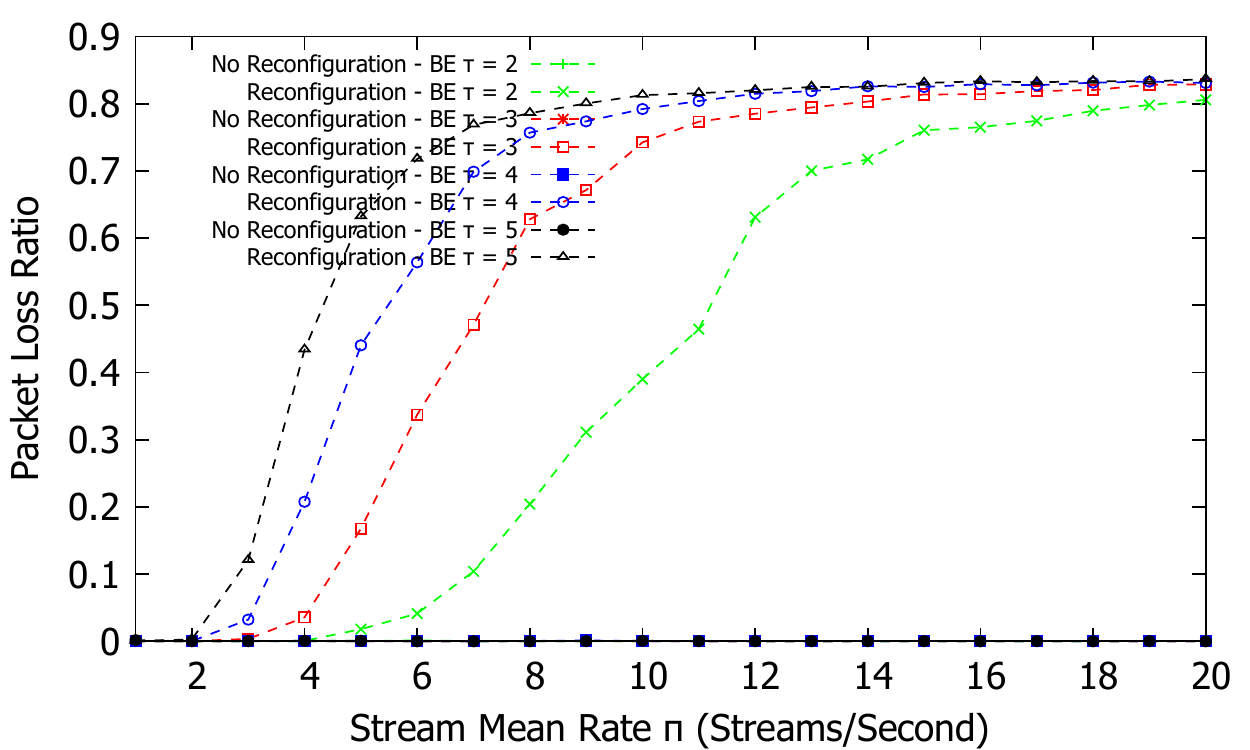}
		\caption{Mid $\rho_{L}$}
	\end{subfigure}
	\begin{subfigure}{\columnwidth} \centering
		\includegraphics[width=3.3in]{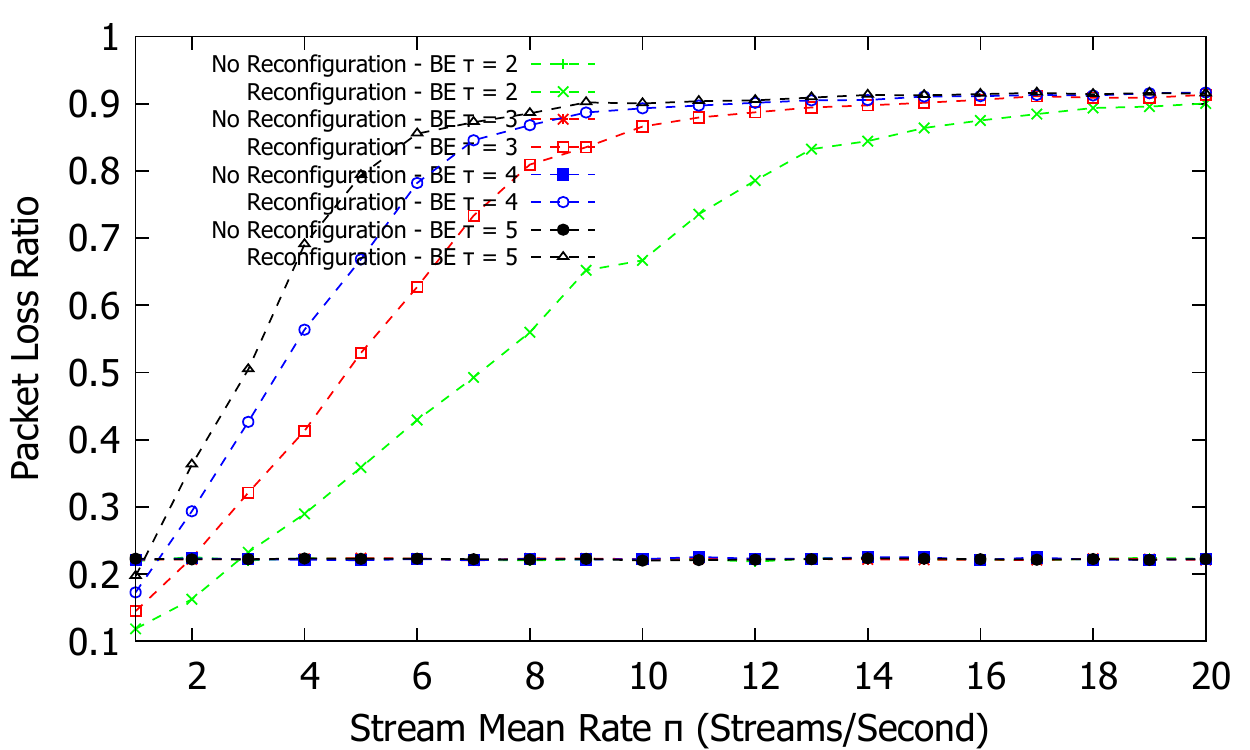}
		\caption{High $\rho_{L}$}
	\end{subfigure}
	\caption{Decentralized Unidirectional Topology: BE Frame loss ratio as a results of TAS.}
	\label{fig_lossProb_BE_dec}
\end{figure}

Similarly, the packet loss rate is nearly similar to the unidirectional centralized model as shown  in Fig.~\ref{fig_lossProb_ST_dec} and Fig.~\ref{fig_lossProb_BE_dec} for ST and BE traffic in the network respectively. The unidirectional topology with either the centralized or decentralized approach generally get bottlenecked much faster compared to the bi-directional results. Therefore, BE traffic quickly suffers as more ST streams request TAS slot reservation. In terms of added improvements, changing the cycle time (GCL time) to different values depending on both ST and BE traffic proportions can generally improve BE traffic whilst still guaranteeing ST streams. In typical industrial environments, ST streams have generally low data rates, are less frequent, and smaller in size than BE traffic, and therefore can be admitted quite easily without much effect on BE traffic.

\subsubsection{Bi-Directional Ring Topology}

\begin{figure} [t!] \centering
	\includegraphics[width=3.3in]{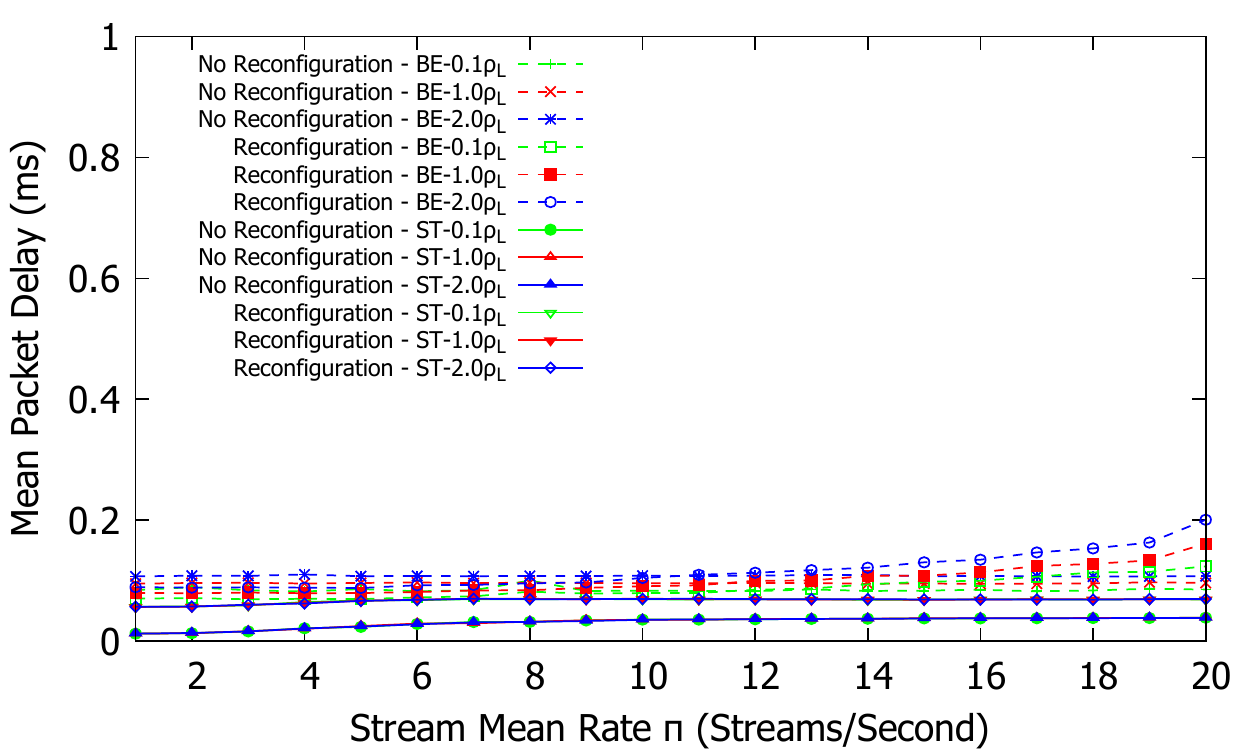}
	\caption{Decentralized Bi-directional Topology: Mean end-to-end delay for ST and BE traffic for $\tau = 2$ under different loads~$\rho_{L}$, mean traffic rates~$\pi$, and initialized gating ratio of $20\%$.}
	\label{fig_delay_2_dec_bi}
\end{figure}

\begin{figure} [t!] \centering
	\includegraphics[width=3.3in]{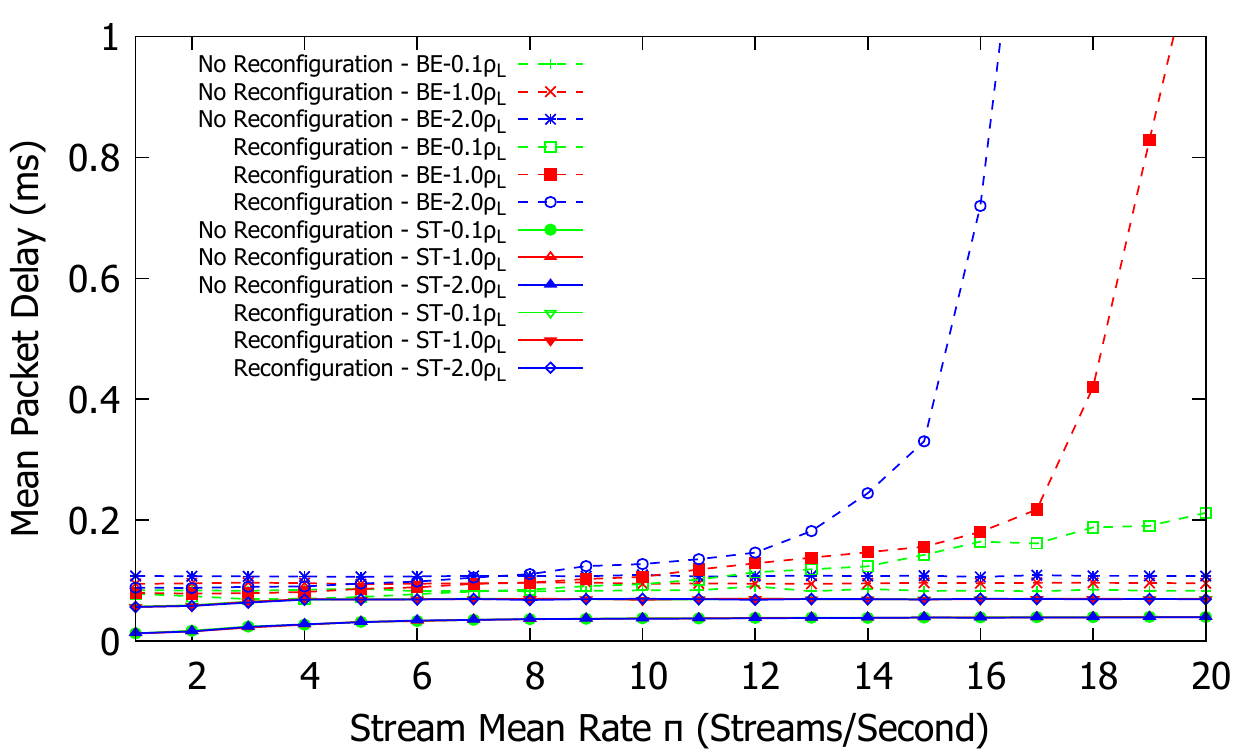}
	\caption{Decentralized Bi-directional Topology: Mean end-to-end delay for ST and BE traffic for $\tau = 3$ under different loads~$\rho_{L}$, mean traffic rates~$\pi$, and initialized gating ratio of $20\%$.}
	\label{fig_delay_3_dec_bi}
\end{figure}

\begin{figure} [t!] \centering
	\includegraphics[width=3.3in]{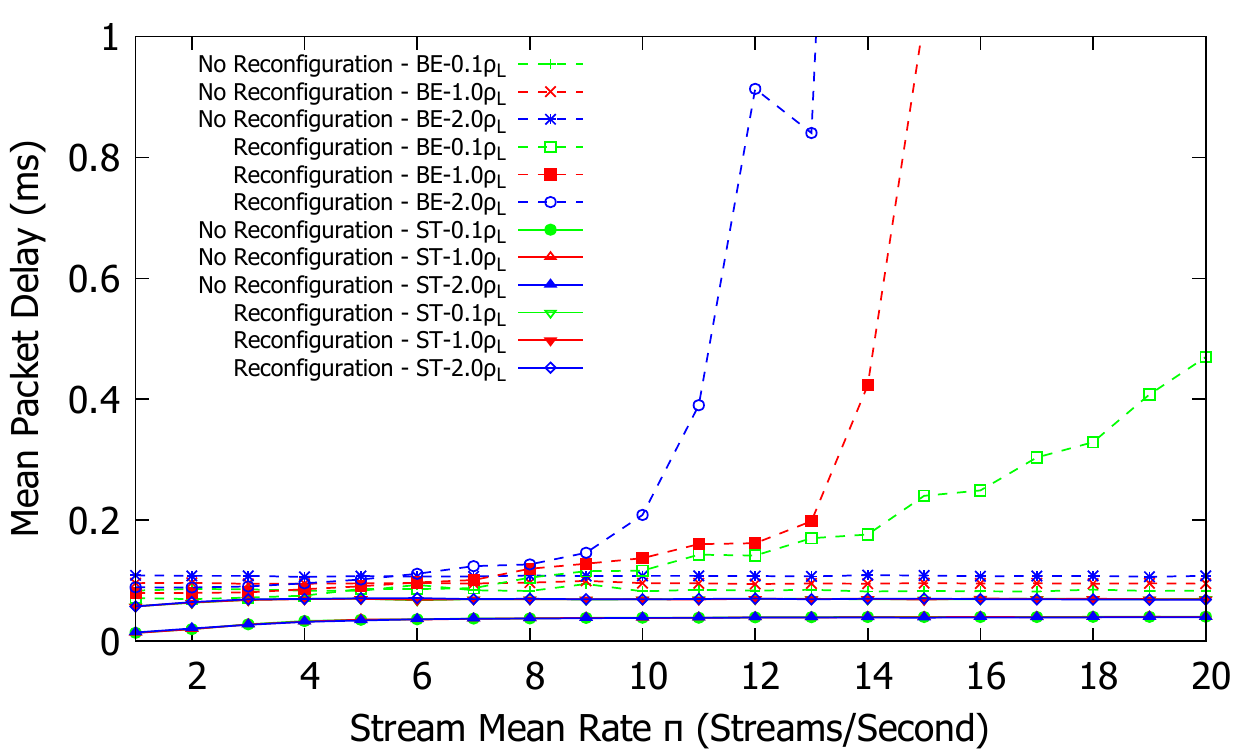}
	\caption{Decentralized Bi-directional Topology: Mean end-to-end delay for ST and BE traffic for $\tau = 4$ under different loads~$\rho_{L}$, mean traffic rates~$\pi$, and initialized gating ratio of $20\%$.}
	\label{fig_delay_4_dec_bi}
\end{figure}

\begin{figure} [t!] \centering
	\includegraphics[width=3.3in]{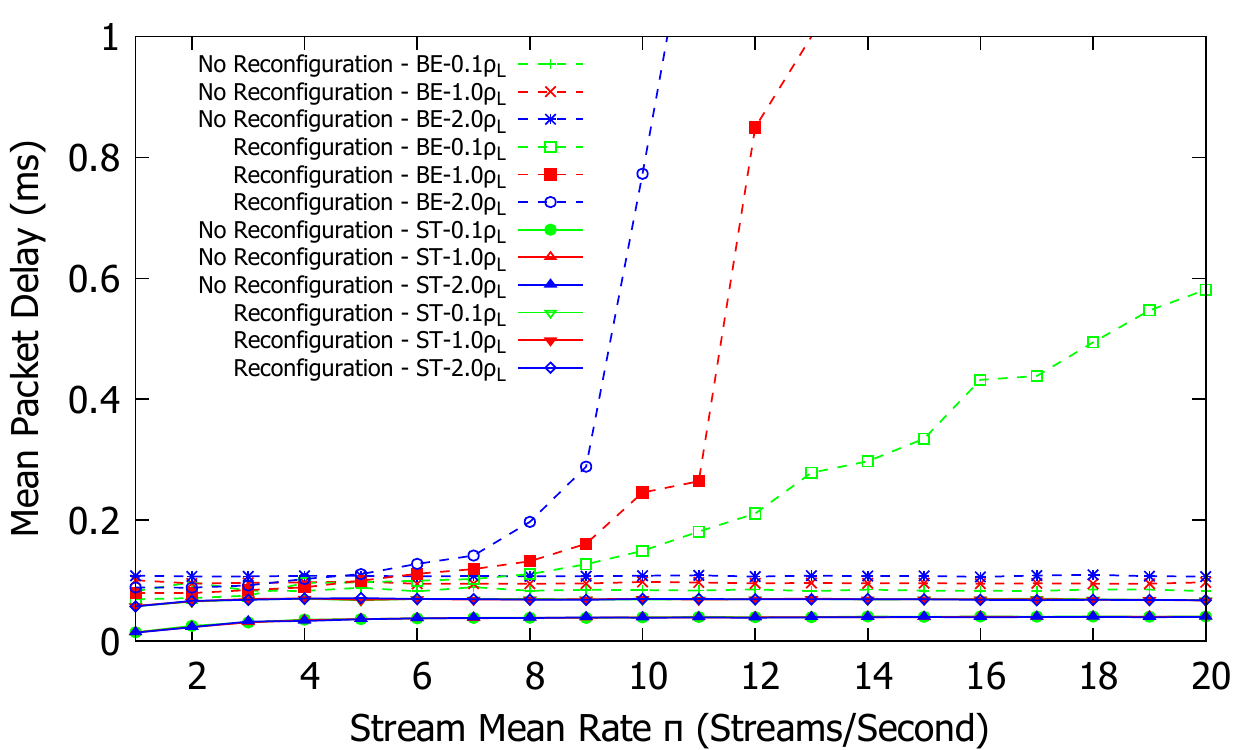}
	\caption{Decentralized Bi-directional Topology: Mean end-to-end delay for ST and BE traffic for $\tau = 5$ under different loads~$\rho_{L}$, mean traffic rates~$\pi$, and initialized gating ratio of $20\%$.}
	\label{fig_delay_5_dec_bi}
\end{figure}

In the bi-directional topology using the decentralized model, in-band CDT traffic affects the data traffic similar to the unidirectional model. Fig.~\ref{fig_delay_2_dec_bi} - \ref{fig_delay_5_dec_bi} shows the average mean delay evaluation for both ST and BE traffic. As $\tau$ is increased, i.e., the number of streams at any time increase, the BE slot reservations are reserved for ST streams which affects BE QoS mean delay.

\begin{figure} [t!] \centering
	\begin{subfigure}{\columnwidth} \centering
		\includegraphics[width=3.3in]{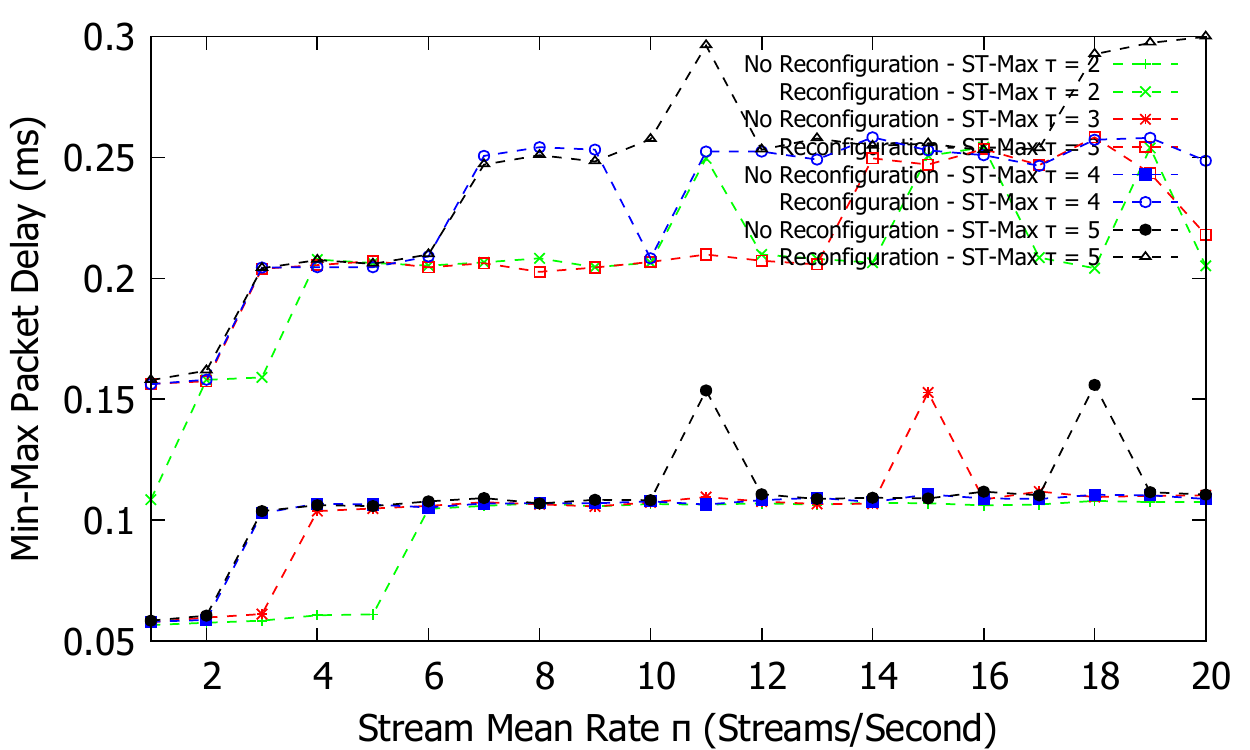}
		\caption{Low $\rho_{L}$}
	\end{subfigure}
	\begin{subfigure}{\columnwidth} \centering
		\includegraphics[width=3.3in]{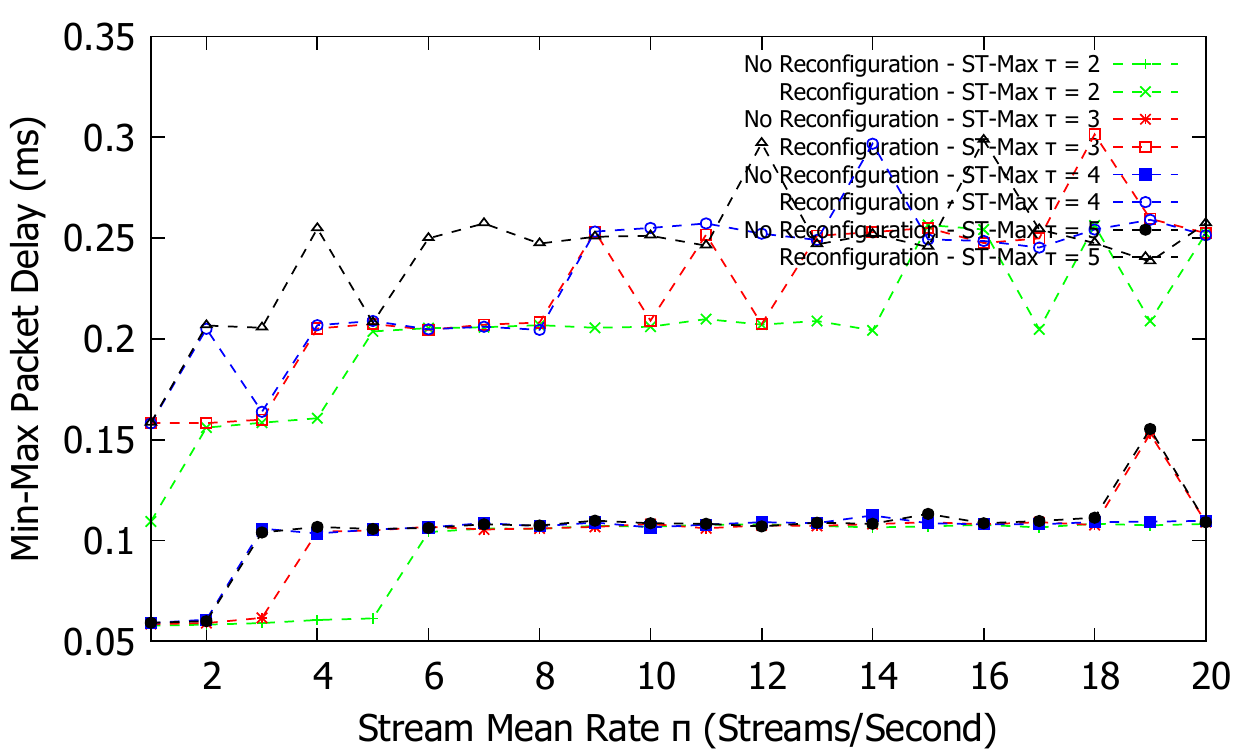}
		\caption{Mid $\rho_{L}$}
	\end{subfigure}
	\begin{subfigure}{\columnwidth} \centering
		\includegraphics[width=3.3in]{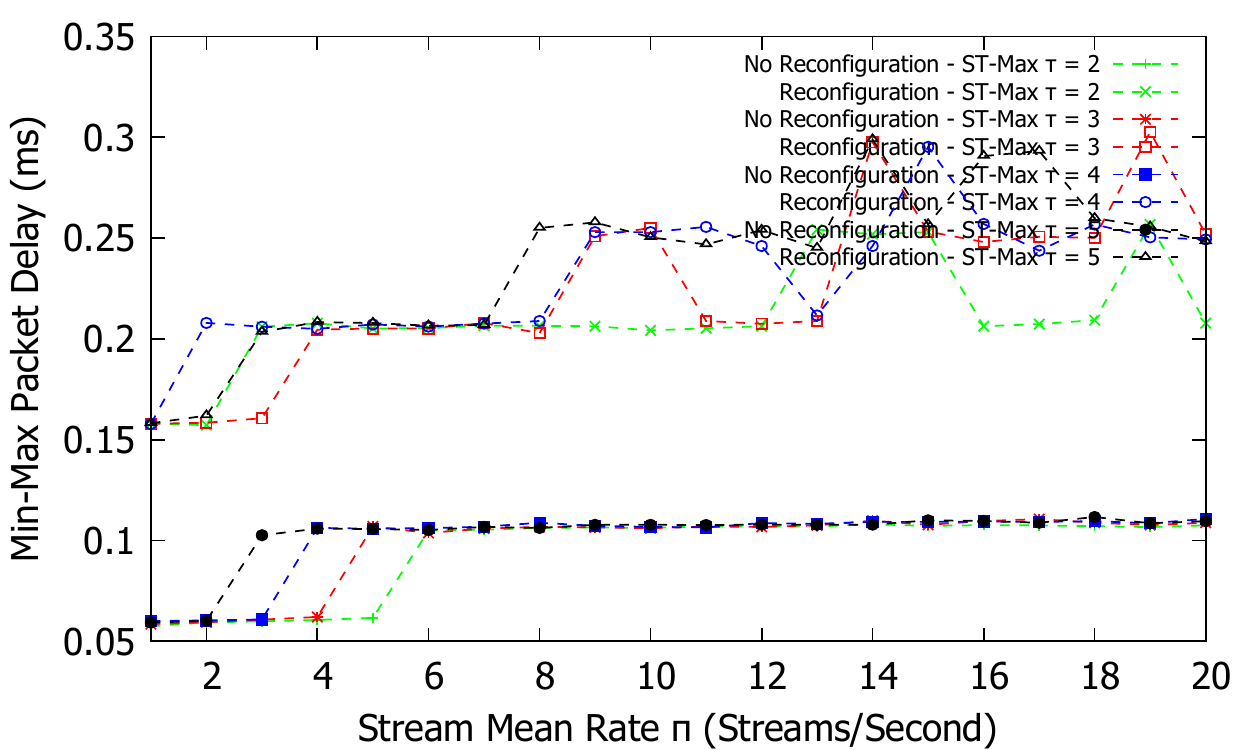}
		\caption{High $\rho_{L}$}
	\end{subfigure}
	\caption{Decentralized Bi-directional Topology: Max delay as a results of TAS.}
	\label{fig_maxDelay_dec_bi}
\end{figure}

In terms of maximum delay, Fig.~\ref{fig_maxDelay_dec_bi} shows the maximum delay evaluation for ST traffic. While the reconfiguration approach looks very similar to the centralized model, the ``no reconfiguration'' approach gets affected by in-band CDT traffic that raises the max delay in some cases to $100~\mu$s.

\begin{figure} [t!] \centering
	\begin{subfigure}{\columnwidth} \centering
		\includegraphics[width=3.3in]{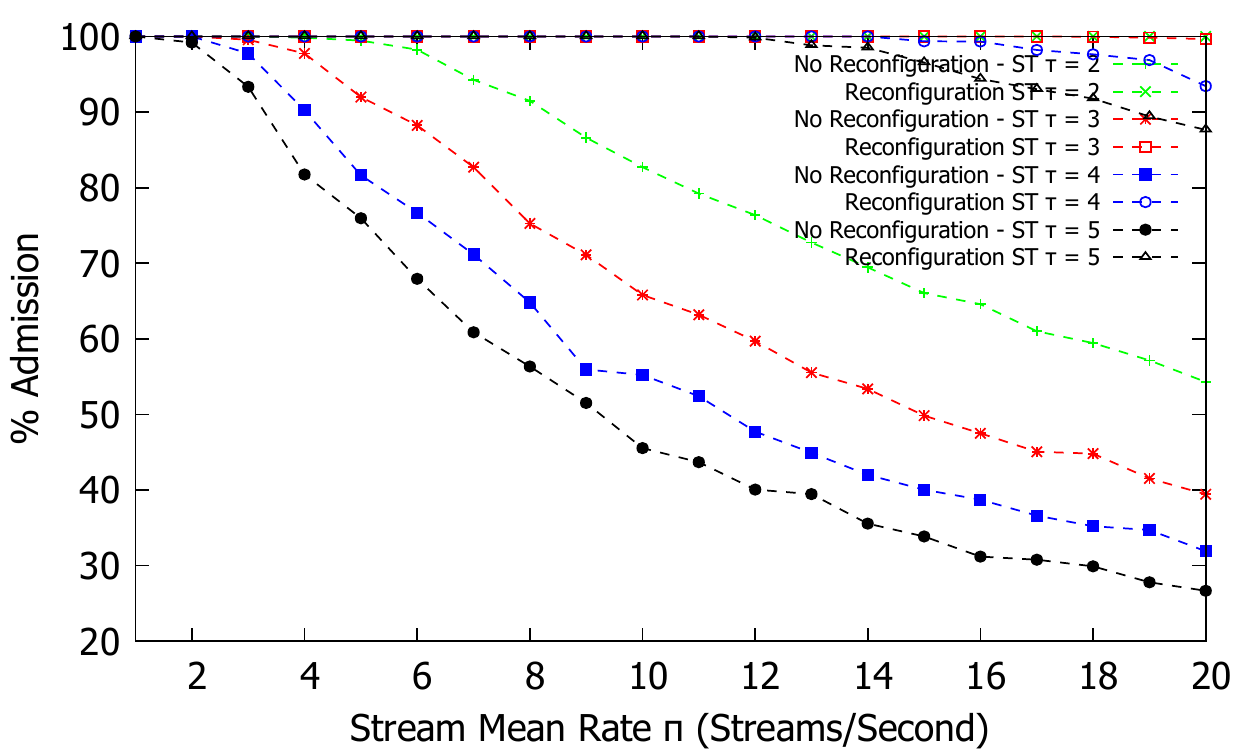}
		\caption{Low $\rho_{L}$}
	\end{subfigure}
	\begin{subfigure}{\columnwidth} \centering
		\includegraphics[width=3.3in]{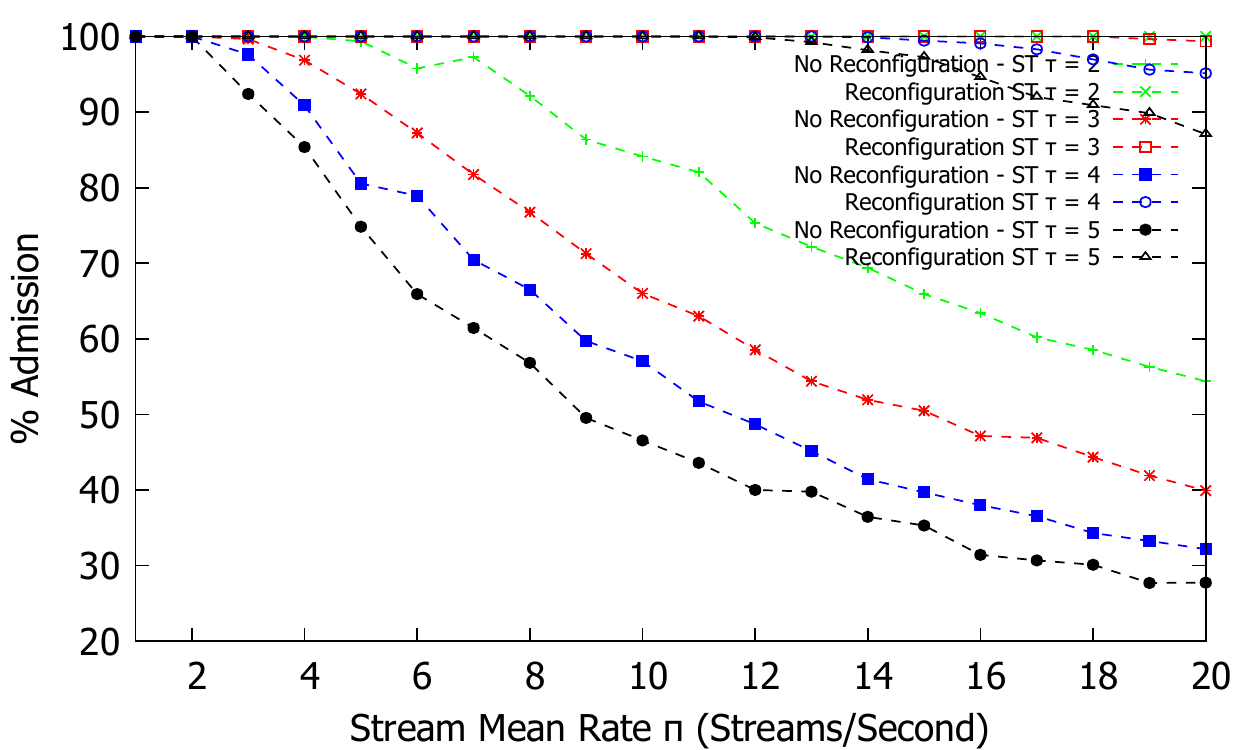}
		\caption{Mid $\rho_{L}$}
	\end{subfigure}
	\begin{subfigure}{\columnwidth} \centering
		\includegraphics[width=3.3in]{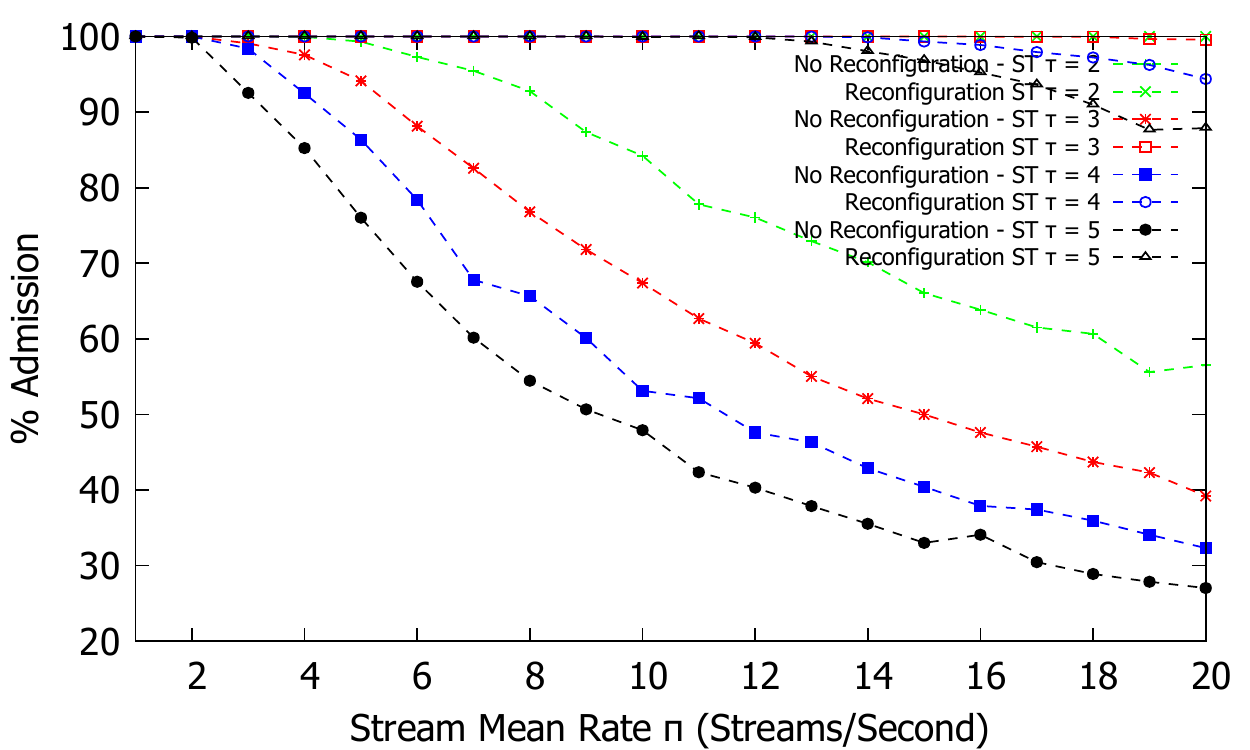}
		\caption{High $\rho_{L}$}
	\end{subfigure}
	\caption{Decentralized Bi-directional Topology: Stream admission as a results of TAS.}
	\label{fig_admin_dec_bi}
\end{figure}

Admission rate is exactly the same as the centralized model as shown in Fig.~\ref{fig_admin_dec_bi} given the same networking parameters.

\begin{figure} [t!] \centering
	\begin{subfigure}{\columnwidth} \centering
		\includegraphics[width=3.3in]{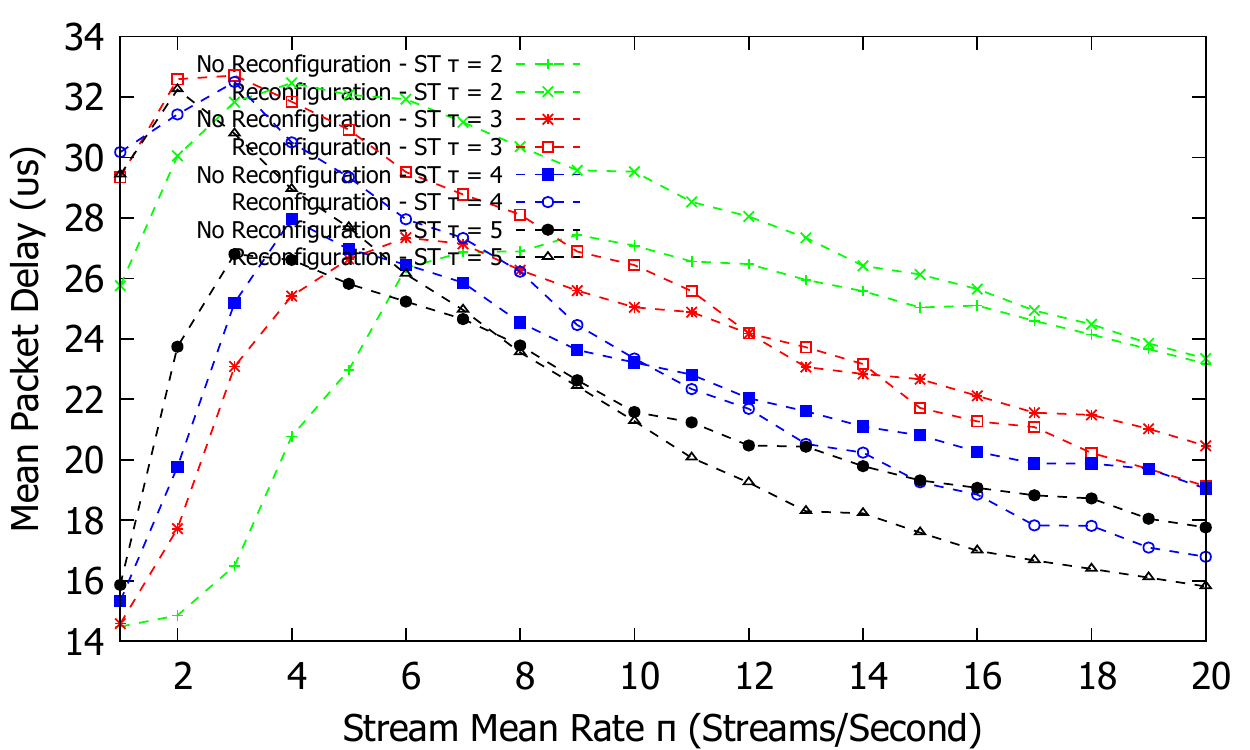}
		\caption{Low $\rho_{L}$}
	\end{subfigure}
	\begin{subfigure}{\columnwidth} \centering
		\includegraphics[width=3.3in]{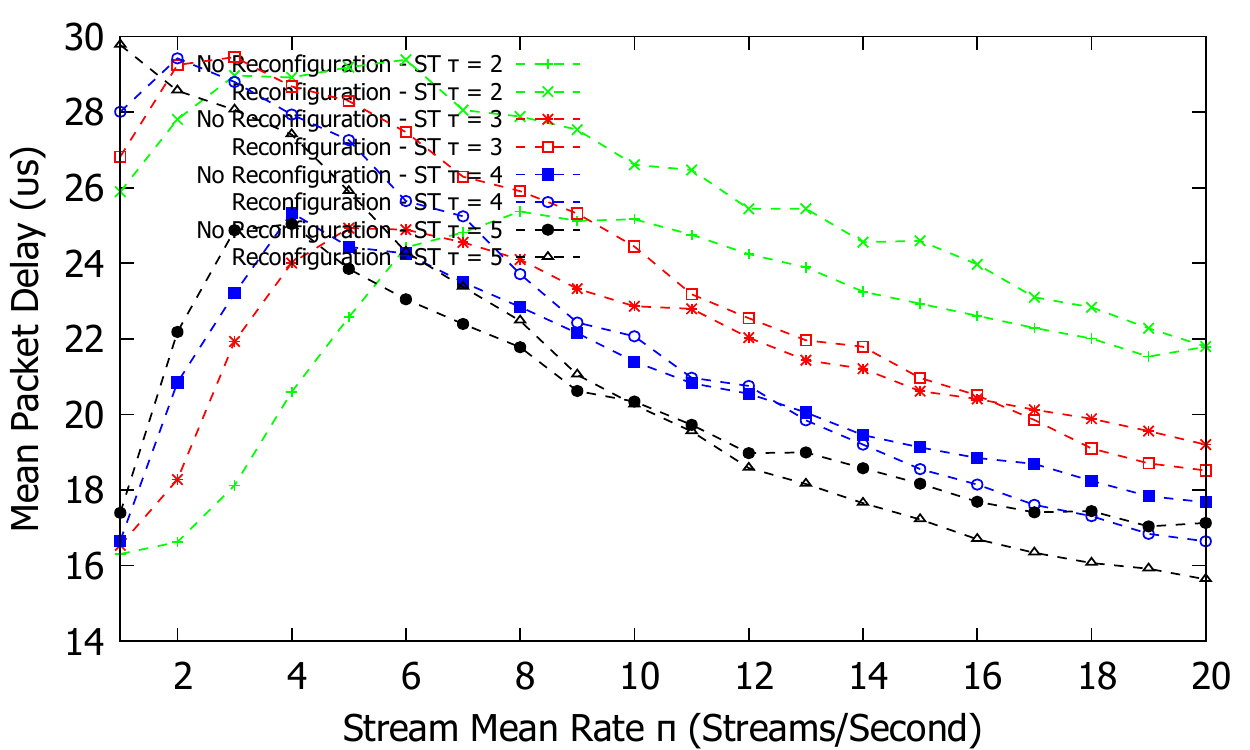}
		\caption{Mid $\rho_{L}$}
	\end{subfigure}
	\begin{subfigure}{\columnwidth} \centering
		\includegraphics[width=3.3in]{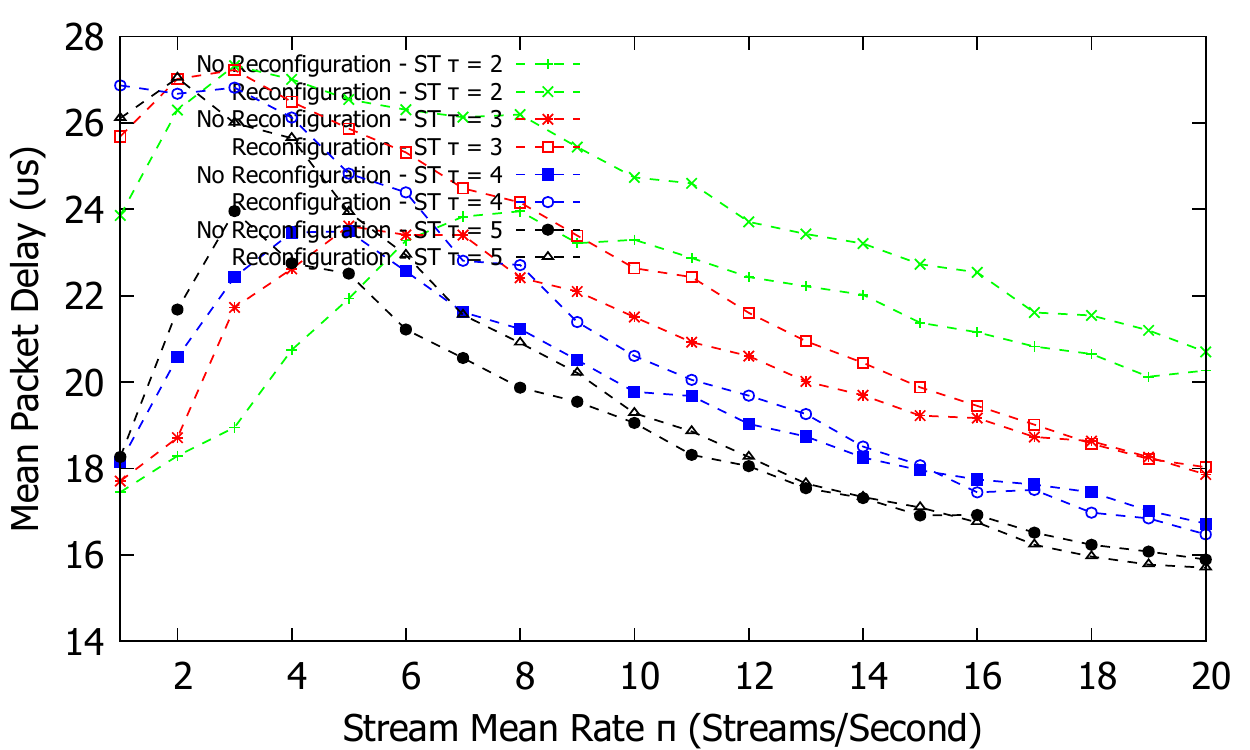}
		\caption{High $\rho_{L}$}
	\end{subfigure}
	\caption{Decentralized Bi-directional Topology: Stream average signaling delay as a results of TAS.}
	\label{fig_signalDelay_dec_bi}
\end{figure}

Fig.~\ref{fig_signalDelay_dec_bi} shows the signaling delay for ST stream registration.It is observed that as the ST stream generation increases, so does the mean signaling delay up to a stream mean rate, $\pi$. The mean delay starts to decrease as the load keeps increasing.

\begin{figure} [t!] \centering
	\begin{subfigure}{\columnwidth} \centering
		\includegraphics[width=3.3in]{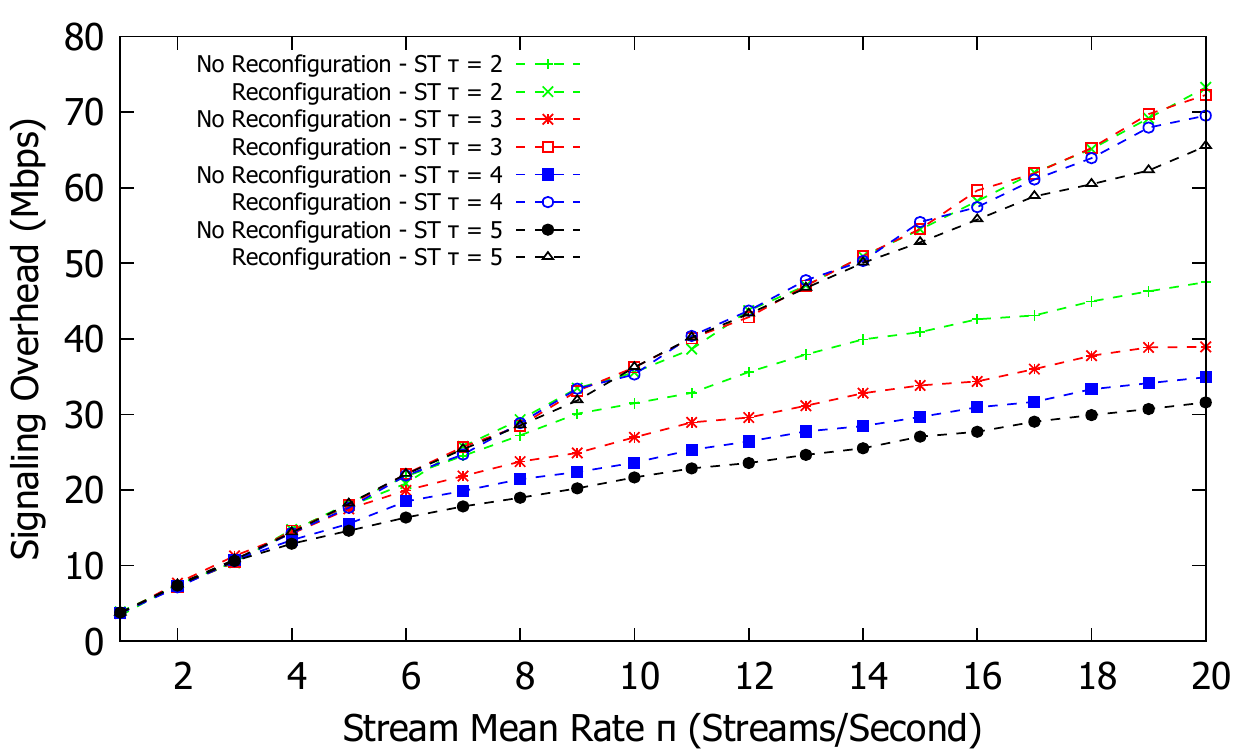}
		\caption{Low $\rho_{L}$}
	\end{subfigure}
	\begin{subfigure}{\columnwidth} \centering
		\includegraphics[width=3.3in]{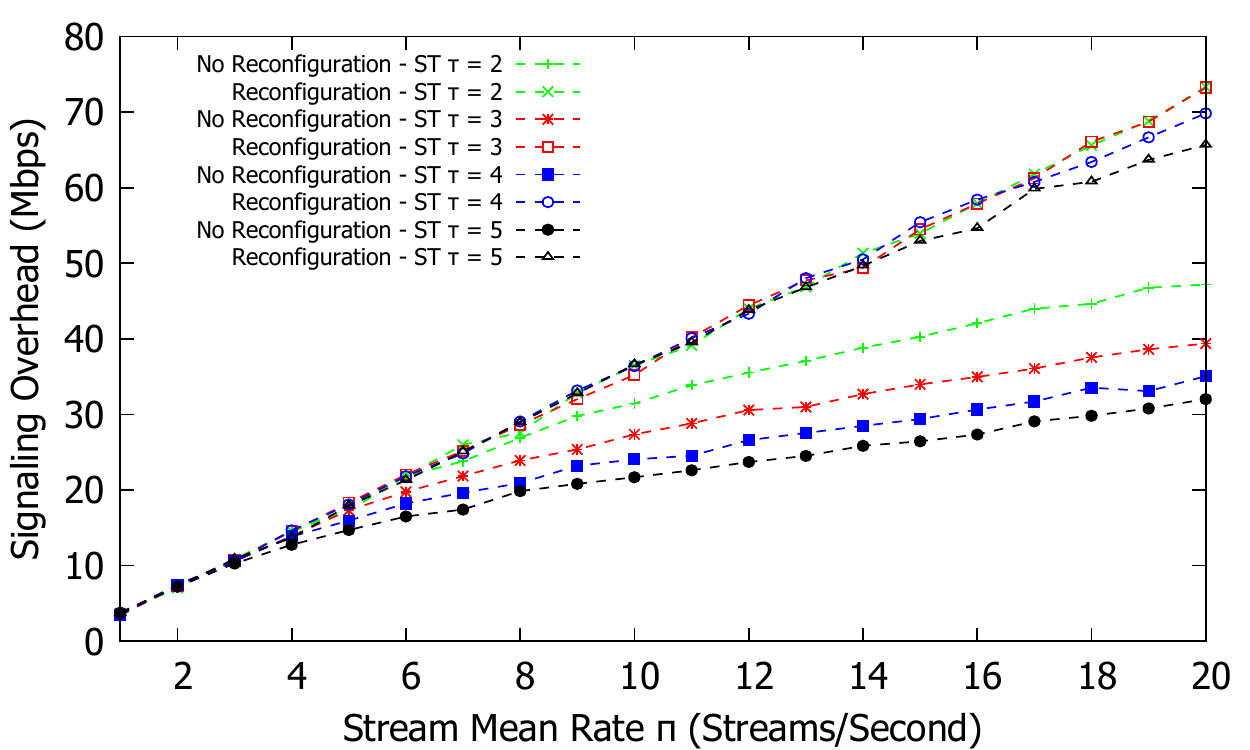}
		\caption{Mid $\rho_{L}$}
	\end{subfigure}
	\begin{subfigure}{\columnwidth} \centering
		\includegraphics[width=3.3in]{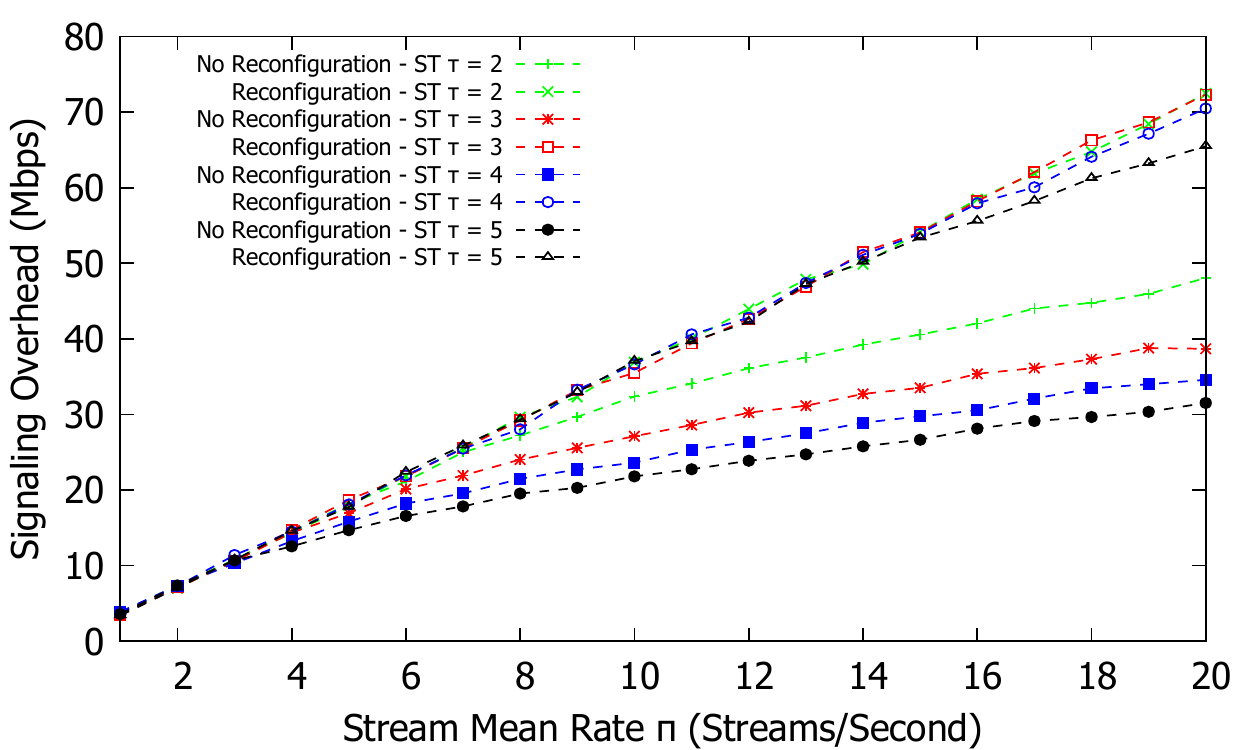}
		\caption{High $\rho_{L}$}
	\end{subfigure}
	\caption{Decentralized Bi-directional Topology: Stream Signaling Overhead as a results of TAS.}
	\label{fig_signalOverhead_dec_bi}
\end{figure}

The stream signaling overhead produces results that are the same as the decentralized model as shown Fig.~\ref{fig_signalOverhead_dec_bi}. In general, the results between different $\tau$ values are close since for any $\tau$ value, almost all the streams are getting accepted generating the same overhead in total.

\begin{figure} [t!] \centering
	\begin{subfigure}{\columnwidth} \centering
		\includegraphics[width=3.3in]{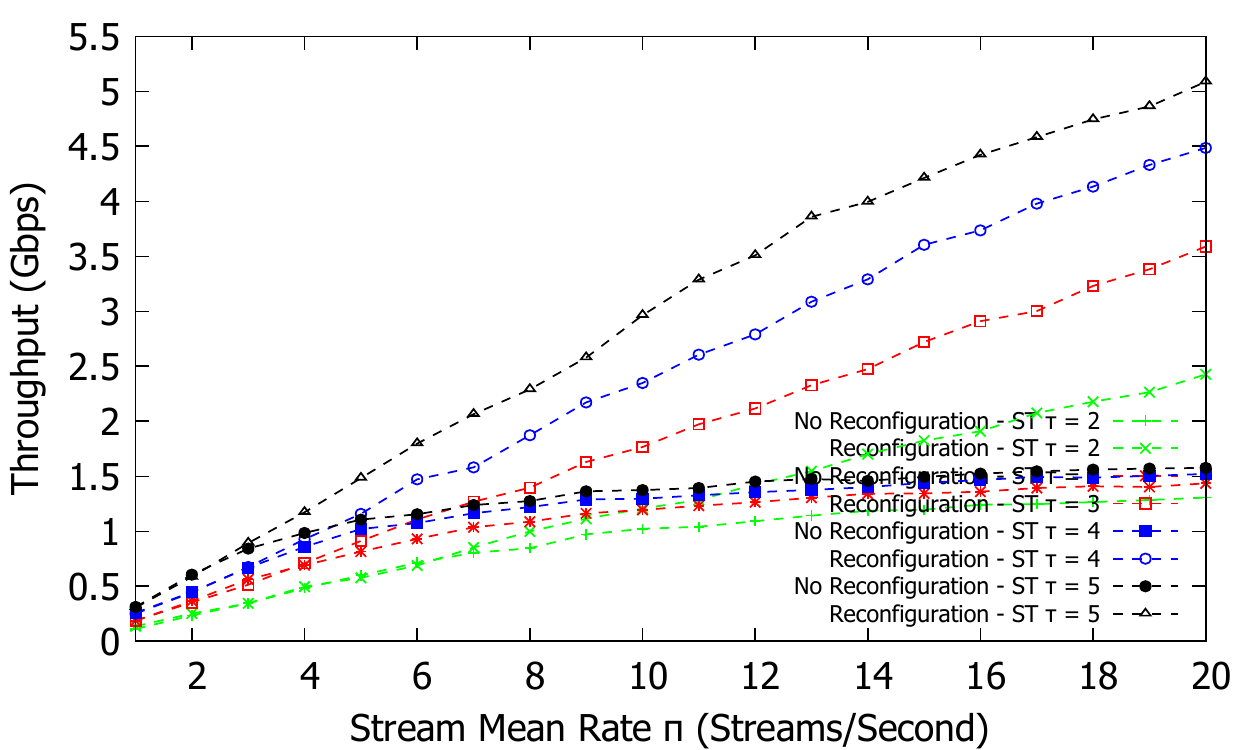}
		\caption{Low $\rho_{L}$}
	\end{subfigure}
	\begin{subfigure}{\columnwidth} \centering
		\includegraphics[width=3.3in]{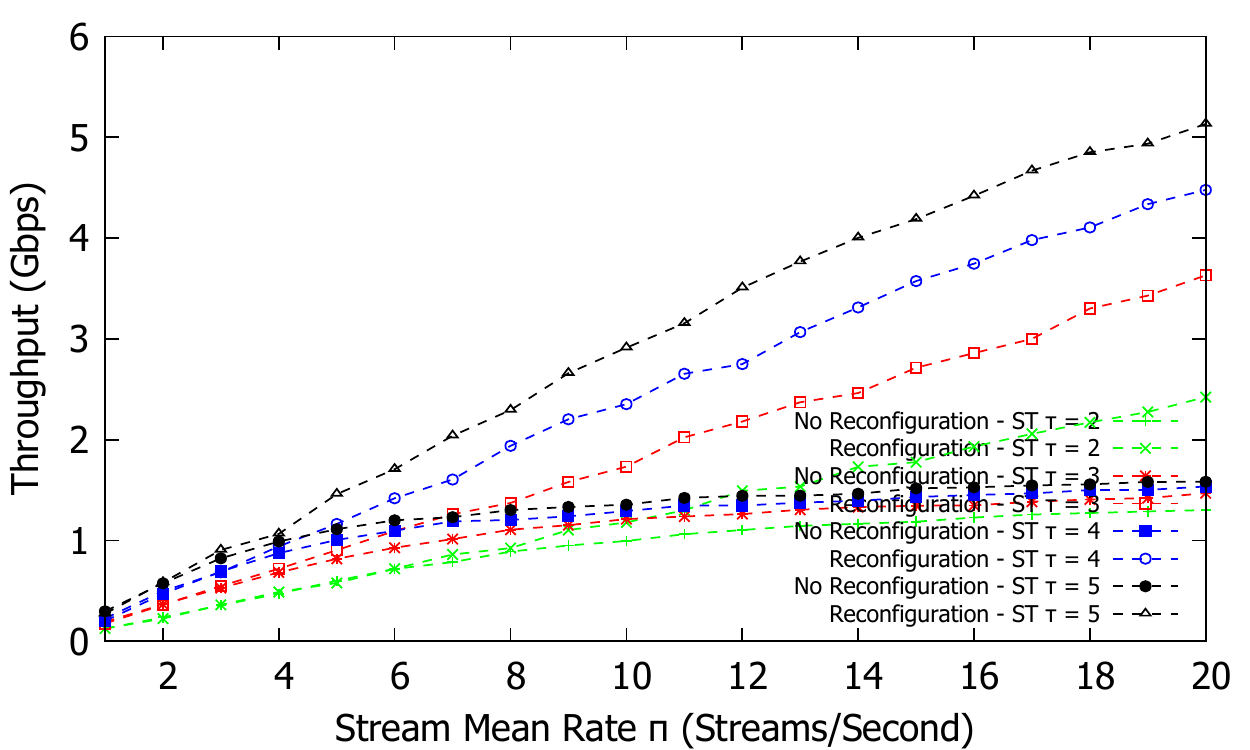}
		\caption{Mid $\rho_{L}$}
	\end{subfigure}
	\begin{subfigure}{\columnwidth} \centering
		\includegraphics[width=3.3in]{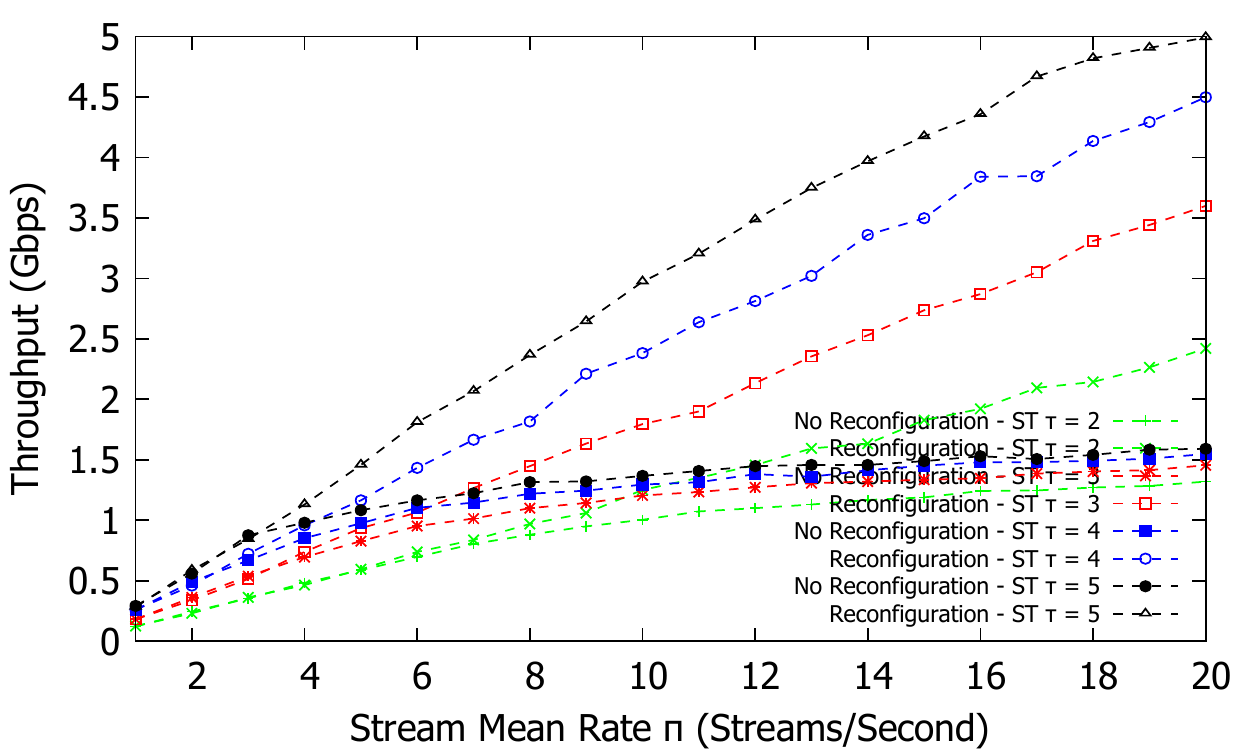}
		\caption{High $\rho_{L}$}
	\end{subfigure}
	\caption{Decentralized Bi-directional Topology: ST Total average throughput measured at the sink as a results of TAS.}
	\label{fig_avgTput_ST_dec_bi}
\end{figure}

\begin{figure} [t!] \centering
	\begin{subfigure}{\columnwidth} \centering
		\includegraphics[width=3.3in]{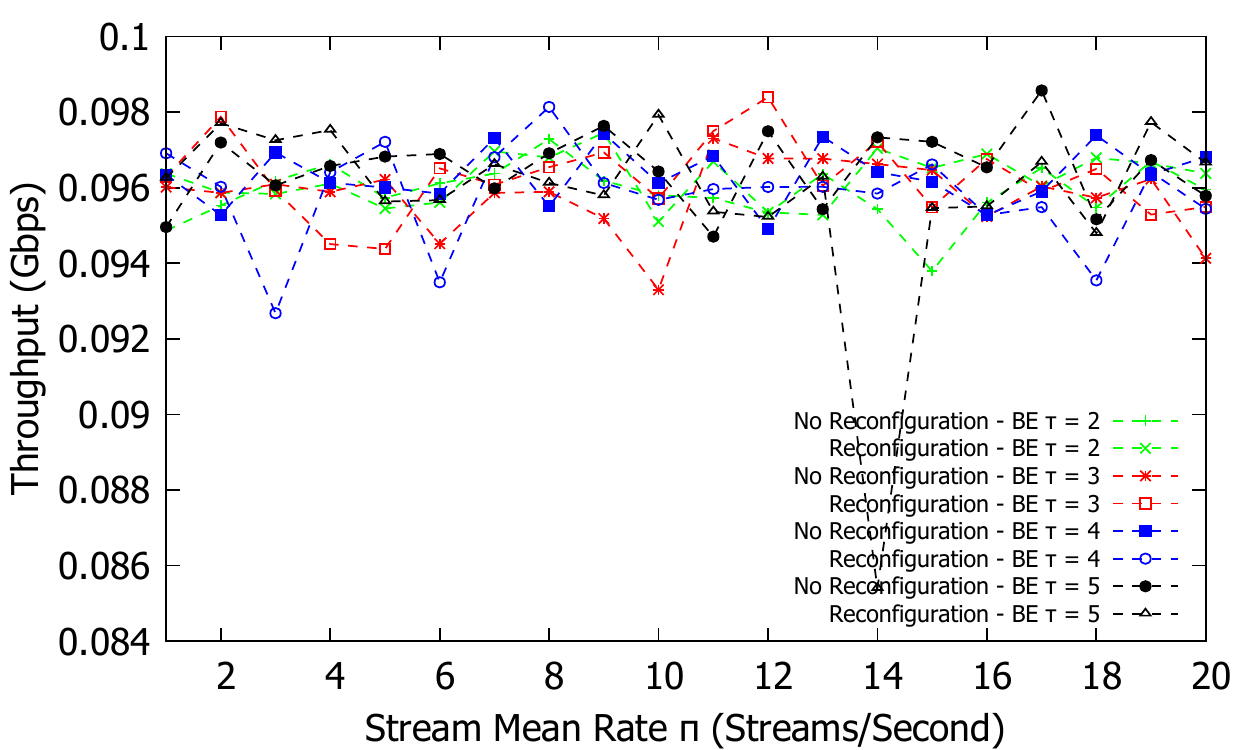}
		\caption{Low $\rho_{L}$}
	\end{subfigure}
	\begin{subfigure}{\columnwidth} \centering
		\includegraphics[width=3.3in]{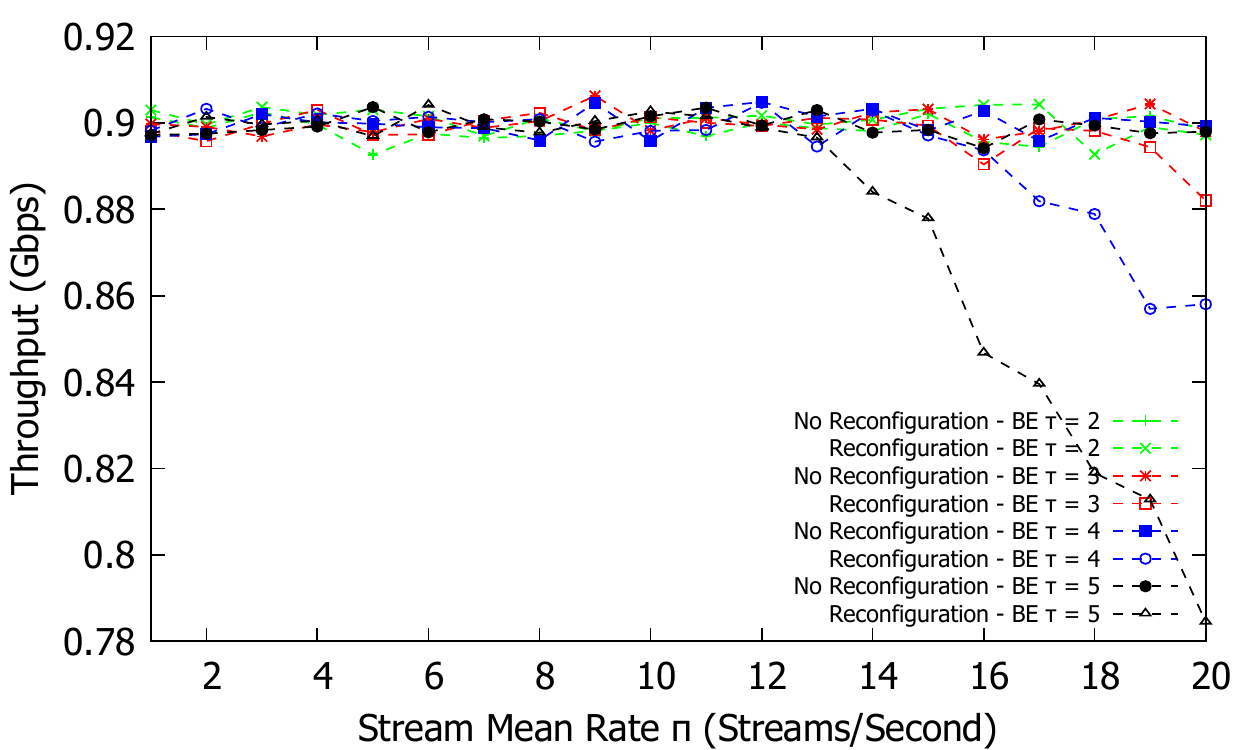}
		\caption{Mid $\rho_{L}$}
	\end{subfigure}
	\begin{subfigure}{\columnwidth} \centering
		\includegraphics[width=3.3in]{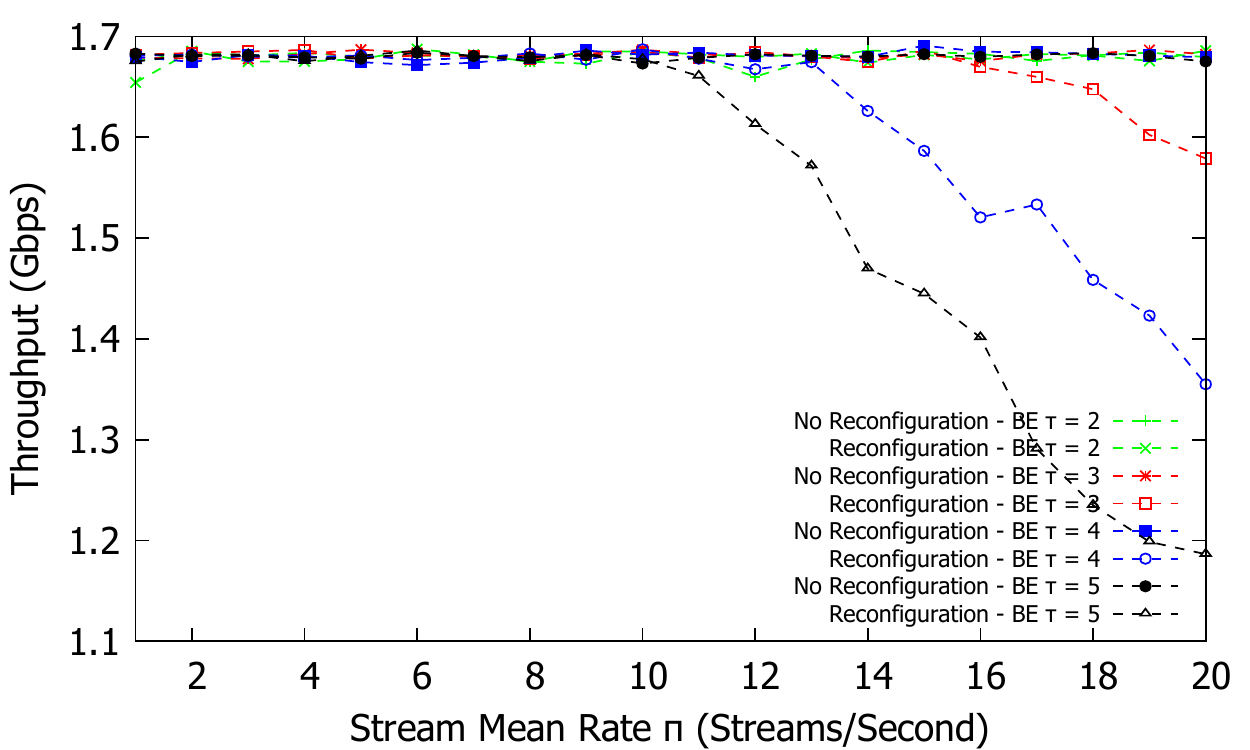}
		\caption{High $\rho_{L}$}
	\end{subfigure}
	\caption{Decentralized Bi-directional Topology: BE Total average throughput measured at the sink as a results of TAS.}
	\label{fig_avgTput_BE_dec_bi}
\end{figure}

Fig.~\ref{fig_avgTput_ST_dec_bi} and Fig.~\ref{fig_avgTput_BE_dec_bi} shows the average throughput measured at the sink for both ST and BE traffic. The average throughput results are nearly identical to the centralized model. While it is difficult to see, the throughput is slightly less than the centralized since we now incorporate in-band control traffic while reduces the link utilization for data traffic.

\begin{figure} [t!] \centering
	\begin{subfigure}{\columnwidth} \centering
		\includegraphics[width=3.3in]{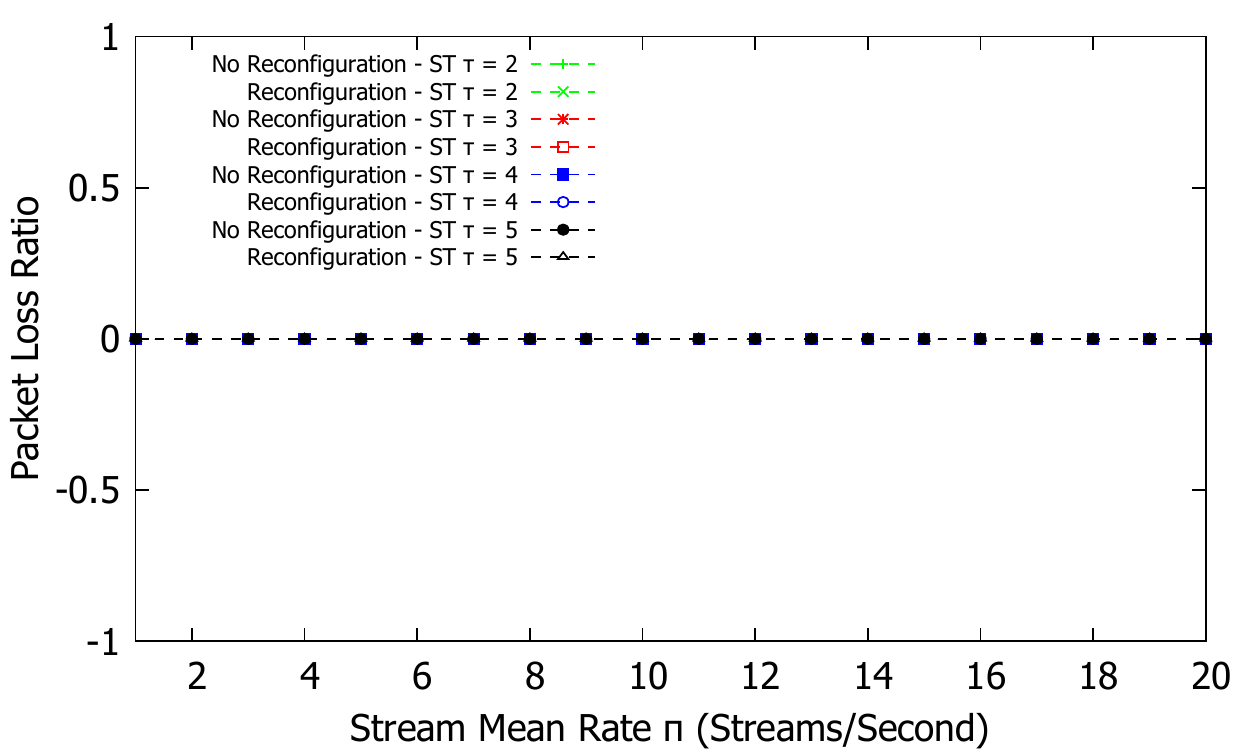}
		\caption{Low $\rho_{L}$}
	\end{subfigure}
	\begin{subfigure}{\columnwidth} \centering
		\includegraphics[width=3.3in]{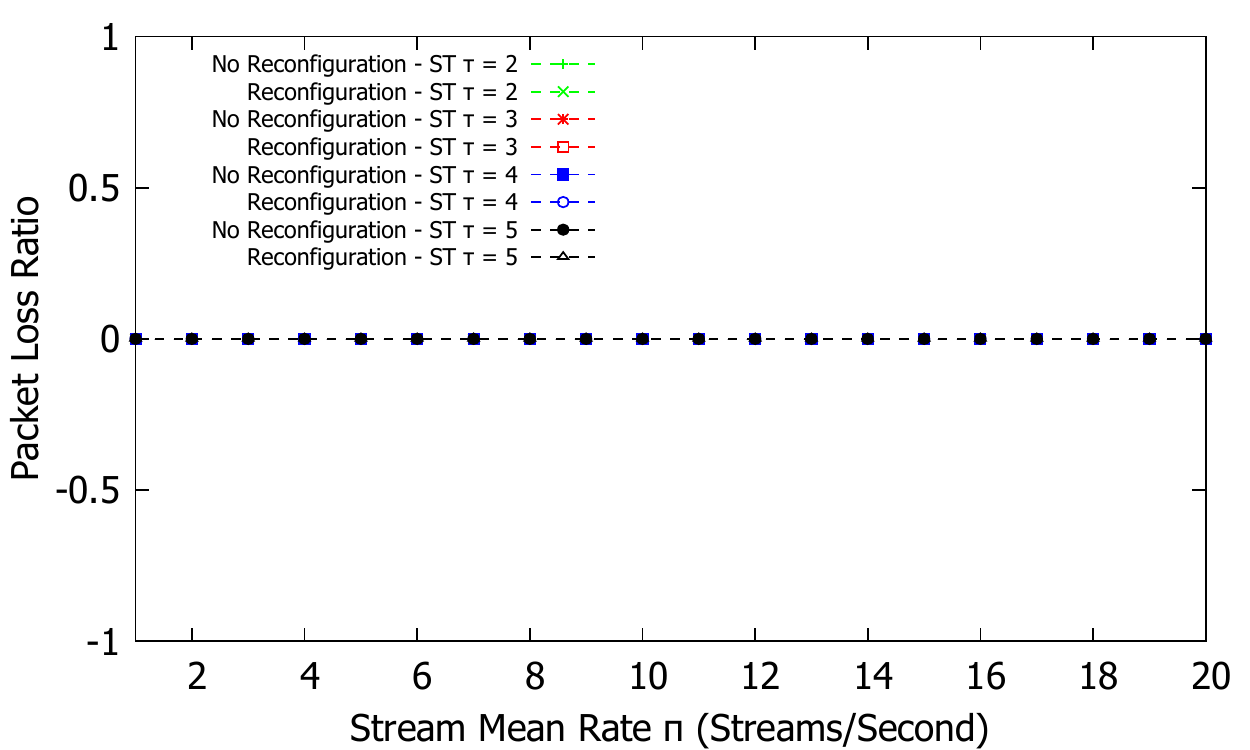}
		\caption{Mid $\rho_{L}$}
	\end{subfigure}
	\begin{subfigure}{\columnwidth} \centering
		\includegraphics[width=3.3in]{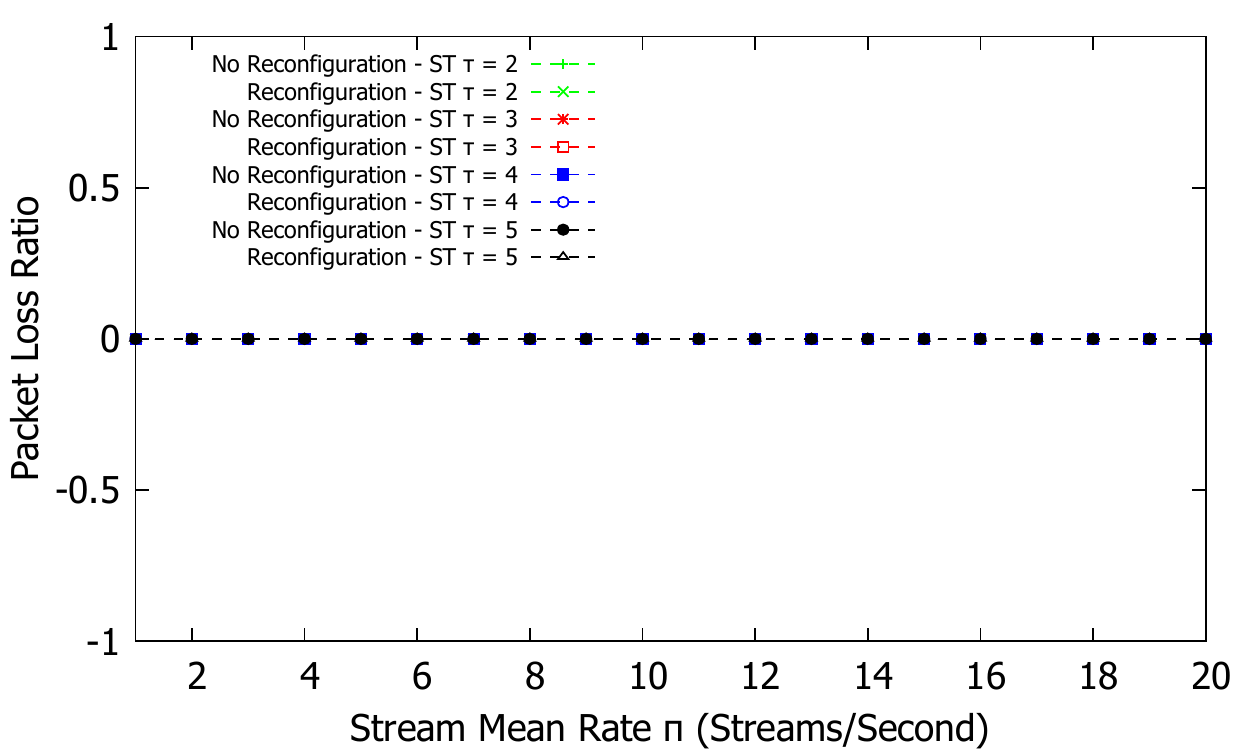}
		\caption{High $\rho_{L}$}
	\end{subfigure}
	\caption{Decentralized Bi-directional Topology: ST Frame loss ratio as a results of TAS.}
	\label{fig_lossProb_ST_dec_bi}
\end{figure}

\begin{figure} [t!] \centering
	\begin{subfigure}{\columnwidth} \centering
		\includegraphics[width=3.3in]{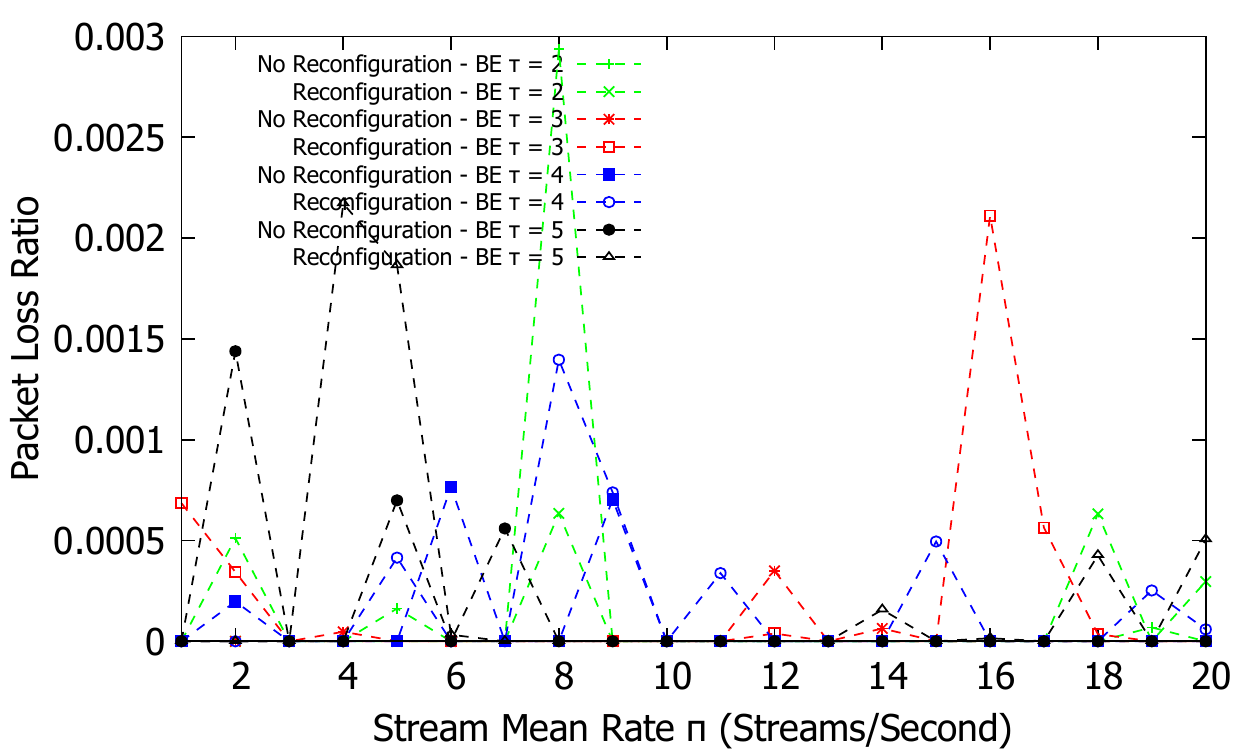}
		\caption{Low $\rho_{L}$}
	\end{subfigure}
	\begin{subfigure}{\columnwidth} \centering
		\includegraphics[width=3.3in]{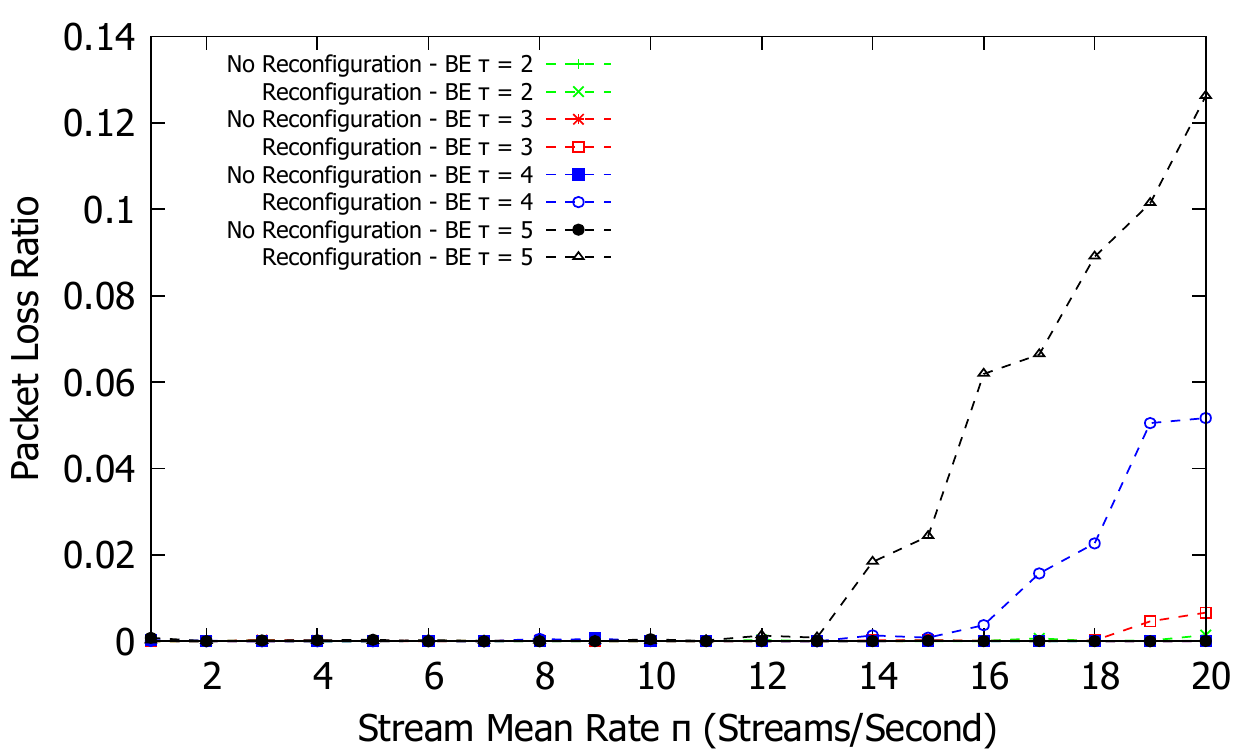}
		\caption{Mid $\rho_{L}$}
	\end{subfigure}
	\begin{subfigure}{\columnwidth} \centering
		\includegraphics[width=3.3in]{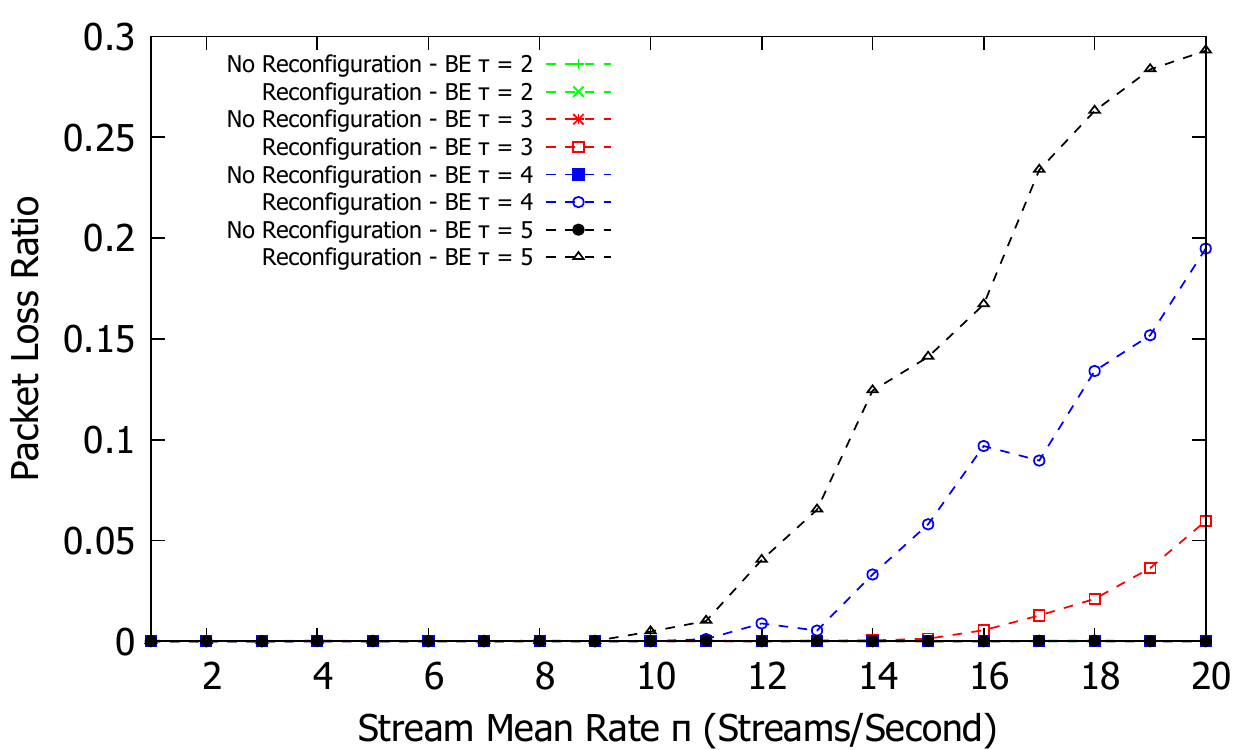}
		\caption{High $\rho_{L}$}
	\end{subfigure}
	\caption{Decentralized Bi-directional Topology: BE Frame loss ratio as a results of TAS.}
	\label{fig_lossProb_BE_dec_bi}
\end{figure}

Fig.~\ref{fig_lossProb_ST_dec_bi} and Fig.~\ref{fig_lossProb_BE_dec_bi} shows the packet loss ratio for ST and BE traffic in the network. Similar to all the different models and topologies, ST streams have zero frame drops as required by TSN. BE traffic results are nearly identical to the centralized model under bi-directional topology. Similarly, the overall performance is largely improved under the bi-directional topology due to the additional port.

While the decentralized model certainly proves to operate nearly identical to the centralized one in terms of QoS metrics and overall admission rate, the main disadvantage is in-band CDT traffic which can lead to delayed ST streams particularly affecting guaranteed maximum delays violations. A work around is to service all the ST streams first, and then service CDT frames before servicing the BE traffic, though this might lead to additional signaling delays depending on the ST load. However, some applications have a more relaxed constraint in guaranteeing maximum delay but require complete segregation of traffic based on the class of service which can be handled using the decentralized model without the overhead complexities of a CNC device. Furthermore, the decentralized model can struggle to find alternate paths without full network visibility, as opposed to the centralized model which can easily reroute streams in the event of failures due to having full network visibility, though this is out of scope for this phase of the project. Moreover, while not tested in our model, adding new devices and removing devices in the decentralized model leads to information flooding across the network that adds complexity and can skew TAS schedules if not handled appropriately. However, in the centralized model, this ``plug and play'' feature can be easily extended to our TSN domain since the CNC has full control and management of the data plane and can adjust and coordinate any scheduling issues in a timely and controllable manner.

\section{Conclusions and Future Work}  \label{concl:sec}
The IEEE 802.1Qcc framework and the 802.1Qbv traffic shaper enable the
implementation of a deterministic forwarding plane that provides
strict guarantees to any scheduled traffic service without any flow or
congestion control mechanism at the source.  Using an automated
network configuration is an imperative tool set to provide a unified
communication platform based on commercial of the shelf (COTS) full-duplex
Ethernet with high bandwidth/low complexity compared to Controller
Area Networks (CANs), Local Interconnect Networks (LINs), and
specialized field-buses in industrial control system applications
(e.g., industrial control, automotive, avionics).  Such network
designs can form a contract with the source to forward mission
critical traffic and to automate the network configuration process using
802.1Qcc for the full lifetime of the stream. Additionally, depending
on the forwarding plane port traffic shaper (e.g., TAS), the required
schedules can be passed to the switch servers using general
user/network information protocols (e.g., TLV, Yang, SNMP).

In this paper, we have investigated the impact of TAS reconfigurations
in response on dynamic network conditions. We have demonstrated the
effectiveness of TAS with and without the CNC, i.e., for centralized
(hybrid) vs. decentralized (fully distributed) models.  We have
examined network QoS traffic characteristics when maximizing stream
admission to the network whilst reserving some BE time slots in the
event of high ST transmission requests.

Based on the insights from the present study we outline the following
future research directions. First, it would be interesting to
judiciously change the GCL time for switches during reconfiguration
whilst satisfying QoS requirements.
The studied reconfiguration techniques should also be examined
in alternate approaches for providing deterministic Qos,
e.g.,~\cite{nas2019tsn,seaman2019pat} as well as in the context of related
QoS oriented routing approaches, e.g.~\cite{chu2018pre,guc2017uni}.

In the wider context of QoS networking and related applications,
deterministic networking should be examined in the context of
emerging multiple-access edge computing
(MEC)~\cite{doa2019pro,gao2019dyn,mar2019mod,sha2018lay},
in particular MEC settings for low-latency
applications~\cite{elb2018tow,xia2019red,zha2018mob}.
As an alternative approach to coordinating the
reconfigurations, emerging softwarized control paradigms, such as
software defined networking can be explored~\cite{ami2018sdn,der2019cou,des2018min,kel2019ada,san2018inf}.
Regarding the reliability aspects, a potential future research direction is to
explore low-latency network coding mechanisms, e.g.,~\cite{ace2018har,coh2019ada,eng2018exp,gab2018cat,luc2018ful,ma2019high,wun2017cat}, to enhance networking protocols targeting reliable low-latency communication.

\bibliographystyle{IEEEtran}
%\bibliography{refs6}

\begin{thebibliography}{10}
\providecommand{\url}[1]{#1}
\csname url@samestyle\endcsname
\providecommand{\newblock}{\relax}
\providecommand{\bibinfo}[2]{#2}
\providecommand{\BIBentrySTDinterwordspacing}{\spaceskip=0pt\relax}
\providecommand{\BIBentryALTinterwordstretchfactor}{4}
\providecommand{\BIBentryALTinterwordspacing}{\spaceskip=\fontdimen2\font plus
\BIBentryALTinterwordstretchfactor\fontdimen3\font minus
  \fontdimen4\font\relax}
\providecommand{\BIBforeignlanguage}[2]{{%
\expandafter\ifx\csname l@#1\endcsname\relax
\typeout{** WARNING: IEEEtran.bst: No hyphenation pattern has been}%
\typeout{** loaded for the language `#1'. Using the pattern for}%
\typeout{** the default language instead.}%
\else
\language=\csname l@#1\endcsname
\fi
#2}}
\providecommand{\BIBdecl}{\relax}
\BIBdecl

\bibitem{finn2018introduction}
N.~Finn, ``Introduction to time-sensitive networking,'' \emph{IEEE
  Communications Standards Magazine}, vol.~2, no.~2, pp. 22--28, 2018.

\bibitem{nas2019ult}
A.~Nasrallah, A.~S. Thyagaturu, Z.~Alharbi, C.~Wang, X.~Shao, M.~Reisslein, and
  H.~ElBakoury, ``Ultra-low latency {(ULL)} networks: The {IEEE TSN} and {IETF
  DetNet} standards and related {5G ULL} research,'' \emph{IEEE Commun. Surv.
  \& Tut.}, vol.~21, no.~1, pp. 88--145, 2019.

\bibitem{nas2019per}
------, ``Performance comparison of {IEEE 802.1 TSN Time Aware Shaper (TAS)}
  and {Asynchronous Traffic Shaper (ATS)},'' \emph{IEEE Access}, vol.~7, pp.
  44\,165--44\,181, 2019.

\bibitem{usecas2018}
R.~Belliardi, J.~Dorr, T.~Enzinger, F.~Essler, J.~Farkas, M.~Hantel, M.~Riegel,
  M.-P. Stanica, G.~Steindl, R.~Wam{\ss}er, K.~Weber, and S.~A. Zuponcic,
  ``{Use Cases IEC/IEEE 60802, V1.3},'' pp. 1--74, Sep. 2018, available from
  http://www.ieee802.org/1/files/public/docs2018/60802-industrial-use-cases-0918-v13.pdf;
  Last accessed Feb. 19, 2019.

\bibitem{IEEE8021Qbv}
``{IEEE Standard for Local and metropolitan area networks -- Bridges and
  Bridged Networks - Amendment 25: Enhancements for Scheduled Traffic},''
  \emph{IEEE Std 802.1Qbv-2015 (Amendment to IEEE Std 802.1Q--- as amended by
  IEEE Std 802.1Qca-2015, IEEE Std 802.1Qcd-2015, and IEEE Std 802.1Q---/Cor
  1-2015)}, pp. 1--57, Mar. 2016.

\bibitem{IEEE8021Qcc}
``{IEEE Draft Standard for Local and metropolitan area networks--Media Access
  Control (MAC) Bridges and Virtual Bridged Local Area Networks Amendment:
  Stream Reservation Protocol (SRP) Enhancements and Performance
  Improvements},'' \emph{IEEE P802.1Qcc/D2.0, October 2017}, pp. 1--207, Jan.
  2017.

\bibitem{chen2017rap}
F.~Chen, ``{Resource Allocation Protocol (RAP) based on LRP for Distributed
  Configuration of Time-Sensitive Streams},'' 2017,
  http://ieee802.org/1/files/public/docs2017/tsn-chen-RAP-whitepaper-0917-v01.pdf.

\bibitem{IEEE8021CS}
N.~Finn, ``{IEEE Draft Standard for Local and metropolitan area networks--Media
  Access Control (MAC) Bridges and Virtual Bridged Local Area Networks
  Amendment: Link-local Registration Protocol},'' \emph{IEEE P802.1CS/D1.2
  December 2017}, Dec. 2017.

\bibitem{IEEE8021Qat}
``{IEEE Standard for Local and Metropolitan Area Networks---Virtual Bridged
  Local Area Networks Amendment 14: Stream Reservation Protocol (SRP)},''
  \emph{IEEE Std 802.1Qat-2010 (Revision of IEEE Std 802.1Q-2005)}, pp. 1--119,
  Sep. 2010.

\bibitem{raagaard2017fog}
M.~Lander, P.~Raagaard, M.~G. Pop, and W.~Steiner, ``Runtime reconfiguration of
  time-sensitive networking {(TSN)} schedules for fog computing,'' in
  \emph{Proc. IEEE Fog World Congress (FWC)}, Oct. 2017.

\bibitem{pop2018enabling}
P.~Pop, M.~L. Raagaard, M.~Gutierrez, and W.~Steiner, ``Enabling fog computing
  for industrial automation through time-sensitive networking (tsn),''
  \emph{IEEE Communications Standards Magazine}, vol.~2, no.~2, pp. 55--61,
  2018.

\bibitem{hackel2019software}
T.~H{\"a}ckel, P.~Meyer, F.~Korf, and T.~C. Schmidt, ``Software-defined
  networks supporting time-sensitive in-vehicular communication,'' \emph{arXiv
  preprint arXiv:1903.08039}, 2019.

\bibitem{herlich2016proof}
M.~Herlich, J.~L. Du, F.~Sch{\"o}rghofer, and P.~Dorfinger, ``Proof-of-concept
  for a software-defined real-time ethernet,'' in \emph{Proc. IEEE
  International Conference on Emerging Technologies and Factory Automation
  (ETFA)}, 2016, pp. 1--4.

\bibitem{nayak2016time}
N.~G. Nayak, F.~D{\"u}rr, and K.~Rothermel, ``Time-sensitive software-defined
  network {(TSSDN)} for real-time applications,'' in \emph{Proc. ACM Int. Conf.
  on Real-Time Networks and Systems}, 2016, pp. 193--202.

\bibitem{nayak2017incremental}
------, ``{Incremental Flow Scheduling \& Routing in Time-sensitive
  Software-defined Networks},'' \emph{IEEE Transactions on Industrial
  Informatics}, vol.~14, no.~5, pp. 2066--2075, May 2017.

\bibitem{nayak2017routing}
N.~G. Nayak, F.~Duerr, and K.~Rothermel, ``{Routing Algorithms for IEEE802.
  1Qbv Networks},'' in \emph{Proc. RTN Workshop, ECRTS}, 2017.

\bibitem{kobzan2018secure}
T.~Kobzan, S.~Schriegel, S.~Althoff, A.~Boschmann, J.~Otto, and J.~Jasperneite,
  ``Secure and time-sensitive communication for remote process control and
  monitoring,'' in \emph{Proc. IEEE Int. Conf. on Emerging Techn. and Factory
  Autom. (ETFA)}, vol.~1, 2018, pp. 1105--1108.

\bibitem{enns2006netconf}
\BIBentryALTinterwordspacing
R.~Enns, M.~Bjorklund, A.~Bierman, and J.~Schönwälder, ``{Network
  Configuration Protocol (NETCONF)},'' RFC 6241, Jun. 2011. [Online].
  Available: \url{https://rfc-editor.org/rfc/rfc6241.txt}
\BIBentrySTDinterwordspacing

\bibitem{bierman2017restconf}
\BIBentryALTinterwordspacing
A.~Bierman, M.~Bjorklund, and K.~Watsen, ``{RESTCONF Protocol},'' RFC 8040,
  Jan. 2017. [Online]. Available: \url{https://rfc-editor.org/rfc/rfc8040.txt}
\BIBentrySTDinterwordspacing

\bibitem{guck2016function}
J.~W. Guck, M.~Reisslein, and W.~Kellerer, ``{Function split between
  delay-constrained routing and resource allocation for centrally managed QoS
  in industrial networks},'' \emph{IEEE Transactions on Industrial
  Informatics}, vol.~12, no.~6, pp. 2050--2061, 2016.

\bibitem{varga2008overview}
A.~Varga and R.~Hornig, ``{An overview of the OMNeT++ simulation
  environment},'' in \emph{Proc. ICST International Conference on Simulation
  Tools and Techniques for Communications, Networks and Systems \& Workshops},
  2008, pp. 1--10.

\bibitem{nas2019tsn}
A.~Nasrallah, V.~Balasubramanian, A.~Thyagaturu, M.~Reisslein, and
  H.~ElBakoury, ``{TSN} algorithms for large scale networks: A survey and
  conceptual comparison,'' \emph{arXiv preprint arXiv:1905.08478}, 2019.

\bibitem{seaman2019pat}
M.~Seaman, ``Paternoster policing and scheduling, {Revision 2.1},'' May 2019,
  available from
  http://www.ieee802.org/1/files/public/docs2019/cr-seaman-paternoster-policing-scheduling-0519-v04.pdf,
  Last accessed May 25, 2019.

\bibitem{chu2018pre}
U.~{Chunduri}, A.~{Clemm}, and R.~{Li}, ``{Preferred Path Routing} - a
  next-generation routing framework beyond {Segment Routing},'' in \emph{Proc.
  IEEE Global Commun. Conf. (GLOBECOM)}, Dec 2018, pp. 1--7.

\bibitem{guc2017uni}
J.~W. Guck, A.~Van~Bemten, M.~Reisslein, and W.~Kellerer, ``Unicast {QoS}
  routing algorithms for {SDN}: A comprehensive survey and performance
  evaluation,'' \emph{IEEE Communications Surveys \& Tutorials}, vol.~20,
  no.~1, pp. 388--415, First Qu. 2018.

\bibitem{doa2019pro}
T.~Doan-Van, A.~Kropp, G.~T. Nguyen, H.~Salah, and F.~H. Fitzek, ``Programmable
  first: Automated orchestration between {MEC} and {NFV} platforms,'' in
  \emph{Proc. IEEE Consumer Commun. \& Netw. Conf. (CCNC)}, 2019, pp. 1--2.

\bibitem{gao2019dyn}
Y.~Gao, W.~Tang, M.~Wu, P.~Yang, and L.~Dan, ``Dynamic social-aware computation
  offloading for low-latency communications in iot,'' \emph{IEEE Internet of
  Things Journal, in print}, 2019.

\bibitem{mar2019mod}
J.~Mart{\'\i}n-P{\'e}rez, L.~Cominardi, C.~J. Bernardos, A.~De~la Oliva, and
  A.~Azcorra, ``Modeling mobile edge computing deployments for low latency
  multimedia services,'' \emph{IEEE Transactions on Broadcasting, in print},
  2019.

\bibitem{sha2018lay}
P.~Shantharama, A.~S. Thyagaturu, N.~Karakoc, L.~Ferrari, M.~Reisslein, and
  A.~Scaglione, ``{LayBack: SDN} management of multi-access edge computing
  {(MEC)} for network access services and radio resource sharing,'' \emph{IEEE
  Access}, vol.~6, pp. 57\,545--57\,561, 2018.

\bibitem{elb2018tow}
M.~S. Elbamby, C.~Perfecto, M.~Bennis, and K.~Doppler, ``Toward low-latency and
  ultra-reliable virtual reality,'' \emph{IEEE Network}, vol.~32, no.~2, pp.
  78--84, 2018.

\bibitem{xia2019red}
Z.~Xiang, F.~Gabriel, E.~Urbano, G.~T. Nguyen, M.~Reisslein, and F.~H. Fitzek,
  ``Reducing latency in virtual machines: Enabling tactile internet for
  human-machine co-working,'' \emph{IEEE Journal on Selected Areas in
  Communications}, vol.~37, no.~5, pp. 1098--1116, 2019.

\bibitem{zha2018mob}
K.~Zhang, S.~Leng, Y.~He, S.~Maharjan, and Y.~Zhang, ``Mobile edge computing
  and networking for green and low-latency {Internet of Things},'' \emph{IEEE
  Communications Magazine}, vol.~56, no.~5, pp. 39--45, 2018.

\bibitem{ami2018sdn}
R.~Amin, M.~Reisslein, and N.~Shah, ``Hybrid {SDN} networks: A survey of
  existing approaches,'' \emph{IEEE Communications Surveys \& Tutorials},
  vol.~20, no.~4, pp. 3259--3306, 2018.

\bibitem{der2019cou}
N.~Deric, A.~Varasteh, A.~Basta, A.~Blenk, R.~Pries, M.~Jarschel, and
  W.~Kellerer, ``Coupling {VNF} orchestration and {SDN} virtual network
  reconfiguration,'' in \emph{Proc. Int. Conf. on Networked Systems (NetSys)},
  2019.

\bibitem{des2018min}
A.~Destounis, S.~Paris, L.~Maggi, G.~S. Paschos, and J.~Leguay, ``Minimum cost
  {SDN} routing with reconfiguration frequency constraints,'' \emph{IEEE/ACM
  Transactions on Networking}, vol.~26, no.~4, pp. 1577--1590, 2018.

\bibitem{kel2019ada}
W.~{Kellerer}, P.~{Kalmbach}, A.~{Blenk}, A.~{Basta}, M.~{Reisslein}, and
  S.~{Schmid}, ``Adaptable and data-driven softwarized networks: Review,
  opportunities, and challenges,'' \emph{Proceedings of the IEEE}, vol. 107,
  no.~4, pp. 711--731, April 2019.

\bibitem{san2018inf}
H.~S{\'a}ndor, B.~Genge, and Z.~Sz{\'a}nt{\'o}, ``Infrastructure and framework
  for response and reconfiguration in {Industry 4.0},'' in \emph{Proc. IEEE
  Int. Symp. on Digital Forensic and Security (ISDFS)}, 2018, pp. 1--6.

\bibitem{ace2018har}
J.~Acevedo, R.~Scheffel, S.~Wunderlich, M.~Hasler, S.~Pandi, J.~Cabrera,
  F.~Fitzek, G.~Fettweis, and M.~Reisslein, ``Hardware acceleration for {RLNC}:
  A case study based on the xtensa processor with the tensilica instruction-set
  extension,'' \emph{Electronics}, vol.~7, no.~9, p. 180, 2018.

\bibitem{coh2019ada}
A.~Cohen, D.~Malak, V.~B. Bracha, and M.~Medard, ``Adaptive causal network
  coding with feedback for delay and throughput guarantees,'' \emph{arXiv
  preprint arXiv:1905.02870}, 2019.

\bibitem{eng2018exp}
A.~Engelmann, W.~Bziuk, A.~Jukan, and M.~M{\'e}dard, ``Exploiting parallelism
  with random linear network coding in high-speed {Ethernet} systems,''
  \emph{IEEE/ACM Transactions on Networking (TON)}, vol.~26, no.~6, pp.
  2829--2842, 2018.

\bibitem{gab2018cat}
F.~Gabriel, S.~Wunderlich, S.~Pandi, F.~H. Fitzek, and M.~Reisslein,
  ``Caterpillar {RLNC} with feedback {(CRLNC-FB)}: Reducing delay in selective
  repeat {ARQ} through coding,'' \emph{IEEE Access}, vol.~6, pp.
  44\,787--44\,802, 2018.

\bibitem{luc2018ful}
D.~E. Lucani, M.~V. Pedersen, D.~Ruano, C.~W. S{\o}rensen, F.~H. Fitzek,
  J.~Heide, O.~Geil, V.~Nguyen, and M.~Reisslein, ``Fulcrum: Flexible network
  coding for heterogeneous devices,'' \emph{IEEE Access}, vol.~6, pp.
  77\,890--77\,910, 2018.

\bibitem{ma2019high}
Z.~Ma, M.~Xiao, Y.~Xiao, Z.~Pang, H.~V. Poor, and B.~Vucetic,
  ``High-reliability and low-latency wireless communication for internet of
  things: Challenges, fundamentals and enabling technologies,'' \emph{IEEE
  Internet of Things Journal, in print}, 2019.

\bibitem{wun2017cat}
S.~Wunderlich, F.~Gabriel, S.~Pandi, F.~H. Fitzek, and M.~Reisslein,
  ``Caterpillar {RLNC (CRLNC)}: A practical finite sliding window {RLNC}
  approach,'' \emph{IEEE Access}, vol.~5, pp. 20\,183--20\,197, 2017.

\end{thebibliography}

% Generated by IEEEtran.bst, version: 1.14 (2015/08/26)

\end{document}